\documentclass[12pt]{report}

\normalsize

\newcounter{sxn}

\newcounter{axn}

\def\br{\begin{thebibliography}{abc}}
\def\er{\end{thebibliography}}

\begin{document}
\title{
FUZZY PHYSICS}
\author{
Badis Ydri\thanks{Email: ydri@physics.syr.edu}\\\\
\bigskip
Physics Department , Syracuse University\\
\bigskip
Syracuse , N.Y , 13244-1130 , U.S.A\\
}
\maketitle
%
\thispagestyle{empty}
\newpage
\thispagestyle{empty}
\begin{center}
{\huge Fuzzy Physics}\\[5em]
by\\[2em]
{\large BADIS YDRI}\\
M. Sc. Physics Department, Syracuse University, Syracuse, NY, 2000.\\[5em]
DISSERTATION\\[2em]
Submitted in partial fulfillment of the requirements for the\\
degree of Doctor of Philosophy in Physics\\
in the Graduate School of Syracuse University\\[1em]
{Friday 21 of September 2001}\\[4em]
\end{center}
\begin{center}
ADVISOR\\
\end{center}
\begin{center}
Professor A. P. BALACHANDRAN\\
\end{center}

\begin{abstract}
Regularization of quantum field theories (QFT's) can be achieved by quantizing the underlying manifold (spacetime or spatial slice) thereby replacing it by a non-commutative matrix model or a ``fuzzy manifold'' . Such discretization by quantization is remarkably successful in preserving symmetries and topological features , and altogether overcoming the fermion-doubling problem . In this thesis, the fuzzification of coadjoint orbits and their QFT's are put forward.
\end{abstract}
\newpage
\begin{center}
\textbf{Acknowledgments}
\end{center}
\begin{quote}
I am immensely grateful to my advisor A. P. Balachandran for showing
tremendous interest and enthusiasm in the development of this work, and for
constant encouragement and guidance. My intellectual debt to him goes
beyond the physics I have learned from him.

I would also like to thank Prof. Joe Schechter and Prof. Rafael Sorkin for many interesting discussions and for their supports over the years .

I would like to thank , Denjoe O'Connor , Giorgio Immirzi , Sachindeo Vaidya , Paulo Teotonio-Sobrinho , T.R.Govindarajan, X.Martin , Garnik Alexanian for many exciting and thought-provoking discussions.

And most of all , I am indebted to my parents for their unquestioning support and love.

\end{quote}
\newpage
\vspace*{2in}
\begin{quote}
\begin{center}
\Large{\it To my Mother and my Father}
\end{center}
\end{quote}

\newpage

\tableofcontents

\chapter{Introduction}

We can find very few fundamental physical models which are amenable to exact treatment. Approximation methods such
as perturbation theory and the $1/N$ expansion remain crucial tools in analyzing different physical systems .
Perturbation theory for example is extremly successful in the case of QED and the $1/N$ expansion turns out to be a
very reliable one for matrix models and large gauge theories. These different approximation schemes are
undoubtedly a part of our physics culture.

Among the important approximation methods for quantum field theories (QFT's) are strong coupling methods of
lattice gauge theories (LGT's) which are based on lattice discretisation of the underlying spacetime or perhaps its
time-slice. They are among the rare effective approaches for the study of confinement in QCD and for the
non-perturbative regularization of QFT's. They enjoyed much popularity in their early days and have retained their
good reputation for addressing certain fundamental problems .

 One feature of naive lattice discretisations however can be
criticised. They do not retain the symmetries of the exact theory except in some rough sense. A related feature is
that topology and differential geometry of the underlying manifolds are treated only indirectly, by limiting the
couplings to ``nearest neighbours''. Thus lattice points are generally manipulated like a trivial topological set,
with a point being both open and closed. The upshot is that these models have  no rigorous representation of
topological defects and lumps like vortices, solitons and monopoles. The complexities in the ingenious solutions
for the discrete QCD $\theta$-term \cite{lattice96} illustrate such limitations. There do exist radical attempts to
overcome these limitations using partially ordered sets \cite{poset}, but their potentials are yet to be adequately
studied.

A new approach to discretisation, inspired by non-commutative geometry (NCG), is being developed since a few years
\cite{4,5,6,7,8,9,10,11,12,13,14,15}. The key remark here is that when the underlying spacetime or spatial cut can
be treated as a phase space and quantized, with a parameter $\hat h$ assuming the role of $\hbar$, the emergent
quantum space is fuzzy, and the number of independent states per (``classical'') unit volume becomes finite. We
have known this result after Planck and Bose introduced such an unltraviolet cut-off and quantum physics later
justified it. ``Fuzzified'' compact manifolds are ultraviolet finite and support only finitely many independent
states. Their continuum limits are the semiclassical $\hat h\rightarrow 0$ limits. This unconventional
dicretization of classical topology is not at all equivalent to the naive one, and we shall see that it does
significantly overcome the previous criticisms .

There are other reasons also to pay attention to fuzzy spaces, be they spacetimes or spatial cuts. There is much
interest among string theorists in matrix models and in describing D-branes using matrices. Fuzzy spaces lead to
matrix models too and their ability to reflect topology better than elsewhere should therefore evoke our curiosity.
They let us devise new sorts of discrete models and are interesting from that perspective. In addition, it has now
been discovered that when open strings end on D-branes which are symplectic manifolds, then the branes can
\cite{16} become fuzzy, in this way one comes across fuzzy tori, ${\bf C}{\bf P}^N$ and many such spaces in string
physics.

The central idea behind fuzzy spaces is discretisation by quantization. It relies on the existence of a suitable
Lagrangian and therefore it does not always work. An obvious limitation is that the parent manifold has to be even
dimensional. (See however ref. \cite{15} for fuzzyfying $RP^3/{\bf Z}_2$ and other non-symplectic manifolds, even
or odd). If it is not, it has no chance of being a phase space. But that is not all. Successful use of fuzzy spaces
for QFT's requires good fuzzy versions of the Laplacian, Dirac equation, chirality operator and so forth, and
their incorporation can make the entire enterprise complicated. The torus ${\bf T}^2$ is compact, admits a
symplectic structure and on quantization becomes fuzzy, or a non-commutative  torus. It supports a finite number
of states if the symplectic form satisfies the Dirac quantization condition. But it is impossible to introduce
suitable derivations without escalating the formalism to infinite dimensions \cite{1,2}.

But we do find a family of classical manifolds elegantly escaping these limitations. They are the co-adjoint orbits
of Lie groups. For semi-simple Lie groups, they are the same as adjoint orbits. It is a theorem that these orbits
are symplectic \cite{last1}. They can often be quantized when the symplectic forms satisfy the Dirac quantization
condition. The resultant fuzzy spaces are described by linear operators on irreducible representations (IRR's) of
the group. For compact orbits, the latter are finite-dimensional. In addition, the elements of the Lie algebra
define natural derivations, and that helps to find Laplacian and the Dirac operator. We can even define chirality
with no fermion doubling and represent monopoles and instantons (See \cite{4,5,6,7,8,9} and the first $3$ papers in
\cite{13}). These orbits therefore are altogether well-adapted for QFT's.

Let us give examples of these orbits:
\begin{itemize}
\item{${\bf S}^2$}:
This is the orbit of $SU(2)$ through the Pauli matrix $\sigma_3$ or any of its multiples $l\,\sigma_3$ ($l\neq 0$).
It is the set $\{l\,g\,\sigma_3\,g^{-1}\,:\, g\in SU(2)\}$. The symplectic form is $l\,d\,{\rm cos}\,\theta\wedge
d\phi$ with $\theta,\phi$ being the usual $S^2$ coordinates, \cite{17}. Quantization gives the spin $l$ $SU(2)$
representations . This case will be  treated in the third and the fourth chapters of this thesis .
\item{${\bf C}{\bf P}^2$}:
${\bf C}{\bf P}^2$ is of particular interest being of dimension $4$. It is the orbit of $SU(3)$ through the
hypercharge $Y=1/3\,\,{\rm diag}(1,1,-2)$ (or its multiples):
\begin{equation}
{\bf C}{\bf P}^2=\{g\,Y\,g^{-1}\,:\, g\in SU(3)\}.
\end{equation}
The associated representations are symmetric products of $3$'s or ${\bar 3}$'s. (See chapter $5$).
\item{$SU(3)/[U(1)\times U(1)]$}:
This 6-dimensional manifold is the orbit of $SU(3)$ through $\lambda_3={\rm diag}(1,-1,0)$ and its multiples.
These orbits give all the IRR's containing a zero hypercharge state.
\end{itemize}
A general class of coadjoint orbits is given by the higher dimensional ${\bf C}{\bf P}^{N}$ spaces defined by
\begin{equation}
{\bf C}{\bf P}^N=SU(N+1)/{U(N)}=\{gY^{(N+1)}g^{-1};g{\in}SU(N+1)\}.
\end{equation}
They are clearly , from this definition , orbits of $SU(N+1)$ through the "hypercharge" like operators
\begin{equation}
Y^{(N+1)}=\frac{1}{N+1}diag(1,1,...,-N).
\end{equation}
For $N=1$ we obtain $Y^{(2)}={\sigma}_3$ and ${\bf C}{\bf P}^1\simeq{\bf S}^2$ . For $N=2$ , on the other hand , we
have $Y^{(3)}=Y$ . From the above definition it is also obvious that the stability group of the hypercharge
operator $Y^{(N+1)}$ is simply the group $U(N)=\{h{\in}SU(N+1):hY^{(N+1)}h^{-1}=Y^{(N+1)}\}$ .
\section{Fuzzy Spaces}
\subsection{The case of ${\bf S}^2$}

As we mentioned earlier the fuzzification of the above ${\bf C}{\bf P}^N$'s is the process of their discretisation
by quantization . One would like to explain in this introduction this concept , which is central to this thesis ,
for the simplest of all the ${\bf C}{\bf P}^N$'s , namely the case of the sphere . To this end we will go into
some details of the fuzzification of ${\bf S}^2$ . The method and the results are generic to all other ${\bf
C}{\bf P}^N$'s .

The starting point is the Wess-Zumino term defined by
\begin{equation}
L={\Lambda}iTr({\sigma}_3g^{-1}\dot{g});g{\in}SU(2),\label{wz}
\end{equation}
whre ${\Lambda}$ is , as of yet , an undetermined real number .

As it was shown in \cite{17} , this Lagrangian arises generally when one tries to avoid the
singularities\footnote{In other words when one tries to find a smooth global system of canonical coordinates for
the phase space . See below and section $4.1.1$ for specific examples.} of the phase space . In such cases a global
Lagrangian can not be found by a simple Legendre transformation of the Hamiltonian and therefore one needs to
enlarge the configuration space . A global Lagrangian over this new extended configuration space can then be shown
to exist and it turns out to contain (\ref{wz}) as a very central piece . Basically (\ref{wz}) reflects the constraints
imposed on the system .

One example which was treated in \cite{17} with great detail is the case of a particle with a fixed spin . For a
free particle one knows that the Lagrangian is given simply by the expression $L=\frac{m}{2}\dot{\vec{x}}^2$ .
However if the particle is constrained to have a fixed spin given by $\vec{S}^2={\lambda}^2$ then the phase space
will be eight dimensional defined by $\{\vec{x},\vec{p},Q,P\}$ , where $Q$ and $P$ describes the two independent
spin degrees of freedom of the particle . Clearly $Q$ and $P$ can not be smooth functions of $\vec{S}$ , they must
show a singularity for at least one value of $\vec{S}$ . Indeed , if they were smooth functions of $\vec{S}$ , there would have been no difference between this particle and the free particle with arbitrary spin . To overcome this difficulty one can enlarge the
configuration space from ${\bf R}^3$ to ${\bf R}^3{\times}SU(2)$ over which the Lagrangian is now given by
$L=\frac{m}{2}{\dot{\vec{x}}}^2+{\Lambda}iTr({\sigma}_3g^{-1}\dot{g})$ . The quantization of this sytem will fix
the number ${\Lambda}$ appropriately . In the Dirac quantization scheme it is fixed to be of the form
${\pm}\sqrt{j(j+1)}$ where $j=0,1/2,1,..$ . In the Gupta-Bleuler quantization approach , on the other hand ,
${\Lambda}$ is fixed to be ${\pm}j$ . Another example for which the above term plays a central role is the system
of a charged particle in the field of a magnetic monopole .

One now defines the sphere ${\bf S}^2=\{{\vec{x}}{\in}{\bf R}^3; \sum_{i=1}^{3}x_i^2={\rho}^2\}$ by the Hopf
fibration
\begin{eqnarray}
{\pi}:SU(2)&{\longrightarrow}&{\bf S}^2\nonumber\\
g&{\longrightarrow}&{\rho}g{\sigma}_3g^{-1}=\vec{x}.\vec{\sigma}\nonumber\\
&\Leftrightarrow&\nonumber\\
{\rho}^2&=&\sum_{i=1}^3x_i^2.
\end{eqnarray}
Clearly the structure group , $U(1)$, of the principal fiber bundle
\begin{equation}
U(1){\longrightarrow}SU(2){\longrightarrow}{\bf S}^2
\end{equation}
leaves the base point ${\vec{x}}$ invariant in the sense that all the elements $gexp(i{\sigma}_3{\theta}/2)$ of
$SU(2)$ are projected onto the same point ${\vec{x}}$ on the base manifold ${\bf S}^2$ . One can then identify the
point $\vec{x}{\in}{\bf S}^2$ with the equivalence class
\begin{equation}
[gexp(i{\sigma}_3{\theta}/2)]{\in}SU(2)/U(1).
\end{equation}

Let us now turn to the quantization of the Lagrangian (\ref{wz}) . First we parametrize the group element $g$ by the
set of variables  $({\xi}_1,{\xi}_2,{\xi}_3)$. The conjugate momenta ${\pi}_i$ are given by the equations
${\pi}_i=\frac{{\partial}L}{{\partial}\dot{\xi}^{i}}={\Lambda}iTr({\sigma}_3g^{-1}\frac{{\partial}g}{{\partial}{\xi}^{i}})$.
${\xi}_i$ and ${\pi}_i$ will satisfy as usual the standard Poisson brackets :
$\{{\xi}_i,{\xi}_j\}=\{{\pi}_i,{\pi}_j\}=0$ and $\{{\xi}_i,{\pi}_j\}={\delta}_{ij}$.

A change in the local coordinates, ${\xi}{\longrightarrow}f(\epsilon)$, which is defined by $
g(f(\epsilon))=exp(i{\epsilon}_i\frac{{\sigma}_i}{2})g(\xi)$ will lead to the identity
$\frac{\partial{g(\xi)}}{{\partial}{\xi}_i} N_{ij}(\xi)=i\frac{{\sigma}_j}{2}g(\xi)$ , where
$N_{ij}(\xi)=\frac{{\partial}f_i(\epsilon)}{{\partial}{\epsilon}_j}|_{{\epsilon}=0}$ . The modified conjugate
momenta , $t_i$ , which are given by
\begin{equation}
t_i=-{\pi}_jN_{ji}=\frac{\Lambda}{\rho}x_i ,\label{modifiedconjugatemomenta}
\end{equation}
will then satify the interesting Poisson's brackets
\begin{eqnarray}
\{t_{i},g\}&=&i\frac{{\sigma}_i}{2}g\nonumber\\
\{t_{i},g^{-1}\}&=&-ig^{-1}\frac{{\sigma}_i}{2}\nonumber\\
\{t_{i},t_{j}\}&=&{\epsilon}_{ijk}t_{k}.\nonumber\\
&&\label{modifiedpoissonbrackets}
\end{eqnarray}
Putting equation (\ref{modifiedconjugatemomenta}) in the last of the equations (\ref{modifiedpoissonbrackets}) , one can derive the following nice result ,
$\{x_{i},x_{j}\}=\frac{\rho}{{\Lambda}}{\epsilon}_{ijk}x_k$ , which is the first indication that we are going to
get a fuzzy sphere under quantization . The classical sphere would correspond to
${\Lambda}{\longrightarrow}{\infty}$ .

However , a more precise treatment would have to start by viewing equations, (\ref{modifiedconjugatemomenta}), as a set of constraints
rather than a set of identities on the phase space $({\xi}_i,t_i)$ . In other words the functions ,
$P_i=t_i-\frac{\Lambda}{\rho}x_i$ , do not vanish identically on the phase space $\{({\xi}_i,t_i)\}$ . However ,
their zeros will define the physical phase space as a submanifold of $\{({\xi}_i,t_i)\}$ . To see that the $P_i$'s
are not the zero functions on the phase space , one can simply compute the Poisson brackets $\{P_i,P_j\}$ . The
answer turns out to be $\{P_i,P_j\}={\epsilon}_{ijk}(P_k-\frac{\Lambda}{\rho}x_k)$ which clearly does not vanish
on the surface $P_i=0$ \footnote{Actually any function $g$ on the phase space $({\xi}_i,t_i)$ for which the Poisson
bracket $\{g,P_i\}$ does not vanish on the surface $P_i=0$ will define a canonical transformation which takes any
point of this surface out of it .} . So the $P_i$'s should only be set to zero after the evaluation of all Poisson
brackets . This fact will be denoted by setting $P_i$ to be weakly zero, i.e
\begin{equation}
P_i\approx0.\label{constraints1}
\end{equation}
Equations (\ref{constraints1}) provide the primary constraints of the system . The secondary constraints of the system are
obtained from the consistency conditions $\{P_i,H\}\approx0$, where $H$ is the Hamiltonian of the system . Since
$H$ is given by $H=v_iP_i$ where $v_i$ are Lagrange multipliers , the requirement $\{P_i,H\}\approx0$ will lead to
no extra constraints on the system . It will only put conditions on the $v$'s \cite{17} .

From equations , (\ref{modifiedpoissonbrackets}) , it is obvious that $t_i$ are generators of the left action of $SU(2)$ on itself . A
right action can also be defined by the generators
\begin{equation}
t_j^{R}=-t_iR_{ij}(g),
\end{equation}
where $R_{ij}(g)$ define the standard $SU(2)$ adjoint representation : $R_{ij}(g){\sigma}_i=g{\sigma}_jg^{-1}$. These right generators satisfy the following Poisson brackets
\begin{eqnarray}
\{t_{i}^R,g\}&=&-ig\frac{{\sigma}_i}{2}\nonumber\\
\{t_{i}^R,g^{-1}\}&=&i\frac{{\sigma}_i}{2}g^{-1}\nonumber\\
\{t_{i}^R,t_{j}^R\}&=&{\epsilon}_{ijk}t_{k}^R.\nonumber\\
&&
\end{eqnarray}
In terms of , $t^R_i$ , the constraints , (\ref{constraints1}), will then take the simpler form
\begin{equation}
t_i^R\approx-{\Lambda}{\delta}_{3i},\label{constraints2}
\end{equation}
These constraints are divided into one independent first class constraint and two independent second class
constraints . $t_3^{R}\approx-{\Lambda}$ is first class because on the surface defined by (\ref{constraints2}) , one have
$\{t_{3}^R,t_i^R\}=0$ for all $i$ . It corresponds to the fact that the Lagrangian (\ref{wz}) is weakly invariant under
the gauge transformations $g{\longrightarrow}gexp(i{\sigma}_3\frac{{\theta}}{2})$ , namely
$L{\longrightarrow}L-{\Lambda}\dot{\theta}$ . The two remaining constraints , $t_{1}^R\approx0$ and
$t_2^R\approx0$ are second class . They can be converted to a set of first class constraints by taking the complex
combinations $t^R_{\pm}=t_1^R{\pm}it_2^R \approx 0$ . We would then have $\{t_3^R,t^R_{\pm}\}={\mp}it^R_{\pm}$ and
therefore all the Poisson brackets $\{t_3^R,t^R_{\pm}\}$ vanish on the surface (\ref{constraints2}) .

Let us now construct the physical wave functions of the system described by the Lagrangian (\ref{wz}) . One starts
with the space ${\bf F}$ of complex valued functions on $SU(2)$ with a scalar product defined by $
({\psi}_1,{\psi}_2)=\int_{SU(2)} d{\mu}(g){\psi}_1(g)^{*}{\psi}_2(g) $ , where $d{\mu}$ stands for the Haar measure
on $SU(2)$ . The physical wave functions are elements of ${\bf F}$ which are also subjected to the constraints
(\ref{constraints2}) . They span a subspace ${\bf H}$ of ${\bf F}$ . For ${\Lambda}<0$ \footnote{If ${\Lambda}$ was positive
the second equation of , (\ref{constraints3}) , should be replaced by $t_{-}^R{\psi}=0$ , and ${\psi}$ would have then been the
lowest weight state of the spin $l={\Lambda}$ representation of the $SU(2)$ group. }, one must then have
\begin{eqnarray}
t_3^R{\psi}&=&-{\Lambda}{\psi}\nonumber\\
t_{+}^R{\psi}&=&0\label{constraints3}
\end{eqnarray}
In other words ${\psi}$ transforms as the highest weight state of the spin $l=|{\Lambda}|$ representation of the
$SU(2)$ group . $|\Lambda|$ is now being quantized to be either an integer or a half integer number . The physical
wave functions are then linear combinations of the form
\begin{equation}
\psi(g)=\sum_{m=-l}^{l}C_{m}<lm|D^{l}(g)|ll>,|{\Lambda}|=l,\label{physicalwaves}
\end{equation}
where $D^{l}(g)$ is the spin $l$ representation of the element $g$ of $SU(2)$ .

Clearly the left action of $SU(2)$ on $g$ will rotate the index $m$ in such a way that , $<lm|D^{l}(g)|ll>$ ,
transform as a basis for the Hilbert space of the $(2l+1)-$dimensional irreducible representation $l=|\Lambda|$ of
$SU(2)$ . Under the right action of $SU(2)$ on $g$ , the matrix element $<lm|D^{l}(g)|ll>$ , will however transform
as the heighest weight state $l=|{\Lambda}|$ , $m=|\Lambda|$ of $SU(2)$ .

Observables of the system will be functions , $f(L_i)\equiv f(L_1,L_2,L_3)$ , of the quantum operators $L_i$ which
are associated with $t_i$ . These functions are the only objects which will have , by construction\footnote{This
is because , by definition , left and right actions of $SU(2)$ commute .}, weakly zero Poisson brackets with the
constraints (\ref{constraints2}) . These observables are linear operators which act on the left of $\psi(g)$ by left
translation , namely
\begin{equation}
[iL_i{\psi}][g]=\Big[\frac{d}{dt}{\psi}(e^{-i\frac{{\sigma}_i}{2}t}g)\Big]_{t=0}
\end{equation}
The operators $f(L_i)$ can be represented by $(2l+1){\times}(2l+1)$ matrices of the form
\begin{equation}
f(L_i)=\sum_{i_1,...,i_k}{\alpha}_{i_1,...,i_k}L_{i_1}...L_{i_k}.\label{expansionoffuzzys2}
\end{equation}
The operators $L_i$ form , by definition , a complete
set of $SU(2)$ generators , namely they satisfy $[L_i,L_j]=i{\epsilon}_{ijk}L_k$ and $\vec{L}^2=l(l+1)$. On the other hand , the summation in (\ref{expansionoffuzzys2}) will clearly terminate because the dimension of the space of all $(2l+1){\times}(2l+1)$ matrices is finite equal to $(2l+1)^2$ . 

In a sense the $L_i$'s provide the fuzzy coordinate functions on the fuzzy sphere ${\bf S}_F^2$ . Fuzzy points are
defined by the eigenvalues of the operators , $L_i$\footnote{The fact that these operators can not be diagonalised
simultaneously is a reflection of the fact that fuzzy points can not be localized .} , whereas fuzzy functions are
given by (\ref{expansionoffuzzys2}) .

The fuzzy sphere ${\bf S}_F^2$ is essentially the algebra ${\bf A}$ of all functions of the form (\ref{expansionoffuzzys2}) . More
precisely ${\bf S}_F^2$ is defined by the $K-cycle$ $({\bf A},{\bf H},D,{\Gamma})$ \cite{1} where ${\bf H}$ is the
Hilbert space spanned by the physical wave functions (\ref{physicalwaves}). We leave the detailed construction of the Dirac
operator $D$ and of the chirality operator ${\Gamma}$ to the third chapter of this thesis . A short discussion of
$D$ and ${\Gamma}$ is however given latter in this introduction .

The last thing one would like to mention concerning the Wess-Zumino term , (\ref{wz}) , is its relation to the
symplectic structure on ${\bf S}^2$ . The first claim is the fact that the symplectic two-form ,
${\epsilon}_{ijk}n_{k}dn_i{\wedge}dn_j$ , on ${\bf S}^2$ can be rewritten in the form
\begin{eqnarray}
{\omega}&=&{\Lambda}id\bigg[Tr{\sigma}_3g(\sigma,t)^{-1}dg(\sigma,t)\bigg]\nonumber\\
&=&-\frac{\Lambda}{2{\rho}^3}{\epsilon}_{ijk}x_{k}dx_i{\wedge}dx_j\nonumber\\
&=&{\Lambda}d cos{\theta}{\wedge}d{\phi}.\label{twoform}
\end{eqnarray}
$t$ in (\ref{twoform}) , is as usual the time variable which goes , say , from $t_1$ to $t_2$ . ${\sigma}$ in the other
hand is a new extra parameter which is chosen to take value in the range $[0,1]$ . By definition $g(1,t)=g(t)$ and
$\vec{n}(1,t)=\vec{n}(t)$ . If one defines the triangle ${\Delta}$ in the plane $(t,\sigma)$ by its boundaries
${\partial}{\Delta}_1=(\sigma,t_1)$ , ${\partial}{\Delta}_2=(\sigma,t_2)$ and ${\partial}{\Delta}_3=(1,t)$, then
it is a trivial exercise to show that
\begin{equation}
S_{WZ}=\int_{\Delta}{\omega}=\int_{t_1}^{t_2} L dt +{\Lambda}i\int_{0}^1
d{\sigma}Tr{\sigma}_3\bigg[g(\sigma,t_1)^{-1}\frac{{\partial}g}{{\partial}{\sigma}}(\sigma,t_1)-g(\sigma,t_2)^{-1}\frac{{\partial}g}{{\partial}{\sigma}}(\sigma,t_2)\bigg]\label{wz1}
\end{equation}
The equations of motion derived from this action are precisely those obtained from the Wess-Zumino term (\ref{wz}) .
This is obvious from the fact that the second term of (\ref{wz1}) , will not contribute to the equations of motion
because it involves the fixed initial and final times , where $g$ is not varied .

\subsection{Higher Coadjoint Orbits}

The above described procedure can be applied , virtually with no modifications, to fuzzify any coadjoint orbit .
It will be explained once more in some detail for the case of ${\bf C}{\bf P}^2$ (see chapter $5$) . The main
ingredient in this activity is the symplectic structure which does exist on any coadjoint orbit . This will be
explained briefly now .

Coadjoint orbits of a compact simple Lie group $G$ are defined by the adjoint action of $G$ on its Lie algebra
${\cal G}$ . The orbit through any element $K{\in}{\cal G}$ is defined by $\{gKg^{-1};g{\in}G\}$ . One can prove
that these orbits are even dimensional and that they admit $G-$invariant symplectic two forms ${\omega}$ . The
proof goes as follows . If the stability group , $H$ , of $K$ is generated by the elements $T_i$ of ${\cal G}$ ,
then the components of the symplectic two-form ${\omega}$ can be defined by
\begin{equation}
{\omega}(S_a,S_b)(K)=TrK[S_a,S_b]\equiv \omega_{ab}(K)\label{stf}
\end{equation}
where $S_a$ generate the orthogonal complement of $H$. From the definition (\ref{stf}), it is clear that the matrix
$[{\omega}]=({\omega}_{ab})$ is antisymmetric and therefore $[{\omega}]^T=-[{\omega}]$. One can also show that
the only solution to the equation ${\omega}_{ab}{\xi}^b=0{\Leftrightarrow}S_a{\xi}^a=0$ is ${\xi}^a=0$, in other
words $[{\omega}]$ is a nondegenerate matrix and therefore the determinat must be non zero . Using now the
following identity , $det[{\omega}]=det[{\omega}]^{T}=det-[{\omega}]=(-1)^{d}det[{\omega}]$ , one trivially deduce
that the dimension of the tangent space at the point $K$ of the orbit , which is generated by the $S_a$'s , is even
dimensional and hence the orbit itself is even dimensional .

At the point $gKg^{-1}$ of the orbit , the tangent space is spanned by $X_a=gS_ag^{-1}$ and therefore the
symplectic structure is given by
\begin{equation}
{\omega}(X_a,X_b)(gKg^{-1})=TrgKg^{-1}[X_a,X_b],
\end{equation}
It is easily shown that ${\omega}(X_a,X_b)(gKg^{-1})={\omega}(S_a,S_b)(K)$ which establishs the invariance of
(\ref{stf}) under the action of the group $G$ .

The ${\omega}_{ab}(K)$ define actually a closed two-form ${\omega}$ given by
\begin{equation}
{\omega}=d\bigg[i{\Lambda}TrKg^{-1}dg\bigg].\label{ctf}
\end{equation}
To show this result one first remarks that the quantity , $g^{-1}dg$ , is in the Lie algebra ${\cal G}$ and
therefore it can be expressed as $g^{-1}dg={\eta}_iT_i+{\xi}_aS_a$ . The above two-form (\ref{ctf}) can then be
rewritten in the form
\begin{equation}
{\omega}=-\frac{i}{2}{\omega}_{ab}(K){\xi}_a\wedge{\xi}_b,\label{ctf1}
\end{equation}
where we have used the identities $TrK[T_i,T_j]=0$ and $TrK[T_i,S_a]=0$ which are consequences of the fact that the
$T_i$'s generate the stability group of $K$ . From (\ref{ctf1}) it is a trivial result that $d{\omega}=0$ .

The quantization of the coadjoint orbit $G/H=\{gKg^{-1}\}$ is equivalent to the quantization of the above closed
two-form (\ref{ctf}) , in other words a constraints analysis of the Wess-Zumino term
\begin{equation}
L=i{\Lambda}TrKg^{-1}\dot{g},
\end{equation}
like we did in the case of ${\bf S}^2$.

\section{Fuzzy Physics}
The remainder of this chapter will be a short , self-contained description of the main results of this thesis .

In order to be able to do physics on fuzzy spaces one needs to have a Dirac operator. On even dimensional spaces ,
the Dirac operator together with the chirality operator define all the differential structure of the space . More
precisely the Dirac operator , as we will see in the next chapter , gives essentially the metric aspects of the
space . On odd dimensional manifolds there are no chirality operators and we are only left with the Dirac operators
to decribe their differential structures .

In chapter $3$ we will derive the Dirac operator on fuzzy ${\bf S}^2_F$ starting from the Dirac operator on
classical ${\bf S}^2$ . Chapter $5$ contains an alternative derivation for the Dirac operator on ${\bf C}{\bf
P}^2$ which is readily generalizable to higher coadjoint orbits. In here we will briefly sketch the results of
chapter $3$ for ${\bf S}^2_F$, so that one can immediately see their applications in the context of the two
following selected examples , a) fuzzy monopoles wave functions (or bundles) and their winding numbers , b) and the
remarkable absence of fermion doubling on fuzzy ${\bf S}^2_F$. A full treatement of these topics is given in
chapter $4$ which also contains a construction of fuzzy sigma fields and fuzzy solitons on ${\bf S}_F^2$. Fuzzy
dynamics and continuum limits of such models are also discussed in chapter $4$.

The last example discussed in this introduction is the construction of fuzzy sections of spinor bundle on fuzzy
${\bf S}_F^2$ . This example together with the above fuzzy monopoles case will provide two concrete examples of the
equivalence between projective modules and sections of vector bundles and their fuzzification [see below for more
explanation]. The case of fuzzy sections of vector bundles will also offer a good example of the use of star
products in noncommutative geometry and fuzzy physics . All of these issues are further expanded in chapter $5$ of
this thesis .

\subsection{Fuzzy Differential Structure} Viewing ${\bf S}^2$ as a submanifold of ${\bf R}^3$ , one can check the
following basic identity
\begin{equation}
{\cal D}_2={\cal D}_3|_{r={\rho}}+\frac{i{\gamma}^3}{\rho}.\label{basicidentity}
\end{equation}
[For an explicit derivation of the above formula see section $3.1.3$] . ${\gamma}^{a}={\sigma}_a$, $a=1,2,3$ , are
the flat gamma matrices in $3-$dimensions . ${\cal D}_2$ , ${\cal D}_3$ are the Dirac operators on ${\bf S}^2$ and
${\bf R}^3$ respectively . ${\cal D}_3|_{r={\rho}}$ is the restriction of the Dirac operator on ${\bf R}^3$ to the
sphere $r={\rho}$ , where ${\rho}$ is the radius of the sphere , namely $\sum_{{a}=1}^3x^2_{a}={\rho}^2$ for any
$\vec{x}{\in}{\bf S}^2$ . The Clifford algebra on ${\bf S}^2$ is two dimensional and therefore at each point
$\vec{n}=\vec{x}/{\rho}$ one has only two independents gamma matrices , they can be taken to be ${\gamma}^1$ and
${\gamma}^2$ . ${\gamma}^3$ should then be identified with the chirality operator ${\gamma}=\vec{\sigma}.\vec{n}$
on ${\bf S}^2$ .

Next by using the canonical Dirac operator ${\cal D}_3=-i{\sigma}_a{\partial}_a$ in (\ref{basicidentity}) one can derive the two
following equivalent expressions for the Dirac operator ${\cal D}_2$ on ${\bf S}_2$ :
\begin{eqnarray}
{\cal D}_{2g}&=&\frac{1}{\rho}(\vec{\sigma}\vec{\cal L}+1)\nonumber\\
{\cal D}_{2w}&=&-\frac{1}{\rho}{\epsilon}_{ijk}{\sigma}_in_j{\cal J}_k.\label{classicaldiracs}
\end{eqnarray}
[see section $3.1.4$ for full details] . ${\cal L}_k=-i{\epsilon}_{kij}x_i{\partial}_j$ is the orbital angular
momentum and ${\cal J}_k={\cal L}_k+\frac{{\sigma}_k}{2}$ is the total angular momentum . $g$ and $w$ in (\ref{classicaldiracs})
stands for Grosse-Klim\v{c}\'{i}k-Pre\v{s}najder and Watamuras Dirac operators respectively . It is not difficult
to check that ${\cal D}_{2w}=i{\gamma}{\cal D}_{2g}={\cal D}_3|_{r=\rho}+\frac{i{\gamma}}{\rho}$ which means that
${\cal D}_{2w}$ and ${\cal D}_{2g}$ are related by a unitary transformation and therefore are equivalent . The
spectrum of these Dirac operators is trivially derived to be given by ${\pm}\frac{1}{\rho}(j+\frac{1}{2})$ where
$j$ is the eigenvalue of $\vec{\cal J}$ , i.e $\vec{\cal J}^2=j(j+1)$ and $j=1/2,3/2,...$ .

As we have already shown in the first part of the introduction , transition from ${\bf S}^2$ to the fuzzy sphere
${\bf S}^2_F$ can be achieved by the replacement
$\vec{n}{\longrightarrow}\vec{n}^F=\frac{\vec{L}^L}{\sqrt{l(l+1)}}$ where ${L}^L_i$ , $i=1,2,3$ ,  are the
generators of the IRR $l$ of $SU(2)$ , i.e $[L_i^L,L_j^L]=i{\epsilon}_{ijk}L_k^L$ . The fuzzy sphere ${\bf S}_F^2$
is defined to be the algebra ${\bf A}$ of $(2l+1){\times}(2l+1)$ matrices of the form (\ref{expansionoffuzzys2}). By definition
$L_i^L$ act on the left of the algebra ${\bf A}$ , namely $L_i^Lf=L_if$ for any $f{\in}{\bf A}$ . The fuzzy
versions of the Dirac operators (\ref{classicaldiracs}) are then
\begin{eqnarray}
D_{2g}&=&\frac{1}{\rho}(\vec{\sigma}.ad\vec{L}+1)\nonumber\\
D_{2w}&=&\frac{1}{\rho}{\epsilon}_{ijk}{\sigma}_in_j^FL_k^R.\label{fuzzydiracs}
\end{eqnarray}
$ad\vec{L}=\vec{L}^L-\vec{L}^R$ is the fuzzy derivation which annihilates the identity matrix in ${\bf A}$ as the
classical derivation $\vec{\cal L}$ annihilates the constant function in ${\cal A}$ . $-\vec{L}^R$ are the
generators of the IRR $l$ of $SU(2)$ which act on the right of the algebra ${\bf A}$ , i.e $-L^R_if=-fL_i$ for any
$f{\in}{\bf A}$ . From this definition one can see that $ad{L}_i$ provide the generators of the adjoint action of
$SU(2)$ on ${\bf A}$ , namely $ad\vec{L}(f)=[\vec{L},f]$ for any $f{\in}{\bf A}$ .

These two fuzzy Dirac operators are not unitarily equivalent anymore . This can be checked by computing their
spectra . The spectrum of $D_{2g}$ is exactly that of the continuum only cut-off at the top total angular momentum
$j=2l+\frac{1}{2}$ . In other words the spectrum of $D_{2g}$ is equal to $\{{\pm}\frac{1}{\rho}(j+\frac{1}{2})$ ,
$j=\frac{1}{2},\frac{3}{2},...2l-\frac{1}{2}\}$ and $D_{2g}(j)=\frac{1}{\rho}(j+\frac{1}{2})$ for
$j=2l+\frac{1}{2}$ . The spectrum of $D_{2w}$ is , however , highly deformed as compared to the continuum spectrum
especially for large values of $j$ . It is given by
$D_{2w}(j)={\pm}\frac{1}{\rho}(j+\frac{1}{2})\sqrt{1+\frac{1-(j+1/2)^2}{4l(l+1)}}$ . From these results it is
obvious that $D_{2g}$ is superior to $D_{2w}$ as an approximation to the continuum .

In the same way one can find the fuzzy chirality operator ${\Gamma}$ by the simple replacement
$\vec{n}{\longrightarrow}\vec{n}^F$ in ${\gamma}=\vec{\sigma}.\vec{n}$ and insisting on the result to have the
following properties  : $1){\Gamma}^2=1$ , ${\Gamma}^{+}={\Gamma}$ and $[{\Gamma},f]=0$ for all $f{\in}{\bf A}$ .
One then finds
\begin{equation}
{\Gamma}=\frac{1}{l+\frac{1}{2}}(-\vec{\sigma}\vec{L}^R+\frac{1}{2}).\label{fuzzychirality0}
\end{equation}
[see section $3.2.4$ for explicit calculations] . Interestingly enough this fuzzy chirality operator anticommutes
with ${\bf D}_{2w}$  and not with ${\bf D}_{2g}$ so ${\bf D}_{2w}$ is a better approximation to the continuum than
$D_{2g}$ from this respect . This is also clear from the spectra above , in the spectrum of $D_{2g}$ the top
angular momentum is not paired to anything and therefore $D_{2g}$ does not admit a chirality operator . A question
then arises naturally , with which Dirac operator should we approximate the continuum? An answer to this question
will be given below when we discuss chiral fermions on ${\bf S}^2_F$ .

\subsection{Fuzzy Non-Trivial Gauge Configurations}

Monopoles are one of the most fundamental non-trivial configurations in field theory. The wave functions of a
particle of charge $q$ in the field of a monopole $g$ , which is at rest at $r=0$ , are known to be given by the
expansion \cite{17}
\begin{equation}
{\psi}^{(N)}(r,g)=\sum_{j,m}c_{m}^{j}(r)<j,m|D^{(j)}(g)|j,-\frac{N}{2}>,\label{mwf}
\end{equation}
where $D^{(j)}:g{\longrightarrow}D^{(j)}(g)$ is the $j$ IRR of $g{\in}SU(2)$ . The integer $N$ is related to $q$
and $g$ by the Dirac quantization condition : $N=\frac{qg}{2{\pi}}$ . $r$ is the radial coordinate of the relative
position $\vec{x}$ of the system , the angular variables of $\vec{x}$ are defined through the element
$g{\in}SU(2)$ by $\vec{\tau}.\vec{n}=g{\tau}_3g^{-1}$ , $\vec{n}=\vec{x}/r$ . It is also a known result that the
precise mathematical structure underlying this physical system is that of a $U(1)$ principal fiber bundle
$SU(2){\longrightarrow}{\bf S}^2$. In other words for a fixed $r=\rho$, the particle $q$ moves on a sphere ${\bf
S}^2$ and its wave functions (\ref{mwf}) are precisely elements of ${\bf {\Gamma}}({\bf S}^2,SU(2))$ , namely sections
of a $U(1)$ bundle over ${\bf S}^2$ . They have the equivariance property
\begin{equation}
{\psi}^{(N)}(\rho,ge^{i{\theta}\frac{{\tau}_3}{2}})=e^{-i{\theta}\frac{N}{2}}{\psi}^{(N)}(\rho,g),\label{equivariance}
\end{equation}
i.e they are not really functions on ${\bf S}^2$ but rather functions on $SU(2)$ because they clearly depend on the
specific point on the $U(1)$ fiber  . In this introduction , we will only consider the case $N={\pm}1$ . The case
$|N|{\neq}1$ being similar and will be treated in great detail in chapter $4$ .

An alternative description of monopoles can be given in terms of K-theory and projective modules . It is based on
the Serre-Swan's theorem \cite{1,last} which states that there is a complete equivalence between vector bundles
over a compact manifold ${\bf M}$ and projective modules over the algebra ${C}(\bf M)$ of smooth functions on
${\bf M}$ . Projective modules are constructed from ${C}(\bf M)^{n}={C}(\bf M){\otimes}{\bf C}^n$ where $n$ is some
integer by the application of a certain projector $p$ in ${\cal M}_{n}({C}(\bf M))$ , i.e the algebra of
$n{\times}n$ matrices with entries in ${C}(\bf M)$ .

In our case ${\bf M}={\bf S}^2$ and $C(\bf M)={\cal A}\equiv$ the algebra of smooth functions on ${\bf S}^2$ . For
a monopole system with winding number $N={\pm}1$ , the appropriate projective module will be constructed from
${\cal A}^2={\cal A}{\otimes}{\bf C}^2$ . It is ${\cal P}^{(\pm 1)}{\cal A}^2$ where ${\cal P}^{(\pm 1)}$ is the
projector
\begin{equation}
{\cal P}^{(\pm 1)}=\frac{1{\pm}\vec{\tau}.\vec{n}}{2}.\label{classicalprojector}
\end{equation}
It is clearly an element of ${\cal M}_{2}(\cal A)$ and satisfies ${\cal P}^{{(\pm 1)} 2}={\cal P}^{(\pm 1)}$ and
${\cal P}^{{(\pm 1)}+}={\cal P}^{(\pm 1)}$. ${\cal P}^{(\pm 1)}{\cal A}^2$ desribes a monopole system with
$N={\pm}1$ as one can directly check by computing its  winding number as follows
\begin{equation}
\pm 1=\frac{1}{2{\pi}i}\int Tr{\cal P}^{({\pm}1)}d{\cal P}^{({\pm}1)}{\wedge}d{\cal P}^{({\pm}1)}.\label{windingnumber1}
\end{equation}
On the contrary to the space of sections ${\bf {\Gamma}}({\bf S}^2,SU(2))$ , elements of ${\cal P}^{(\pm 1)}{\cal
A}^2$ are by construction invariant under the action $g{\longrightarrow}gexp(i{\theta}\frac{{\tau}_3}{2})$ . The
other advantage of ${\cal P}^{(\pm 1)}{\cal A}^2$ as compared to ${\bf {\Gamma}}({\bf S}^2,SU(2))$ is the fact
that its fuzzification is much more straight forward . The fuzzification of the space of sections ${\bf
{\Gamma}}({\bf S}^2,SU(2))$ will be the subject of the last example of this introduction .

Before we start the fuzzification of ${\cal P}^{(\pm 1)}{\cal A}^2$ , let us first comment on the relation between
the wave functions ${\psi}^{(\pm 1)}$ given in equation (\ref{mwf}) and those belonging to ${\cal P}^{(\pm 1)}{\cal
A}^2$ . The projector ${\cal P}^{(\pm 1)}$ can be rewritten as ${\cal P}^{(\pm 1)
}=D^{(\frac{1}{2})}\frac{1{\pm}{\tau}_3}{2}D^{(\frac{1}{2})+}(g)$ where
$D^{(\frac{1}{2})}:g{\longrightarrow}D^{(\frac{1}{2})}(g)=g$ is the $\frac{1}{2}$ IRR of $SU(2)$ . Hence ${\cal
P}^{(\pm 1)}D^{(\frac{1}{2})}(g)|\pm>=D^{(\frac{1}{2})}(g)\frac{1{\pm}{\tau}_3}{2}|\pm>=D^{(\frac{1}{2})}(g)|\pm>$
, where $|\pm>$ are defined by ${\tau}_3|\pm>=\pm|\pm>$ . In the same way one can show that ${\cal P}^{(\pm
1)}D^{(\frac{1}{2})}(g)|\mp>=0$ . This last result means that
\begin{eqnarray}
{\cal P}^{(\pm 1)}&=&D^{(\frac{1}{2})}(g)|\pm><\pm|D^{(\frac{1}{2})+}(g)\nonumber\\
&{\rm or}&\nonumber\\
({\cal P}^{(\pm 1)})_{ij}&=&D^{(\frac{1}{2})}_{i\pm}(g)D^{(\frac{1}{2})+}_{{\pm}j}(g).
\end{eqnarray}
$<\pm|D^{(\frac{1}{2})+}(g)$ defines a map from ${\cal P}^{(\pm 1)}{\cal A}^2$  into ${\bf {\Gamma}}({\bf
S}^2,SU(2))$ as follows
\begin{eqnarray}
<\pm|D^{(\frac{1}{2})+}(g)&:&{\cal P}^{(\pm 1)}{\cal A}^2{\longrightarrow}{\bf {\Gamma}}({\bf S}^2,SU(2))\nonumber\\
&&|\psi>{\longrightarrow}<\pm|D^{(\frac{1}{2})+}(g)|\psi>={\psi}^{(\pm 1)}(\rho,g).\label{mwf1}
\end{eqnarray}
$<\pm|D^{(\frac{1}{2})+}(g)|\psi>$ has the correct transformation law (\ref{equivariance}) under
$g{\longrightarrow}gexp(i{\theta}\frac{{\tau}_3}{2})$ as one can check by using the basic equivariance property
\begin{equation}
D^{(\frac{1}{2})}(ge^{i{\theta}\frac{{\tau}_3}{2}})|\pm>=e^{\pm i \frac{\theta}{2}}D^{(\frac{1}{2})}(g)|\pm>.
\end{equation}
In the same way $D^{(\frac{1}{2})}(g)|\pm>$ defines a map , ${\bf {\Gamma}}({\bf S}^2,SU(2)){\longrightarrow}{\cal
P}^{(\pm 1)}{\cal A}^2$, which takes the wave functions ${\psi}^{(\pm 1)}$ to the two components elements
${\psi}^{(\pm 1)}D^{(\frac{1}{2})}(g)|\pm>$ of ${\cal P}^{(\pm 1)}{\cal A}^2$ . Under
$g{\longrightarrow}gexp(i{\theta}\frac{{\tau}_3}{2})$ , the two phases coming from ${\psi}^{(\pm 1)}$ and
$D^{(\frac{1}{2})}(g)|\pm>$ cancel exactly so that their product is a function over ${\bf S}^2$ .

Towards fuzzification one rewrites the winding number (\ref{windingnumber1}) in the form
\begin{eqnarray}
\pm 1 &=& -\frac{1}{4 \pi}\int d (\cos{\theta}){\wedge}d{\phi}\;{\rm Tr}\;{\gamma}{\cal P}^{(\pm 1)}\;[{\cal
D}, {\cal P}^{(\pm 1)}]\;[{\cal D}, {\cal P}^{(\pm 1)}](\vec{n})\nonumber\\
&=& -Tr_{\omega} \left( \frac{1}{|{\cal D}|^2}\gamma \;{\cal P}^{(\pm 1)}\; [{\cal D}, {\cal P}^{(\pm 1)}] \;[{\cal
D}, {\cal P}^{(\pm 1)}]\; \right).\label{windingnumber2}
\end{eqnarray}
The first line is trivial to show starting from (\ref{windingnumber1}) , whereas the second line is essentially Connes trace
theorem which we will prove in the next chapter. $|{\cal D}|=$ positive square root of ${\cal D}^{\dagger}{\cal
D}$ while $Tr_{\omega}$ is the Dixmier trace[see the next chapter for the definition] . In the fuzzy setting , this
Dixmier trace will be replaced by the ordinary trace because the algebra of functions on fuzzy ${\bf S}^2_F$ is
finite dimensional .

${\cal D}$ in (\ref{windingnumber2}) is either ${\cal D}_{2g}$ or ${\cal D}_{2w}$ which are given in equation (\ref{classicaldiracs}) . They
both give the same answer ${\pm}1$ . The fuzzy analogues of ${\cal D}_{2g}$ and ${\cal D}_{2w}$ are respectively
$D_{2g}$ and $D_{2w}$ given by equation (\ref{fuzzydiracs}) . These latter operators were shown to be different and therefore
one has to decide which one should we take as our fuzzy Dirac operator . $D_{2g}$ does not admit as it stands a
chirality operator and therefore its use in the computation of winding numbers requires more care which we will do
in the next section on chiral fermions . $D_{2w}$ admits the fuzzy chirality operator (\ref{fuzzychirality0}) which will be used
instead of the continuum chirality ${\gamma}=\vec{\sigma}.\vec{n}$. However $D_{2w}$ has a zero eigenvalue for
$j=2l+\frac{1}{2}$ so it must be regularized for its inverse in (\ref{windingnumber2}) to make sense. This will be understood
but not done explicitly in this introduction , a careful treatement is given in chapter $4$.

Finally the projector ${\cal P}^{(\pm 1)}$ will be replaced by a fuzzy projector $p^{(\pm 1)}$ which we will now
find  . We proceed like we did in finding the chirality operator ${\Gamma}$ , we replace $\vec{n}$ in (\ref{classicalprojector}) by
$\vec{n}^F={\vec{L}^L}/{\sqrt{l(l+1)}}$ and insist on the result to have the properties $p^{(\pm 1)2}=p^{(\pm 1)}$
and $p^{(\pm 1)+}=p^{(\pm 1)}$ . We also require this projector to commute with the chirality operator ${\Gamma}$ ,
the answer for winding number $N=+1$ turns out to be
\begin{equation}
p^{(+1)}=\frac{1}{2}+\frac{1}{2l+1}\big[\vec{\tau}.\vec{L}^L+\frac{1}{2}\big].
\end{equation}
This can be rewritten in the following useful form
\begin{equation}
p^{(+1)}=\frac{\vec{K}^{(1)2}-(l-\frac{1}{2})(l+\frac{1}{2})}{(l+\frac{1}{2})(l+\frac{3}{2})-(l-\frac{1}{2})(l+\frac{1}{2})},\label{fuzzyprojector1}
\end{equation}
where $\vec{K}^{(1)}=\vec{L}^L+\frac{{\vec{\tau}}}{2}$ . This allows us to see immediately that $p^{(+1)}$ is the
projector on the subspace with the maximum eigenvalue $l+\frac{1}{2}$ . Similarly , the projector $p^{(-1)}$ will
correspond to the subspace with minimum eigenvalue $l-\frac{l}{2}$ , namely
\begin{equation}
p^{(-1)}=\frac{\vec{K}^{(1)2}-(l+\frac{1}{2})(l+\frac{3}{2})}{(l-\frac{1}{2})(l+\frac{1}{2})-(l+\frac{1}{2})(l+\frac{3}{2})}.\label{fuzzyprojector2}
\end{equation}
By construction (\ref{fuzzyprojector2}) as well as (\ref{fuzzyprojector1}) have the correct continuum limit (\ref{classicalprojector}), and they are in the
algebra ${\cal M}_2(\bf A)$ where ${\bf A}$ is the fuzzy algebra on fuzzy ${\bf S}^2_F$ , i.e
$2(2l+1){\times}2(2l+1)$ matrices . Fuzzy monopoles with winding number ${\pm}1$ are then desribed by the
projective modules $p^{(\pm 1)}{\bf A}^2$ .

If one include spin , then ${\bf A}^2$ should be enlarged to ${\bf A}^4$ . It is on this space that the Dirac
operator $D_{2w}$ as well as the chirality operator ${\Gamma}$ are acting . In the fuzzy the left and right actions
of the algebra ${\bf A}$ on ${\bf A}$ are not the same , For each $a \in {\bf A}$ , we thus have operators $a^{L,
R} \in {\bf A}^{L,R}$ acting on $\xi \in {\bf A}$ according to $a^L \xi = a \xi, a^R \xi = \xi a$. [Note that $a^L
b^L = (ab)^L $ while $a^R b^R = (ba)^R$] . The left action is genearted by $L_i^L$ whereas the right action is
genearted by $-L_i^R$  so that we are effectively working with the algebra ${\bf A}^L{\otimes}{\bf A}^R$ . A
representation ${\Pi}$ of this algebra is provided by ${\Pi}(\alpha)={\alpha}{\otimes}{\bf 1}_{2{\times}2}$ for any
${\alpha}{\in}{\bf A}^L{\otimes}{\bf A}^R$ . It acts on the Hilbert space ${\bf A}^4{\oplus}{\bf A}^4$ .

With all these considerations , one might as well think that one must naively replace
$Tr_{\omega}{\longrightarrow}Tr$ , ${\gamma}{\longrightarrow}{\Gamma}$ , ${\cal D}{\longrightarrow}D_{2w}$ and
${\cal P}^{(\pm 1)}{\longrightarrow}p^{(\pm 1)}$ in (\ref{windingnumber2}) to get its fuzzy version . This is not totally correct
since the correct discrete version of (\ref{windingnumber2}) turns out to be
\begin{equation}
c(\pm 1)=-Tr{\epsilon}P^{(\pm 1)}[F_{2w},P^{(\pm 1)}][F_{2w},P^{(\pm 1)}],\label{windingnumber3}
\end{equation}
with
\begin{equation}
{\bf F}_{2w} = \left( \begin{array}{cc}
             0  & \frac{D_{2w}}{|D_{2w}|}\\
             \frac{D_{2w}}{|D_{2w}|}& 0
           \end{array} \right),
           \quad \epsilon=\left( \begin{array}{cc}
                                                    {\Gamma} & 0 \\
                                                        0  & {\Gamma}
                                                 \end{array} \right).\label{fredholm}
\end{equation}
and
\begin{equation}
P^{(\pm 1)}=\left( \begin{array}{cc}
           \frac{1+\Gamma}{2} p^{(\pm)} & 0 \\
                    0  & \frac{1-\Gamma}{2} p^{(\pm)}
         \end{array} \right). \label{projector}
\end{equation}
[For a complete proof of (\ref{windingnumber3}) using index theory see chapter $4$] . For $p^{(+1)}$ one finds that
$c(+1)=+1+[2(2l+1)+1]$ while for $p^{(-)}$ we find $c(-1)=-1+[2(2l)+1]$. They are both wrong if compared to (\ref{windingnumber2})!

The correct answer is obtained by recognizing that $c(\pm 1)$ is nothing but the index of the operator
\begin{equation}
\hat{f}^{(+)}=\frac{1-\Gamma}{2}p^{(\pm 1)}\frac{D_{2w}}{|D_{2w}|}p^{(\pm 1)}\frac{1+\Gamma}{2}.
\end{equation}
This index counts the number of zero modes of $\hat{f}^{(+)}$ . The proof starts by remarking that , by
construction , only the matrix elements $<p^{(\pm 1)}U_{-}|{\hat{f}}^{(+)}|p^{(\pm 1)}U_{+}>$ where
$U_{\pm}=\frac{1{\pm}\Gamma}{2}{\bf A}^4$ , exist and therefore ${\hat{f}}^{(+)}$ is a mapping from
$\hat{V}_{+}=p^{(\pm 1)}U_{+}$ to $\hat{V}_{-}=p^{(\pm 1 )}U_{-}$ . Hence
$Index{\hat{f}}^{(+)}=dim\hat{V}_{+}-dim\hat{V}_{-}$ .

Since one can write the chirality operator ${\Gamma}$ in the form
${\Gamma}=\frac{1}{l+1/2}\Big[j(j+1)-(l+1/2)^2\Big]$ where $j$ is the eigenvalue of $(-\vec{L}^R +
\frac{\vec{\sigma}}{2})^2$ , $j=l\pm 1/2$ for which $\Gamma|_{j=l\pm 1/2}=\pm 1$ defines the subspace $U_{\pm}$
with dimension $2(l\pm 1/2)+1$ . On the other hand , for $p^{(+1)}$ which projects down to the subspace with
maximum eigenvalue $k_{max}=l+\frac{1}{2}$ of the operator $\vec{K}^{(1)}=\vec{L}+\frac{{\vec{\tau}}}{2}$ ,
$\hat{V}_{\pm}$ has dimension $[2(l{\pm}1/2)+1][2(l+1/2)+1]$ and so the index is
$Index{\hat{f}}^{(+)}=c(+1)=2(2l+2)$.

This result signals the existence of zero modes of the operator ${\hat{f}}^{(+)}$ . Indeed for ${\Gamma}=+1$ one
must couple $l+\frac{1}{2}$ to $l+\frac{1}{2}$ and obtain $j=2l+1,2l,..0$, whereas for ${\Gamma}=-1$ we couple
$l+\frac{1}{2}$ to $l-\frac{1}{2}$ and obtain $j=2l,...,1$ . $j$ here denotes the total angular momentum
$\vec{J}=\vec{L}^L-\vec{L}^R+\frac{\vec{\sigma}}{2}+\frac{\vec{\tau}}{2}$ . Clearly the eigenvalues
$j^{(+1)}=2l+1$ and $0$ in $\hat{V}_+$ are not paired to anything . The extra piece in $c(+1)$ is therefore
exactly equal to the number of the top zero modes , namely $2j^{(+1)}+1=2(2l+1)+1$ . These modes do not exist in
the continuum and therefore they are of no physical relevance and must be projected out . This can be achieved by
replacing the projector $p^{(+1)}$ by a corrected projector ${\pi}^{(+1)}=p^{(+1)}[1-{\pi}^{(j^{(+1)})}]$ where
${\pi}^{(j^{(+1)})}$ projects out the top eigenvalue $j^{(+1)}$ , it can be easily written down explicitly [see
equation $(4.35)$] . Putting ${\pi}^{(+1)}$ in (\ref{windingnumber3}) gives exactly $c(+1)=+1$ which is the correct answer .

The same analysis goes for $p^{(-1)}$ . Since it projects down to the subspace with minimum eigenvalue
$k_{max}=l-\frac{1}{2}$ of the operator $\vec{K}^{(1)}=\vec{L}+\frac{\vec{\tau}}{2}$ , $\hat{V}_{\pm}$ has now
dimension $[2(l{\pm}1/2)+1][2(l-1/2)+1]$ and so the index is $Index{\hat{f}}^{(+)}=c(-1)=4l$. By coupling
$l-\frac{1}{2}$ to $l+\frac{1}{2}$ one obtains $j=2l,2l-1,..1$ which are all ${\Gamma}=+1$ states, whereas by
coupling $l-\frac{1}{2}$ to $l-\frac{1}{2}$ one obtains $j=2l-1,...,1,0$ with ${\Gamma}=-1$ . Clearly the top
eigenvalue $j^{(-1)}=2l$ in $\hat{V}_+$ and $0$ in $\hat{V}_-$ are not paired to anything. $c(-1)$ can then be
rewritten as $c(-1)=[2j^{(-1)}+1]-[+1]$. Again the top modes do not exist in the continuum and therefore are of no
physical relevance and must be projected out . This can be achieved by replacing the projector $p^{(-1)}$ by a
corrected projector ${\pi}^{(-1)}=p^{(-1)}[1-{\pi}^{(j^{(-1)})}]$ where ${\pi}^{(j^{(-1)})}$ projects out the top
eigenvalue $j^{(-1)}$. Putting ${\pi}^{(-1)}$ in (\ref{windingnumber3}) gives exactly $c(-1)=-1$ which is what we want .

This idea of obtaining fuzzy projectors describing fuzzy monoples with all the correct properties and with the
right continuum limit generalizes easily to higher charges . For winding number ${\pm}N$ one introduces $N$ set of
Pauli matrices $\vec{\tau}^{(i)}$ and define $\vec{K}^{(N)}=\vec{L}^L+\sum_{i=1}^N\frac{{\vec{\tau}}^{(i)}}{2}$ .
Fuzzy Monopole of charge ${\pm}N$ is described by the projective module $p^{(\pm N)}{\bf A}^{2^{N}}$ , where
$p^{(\pm N)}$ is the projector on the subspace with maximum (minimum) eigenvalue , $l+\frac{N}{2}(l-\frac{N}{2})$,
of the operator $\vec{K}^{(N)}$. They have the right winding number $\pm N$ given by formulae like (\ref{windingnumber3}). These
projectors reduce in the continuum limit to ${\cal P}^{\pm N}=\prod_{i=1}^N
\frac{1{\pm}\vec{\tau}^{(i)}.\vec{n}}{2}$ as one can check , and in terms of sections of vector bundles the
projective module , ${\cal P}^{(\pm N)}{\cal A}^{2^N}$, corresponds to the tensor product of $N$ wave functions of
the form (\ref{mwf1}).

\subsection{Chiral Fermions on Fuzzy Spaces}
The Dirac operator $D_{2w}$ was shown to admit a chirality operator ${\Gamma}$ , i.e $\{D_{2w},{\Gamma}\}=0$ .
However there was one obvious problem with this Dirac operator , its spectrum. It has very small eigenvalues
$D_{2w}(j)$ for large values of $j$\footnote{As usual $j$ denotes the eigenvalues of the total angular momentum
which is , in the absence of monopoles , given by $\vec{J}=\vec{L}^L-\vec{L}^R+\frac{\vec{\sigma}}{2}$ .}, in
particular its top eigenvalue corresponding to $j=2l+\frac{1}{2}$ is identically zero . The spectrum of the Dirac
operator $D_{2g}$ was , on the other hand , identical to the continuum spectrum
${\pm}\frac{1}{\rho}(j+\frac{1}{2})$ only truncated at the top eigenvalue $j=2l+\frac{1}{2}$ . But this $D_{2g}$
does not commute with ${\Gamma}$ as the top mode is not paired to anything .

One would like to be able to define chiral fermions on fuzzy ${\bf S}^2_F$ with the above truncated spectrum . To
this end , one first projects out the top mode , which is a source of problem for both Dirac operators , and
define the space $V$ through the projector $P$ given by $P|j,j_3>=|j,j_3>$ for any $j{\leq}2l-\frac{1}{2}$ and
$P|j=2l+\frac{1}{2},j_3>=0$. Next recall that in the continuum we had the identity ${\cal D}_{2w}=i{\gamma}{\cal
D}_{2g}$ which can be put in the form $|{\cal D}_{2g}|^{-1}{\cal D}_{2g}{\cal D}_{2w}|{\cal D}_{2w}|^{-1}=i|{\cal
D}_{2g}|^{-1}{\cal D}_{2g}{\gamma}{\cal D}_{2g}|{\cal D}_{2w}|^{-1}$ . Then by using the facts $|{\cal
D}_{2g}|=|{\cal D}_{2w}|$ and $[{\gamma},|{\cal D}_{2g}|]=[{\gamma},|{\cal D}_{2w}|]=0$ one can see that
${\gamma}=\vec{\sigma}.\vec{n}$ can be rewritten
\begin{equation}
\gamma~=~i{\frac{{\cal D}_{2g}}{|{\cal D}_{2g}|}}{\frac{{\cal D}_{2w}} {|{\cal D}_{2w}|}}\label{gama}.
\end{equation}
Naive fuzzification would give
\begin{equation}
{\Gamma}^{'}~=~iF_{2g}F_{2w}.
\end{equation}
where $F_{2g}$ and $F_{2w}$ are given on the space $V$ by $\frac{D_{2g}}{|D_{2g}|}$ and $\frac{D_{2w}}{|D_{2w}|}$
respectively. On the subspace corresponding to $j=2l+\frac{1}{2}$ both $F_{2g}$ and $F_{2w}$ are set to zero .

It is a very interesting exercise to check that ${\Gamma}^{'}$ is indeed a chirality operator which is different
from ${\Gamma}$ . By construction , it also has the correct continuum limit . Further it anticommutes with both
Dirac operators $PD_{2g}P$ and $PD_{2w}P$ . [see chapter $4$ for the explicit proof].

One can now replace the chirality operator ${\Gamma}^{'}$ in the chiral pair $(PD_{2g}P,{\Gamma}^{'})$ by
${\Gamma}$ as follows . The triplet $e_1=F_{2g}$ , $e_2=F_{2w}$ and $e_3={\Gamma}^{'}$ can be shown to form a
Clifford algebra , namely $\{e_a,e_b\}={\delta}_{ab}{\bf 1}_{V}$ . In other words $e_a$ together with the identity
on $V$ will generate all linear operators on $V$ , in particular ${\Gamma}$ can be written as a linear combination
of them . But since ${\Gamma}$ is the chirality operator of $e_2=F_{2w}$ , its expansion in terms of $e_a$ and
${\bf 1}_V$ will only contain $e_1$ and $e_3$ , and therefore it must be in the plane generated by $e_1=F_{2g}$ and
$e_3={\Gamma}^{'}$ . Hence ${\Gamma}^{'}$ can be rotated to ${\Gamma}$ by a unitary transformation
$U_{e_2}(\theta)$ with an angle ${\theta}$ around $e_2$, i.e
${\Gamma}=U_{e_2}(\theta)^{+}{\Gamma}^{'}U_{e_2}(\theta)$. Similarly the Dirac operator $PD_{2g}P$ will be rotated
to a new Dirac opertaor $D=U_{e_2}(\theta)^{+}PD_{2g}PU_{e_2}(\theta)$ . [see chapter $4$ for the explicit form of
the rotation $U_{e_2}$] .

Using this new pair $(D,{\Gamma})$ in the winding number formula (\ref{windingnumber3}) will give exactly the correct charges
${\pm}1$ with no need to modify the projectors $p^{(\pm 1)}$ , namely
\begin{equation}
\pm 1=-Tr{\epsilon}P^{(\pm 1)}[F,P^{(\pm 1)}][F,P^{(\pm 1)}],
\end{equation}
where now $F$ is given by (\ref{fredholm}) with the substitution $\frac{D_{2w}}{|D_{2w}|}{\longrightarrow}\frac{D}{|D|}$ .

\subsection{Fuzzy Spinors and the Star Product}
This is another example in which the relation of projective modules to sections of vector bundles is worked out
explicitly in both the continuum and the fuzzy case. Fuzzification here is elegantly achieved by the means of star
product . The considerations of this last section are relevant for spin $\frac{1}{2}$ particles moving on ${\bf
S}^2$ in the absence of monopoles . The case of ${\bf C}{\bf P}^2$ and/or the inclusion of monopoles can be
treated similarly and is done in chapter $5$.

${\cal D}_{2g}$ acts on ${\cal A}{\otimes}{\bf C}^2{\equiv}{\cal A}^2=\{|\psi>=\left(
\begin{array}{c}
{\psi}_1\\
{\psi}_2
\end{array}\right)\ ,{\psi}_i{\in}{\bf C}^{\infty}({\bf
S}^2)\}$ and it anticommutes with the chirality operator ${\gamma}=\vec{\sigma}.\vec{n}$ . To write down explicitly
a general element of the projective module ${\cal A}^2$ , one can for example expand it in terms of the
eigenfunctions of ${\gamma}$ . To this end one first remark that a general point $\vec{n}$ of ${\bf S}^2$ is given
in terms of $g=D^{(\frac{1}{2})}(g)$ by $\vec{\sigma}.\vec{n}=D^{(\frac{1}{2})}(g){\sigma}_3D^{(\frac{1}{2})-1}(g)$
and hence ${\gamma}D^{(\frac{1}{2})}(g)=D^{(\frac{1}{2})}(g){\sigma}_3$ . Now by taking the standard basis
$\{|+>,|->\}$ defined by the equations ${\sigma}_3|\pm>=\pm|\pm>$ , one have the following identity
\begin{equation}
{\gamma}D^{(\frac{1}{2})}(g)|\pm>={\pm}D^{(\frac{1}{2})}(g)|\pm>.
\end{equation}
In other words $D^{(\frac{1}{2})}(g)|+>$ and $D^{(\frac{1}{2})}(g)|->$ are the eigenfunctions of ${\gamma}$ with
$+1$ and $-1$ helicity respectively . Let us write these spinors as
\begin{equation}
|{\psi}^{(\frac{1}{2})}_{\pm}(g)>\equiv \left(
\begin{array}{c}
<+|D^{(\frac{1}{2})}(g)|\pm>\\
<-|D^{(\frac{1}{2})}(g)|\pm>
\end{array}\right)\,\label{spinor}
\end{equation}
By construction , these spinors $|{\psi}^{(\frac{1}{2})}_{\pm}(g)>$ have the equivariance property
\begin{equation}
|{\psi}^{(\frac{1}{2})}_{\pm}(ge^{i\frac{{\sigma}_3}{2}{\theta}})>=|{\psi}^{(\frac{1}{2})}_{\pm}(g)>e^{{\pm}i\frac{{\theta}}{2}}.\label{equivariance2}
\end{equation}
The elements of the projective module ${\cal A}^2$ , and hence too its chirality ${\pm}1$ subspaces
$\frac{1{\pm}{\sigma}.\hat{n}}{2}{\cal A}^2$ , are by construction invariant under $g\rightarrow
ge^{i\frac{{\sigma}_3}{2}{\theta}}$ . This is because they are functions on ${\bf S}^2$ and not on $SU(2)$ . The
expansion of elements of these subspaces using (\ref{spinor}) must thus have another factor in each term transforming
with the opposite phase to that in (\ref{equivariance2}) . Accounting for this fact, we can write for $|{\psi}>\in {\cal A}^2$ ,
\begin{eqnarray}
|\psi(g)>&=&|{\psi}^{+}(g)>+|{\psi}^{-}(g)>,\nonumber\\
|{\psi}^{\pm}(g)>&=&\Big[\sum_{j,m}{\xi}_m^{j\pm}<jm|{\psi}^{(j)}_{\mp}(g)>\Big]|{\psi}^{(\frac{1}{2})}_{\pm}(g)>~,~{\xi}_m^{j\pm}{\in}{\bf
C} .\label{expansionspinors}
\end{eqnarray}
$|{\psi}^{+}(g)>$ and $|{\psi}^{-}(g)>$ belong to the subspaces $\frac{1+{\sigma}.\hat{n}}{2}{\cal A}^2$ and
$\frac{1-{\sigma}.\hat{n}}{2}{\cal A}^2$ respectively and hence they represent left handed spinors and right handed
spinors on the sphere ${\bf S}^2$  . $|{\psi}^{(j)}_{\mp}(g)>$ in equation (\ref{expansionspinors}) is , on the other hand ,
defined by
\begin{equation}
|{\psi}^{(j)}_{\mp}(g)>=D^{(j)}(g)|j,{\mp}\frac{1}{2}>
\end{equation}
$j$ here are the eigenvalues of the total angular momentum $\vec{\cal J}=\vec{\cal L}+\frac{\vec{\sigma}}{2}$  . It
is half integer equal to $l+\frac{1}{2}$ or $l-\frac{1}{2}$ and therefore the states $|j,{\mp}\frac{1}{2}>$ do
always exist. $l$ are of course the eigenvalues of the orbital angular momentum $\vec{\cal L}$ . $D^{(j)}(g)$ is
the $j$ IRR of the element $g{\in}SU(2)$ . It is now not difficult to see that under the transformation
$g{\longrightarrow}gexp(i\frac{{\sigma}_3}{2}{\theta})$ we have
\begin{eqnarray}
|{\psi}^{(j)}_{\mp}(ge^{i\frac{{\sigma}_3}{2}{\theta}})>&=&|{\psi}^{(j)}_{\mp}(g)>e^{{\mp}i\frac{\theta}{2}}.\label{equivariance3}
\end{eqnarray}
From (\ref{equivariance2}) and (\ref{equivariance3}) it is then obvious that (\ref{expansionspinors}) are invariant under $g\rightarrow
ge^{i\frac{{\sigma}_3}{2}{\theta}}$ .

To find sections of the spinor bundle over ${\bf S}^2$ one simply acts with the Dirac operator ${\cal D}_{2g}$ on
the elements (\ref{expansionspinors}) of ${\cal A}^2$ . A non-trivial calculation gives

\begin{eqnarray}
{\cal D}_{2g}|{\psi}(g)>&=&-\frac{1}{\rho}\sum_{j,m}{\xi}_m^{j+}<jm|{\psi}_{+}^{(j)}(g)><j,\frac{1}{2}|{\cal
J}_{+}^{(j)}|j,-\frac{1}{2}>|{\psi}_{-}^{(\frac{1}{2})}(g)>\nonumber\\
&-&\frac{1}{\rho}\sum_{j,m}{\xi}_m^{j-}<jm|{\psi}^{(j)}_{-}(g)><j,-\frac{1}{2}|{\cal
J}_{-}^{(j)}|j,+\frac{1}{2}>|{\psi}_{+}^{(\frac{1}{2})}(g)>,\nonumber\\
&&\label{action1}
\end{eqnarray}
[see chapter $5$ for the complete proof] . ${\cal J}_{\pm}^{(j)}={\cal J}_{1}^{(j)}{\pm}i{\cal J}_{2}^{(j)}$ are
the lowering and raising generators of the $j$ IRR of $SU(2)$ . (\ref{action1}) can also be rewritten in the
"Dirac-K{\"a}hler" form
\begin{eqnarray}
{\cal D}_{2g}^{D-K}{\psi}^{D-K}\equiv-\frac{1}{\rho}\sum_{j}
                  && \left[ \begin{array}{cc}
                     0  & <j,-\frac{1}{2}|{\cal J}^{(j)}_{-}|j,+\frac{1}{2}>\\
                     <j,\frac{1}{2}|{\cal J}_{+}^{(j)}|j,-\frac{1}{2}>  &
                     0
                           \end{array}
                    \right]{\times}\nonumber\\
&&\left[
                            \begin{array}{c}
                  \sum_{m}\xi^{j+}_m<jm|{\psi}_{+}^{(j)}>\\
                  \sum_{m}\xi^{j-}_m <jm|{\psi}_{-}^{(j)}>
                              \end{array}
                            \right],\nonumber\\
&&
\end{eqnarray}
with
\begin{equation}
{\psi}^{D-K}=\left[
                            \begin{array}{c}
                  \sum_{j,m}\xi^{j+}_m<jm|{\psi}_{-}^{(j)}>\\
                  \sum_{j,m}\xi^{j-}_m <jm|{\psi}_{+}^{(j)}>
                              \end{array}
                            \right].\label{kahler}
\end{equation}
These are by definition the sections of the spinor bundle over ${\bf S}^2$ . As expected they are essentially
given in terms of the rotation matrices $<j,m|{\psi}_{\pm}^{(j)}>=D^{(j)}_{m,{\pm}\frac{1}{2}}$.

Towards the fuzzification of (\ref{expansionspinors}) and (\ref{kahler}) one introduces the coherent states
\begin{eqnarray}
|g ,l\rangle &=&U^{(l)}(g)|l,l\rangle \nonumber\\
|g ,l+\frac{1}{2}\rangle &=&U^{(l+\frac{1}{2})}(g)|l+\frac{1}{2},l+\frac{1}{2}\rangle \label{coherentstates} 
\end{eqnarray}
induced from the highest weight vectors $|l;l\rangle$ and $|l+\frac{1}{2};l+\frac{1}{2}>$ respectively.
$g{\rightarrow}U^{(k)}(g)$ , $k=l,l+\frac{1}{2}$ , is the angular momentum $k$ IRR of $SU(2)$. They have the
equivaraince property $|ge^{i{{\sigma}_3\over 2}{\theta}},k\rangle=e^{ik{\theta}}|g,k\rangle $.

It is a theorem \cite{ref21} that the diagonal matrix elements $\langle g,k|a|g,k\rangle$ completely determine the
operator $a$. Further $\langle  ge^{i{{\sigma}_3\over 2}{\theta}},k|a|ge^{i{{\sigma}_3\over 2}{\theta}},k\rangle
=\langle g,k|a|g,k\rangle$ so that $\langle g,k|a|g,k\rangle$ depends only on
$g{\sigma}_3g^{-1}=\vec{\sigma}.\vec{n}$ . In this way, we have the map
\begin{eqnarray}
{\bf A}&{\rightarrow}&{\cal A},\nonumber\\
a&{\rightarrow}&{\tilde a}=\langle g,k|a|g,k\rangle.
\end{eqnarray}
In particular the spherical harmonic $Y_{KM}$ is the image of an operators $T^{K}_{M}$ defined by
$Y_{KM}(\vec{n})=\langle g,k|T^{K}_{M}|g,k\rangle $ . ``Fuzzy spherical harmonics'' $T^{K}_{M}$ provide a basis of
matrices for ${\bf A}=Mat_{2l+1}$ and can only occur with $K{\leq}2l$ . The proof goes as follows , $SU(2)$ has two
different actions on ${\bf A}$ , the left action given by the generators $\vec{L}^L$ of the $l$ IRR of $SU(2)$ and
the right action given by the generators $-\vec{L}^R$ of the $l$ IRR of $SU(2)$ . The total action is then given
by the orbital angular momentum $\vec{\cal L}=\vec{L}^L-\vec{L}^R$ which has eigenvalues $K=0,...,2l$ .
$T^{K}_{M}$ can then be chosen to be such that $\vec{\cal L}^2T^K_M=K(K+1)T^K_M$ and ${\cal L}_3T^K_M=MT^K_M$.
 Elements of ${\bf A}$ can then be thought of as ordinary functions on ordinary ${\bf S}^2$ with expansions ( in
terms of $Y_{KM}$ ) cut-off at angular momenta $2l$ . For their product to not have terms with $K>2l$ , the
multiplication rule should be changed to that given by the star product. Thus consider $\langle
g,k|T^{K}_{M}T^{L}_{N}|g,k\rangle$. The functions $Y_{KM}$ and $Y_{LN}$ completely determine $T^{K}_{M}$ and
$T^{L}_{N}$, and for that reason also this matrix element . Hence it is the value of the function $Y_{KM}*Y_{LN}$,
linear in each factor, at $\vec{n}$:
\begin{equation}
\langle g,k|T^{K}_{M}T^{L}_{N}|g,k\rangle = [Y_{KM}*Y_{LN}](\vec{n}).
\end{equation}
The star product $*$ extends by linearity to all functions with angular momenta ${\leq}2l$. The resultant algebra
is $({\cal A},*)$ and it is isomorphic to the algebra ${\bf A}$. The explicit formula for $*$ on fuzzy ${\bf
S}^2_F$ has been found by Pre\v{s}najder \cite{9} [see also \cite{15,badis1} and references therein].

$|g,l>$ and $|g,l+\frac{1}{2}\rangle$ given in equation (\ref{coherentstates}) span vector spaces $V_{l}$ and $V_{l+\frac{1}{2}}$
respectively . Let $T^{j}_{m+}\footnote{Similraly these $T^{j}_{m+}$ are defined by
$\vec{J}^2T^j_{m+}=j(j+1)T^j_{m+}$ and $J_3T^j_{m+}=mT^j_{m+}$ where
$\vec{J}=\vec{L}^L-\vec{L}^R+\frac{\vec{\sigma}}{2}$ and $j=2l+\frac{1}{2},...,\frac{1}{2}$ . The index $+$ , on
the other hand , is simply denoting the fact that $T^{j}_{m+}$ is a linear operator from
$V_{l+1/2}{\longrightarrow}V_l$.}{\in}Hom(V_{l+1/2},V_l)$ be a linear operator from $V_{l+1/2}$ to $V_l$ . It is a
$(2l+1){\times}(2(l+\frac{1}{2})+1)$ matrix in a basis of $V_{l+\frac{1}{2}}$ and $V_{l}$ . This operator has the
transformation property
$<g,l|T^{j}_{m+}|g,l+\frac{1}{2}>{\longrightarrow}e^{i\frac{\theta}{2}}<g,l|T^{j}_{m+}|g,l+\frac{1}{2}>$ under
$g{\longrightarrow}ge^{i\frac{{\sigma}_3}{2}{\theta}}$ . One also has the transformation property

\begin{equation}
U^{(l)}(g)^{\dagger}T^{j}_{m+}U^{(l+\frac{1}{2})}(g)=\sum_{m^{'}}D^{(j)}_{mm^{'}}(g) T^{j}_{m^{'}+},
\end{equation}
where now $j<2l+\frac{1}{2}$ since we have to project out the top mode $j=2l+\frac{1}{2}$ as explained in the last
section , and $D^{(j)}_{mm^{'}}(g)$ is $ =<j,m|D^{(j)}(g)|j,m^{'}>$. Then one can make the identification
\begin{equation}
<j,m|{\psi}_{+}^{(j)}(g)>\equiv D^{(j)}_{m,+\frac{1}{2}}(g)=\langle g,l|T^{j}_{m+}|g,l+\frac{1}{2}\rangle,
\end{equation}
since from equation (\ref{equivariance3}) one can easily see that
$<j,m|{\psi}_{+}^{(j)}(g)>{\longrightarrow}e^{i\frac{\theta}{2}}<j,m|{\psi}_{+}^{(j)}(g)>$ under
$g{\longrightarrow}ge^{i\frac{{\sigma}_3}{2}{\theta}}$ . In other words $T^{j}_{m+}$ is the fuzzy version of
$<j,m|{\psi}^{(j)}_{+}(g)>$ and will then be associated with the negative helicity part of the fuzzy wave function
[see equation (\ref{expansionspinors})].

For helicity $+$ we have to consider $T^{j}_{m-}{\in}Hom(V_l,V_{l+1/2})$, with
\begin{equation}
U^{(l+\frac{1}{2})}(g)^{\dagger}T^{j}_{m-}U^{(l)}(g)=\sum_{m^{'}}D^{(j)}_{mm^{'}}(g)T^{j}_{m^{'}-}.
\end{equation}
Of course now ,
\begin{equation}
<j,m|{\psi}_{-}^{(j)}(g)>\equiv D^{(j)}_{m,-\frac{1}{2}}(g)=<g,l+\frac{1}{2}|T^{j}_{m-}|g,l>,
\end{equation}
where both sides will acquire now a phase $exp(-i\frac{\theta}{2})$ under the right $U(1)$ action , namely under
$g{\longrightarrow}gexp(i\frac{{\sigma}_3}{2}{\theta})$ . $T^j_{m-}$ is then clearly the fuzzy version of
$<j,m|{\psi}^{(j)}_{-}(g)>$ .

We can restore spin parts to fuzzy wave functions. The spin wave functions for helicity $\pm$ are
$T^{\frac{1}{2}}_{m_{s}{\pm}}$ , where $m_s$ denotes the two components of the spinor . The positive chirality
spinors are defined by
\begin{equation}
<\frac{1}{2},m_s|{\psi}_{+}^{(\frac{1}{2})}(g)>\equiv D^{(\frac{1}{2})}_{m_s+\frac{1}{2}}(g)=
<g,l|T^{\frac{1}{2}}_{m_s+}|g,l+\frac{1}{2}>,
\end{equation}
while the negative chirality spinors are defined by
\begin{equation}
<\frac{1}{2},m_s|{\psi}_{-}^{(\frac{1}{2})}(g)>\equiv D^{(\frac{1}{2})}_{m_s-\frac{1}{2}}(g)=
<g,l+\frac{1}{2}|T^{\frac{1}{2}}_{m_s-}|g,l>.
\end{equation}
So the two components of the total fuzzy wave functions for helicity $\pm$ are
\begin{equation}
<\frac{1}{2},m_s|{\psi}^{\pm}_F>=\Big[\sum_{j,m} {\xi}^{j\pm}_{m} T^{j}_{m{\mp}}\Big]T^{\frac{1}{2}}_{m_s{\pm}},\ \
\ {\xi}^{j\pm}_{m}{\in}{\bf C},m_s=+\frac{1}{2},-\frac{1}{2}.
\end{equation}
This is the fuzzy version of equation (\ref{expansionspinors}) .

The Dirac operator ${\bf D}_{2g}$ is given by the truncated version of (\ref{action1}) :
\begin{eqnarray}
&&{\rho}\sum_{m_s}({\bf D}_{2g})_{m_s^{'}m_s}\{\sum_{j,m}\xi^{j+}_{m} T^{j}_{m-}T^{1/2}_{m_s+}+\sum_{j,m}\xi^{j-}_{m}T^{j}_{m+}T^{1/2}_{m_s-}\} =\nonumber\\
&&-\{\sum_{j,m}\xi^{j+}_{m}T^{j}_{m+}(J^{(j)}_{+})_{+1/2,-1/2}\} \{T^{1/2}_{m_s^{'}-}\}
-\{\sum_{j,m}\xi^{j-}_{m}T^{j}_{m-}(J^{(j)}_{-})_{-1/2,+1/2}\}
\{T^{1/2}_{m_s^{'}+}\},\nonumber\\
&&
\end{eqnarray}
$J^{(j)}_{i}$ being the angular momentum $j$ images of $\frac{{\sigma}_{i}}{2}$. Similarly fuzzy sections are
given by
\begin{equation}
{\psi}^{D-K}_F=\left[
                            \begin{array}{c}
                  \sum_{j,m}\xi^{j+}_mT^j_{m-}\\
                  \sum_{j,m}\xi^{j-}_m T^j_{m+}
                              \end{array}
                            \right].
\end{equation}
This is the fuzzy analogue of (\ref{kahler}) .
\subsubsection{Organization of the Thesis}
This thesis is essentially based on \cite{12} , the first two papers of \cite{13} and the reference \cite{badis} .
It is organized as follows : chapter two is a short , but fairly technical introduction to the subject of noncommutative
geometry , which can be skipped in a first reading . For us , the most pertinent results of this chapter are Connes's trace theorem , the stability of cyclic cocycle
and the definition of cyclic cohomology.

Chapter $3$ is devoted to the construction of the K-cycle $({\bf A},{\bf H},D,{\Gamma})$ describing the fuzzy
sphere ${\bf S}^2_F$ and its fuzzy geometry . The core of the thesis is chapter $4$ which contains the formulation
of several physical problems on ${\bf S}^2_F$ . First one construct fuzyy monopoles  with the right charges . Fuzzy
solitons and ${\sigma}-$ models are also written down on ${\bf S}^2_F$ with the correct winding numbers . Their
dynamics and continuum limits are also discussed . Finally chiral fermions are elegantly defined on fuzzy ${\bf
S}^2_F$ with no doubling .

Chapter $5$ deals with ${\bf C}{\bf P}^2$ , its fuzzification and its Dirac operator. As ${\bf C}{\bf P}^2$ is not
a spin manifold but a spin$_c$ manifold , fermions are not symmetric between left and right , neverthless they can
be defined both in the continuum and in the fuzzy case. This chapter contains also the fuzzification of sections of
vector bundles using the star product and coherent states representations. This gives us an alternative approach
to desribe fermions , monopoles and solitons on ${\bf S}^2_F$ and ${\bf C}{\bf P}^2_F$ . The approach used in
chapter $4$ was based on projective modules and K-theory .

We conclude the thesis in the last chapter by some general remarks concerning fuzzy spaces , noncommutative
geometry and fuzzy quantum field theories .

%

\chapter{A Few Elements of NCG}
This chapter contains a very brief introduction to the subject of noncommutative geometry (NCG) . In particular ,
many concepts of NCG will be built by comparison with those of ordinary differential geometry through working out
explicitly several examples . This introduction is far from being complete and the level of rigour is also not that
of \cite{1} . However in writing it , extensive use of \cite{1,2,3.1,3.4,4} has been made. Our central aim is to Introduce the technology of NCG without actually going too much into the technical details . The
other goal is to develop some necessary tools which will be crucially used in the bulk of the thesis . Examples of
such tools are the Dixmier trace , K-cycles , Dirac operator , cyclic cohomology and Fredholm modules . The most
important results of this chapter are : the explicit proof of Connes trace theorem which allows us to directly
verify , in the case of Riemannian manifolds , the axiom that the Dirac operator gives the metric . The
other important result is the stability of cyclic cocycle which allows us to define
topological numbers for arbitrary spaces .

However , because those results of this chapter which will be actually used in the remaining chapters are fairly simple and elementary , the reader can skip it in a first reading and only consult it when needed in the bulk of the thesis .

\section{First Example : Quantum Mechanics a la Heisenberg }

Quantum Mechanics is the very first example in which the space , which is the phase space in this case ,
becomes noncommutative . This is obvious from the fact that the phase space coordinates $q$ and $p$ , in the
quantum theory , will satisfy the nontrivial commutation relation $[q,p]=i\hbar$ . Phase space acquires a cell-like
structure with minimum volume given roughly by $\hbar$. In this section we will rederive this result in an
algebraic form , i.e a form in which the noncommutativity is established at the level of the underlying algebra of
functions.

It is a textbook result that the classical atom can be characterized by a set of positive real numbers , ${\nu}_i$
, called the fundamental frequencies . This atom, if viewed as a classical system , radiates via its dipole moment
interaction until it collapses . The intensity of this radiation is given by
\begin{eqnarray}
I_n &\propto&|<\nu,n>|^4\nonumber\\
&where&\nonumber\\
<\nu,n>&=&\sum_in_i{\nu}_i,n_i{\in}Z.
\end{eqnarray}
It is clear that all possible emitted frequencies , $<\nu,n>$ , form a group ${\Gamma}$ under the addition
operation of real numbers
\begin{equation}
{\Gamma}=\{<n,\nu>;n_i{\in}Z\},
\end{equation}
indeed given two frequencies $<\nu,n>=\sum_in_i{\nu}_i$ and $<\nu,n^{'}>=\sum_in_i^{'}{\nu}_i$ in ${\Gamma}$, it is
obvious that $<\nu,n+n^{'}>=\sum_i(n_i+n_i^{'}){\nu}_i$ is also in ${\Gamma}$ .

The algebra of classical observables of this atom can be obtained as the convolution algebra of the the abelian
group $\Gamma$ . To see how this works exactly, one first recalls that any function on the phase space $X$ of this
atom can be expanded as ( an almost ) periodic series
\begin{equation}
f(q,p;t)=\sum_{n}f(q,p;n)e^{2{\pi}i<n,\nu>t} ; n\equiv(n_1,...,n_k) .
\end{equation}
The Fourier coefficients $f(q,p;n)$ are labelled by the elements $n{\in}\Gamma$ . One can then check that the
convolution product defined by
\begin{equation}
f*g(q,p;t;n)=\sum_{n_1+n_2=n}f(q,p;t;n_1)g(q,p;t;n_2),\label{convolutionproduct}
\end{equation}
where $f(q,p;t;n)=f(q,p;n)exp(2{\pi}i<n,\nu>t)$ , leads to the ordinary commutative pointwise multiplication of
the corresponding functions $f(q,p;t)$ and $g(q,p;t)$, namely
\begin{equation}
fg(q,p;t)\equiv f(q,p;t)g(q,p;t)=\sum_{n}f_1*f_2(q,p;t;n).
\end{equation}
The key property leading to this result is the fact that ${\Gamma}$ is an abelian group.

If we take experimental facts in our account then we know that the atom must obey the Ritz-Rydberg combination
principle which says that

a)Rays in the spectrum are labeled with two indices .

b)Frequencies of these rays obey the law of composition

\begin{eqnarray}
{\nu}_{ij}&=&{\nu}_{ik}+{\nu}_{kj}\nonumber\\
\end{eqnarray}
which we write as
\begin{eqnarray}
(i,j)&=&(i,k)\circ(k,j).
\end{eqnarray}
The emitted frequencies ${\nu}_{ij}$ are therefore not parametrized by the group $\Gamma$ but rather by the
groupoid $\Delta$ of all pairs $(i,j)$ . It is a groupoid since not all frequencies can be composed to give
another allowed frequency , every element $(i,j)$ has an inverse $(j,i)$ , and $\circ$ is associative.

The quantum algebra of observables is then the convolution algebra of the groupoid $\Delta$, and it turns out to
be a noncommutative (matrix) algebra as one can see by rewriting (\ref{convolutionproduct}) in the form
\begin{equation}
F_1F_{2(i,j)}=\sum_{(i,k)\circ(k,j)=(i,j)}F_{1(i,k)}F_{2(k,j)}.
\end{equation}
One can easily check that $F_1F_2{\neq}F_2F_1$ , so $F's$ fail to commute .

To implement the element of the quantum algebra as matrices one should replace $
f(q,p;t;n)=f(q,p;n)e^{2{\pi}i<n,\nu>t}$ by
\begin{equation}
F(Q,P;t)_{(i,j)}=F(Q,P)_{(i,j)}e^{2{\pi}i{\nu}_{ij}t}.
\end{equation}
From here , Heisenberg's equation of motion , phase space canonical commutation relations and Heisenberg's
uncertainty relations follow in the usual way .
\section{Compact Operators as Noncommutative Infinitesimals}

It is a set of deep results due to Gelfand , Naimark , Connes and others that all the properties of a space $X$, namely its topology , measure
theory , De Rham theory, K-theory,... can be coded in the algebra of functions $C(X)$ on this space . 

Connes's noncommutative geometry (NCG) is a very precise construction in which the above basic theorem is explicitly
implemented , not only for smooth differentiable manifolds , but also for general spaces . One fruitful way of
introducing NCG is by stating axioms of diferential geometry in a form suitable for generalization . Differential
geometry will appear therefore as only a very special case of NCG .

Before we state the axioms , we need first to write down the spectral or quantized calculus , which is a
generalization of the usual calculus on manifolds. It can be summarized in the following table
\begin{eqnarray}
~Complex ~variables &{\longrightarrow}&~Operators ~on ~a
~Hilbert ~space\nonumber\\
~Real ~variables&\longrightarrow& ~Selfadjoint ~operators\nonumber\\
~Infinitesimal &\longrightarrow& ~Compact
~operators\nonumber\\
~Integral &\longrightarrow& ~Dixmier ~trace\nonumber\\
&&
\end{eqnarray}
The first two lines are essentially borrowed from QM , whereas the third line will be explained in this section .
Next section will be devoted to the last line .

{{\bf{Definition1}}}

An operator $T$ on a Hilbert space $H$ is said to be compact if it can be approximated in norm by finite rank
operators . More precisely

\begin{eqnarray}
&{\forall}&{\epsilon}{>}0 , \exists ~a ~finite ~dimensional ~space
~E{\in}H : ||T_{E^{\bot}}||<\epsilon.\nonumber\\
\end{eqnarray}
With this definition , it is clear that compact operators are in a sense small .

{{\bf{Definition2}}}

One can alternatively define compact operators as follows : They admit a uniformly convergent (in norm) expansion
of the form
\begin{equation}
T=\sum_{n{\geq}0}{\mu}_n(T)|{\psi}_n><{\phi}_n|
\end{equation}
where $ 0{\leq}{\mu}_{i+1}(T){\leq}{\mu}_{i}(T)$ , $ \{|{\psi}_n>\}and\{|{\phi}_n>\} $ are orthonormal (not necessarily
complete) sets . One can now make the following remarks : $1)$The size of the compact operator $T$ (infinitesimal)
is governed by the rate of decay of the sequence $\{{\mu}_n(T)\}$ , as $ n{\longrightarrow}{\infty}$ . $2)$If we
polar decompose , $T=U|T|$ where $|T|=\sqrt{T^{*}T}$ and $U$ is the phase of $T$ , then one can show that the
characteristic values of $T$ , ${\mu}_n(T)$ , are basically the eigenvalues of $|T|$ with eigenvectors
$|{\phi}_n>$.

The characteristic values , ${\mu}_n(T)$ , satisfy

\begin{eqnarray}
{\mu}_{n+m}(T_1+T_2)&{\leq}&{\mu}_n(T_1)+{\mu}_m(T_2)\nonumber\\
{\mu}_{n+m}(T_1T_2)&{\leq}&{\mu}_n(T_1){\mu}_m(T_2)\nonumber\\
{\mu}_n(TT_1)&{\leq}&||T||{\mu}_n(T_1)\nonumber\\
{\mu}_n(T_1T)&{\leq}&||T||{\mu}_n(T_1).\label{characteristicvalues}
\end{eqnarray}
where $T_1$ , $T_2$ are compact operators and $T$ is a bounded operator . For $n=m=0$, the above inequalities can
be shown by using the fact that
\begin{eqnarray}
{\mu}_0(T)&=&sup\{||T|\chi>||:|\chi>{\in}H,|||\chi>||{\leq}1\}\nonumber\\
&=&||T||\equiv ~the ~operator ~norm ,
\end{eqnarray}
which trivially satisfies all of (\ref{characteristicvalues}) . One can also show that ${\mu}_n(T)$ , $n{\neq}0$, behaves as a norm as
follows . First let ${\cal L}(H)$ be the set of all bounded operators on $H$ and $R_n$ the set of all operators
with rank less than $n$ , i.e $R_{n}=\{S{\in}{\cal L}(H):dim(ImS){\leq}n\}$. Then from the above first definition
of compact operators , one can write
\begin{eqnarray}
{\mu}_{n}(T)&=&dist(T,R_n),\forall n{\in}N\nonumber\\
&with&\nonumber\\
Lim{\mu}_n(T)&{\longrightarrow}&0~when~n{\longrightarrow}{\infty}.
\end{eqnarray}
This is another way of writing that the compact operator $T$ is a norm limit of operators with finite rank . From
this definition and the obvious inclusions $R_{n}+R_{m}\subset R_{n+m}$ the inequalities (\ref{characteristicvalues}) follow easily .
In showing (\ref{characteristicvalues}) we need also to use the fact that the set of compact operators forms a two-sided ideal in
${\cal L}(H)$, i.e $R_n{\cal L}(H)={\cal L}(H)R_n=R_n$. This is obvious since compact operators among bounded
operators are like infinitesimal numbers among numbers .

{\bf{Definition 3}: Order of a Compact Operator}

A compact operator $T$  is of order ${\alpha}{\in}R^{+}$ iff
\begin{eqnarray}
&\exists& C <{\infty} : {\mu}_n(T){\leq}Cn^{-\alpha} ,
{\forall}n{\geq}1\nonumber\\
&\Leftrightarrow&{\mu}_{n}=O(n^{-\alpha}),n{\longrightarrow}{\infty}.
\end{eqnarray}

{\bf{Example $1$}}

Let us check that some of the intuitive rules of calculus of infinitesimals are still valid for compact operators
. For example if $T_1$ , $T_2$ are of orders ${\alpha}$, ${\beta}$ then $T_1T_2$ is of order ${\alpha}+{\beta}$ .
we start with
\begin{equation}
{\mu}_{n+m}(T_1T_2){\leq}{\mu}_{n}(T_1){\mu}_{m}(T_2).
\end{equation}
But ${\mu}_{n}(T_1)=O(n^{-{\alpha}})$ , ${\mu}_{m}(T_2)=O(m^{-{\beta}})$ and ${\mu}_{p}(T_1T_2)=O(p^{-{\gamma}})$
. ${\alpha}$ , $\beta$ and ${\gamma}$ are the orders of $T_1$ , $T_2$ and $T_1T_2$ respectively . Then

\begin{eqnarray}
O((n+m)^{-{\gamma}})&{\leq}&O(n^{-{\alpha}})O(m^{-{\beta}})\nonumber\\
O(n^{{\alpha}})O(m^{{\beta}})&{\leq}&O((n+m)^{{\gamma}})\nonumber\\
&{\Longrightarrow}&\nonumber\\
O(n^{{\alpha}+{\beta}-{\gamma}})&{\leq}&1.
\end{eqnarray}
Where we have assumed that $n=m$ . One then concludes that ${\gamma}={\alpha}+{\beta}$.

{{\bf{Example $2$}}}

The volume $d^dx$ in $d$ dimensions is an infinitesimal of order $1$ and therefore the differential $dx$ is of order
$1/d$\footnote{This is because by definition the integral has as a domain the set of compact operators of order $1$ , in other words $d^dxf(x)$ must be a compact operator of order $1$ and therefore $d^dx$ is a compact operator of order $1$ . Remember that a bounded operator $f(x)$ times a compact operator $d^dx$ is still a compact operator of the same order. } .

{\bf{Example $3$}}

On a $d$-dimensional manifold $M$ , the Dirac operator $D=D^{+}=|D|$ has the eigenvalues (Weyl formula)
\begin{equation}
{\mu}_j(D)\simeq2{\pi}(\frac{d}{{\Omega}_d volM})^{1/d}j^{1/d}
\end{equation}
for large $j$ . So $D^{-1}$ is infintesimal of order $1/d$ and therefore $D^{-d}$ is an infinitesimal of order $1$ .

\section{Dixmier Trace as the Noncommutative Integral}

One starts with the usual trace , which has as a domain the space ${\cal L}^1$ of trace class operators . Let
$T{\in}{\cal L}^1$ be a positive and compact operator of order $1$, then one can compute
\begin{eqnarray}
{\sigma}_N(T)\equiv TrT|_{N}=\sum_{n=0}^{N-1}{\mu}_n(T){\leq}ClnN+\frac{C^{'}}{N}+~finite~terms.
\end{eqnarray}
In other words the ordinary trace is at most logarithmically divergent and should be replaced by
\begin{eqnarray}
&&TrT{\longrightarrow}Lim_{N{\longrightarrow}{\infty}}{\gamma}_N(T)\nonumber\\
&&where\nonumber\\
&&{\gamma}_N(T)=\frac{{\sigma}_N(T)}{lnN}=\frac{1}{lnN}\sum_{n=0}^{N-1}{\mu}_{n}(T).\label{dixmiertrace1}
\end{eqnarray}
The sequence $\{{\gamma}_N(T)\}$ satisfies
\begin{eqnarray}
{\gamma}_N(T_1+T_2)&{\leq}&{\gamma}_{N}(T_1)+{\gamma}_N(T_2){\leq}{\gamma}_{2N}(T_1+T_2)(1+\frac{ln2}{lnN}).\nonumber\\
&&
\end{eqnarray}
We can see immediately That ${\gamma}_N$ is not linear , and that linearity will be recovered if the sequence
$\{{\gamma}_N\}$ converges . One needs then to replace (\ref{dixmiertrace1}) by something else , namely

\begin{equation}
Tr_{\omega}(T)=Lim_{\omega}{\gamma}_N(T).
\end{equation}
This is the Dixmier trace . $Lim_{\omega}$ is a linear form on the space of bounded sequences $\{{\gamma}_N\}$ .
It is positive , linear , scale invariant and it converges to the ordinary limit if the sequence on which it is
evaluated converges. Explicitly, it satisfies:
\begin{eqnarray}
Tr_{\omega}(T)&{\geq}&0 \nonumber\\
Tr_{\omega}({\lambda}_1T_1
+{\lambda}_2T_2)&=&{\lambda}_1Tr_{\omega}T_1+{\lambda}_2Tr_{\omega}T_2\nonumber\\
Tr_{\omega}(BT)&=&Tr_{\omega}(TB),B ~is ~a ~bounded ~operator\nonumber\\
Tr_{\omega}(T)&=&0 , ~if ~T ~is ~of ~order ~higher ~than
~1.\nonumber\\
\end{eqnarray}
The $4th$ equation means that infinitesimals of order $1$ are in the domain of the Dixmier trace , while those of
order higher than $1$ have vanishing trace . The proof is pretty obvious from the above construction.

The Dixmier trace can be extended to the whole space ${\cal L}^{1,\infty}$ , the space of trace class compact
operators of order $1$ , because of the linearity of the trace and the fact that ${\cal L}^{1,\infty}$ is generated
by its positive part .

{\bf{Example $4$}}

The Laplacian on a $d$-dimensional torus $T^d=R^d/(2{\pi}Z)^d$ (and its eigenvalues) is (are)
\begin{eqnarray}
{\Delta}&=&-(\frac{\partial}{{\partial}x^1})^2-.....-(\frac{\partial}{{\partial}x^d})^2\nonumber\\
\vec{p}^2&=&p_1^2+...+p_d^2 ,
\end{eqnarray}
One would like to compute $Tr_{\omega}{\Delta}^{-d/2}$ . The eigenvalues of ${\Delta}^{-d/2}$ are
${\mu}_p({\Delta}^{-d/2})=|\vec{p}|^{-d}$ . The multiplicity of this eigenvalue is the number of points in $Z^d$
of length $|\vec{p}|$ which is proportional to the volume
\begin{equation}
N_{p+dp}-N_{p}={\Omega}_dp^{d-1}dp
\end{equation}
${\Omega}_d$ is a $d-1$ dimensional sphere . Therefore
\begin{eqnarray}
\frac{1}{lnN}\sum_{n=0}^{N-1}{\mu}_n(T)&=&\frac{1}{lnN_{k}}\sum_{p{\leq}k}p^{-d}\nonumber\\
&\sim&\frac{1}{dlnk}\int_{1}^{k}p^{-d}({\Omega}_dp^{d-1}dp)\nonumber\\
&\sim&\frac{{\Omega}_d}{d}\nonumber\\
&{\Longrightarrow}&\nonumber\\
Tr_{\omega}({\Delta}^{-d/2})&=&\frac{{\Omega}_d}{d}.
\end{eqnarray}
Since on the torus , $|D|^2={\Delta}$ , this result can be written as
\begin{equation}
Tr_{\omega}({|D|}^{-d})=\frac{{\Omega}_d}{d}
\end{equation}
The Dirac operator here seems to play the role of the metric .

\section{Spectral Triples or K-cycles as Noncommutative Spaces}

An arbitrary space $X$ can be always defined by a set $(A,H,D)$ where $A$ is an involutive algebra of bounded
operators on the Hilbert space $H$ , and $D=D^{+}$ is an operator acting on $H$ with the properties
\begin{eqnarray}
&D^{-1}& ~is ~a ~compact ~operator ~on ~H_{\bot}\nonumber\\
&[D,a]& ~is ~a ~bounded ~operator ~for ~any ~a{\in}A.\nonumber\\
&&\label{propertiesofD}
\end{eqnarray}
$H_{\bot}$ is the orthogonal complement of the finite dimensional kernel of $D$ .

The above space is compact in the sense that the spectrum of $D$ is by construction discrete with finite
multiplicity . A noncompact space will be obtained if the algebra $A$ has no unit , more precisely ,
we need to replace the first line in (\ref{propertiesofD}) by the following condition : For any $a{\in}A$ and ${\lambda}$ not
in $R$ , $a(D-{\lambda})^{-1}$ is a compact operator . $(A,H,D)$ is called the spectral triple or K-cycle and
contains everything that is to know about our space .

\subsubsection{States as Noncommutative Points}

A point in the above K-cycle , or noncommutative space , is a state on the $C^{*}$ algebra ${A}$ , in other words a
linear functional
\begin{eqnarray}
{\psi}&:&A{\longrightarrow}C\nonumber\\
&where&\nonumber\\
{\psi}(a^*a)&{\geq}&0,{\forall}a{\in}A\nonumber\\
||{\psi}||&=&sup\{|{\psi}(a)|:||a||{\leq}1\}.
\end{eqnarray}
One can check that $||\psi||={\psi}(1)=1$ . The set ${\cal S}(A)$ of all states is a convex space, in other words
: given any two states ${\psi}_1$ and ${\psi}_2$ and a real number $0{\leq}{\lambda}{\leq}1$ then
${\lambda}{\psi}_1+(1-{\lambda}){\psi}_2~is~{\in}{\cal S}(A)$ . The boundary of ${\cal S}(A)$ is generated by pure
states .
\subsubsection{Even Spectral Triple}

The spectral triple $X=(A,H,D)$ is said to be even (otherwise it is said to be odd) if there is a $Z_2$ grading
${\Gamma}$ of $H$ , satisfying
\begin{eqnarray}
{\Gamma}^2&=&1,\nonumber\\
{\Gamma}^{+}&=&{\Gamma}\nonumber\\
\{{\Gamma},D\}&=&[{\Gamma},a]=0,\forall a{\in}A\nonumber\\
&&
\end{eqnarray}

\subsubsection{The Real Structure $J$ as the Noncommutative CP operation}

The spectral triple $X=(A,H,D)$ is said to be  real (otherwise it is said to be complex) if there is an antilinear
isometry , $J : H{\longrightarrow}H$ , which satisfy the following
\begin{eqnarray}
J^2&=&{\epsilon}(d)1,\nonumber\\
JD&=&{\epsilon}^{'}(d)DJ ,\nonumber\\
J{\Gamma}&=&(i)^d{\Gamma}J\nonumber\\
J^{+}&=&J^{-1}={\epsilon}(d)J.\label{realstructure}
\end{eqnarray}
The mod $8$ periodic functions ${\epsilon}(d)$ and ${\epsilon}^{'}(d) $ are given by
\begin{eqnarray}
{\epsilon}(d)&=&(1,1,-1,-1,-1,-1,1,1)\nonumber\\
{\epsilon}^{'}(d)&=&(1,-1,1,1,1,-1,1,1).
\end{eqnarray}
If the space $X$ is a Riemannian spin manifold $M$ , then the real structure is exactly the $CP$ operation
\begin{equation}
J{\psi}=C{\bar{\psi}}.
\end{equation}
$C$ being the charge conjugation operator .
\section{The Dirac Operator as the Noncommutative Metric}
The Dirac operator of the K-cycle $X=(A,H,D)$ can be used to define a distance formula on the space ${\cal S}(A)$ :
the space of states on the algebra $A$ . Given two states ( points ) on $A$ ( of $X$ ) , ${\psi}_1$ and ${\psi}_2$
the distance between them is given by
\begin{eqnarray}
d({\psi}_1,{\psi}_2)&=&sup_{a{\in}A}\{|{\psi}_1(a)-{\psi}_2(a)|:||[D,a]||{\leq}1\},\nonumber\\
&&\label{distanceformula}
\end{eqnarray}
$D$ essentially contains all the metric informations of the space $X$ .

{\bf{{Example $5$}}}

Let us check that the distance formula (\ref{distanceformula}) will reduce to the ordinary distance when the space $X$ is an
ordinary manifold $M$ . In this case , $A=C^{\infty}(M)$ , $D={\gamma}^{\mu}{\partial}_{\mu}$ . The space of
states ${\cal S}(C^{\infty}(M))$ is now the space of characters $M(C^{\infty})$ which can be identified with the
manifold itself as follows
\begin{eqnarray}
x{\in}M&{\longrightarrow}&{\psi}_x{\in}M(C^{\infty})\nonumber\\
&such ~that&\nonumber\\
{\psi}_x(f)&=&f(x),\forall f{\in}C^{\infty}(M)\nonumber\\
&&
\end{eqnarray}
The distance (\ref{distanceformula}) takes then the form
\begin{eqnarray}
d(x_1,x_2)&=&sup_{f{\in}C^{\infty}(M)}\{|f(x_1)-f(x_2)|:||[D,f]||{\leq}1\}.\nonumber\\
&&
\end{eqnarray}
Next since $[D,f]={\gamma}^{\mu}{\partial}_{\mu}f$ one has $||[D,f]||=sup_{x{\in}M}||\vec{\partial}f||$ , and hence

\begin{eqnarray}
|f(x_2)-f(x_1)|&=&|\int_{x_1}^{x_2}\vec{\partial}f.d\vec{x}|\nonumber\\
&{\leq}&\int_{x_1}^{x_2}|\vec{\partial}f.d\vec{x}|\nonumber\\
&{\leq}&\int_{x_1}^{x_2}|\vec{\partial}f|ds\nonumber\\
&{\leq}&\int_{x_1}^{x_2}||[D,f]||ds\nonumber\\
&{\leq}&\int_{x_1}^{x_2}ds.\nonumber\\
&{\Longrightarrow}&\nonumber\\
d(x_1,x_2)&=&Inf(\int_{x_1}^{x_2}ds).\label{distance}
\end{eqnarray}
In the above proof we have assumed for simplicity that the functions , $f{\in}C^{\infty}(M)$, are real valued  .
However the result (\ref{distance}) will also hold if $f$'s are complex valued functions on $M$ . The only difference is
that one finds now that the norm of the bounded operator $[D,f]$ is equal to the Lipschitz norm of $f$ ,
\begin{equation}
||[D,f]||=||f||_{Lip}=sup_{x_1{\neq}x_2}\frac{|f(x_1)-f(x_2)|}{Inf ( \int_{x_1}^{x_2}ds)}.
\end{equation}
\subsection{Example $6$ : Connes Trace Theorem}

Before we state the first axiom of NCG , one needs to do one more computation in which one sees once more that the
Dirac operator $D=i{\gamma}^{\mu}{\partial}_{\mu}$ of a $d$ dimensional spin manifold is intimately related to the
metric . More precisely, one would like to show that the Riemannian measure on $M$ is given by
\begin{equation}
Tr_{\omega}f|D|^{-d}=\int_{M} f(x) \sqrt{detg(x)}dx^1{\wedge}dx^2{\wedge}..{\wedge}dx^{d}.
\end{equation}
The first step is to recognize that the Dirac operator $D$ is a first order elliptic pseudodifferential operator .
The statement that $D$ is a first order is trivial as one can see from its expression . Being pseudodifferential
operator means that , it is an operator between two Hilbert spaces $H_1$ and $H_2$ of sections of Hermitian vector
bundles over $M$ which can be written in local coordinates as
\begin{eqnarray}
D{\psi}(x)&=&\frac{1}{(2{\pi})^d}\int
e^{ip(x-y)}a(x,p){\psi}(y)d^dyd^dp\nonumber\\
&where&\nonumber\\
a(x,p)&=&-p_{\mu}{\gamma}^{\mu}.
\end{eqnarray}
In this case $H_1=H_2=L^2(M,S)$ : the Hilbert space of square integrable sections of the irreducible spinor bundle
over $M$  . $a(x,p)$ is called the principal symbol of the operator $D$ and since it is invertible for $p{\neq}0$
, the Dirac operator is called elliptic .

It is not difficult to check that the principal symbol of the second order operator, $D^2$, will be given by
$p^2{\bf 1}={\eta}^{{\mu}{\nu}}p_{\mu}p_{\nu}{\bf 1}$ where we have assumed that we are in a locally flat metric (
which can always be done ). One can then compute the principal symbol of $D^{-2}$ as follows
\begin{eqnarray}
D^{-2}{\psi}(y)&=&\frac{1}{(2{\pi})^d}\int
e^{ip(y-x)}a(x,p){\psi}(x)d^dxd^dp\nonumber\\
&where ~now&\nonumber\\
a(x,p)&=&\frac{1}{|p|^2}{\bf 1}.
\end{eqnarray}
This operator is of order $-2$ .  From here one can directly conclude that the operator $|D|^{-d}=(D^2)^{-d/2}$ is
also a pseudodifferential operator of order $-d$ and its principal symbol is given by $|p|^{-d}{\bf 1}$ .

Now given any $f{\in}C^{\infty}(M)$ , it will act as a bounded multiplicative operator on the Hilbert space , and
therefore the operator $f|D|^{-d}$ will be also a pseudodifferential operator of order $-d$ , with a principal
symbol given by $f(x)|p|^{-d}{\bf 1}$ .

The identity $\bf 1$ which appears above is an $N{\times}N$ unit matrix which acts on the spinor bundle , so
$N=2^{d/2}$ for even dimensional manifolds and $N=2^{(d-1)/2}$ for odd dimensional manifolds .

The next step is to use the famous Connes trace theorem , which asserts that the Dixmier trace of a
pseudodifferential operator (of order $-d$ ) over a $d$ dimensional Riemannian manifold is proportional to the
Wodzicki residue of that operator , more precisely
\begin{eqnarray}
Tr_{\omega}A&=&\frac{1}{d(2{\pi})^d}WresA\nonumber\\
&=&\frac{1}{d(2{\pi})^d}\int_{S^*M}Tr[a(x,p)]{\sigma}_pdx^{1}{\wedge}dx^{2}...{\wedge}dx^{d}.\nonumber\\
&where&\nonumber\\
{\sigma}_p&=&\sum_{j=1}^{d}(-1)^{j-1}p_jdp_1{\wedge}..{\wedge}{\hat{dp_j}}{\wedge}...{\wedge}dp_{d}.
\nonumber\\
&and&\nonumber\\
S^*M&=&\{(x,p){\in}T^*M:|p|=1\}.
\end{eqnarray}
Therefore it is straight forward to see that
\begin{eqnarray}
Tr_{\omega}f|D|^{-d}&=&\frac{1}{d(2{\pi})^d}Wresf|D|^{-d}\nonumber\\
&=&\frac{1}{d(2{\pi})^d}\int_{S^*M}Tr[f(x)|p|^{-d}{\bf 1}]{\sigma}_pdx^{1}{\wedge}dx^{2}...{\wedge}dx^{d}.\nonumber\\
&=&\frac{N}{d(2{\pi})^d}\int_{S^{d-1}}{\sigma}_p\int_{M}f(x)dx^{1}{\wedge}dx^{2}...{\wedge}dx^{d}.\nonumber\\
&=&\frac{N{\Omega}_{d-1}}{d(2{\pi})^d}\int_{M}f(x)dx^{1}{\wedge}dx^{2}...{\wedge}dx^{d}.
\end{eqnarray}

\section{Axioms of NCG}

{\subsubsection{Axiom $1$ : Dimension}}

We are now ready to state the first axiom of NCG. It is written as

\begin{eqnarray}
|D|^{-2}&=&\sum_{{\mu},{\nu}=1}^{d}[F,X^{\mu}]^{*}{\eta}_{{\mu}{\nu}}[F,X^{\nu}],\nonumber\\
\end{eqnarray}
$F$ is the sign of the Dirac operator , $F=\frac{D}{|D|}$ , and it is assumed to satisfy $[F,|D|^{-2}]=0$ .
$X^{\mu}$ are the generators of $A$ while ${\eta}=({\eta}_{{\mu}{\nu}})$ is in $M_d(A)$ , i.e $d{\times}d$ matrices with entries in
$A$. The positive compact operator $|D|^{-2}$ can be thought of as the square of the infinitesimal length element
over the space $X=(A,H,D)$.

Usually this axiom is formulated as follows : $D^{-1}$ is a compact operator of order $1/d$ , where $d$ will be by
definition the dimension of the above space (K-cycle) $X\equiv(A,H,D)$ . This is called the Dimension Axiom .

{\bf Example $7$}

Let us compute the dimension of $S^2$ from its Dirac operator . The Dirac operator on $S^2$ is known to have the
form
\begin{equation}
D=\vec{{\sigma}}.\vec{L}+1
\end{equation}
Its square , $D^2$ , has the spectrum
\begin{equation}
k^2=(j+\frac{1}{2})^2 , j=l+\frac{1}{2} ~or ~j=l-\frac{1}{2}.
\end{equation}
$k^2=(j+\frac{1}{2})^2$ is an eigenvalue of $D^2$ with a total multiplicity equal to :
$4(j+\frac{1}{2})=4k$. One can then compute
\begin{eqnarray}
Tr_{\omega}|D|^{-2}&=&Lim_{\omega}\frac{1}{LnN_{M}}\sum_{k=1}^{M}\frac{1}{k^2}{\times}{4k}\nonumber\\
&=&Lim_{\omega}\frac{4}{lnN_M}ln M.
\end{eqnarray}
But , $N_M=\sum_{k=1}^{M}4k=2M(M+1)$ , and therefore
\begin{equation}
Tr_{\omega}|D|^{-2}=2\label{euler}
\end{equation}
This equation means that $|D|^{-2}$ is in the domain of the Dixmier trace and therefore it is a compact operator
of order $1$ . $D^{-1}$ is then a compact operator of order  $1/2$ which leads to the conclusion that the
dimension of $S^2$ is $2$ . (\ref{euler}) is exactly the Euler character of $S^2$ .

\subsubsection{Axiom $2$ : Reality}

One can use the real structure $J$ , introduced in equation (\ref{realstructure}) , to define the opposite algebra $A^0$ . Its
elements $a^0$ are defined by
\begin{equation}
a^0=Ja^{*}J^{+}
\end{equation}
with product
\begin{equation}
(ba)^0=a^0b^0.
\end{equation}
The two algebras $A$ and $A^0$ are also required to commute with each other , i.e
\begin{equation}
[a,b^0]=0 , ~\forall a , b{\in}A .\label{reality}
\end{equation}
Equation (\ref{reality}) is the reality axiom of noncommutative geometry .

\subsubsection{Axiom $3$ : First Order}

The real structure $J$ and the Dirac operator $D$ should also satisfy one more condition known as the first order
axiom of noncommutative geometry. For all $a$ and $b{\in}A$ , one must have
\begin{equation}
[[D,a],b^0]=0\label{firstorder}
\end{equation}
{\bf Example $8$}

In the case of ordinary commutative manifold $M$ with a Dirac operator $D=i{\gamma}^{\mu}{\partial}_{\mu}$ , the
opposite algebra $A^{0}$ coincides with the algebra $A$ itself , since $ (ba)^0=a^0b^0=b^0a^0 $ , and therefore
the condition (\ref{firstorder}) will simply mean that the Dirac operator is a first order operator .

\subsubsection{Axiom $4$ : Regularity}

For all $a{\in}A$ , the operator $[D,a]$ is a bounded operator on the Hilbert space $H$, and both $a$ and $[D,a]$
are in the domain of ${\delta}^m$ for all integers $m$ . ${\delta}$ being the derivation defined by :
${\delta}(a)=[|D|,a]$. This is the algebraic formulation of the smoothness of the elements of $A$.

There are three more axioms regarding orientability , finiteness of the K-cycle and Poincare duality and K-theory , whose discussions will take us out of the scope of this thesis , so we stop here .

The subject of the next chapter will be the construction of noncommutative differential calculus .
\section{Fredholm Module}
\subsection{Definition}

Given the K-cycle $X=(A,H,D)$ , One way to introduce a Fredholm module structure is by defining the operator
\begin{equation}
F=\left(
\begin{array}{cc}
0 & D^{-1} \\
D & 0
\end{array}
\right),\label{fredholmmodule}
\end{equation}
on $H_2=H{\oplus}H$ . By construction (\ref{fredholmmodule}) is such that $ F^{2}=1$ . Let $ {\pi} $ be an involutive
representation of the algebra $A$ on the Hilbert space $H_2=H{\oplus}H $ given by :

\begin{equation}
{\forall}  f  {\in}A  : {\pi}(f) = \left(
\begin{array}{cc}
f & 0 \\
0 & f
\end{array}
\right).\label{representationh2}
\end{equation}
The exterior derivative $ df $ of any element $f$ of $A$ is defined by $df=i[F,{\pi}(f)]$ or

\begin{equation}
df=i\left(
\begin{array}{cc}
0       & {\lbrack}D^{-1},f{\rbrack} \\
{\lbrack}D,f{\rbrack} & 0
\end{array}
\right).\label{exterior1}
\end{equation}
We will assume that $[F,{\pi}(f)]$ is a compact operator on $H_2$ for any $f$ in $A$: in other words $df$ is an
infintesimal variable . we will also assume that $ {\pi}(f)( F - F^{+} ) $ is a compact operator. The pair
$(H_2,F)$ define a Fredholm module , it is an even Fredholm module if the Hilbert space $H$ admits a $ Z/2 $
grading $ {\Gamma} $ . On $H_2$ the chirality operator is therefore given by

\begin{equation}
{\Gamma}_2=\left(
\begin{array}{cc}
{\Gamma} & 0 \\
0 & {\Gamma}
\end{array}
\right).\label{chiralityh2}
\end{equation}
By construction we have $ {\Gamma}_2^{2} =1 $ , $ {\Gamma}^{+}_2={\Gamma}_2$ ,$ {\Gamma}_2F=-F{\Gamma}_2$ and $
{\Gamma}_2{\pi}(f) = {\pi}(f){\Gamma}_2 $ for any element $f$ of the algebra $A$ .

\subsection{Cyclic Cohomology}

\subsubsection{The Schatten-Von Neumann classes}

The Schatten-Von Neumann ideal of compact operators $ {\cal L}^{p} $ ( where $p$  is a real number $ {\geq} 1 $ )
is defined as the space of all bounded operators $T$ on $H_2$ such that the trace of $ |T|^{p} $ is finite , in
other words :

\begin{equation}
\sum_{n=0}^{\infty} ( {\mu}_{n}(T) )^{p} < {\infty},
\end{equation}
where $ {\mu}_{n}(T) $ is the $ nth $ eigenvalue of $|T|$. The above condition simply means that the eigenvalues of
$T$ must decrease fast enough at infinity .These classes are used to measure the size of the differential $ [
F,{\pi}(f) ] $ .

One last remark is that ${\cal L}^{p} {\subset} {\cal L}^{q}$  if  $p {\leq}q$  which can be written as :

\begin{equation}
{\cal L}^{1}{\subset}{\cal L}^{2}{\subset}...{\subset}{\cal L}^{p}{\subset}...{\subset}{\cal L}^{\infty},
\end{equation}
where $ {\cal L}^{1} $ are trace-class operators , $ {\cal L}^{2} $ are Hilbert-Schmidt operators and $ {\cal
L}^{\infty} $ are the compact operators .

\subsubsection{p-summable Fredholm module}

A Fredholm module $ (H_2 , F )$ is called  p-summable if :

\begin{equation}
[ F,{\pi}(f) ]  {\in} {\cal L}^{p}(H_2)  ,  {\forall} f {\in} A.
\end{equation}

\subsubsection{The differential envelope  $ {\Omega} $ of  $A$}

Let $n$ be an even  integer , $ n{\geq}0 $ ,and let us assume that our Fredholm module $ (H_2,F) $  is (
n+1)-summable , in other words :

\begin{equation}
[ F,{\pi}(f) ] {\in} {\cal L}^{n+1}(H_2)  ,  {\forall} f {\in}A.\label{n+1summability}
\end{equation}
We can  associate to the algebra $A$ a bigger algebra $ {\Omega} $ called the differential envelope of $A$ in the
following way :
\begin{equation}
{\Omega}= {\bigoplus}_{k=0}^{n}  {\Omega}_{k},\label{envelope}
\end{equation}
where $ {\Omega}_{0} =A $ and ${\Omega}_{k}$ , $k>1$ , is the space generated by the operators :

\begin{equation}
{\omega}={\pi}(f_{0})[F,{\pi}(f_{1})][F,{\pi}(f_{2})]....[F,{\pi}(f_{k})],\label{kform}
\end{equation}
where $ f_0 $,$ f_1 $....,$ f_k $ are elements of $A$ . In fact , the operators $ {\omega} $ define the space of
k-forms over the algebra $A$ , in particular ${\Omega}_{1}$ is the space of one forms and ${\Omega}_{2}$ is the
space of two-forms . By using the so called Holder inequality  which can be stated as :

\begin{equation}
{\cal L}^{p_{1}}{\cal L}^{p_{2}}.....{\cal L}^{p_{k}} {\subset} {\cal L}^{p}  ,
 for \frac{1}{p}=\sum_{j=1}^{k}\frac{1}{p_{j}},
\end{equation}
one can see that  : $ {\Omega}_{k} {\subset} {\cal L}^{\frac{n+1}{k}} $ . The product in $ {\Omega} $ is the
product of operators given by :

\begin{equation}
{\forall}{\psi}{\in}{\Omega}_{k}  , {\forall}{\phi}{\in}{\Omega}_{p}:{\psi}{\phi} {\in} {\Omega}_{k+p}.
\end{equation}
\subsubsection{The exterior derivative}
The differential envelope ${\Omega}$ is a graded algebra in the following sense . In general one can define the
exterior derivative $d$ as a map from $ {\Omega} $ into $ {\Omega} $ given by :

\begin{eqnarray}
d {\omega}&=&i( F{\omega}- (-1)^{k}{\omega}F ) , {\forall}{\omega}{\in}{\Omega}_{k}\nonumber\\
&&\label{exterior2}
\end{eqnarray}
For $k=0$ , (\ref{exterior2}) reduces precisely to (\ref{exterior1}) . More precisely this exterior derivative $d$ maps $k-$forms
into $(k+1)-$forms , in other words given a $k-$form ${\omega}$, one can compute

\begin{eqnarray}
d {\omega} &=&i\bigg[F{\omega}- (-1)^{k}{\omega}F\bigg]\nonumber\\
&=&i\bigg[F{\pi}(f_{0})[F,{\pi}(f_{1})][F,{\pi}(f_{2})]....[F,{\pi}(f_{k})] - (-1)^{k}{\pi}(f_{0})[F,{\pi}(f_{1})][F,{\pi}(f_{2})]....[F,{\pi}(f_{k})]F\bigg]\nonumber\\
&=&i[F,{\pi}(f_{0})][F,{\pi}(f_{1})][F,{\pi}(f_{2})]....[F,{\pi}(f_{k})],\nonumber\\
&&\label{dkform}
\end{eqnarray}
where we have used the expression (\ref{kform}) of ${\omega}{\in}{\Omega}_k$ and the identity
$F[F,{\pi}(f)]=-[F,{\pi}(f)]F$ . From (\ref{dkform}) , it is very clear that $ d{\omega} $ is in ${\Omega}_{k+1} $ ,
which means that

\begin{equation}
d : {\Omega}_{k}{\longrightarrow}{\Omega}_{k+1}.
\end{equation}
Now let us check that $d$ is a graded derivation as follows :

\begin{eqnarray}
d({\omega}_{1}{\omega}_{2})&=&i\bigg[F{\omega}_{1}{\omega}_{2}-(-1)^{k}{\omega}_{1}{\omega}_{2}F\bigg]\nonumber\\
&=&i\bigg[\big[F{\omega}_{1}-(-1)^{k_{1}}{\omega}_{1}F+(-1)^{k_{1}}{\omega}_{1}F\big]{\omega}_{2}-(-1)^{k}{\omega}_{1}{\omega}_{2}F\bigg]\nonumber\\
&=&(d{\omega}_{1}){\omega}_{2} + i{\omega}_{1}\bigg[(-1)^{k_{1}}F{\omega}_{2}-(-1)^{k_{1}+k_{2}}{\omega}_{2}F\bigg]\nonumber\\
&=&(d{\omega}_{1}){\omega}_{2} +(-1)^{k_{1}}{\omega}_{1}(d{\omega}_{2}).\nonumber\\
&&\label{leibnitz}
\end{eqnarray}
For $k_1=$even , (\ref{leibnitz}) is exactly Leibnitz's rule .

One can write the definition (\ref{exterior2}) of the exterior derivative as follows: $ d {\omega} = i(F{\omega}-
(-1)^{k}{\omega}F)=i( F{\omega} - {\Gamma}_2{\omega}{\Gamma}_2F) $ . The proof is simple and consists in the
observation that ${\omega}$ contains $( k+1 )$ elements of the algebra $A$ which all commute with $ {\Gamma}_2 $ ,
while the $k$ operators $F$ anticommute with $ {\Gamma}_2 $ and therefore the extra sign $ (-1)^{k} $ . So

\begin{eqnarray}
d{\omega}&=&i(F{\omega}-{\Gamma}_2{\omega}{\Gamma}_2F)\nonumber\\
&=&i{\Gamma}_2({\Gamma}_2F{\omega}-{\omega}{\Gamma}_2F)\nonumber\\
&=&i{\Gamma}_2[{\Gamma}_2F,{\omega}].\nonumber\\
&&\label{exterior3}
\end{eqnarray}
From (\ref{exterior3}) it is easily verified that $d$ satisfies $ d^{2}=0 $ , indeed :

\begin{eqnarray}
d^{2}{\omega}&=&i{\Gamma}_2[{\Gamma}_2F,d{\omega}]=i{\Gamma}_2[{\Gamma}_2F,i{\Gamma}_2[{\Gamma}_2F,{\omega}]]\nonumber\\
&=&-{\Gamma}_2[{\Gamma}_2F,F{\omega}-{\Gamma}_2{\omega}{\Gamma}_2F]=-F(F{\omega}-{\Gamma}_2{\omega}{\Gamma}_2F)+{\Gamma}_2(F{\omega}-{\Gamma}_2{\omega}{\Gamma}_2F){\Gamma}_2F\nonumber\\
&=&-F^{2}{\omega}+F{\Gamma}_2{\omega}{\Gamma}_2F-F{\Gamma}_2{\omega}{\Gamma}_2F+{\omega}F^{2}\nonumber\\
&=&0.\nonumber\\
\end{eqnarray}
The pair $ ( {\Omega},d ) $  defines a graded differential algebra .

\subsubsection{The cycle $({\Omega},d,Tr_s)$}

One can define a closed graded trace of degree $n$ , recalling that $n$ is an even integer introduced in equations
(\ref{n+1summability}) and (\ref{envelope}) , by :

\begin{eqnarray}
Tr_{s} :& {\Omega}_{n}& {\longrightarrow}{\bf C}\nonumber\\
&{\omega}&{\longrightarrow}Tr_{s}({\omega})=Tr^{'}({\Gamma}_2{\omega}).\nonumber\\
&&\label{tracecycle}
\end{eqnarray}
$Tr^{'}(x) $ is defined only for those $x$'s which are such that the combination  $Fx+xF$ is in ${\cal L}^{1}(H_2)
$ , i.e belongs to the space of trace class operators , it is given by :

\begin{equation}
Tr^{'}(x)=\frac{1}{2}Tr(F(Fx+xF)).
\end{equation}
To prove that the combination $Fx+xF$ , for $x={\Gamma}_2{\omega}$, is in fact in ${\cal L}^{1}(H_2)$, we simply
compute $F{\Gamma}_2{\omega}+{\Gamma}_2{\omega}F=i{\Gamma}_2d{\omega}$ where we have used the fact that
${\omega}{\in}{\Omega}_n$ and that $n$ is even . $d{\omega}$ is clearly in ${\Omega}_{n+1}$ but by using Holder
inequality we can check that ${\Omega}_{n+1}{\subset}{\cal L}^{1}(H_2)$ , hence $d{\omega}$ and therefore
$F{\Gamma}_2{\omega}+{\Gamma}_2{\omega}F$ are ${\in}{\cal L}^{1}(H_2)$ .

Since $Tr_s(\omega)$ depends only on $d{\omega}$ and $d^2=0$ , it is trivial to see that $Tr_s(d{\omega})=0$ , i.e
the trace (\ref{tracecycle}) is closed . The trace (\ref{tracecycle}) is also a graded trace because:
${\forall}{\omega}_{1}{\in}{\Omega}_{k_{1}}$ and ${\forall} {\omega}_{2}{\in}{\Omega}_{k_{2}}$ such that : $
k_{1}+k_{2} = n $ we find :

\begin{eqnarray}
Tr_{s}({\omega}_{1}{\omega}_{2})&=&-\frac{i}{2}Tr{\Gamma}_2Fd({\omega}_{1}{\omega}_{2})\nonumber\\
&=&-\frac{i}{2}Tr{\Gamma}_2F\bigg[d{\omega}_{1}{\omega}_{2}+(-1)^{k_1}{\omega}_{1}d{\omega}_{2}\bigg]\nonumber\\
&=&-\frac{i}{2}Tr{\Gamma}_2\bigg[(-1)^{k_1+1}d{\omega}_1F{\omega}_2+(-1)^{k_1}F{\omega}_1d{\omega}_2\bigg],
\end{eqnarray}
where we have used the identity $Fd{\omega}_1+(-1)^{k_1}d{\omega}_1F=0$ .

The triplet $ ( {\Omega},d,Tr_{s} ) $ defines a cycle with dimension $n$ over the the algebra $A$. It is a
theorem that this cycle is essentially determined by its character ${\tau}$, i.e the $(n+1)-$linear function defined
by :

\begin{eqnarray}
{\tau}_{n}({\pi}(f_0),{\pi}(f_1),{\pi}(f_2)....,{\pi}(f_n))&=&\nonumber\\
Tr^{'}{\Gamma}_2{\pi}(f_0)[F,{\pi}(f_1)][F,{\pi}(f_2)].....[F,{\pi}(f_n)]&.&\nonumber\\
\end{eqnarray}
${\tau}_{n}$ is called the character or the cocycle of the cycle $( {\Omega},d,Tr_{s} )$ .
\subsection{The Hochschild Complex and its Cyclic Subcomplex}

Let $A^{*} $ be  the algebraic dual of $A$ , i.e the space of all linear functionals ${\phi}$ on $A$ :
\begin{eqnarray}
{\phi}:&A&{\longrightarrow}{\bf C}\nonumber\\
&f&{\longrightarrow}{\phi}(f).
\end{eqnarray}
$A^{*}$ is a bimodule in the sense that for any $a$ and $b$ in $A$ and ${\phi}{\in}A^{*} $ , the object $a{\phi}b$
is in $A^{*} $ defined by :

\begin{equation}
a{\phi}b(c)={\phi}(bca)
\end{equation}

Let now $ C^{p}=C^{p}(A,A^{*}) $ be the space of all $p-$linear maps from $A$ to $A^{*} $. Any element $T$ of
$C^{p} $ can be viewed as a $(p+1)-$linear functional $ {\tau} $ on $A$ given by :

\begin{equation}
{\tau}({\pi}(f_0),{\pi}(f_1),...{\pi}(f_p))=[T({\pi}(f_1),...{\pi}(f_p))]({\pi}(f_0)) {\in} C.
\end{equation}
The Hochschild coboundary map $b$ is defined as follows . To the boundary $bT$ corresponds a $(p+2)-$linear
functional $b{\tau}$ given by :

\begin{eqnarray}
[b{\tau}]({\pi}(f_0),....{\pi}(f_{p+1}))&=&{\tau}({\pi}(f_0){\pi}(f_1),....{\pi}(f_{p+1}))\nonumber\\
&+&\sum_{i=1}^{p}(-1)^{i}{\tau}({\pi}(f_0),...,{\pi}(f_i){\pi}(f_{i+1}),...,{\pi}(f_p))\nonumber\\
&+&(-1)^{p+1}{\tau}({\pi}(f_{p+1}){\pi}(f_0),...,{\pi}(f_p)).\nonumber\\
\end{eqnarray}
Hochschild cochains of degree $p$ are defined to be those elements $ T{\in}C^{p} $ which are also linear
functionals $\hat{\tau}$ on ${\Omega}_p$ defined by
\begin{eqnarray}
{\hat{\tau}}({\omega})&=&{\tau}({\pi}(f_0),{\pi}(f_1),...,{\pi}(f_p))\nonumber\\
&where&\nonumber\\
{\omega}&=&{\pi}(f_0)d{\pi}(f_1)....d{\pi}(f_p) {\in} {\Omega}^{p},\nonumber\\
\end{eqnarray}
and which vanish on the $d{\Omega}_{p-1}$ part of ${\Omega}_p$ , i.e
\begin{equation}
{\hat{\tau}}(d{\omega}^{'})=0 ,
\end{equation}
where ${\omega}^{'}{\in}{\Omega}_{p-1}$ .

Now we define the Hochschild cocycles as all those Hochschild cochains which satisfy the extra following condition
:

\begin{equation}
[b\hat{{\tau}}]({\omega})=0,
\end{equation}
where $[b\hat{\tau}](\omega)=[b{\tau}]({\pi}(f_0),...,{\pi}(f_{p+1}))$ and
${\omega}={\pi}(f_0)d{\pi}(f_1)...d{\pi}(f_{p+1}){\in}{\Omega}_{p+1}$. By definition , the $p-$th cohomology group
of the algebra $A$ with coefficients in $A^{*}$ is the cohomology $ H^{p}=H^{p}(A,A^{*}) $ of the Hochschild
complex $ (C^{p}(A,A^{*}),b) $ .

Finally we define the cyclic cocycles as those Hochschild cocycles which satisfy:

\begin{equation}
{\tau}^{\gamma}={\epsilon}({\gamma}){\tau},\label{cc}
\end{equation}
where $ {\gamma} $ denotes  any cyclic permutation of $ \{0,1,...p\}$\footnote{${\tau}^{\gamma}({\pi}(f_0),{\pi}(f_1),...,{\pi}(f_p))={\tau}({\pi}(f_{n_0}),{\pi}(f_{n_1}),...,{\pi}(f_{n_p}))$ where $\{n_0,n_1,...,n_p\}$ is the permutation ${\gamma}{\in}{\Gamma}$ of $\{0,1,...,p\}$.} , and ${\epsilon}(\gamma)$ is the corresponding sign , for even permutations it is plus whereas for odd permutations it is minus. Given an arbitrary Hochschild cocycle $
{\tau} $ , we can associate to it a cyclic cocycle as follows :

\begin{equation}
A{\tau} = \sum_{{\gamma}{\in}{\Gamma}}{\epsilon}({\gamma}){\tau}^{\gamma},\label{cycliccocycle}
\end{equation}
where $ {\Gamma} $ stands for the group of cyclic permutations of $ \{ 0,1,...,p\} $ , and $A$ is a linear map from
$ C^{p} $ into $ C^{p} $ given by the above equation, i.e (\ref{cycliccocycle}). Obviously the range of $A$ is the subspace $
C^{p}_{\lambda} $ of $ C^{p} $ , namely the space of Hochschild cocycles which satisfy equation (\ref{cc}). Although
the Hochschild coboundary operator $b$ does not commute with cyclic permutations , it can be proven that it maps
cyclic cocycles to cyclic cocycles . By definition , the $ p-$th cyclic cohomology group of the algebra $A$ with
coefficients in $ A^{*} $ is the cohomology  $ HC^{p}=HC^{p}(A,A^{*}) $ of the cyclic complex $ (
C^{p}_{\lambda}(A,A^{*},b) $. Clearly $ (C^{p}_{\lambda},b) $ is a subcomplex of the Hochschild complex .

\subsubsection{Fredholm module's character as cyclic cocycle}

In section $(2.7.1)$ we associated to the even K-cycle $X=(A,H,D)$ an $(n+1)-$summable Fredholm module structure
$(H_2,F)$ . Then , in section $(2.7.2)$, this Fredholm module was completely charcaterized by the charcacter
${\tau}_n$ of its cycle $({\Omega},d,Tr_s)$. This character ${\tau}_n$ is explicitly given by

\begin{equation}
{\tau}_n({\pi}(f_0),{\pi}(f_1),...,{\pi}(f_n))=\frac{1}{{i}^{n}} Tr{\Gamma}_2{\pi}(f_0)d{\pi}(f_1)...d{\pi}(f_n).
\end{equation}
Recall that $n$ was taken to be even . This character is clearly an $(n+1)-$linear map from the algebra $A$ into
the complex numbers . It can be associated with an element $T_n{\in}C^n(A,A^{*})$ in the following way
\begin{equation}
T_n[({\pi}(f_1),...,{\pi}(f_n))]({\pi}(f_0))={\tau}_n({\pi}(f_0),{\pi}(f_1),...,{\pi}(f_n)),
\end{equation}
of course $T_n$ is an $n-$linear map from $A$ to $A^{*}$ . In the same way , one can associate to ${\tau}_n$ a map
$\hat{\tau}_n$ from ${\Omega}_n$ into ${\bf C}$ by the equation
\begin{eqnarray}
{\hat{\tau}}_n({\omega})&=&{\tau}_n({\pi}(f_0),{\pi}(f_1),...,{\pi}(f_n))=\frac{1}{i^n}Tr{\Gamma}_2{\omega}\nonumber\\
&where&\nonumber\\
{\omega}&=&{\pi}(f_0)d{\pi}(f_1)....d{\pi}(f_n){\in}{\Omega}_n.\nonumber\\
\end{eqnarray}
Let us now check that this character , $\hat{\tau}_n$ , is a cyclic cocycle . First one needs to check that it is a
Hochschild cochain , in other words for any ${\omega}^{'}{\in}{\Omega}_{n-1}$ we must have
$\hat{\tau}_n(d{\omega}^{'})=0$ . Indeed for ${\omega}^{'}{\in}{\Omega}_{n-1}$ we have

\begin{equation}
d{\omega}^{'}=d{\pi}(f_1)d{\pi}(f_2)....d{\pi}(f_{n}),
\end{equation}
and therefore
\begin{eqnarray}
{\hat{\tau}}_n(d{\omega}^{'})&=&{\tau}_n(1,{\pi}(f_1),...{\pi}(f_{n}))\nonumber\\
&=&Tr{\Gamma}_2[F,{\pi}(f_1)]....[F,{\pi}(f_{n})]\nonumber\\
&=&Tr{\Gamma}_2F{\pi}(f_1)[F,{\pi}(f_2)]...[F,{\pi}(f_{n})]\nonumber\\
&-&Tr{\Gamma}_2{\pi}(f_1)F[F,{\pi}(f_2)]...[F,{\pi}(f_{n})]\nonumber\\
&=&Tr{\Gamma}_2F{\pi}(f_1)[F,{\pi}(f_2)]...[F,{\pi}(f_{n})]\nonumber\\
&-&(-1)^{n-1}TrF{\Gamma}_2{\pi}(f_1)[F,{\pi}(f_2)]...[F,{\pi}(f_{n})]\nonumber\\
&=&0,\nonumber\\
\end{eqnarray}
where we have used the identity  $F[F,{\pi}(f)]=-[F,{\pi}(f)]F$ .

Next one must check that , $\hat{\tau}_n$ , is a Hochschild cocycle , i.e  $ b\hat{\tau}_n=0 $ or more precisely

\begin{eqnarray}
[b{\tau}_n]({\pi}(f_0),...,{\pi}(f_{n+1}))&=&{\tau}_n({\pi}(f_0){\pi}(f_1),...,{\pi}(f_{n+1}))\nonumber\\
&+&\sum_{i=1}^{n}(-1)^{i}{\tau}_n({\pi}(f_0),...,{\pi}(f_i){\pi}(f_{i+1}),...,{\pi}(f_n))\nonumber\\
&+&(-1)^{n+1}{\tau}_n({\pi}(f_{n+1}){\pi}(f_0),...,{\pi}(f_n)).\nonumber\\
&&\label{hochschild}
\end{eqnarray}
To prove (\ref{hochschild}) , let us simply compute the second term above :

\begin{eqnarray}
\sum_{i=1}^{n}(-1)^{i}{\tau}_n({\pi}(f_0),...,{\pi}(f_i){\pi}(f_{i+1}),...,{\pi}(f_n))&=&\nonumber\\
\sum_{i=1}^{n}(-1)^{i}Tr{\Gamma}_2{\pi}(f_0)[F,{\pi}(f_1)]...[F,{\pi}(f_i){\pi}(f_{i+1})]...[F,{\pi}(f_n)]&=&\nonumber\\
\sum_{i=1}^{n}(-1)^{i}Tr{\Gamma}_2{\pi}(f_0)[F,{\pi}(f_1)]...[F,{\pi}(f_i)]{\pi}(f_{i+1})...[F,{\pi}(f_n)]&+&\nonumber\\
\sum_{i=1}^{n}(-1)^{i}Tr{\Gamma}_2{\pi}(f_0)[F,{\pi}(f_1)]...{\pi}(f_i)[F,{\pi}(f_{i+1})]...[F,{\pi}(f_n)]&=&\nonumber\\
(-1)^nTr{\Gamma}_2{\pi}(f_0)[F,{\pi}(f_1)]...[F,{\pi}(f_n)]{\pi}(f_{n+1})&+&\nonumber\\
\sum_{i=1}^{n-1}(-1)^{i}Tr{\Gamma}_2{\pi}(f_0)[F,{\pi}(f_1)]...[F,{\pi}(f_{i})]{\pi}(f_{i+1})...[F,{\pi}(f_n)]&+&\nonumber\\
(-1)^{1}Tr{\Gamma}_2{\pi}(f_0){\pi}(f_1)[F,{\pi}(f_2)]...[F,{\pi}(f_n)]&+&\nonumber\\
\sum_{i=2}^{n}(-1)^{i}Tr{\Gamma}_2{\pi}(f_0)[F,{\pi}(f_1)]...{\pi}(f_i)[F,{\pi}(f_{i+1})]...[F,{\pi}(f_n)]&=&\nonumber\\
{\tau}({\pi}(f_{n+1}){\pi}(f_{0}),...,{\pi}(f_n))&-&\nonumber\\
{\tau}({\pi}(f_0){\pi}(f_1),...,{\pi}(f_{n+1}))&.&\nonumber\\
\end{eqnarray}
From this last result , one can easily see that , $b{\tau}_n=0$ , as desired .

Finally one must check that , ${\tau}_n$ , is a cyclic cocycle , in other words :

\begin{equation}
{\tau}_n({\pi}(f_0),{\pi}(f_1),...,{\pi}(f_n))={\tau}_n({\pi}(f_1),...,{\pi}(f_{n}),{\pi}(f_0)).
\end{equation}
Indeed ,

\begin{eqnarray}
{\tau}_n({\pi}(f_0),{\pi}(f_1),...,{\pi}(f_n))&=&Tr{\Gamma}_2{\pi}(f_0)[F,{\pi}(f_1)]...[F,{\pi}(f_n)]\nonumber\\
&=&Tr{\Gamma}_2{\pi}(f_0)F{\pi}(f_1)[F,{\pi}(f_2)]...[F,{\pi}(f_n)]\nonumber\\
&-&Tr{\Gamma}_2{\pi}(f_0){\pi}(f_1)F[F,{\pi}(f_2)]...[F,{\pi}(f_n)]\nonumber\\
&=&Tr{\Gamma}_2{\pi}(f_0)F{\pi}(f_1)[F,{\pi}(f_2)]...[F,{\pi}(f_n)]\nonumber\\
&+&Tr{\Gamma}_2{\pi}(f_0){\pi}(f_1)[F,{\pi}(f_2)]...[F,{\pi}(f_n)]F\nonumber\\
&=&Tr{\Gamma}_2{\pi}(f_1)[F,{\pi}(f_2)]...[F,{\pi}(f_n)][F,{\pi}(f_0)]\nonumber\\
&=&{\tau}_n({\pi}(f_1),...,{\pi}(f_n),{\pi}(f_0)).\nonumber\\
\end{eqnarray}
With this result one concludes the proof that the character ${\tau}_n$ of the Fredholm module $ (H_2,F) $ is a
cyclic cocycle . The integer $n$ in all the above equations is by construction the smallest integer compatible
with the $ (n+1)-$ summability of the Fredholm module $ (H_2,F) $ . For example it is equal to $2 $ in the case of
the sphere .

\subsubsection{Stability of cyclic cocycles}

The cyclic cocycle $ {\tau}_{2} $ of the two dimensional cycle $ ({\Omega},d,Tr_{s}) $ is given by :

\begin{eqnarray}
{\tau}_{2}({\pi}(f_0),{\pi}(f_1),{\pi}(f_2))
&=&Tr{\Gamma}_2{\pi}(f_0)[F,{\pi}(f_1)][F,{\pi}(f_2)].\nonumber\\
\end{eqnarray}
By using equations (\ref{fredholmmodule}) , (\ref{representationh2}) and (\ref{chiralityh2}) we obtain :

\begin{eqnarray}
{\tau}_{2}({\pi}(f_0),{\pi}(f_1),{\pi}(f_2))&=&Tr{\Gamma}f_0[D^{-1},f_1][D,f_2] + Tr{\Gamma}_2f_0[D,f_1][D^{-1},f_2].\nonumber\\
&&
\end{eqnarray}
This last equation can be rewritten using the identity , $ [D^{-1},f]=-D^{-1}[D,f]D^{-1} $, as :
\begin{eqnarray}
{\tau}_{2}({\pi}(f_0),{\pi}(f_1),{\pi}(f_2))&=&-Tr{\Gamma}D^{-1}[D,f_0]D^{-1}[D,f_1]D^{-1}[D,f_2].\nonumber\\
&&\label{fermionicloop}
\end{eqnarray}
From this last result , one sees that the cyclic cocycle , $ {\tau}_2 $ , can be interpreted as a fermionic one
loop Feynman diagram with one insertion of the helicity operator $ {\Gamma} $ . For example , if we consider $A$ to
be the algebra of smooth functions on the sphere , then (\ref{fermionicloop}) takes the form

\begin{equation}
{\tau}_{2}({\pi}(f_0),{\pi}(f_1),{\pi}(f_2))=-\int_{S^{2}}Tr{\Gamma}D^{-1}[D,f_0]D^{-1}[D,f_1]D^{-1}[D,f_2]d^{2}x,
\end{equation}
where now the trace is only over the spin indices . The element $f_i$ is a superposition of exponentials $
exp(ik_{i}x) $ , and therefore in the Fourier space $ [D,f_i] $ will appear as an insertion of $
{\gamma}_{\mu}k^{\mu}_{i} $ at the vertex $ i $ . $ D^{-1} $ , on the other hand , appears as a propagator .
Finally the overall conservation of momentum : $ {\delta}( k_0+k_1+k_2 ) $ reduces the number of variables to two .

Next one would like to extend the cyclic cocycle , ${\tau}_2$ , which is defined over the algebra $A$ to a cyclic
cocycle , ${\tau}_2^e$ , which is defined over the algebra $ M_{2}(A) $ , i.e the algebra of $2{\times}2$ matrices
with entries in the algebra $A$ . The extended cyclic cocycle $ {\tau}_2^{e} $ should also satisfy the two
conditions satisfied by the original cyclic cocycle ${\tau}_2$ , namely :

\begin{eqnarray}
a&)&{\tau}_2^{e}({\pi}(f_0),{\pi}(f_1),{\pi}(f_2))={\tau}_2^{e}({\pi}(f_2),{\pi}(f_0),{\pi}(f_1))\nonumber\\
b&)&b{\tau}_2^{e}=0\nonumber\\
&&\label{extendedcc}
\end{eqnarray}
This extension is given by : $ {\forall}f_0, f_1 , f_2 $ in  $ M_{2}(A) $ we write

\begin{equation}
{\tau}_2^{e}({\pi}(f_0),{\pi}(f_1),{\pi}(f_2))
{\equiv}Tr{\sigma}_{i}{\sigma}_{j}{\sigma}_{k}.{\tau}_2({\pi}(f_0^i),{\pi}(f_1^j),{\pi}(f_2^k)),
\end{equation}
where $ f_0={\sigma}_{i}f_{0}^{i} $ , $ f_1={\sigma}_{i}f_{1}^{i} $ and $ f_2={\sigma}_{i}f_{2}^{i} $ . It is
obvious that $ {\tau}_2({\pi}(f_0^i),{\pi}(f_1^j),{\pi}(f_2^k)) $ is well defined since $ f_0^i $ , $ f_1^j $ and $
f_2^k $ are all elements of the algebra $A$ .  We can also (easily) check that the two properties $ a) $ and $ b) $
given in equation (\ref{extendedcc}) are both satisfied for this $ {\tau}_{2}^{e} $ . This extension is very useful
because one of the central object of this thesis is
\begin{equation}
{\tau}_2^e(P)\equiv{\tau}_2^e(P,P,P),\label{centralthesis}
\end{equation}
where $P$ is an arbitrary idempotent of $M_2(A)$ , in other words a selfadjoint element of $M_2(A)$ which also
satisfies $P^2=P$ . In the next chapters , (\ref{centralthesis}) will be interpreted as the Chern character of some bundle .

The last thing one needs to do in this chapter is to check the stability of (\ref{centralthesis}) under the deformation of $P$
among the idempotents of $ M_2(A) $ . In other words , under
\begin{equation}
P{\longrightarrow}P^{`}=UPU^{-1},
\end{equation}
where $ U $ is a unitary transformation , $ U^{+}=U^{-1} $ , one must have $ {\tau}^e_2(P^{`})={\tau}^e_2(P) $.

For infintesimal transformations , $ U = 1 + T $ , we have $P^{`}=P+{\delta}P$ where ${\delta}P=[T,P]$ . If we
write $ P^{`}={\sigma}_{i}f^{'i} $ and $ P={\sigma}^{i}f^i $ then $ {\delta}P={\sigma}_i{\delta}f^i $ where $
{\delta}f^i=f^{'i}-f^i $ . Hence

\begin{eqnarray}
{\tau}_2^e(P^{`})&=&{\tau}_{2}^{e}(P^{`},P^{`},P^{`}){\equiv}Tr{\sigma}_i{\sigma}_j{\sigma}_k.
{\tau}_{2}({\pi}(f^{'i}),{\pi}(f^{'j}),{\pi}(f^{'k}))\nonumber\\
&=&-Tr{\sigma}_i{\sigma}_j{\sigma}_k.Tr{\Gamma}D^{-1}[D,f^{'i}]D^{-1}[D,f^{'j}]D^{-1}[D,f^{'k}]\nonumber\\
&=&-Tr{\sigma}_i{\sigma}_j{\sigma}_k.Tr{\Gamma}D^{-1}[D,f^i+{\delta}f^i]D^{-1}[D,f^j+{\delta}f^j]D^{-1}[D,f^k+{\delta}f^k]\nonumber\\
&=&{\tau}_{2}^{e}(P,P,P)+{\tau}_{2}^{e}(P,P,{\delta}P)+{\tau}_{2}^{e}(P,{\delta}P,P)+{\tau}_{2}^{e}({\delta}P,P,P)\nonumber\\
&{\Longrightarrow}&\nonumber\\
{\delta}{\tau}_{2}^{e}&=&3{\tau}_{2}^{e}({\delta}P,P,P),\nonumber\\
\end{eqnarray}
where we have used , in the last line , the fact that the extended cyclic cocycle, ${\tau}_2^e$, is symmetric
under cyclic permutations .

Now from the fact that , $ b{\tau}_{2}^{e}=0 $ , we have :

\begin{eqnarray}
{\tau}_{2}^{e}(TP^1,P^2,P^3) - {\tau}_{2}^{e}(T,P^1 P^2,P^3)+
{\tau}_{2}^{e}(T,P^1,P^2P^3) - {\tau}_{2}^{e}(P^3T,P^1,P^2)= 0,\nonumber\\
\end{eqnarray}
we get for $P^1=P^2=P^3$
\begin{eqnarray}
{\tau}_{2}^{e}(TP,P,P) - {\tau}_{2}^{e}(PT,P,P)&=&0\nonumber\\
&{\Longrightarrow}&\nonumber\\
{\tau}_2^e({\delta}P,P,P)&=&0,
\end{eqnarray}
since ${\delta}P=TP-PT$ . Hence ${\delta}{\tau}_2^e=0$ and therefore the cyclic cocycle , ${\tau}_2^e$ , is stable
.

%

\chapter{Fuzzy ${\bf S}^2$}

This chapter is entirely devoted to the construction of the formalism needed to describe the fuzzy sphere ${\bf
S}^2_F$ . A K-cycle $({\bf A},{\bf H},D,{\Gamma})$ describing ${\bf S}^2_F$ will be obtained from the K-cycle
$({\cal A},{\cal H},{\cal D},{\gamma})$ describing the classical sphere ${\bf S}^2$ by quantizing the underlying
symplectic structure of ${\bf S}^2$ , namely ${\omega}=ld(cos{\theta}){\wedge}d{\phi}$ . It is a theorem due to
Connes \cite{1} that this K-cycle will code all the geometrical properties of the space . For even dimensional
spaces , the K-cycle consists of an algebra of operators , a representation space on which the algebra acts , and a
Dirac operator as well as a chirality operator defining the differential structure of the space . For odd
dimensional spaces the chirality operator does not exist .

The method given here for the case of , ${\bf S}^2={\bf C}{\bf P}^1$ , will work for all other ${\bf C}{\bf P}^N$
manifolds so that generalization is straightforward . Because ${\bf C}{\bf P}^N$ manifolds with $N$ even , starting with ${\bf C}{\bf P}^2$ , do not admit spin structure but only spin$_c$ structure, their case present more
complications and will be treated in chapter $5$ .

\section{Continuum Considerations}
One starts this section by briefly reviewing the ordinary differential geometry of the two sphere ${\bf S}^{2}$.
We will also try to reformulate all the relevant aspects of this geometry in algebraic terms so that
generalization to the fuzzy sphere can be made.

\subsection{The Algebra ${\cal A}$ of Functions on ${\bf S}^2$}

The sphere is a two dimensional compact manifold defined by the set of all points $(x_{1},x_{2},x_{3})$ of ${\bf
R}^{3}$ which satisfy :

\begin{equation}
x_{1}^{2}+x_{2}^{2}+x_{3}^{2}={\rho}^{2}\label{calgebra1}
\end{equation}
The algebra ${\cal A}$ of smooth , complex valued and square integrable functions on the sphere is of course
commutative with respect to the pointwise multiplication of functions . A basis for this algebra can be chosen to
be provided by the spherical harmonics $ Y_{lm}$ , namely

\begin{eqnarray}
f(x)=f({\theta},{\phi})&=&\sum_{i_1,...,i_k}f_{i_1...i_k}x_{i_1}...x_{i_k}\nonumber\\
&=&\sum_{lm}c_{lm}Y_{lm}({\theta},{\phi}).\label{calgebra2}
\end{eqnarray}
When the coordinates $x_i$  are quantized , they become operators realized as matrices on a certain Hilbert space
${\bf H}$ and therefore the functions $f$ become also matrices acting on this ${\bf H}$ . The set of all these
operators , $f$ , form an algebra ${\bf A}$ . The numbers $f_{i_1...i_k}$ will clearly preserve their meaning as
the coeficients of the expansion of these operators . We would like to reformulate the geometry of ${\bf S}^{2}$ in
terms of the algebra ${\cal A}$ defined on it . One crucial step towards the definition of the fuzzy sphere will
be then the simple replacement ${\cal A}{\longrightarrow}{\bf A}$ and ${\cal H}{\longrightarrow}{\bf H}$ , where
${\cal H}$ is the Hilbert space on which the algebra ${\cal A}$ is represented . A basis for ${\cal H}$ is
provided by the standard infinite dimensional set of kets $\{|\vec{x}>\}$ , the action of an element $f$ of ${\cal
A}$ on $|\vec{x}>$ will give the value of this function at the point $\vec{x}$ .

An alternative , manifestly $SU(2)-$invariant , description of ${\cal A}$ can be given as follows \cite{last3}.
The algebra ${\cal A}$ is the quotient of the algebra ${\bf C}^{\infty}({\bf R}^3)$ of all smooth functions of
${\bf R}^{3}$ by its ideal ${\cal I}$ consisting of all functions of the form : $h(x)(x_{i}x_{i}-{\rho}^{2})$. A
scalar product on ${\cal A}$ is then given by : $ (f,g)={\frac{1}
{2{\pi}{\rho}}}{\int}d^{3}x{\delta}(x_{i}x_{i}-{\rho}^{2})f^{*}(x)g(x)$. Here $f,g{\in}{\cal A}$ and $f(x)$,$g(x)$
are their representatives in ${\bf C}^{\infty}({\bf R}^3)$ respectively . For example the norms of the generators
$x_i$ of the algebra ${\cal A}$ are computed to be
${\parallel}x_{i}{\parallel}^2=(x_{i},x_{i})={\frac{{\rho}^2}{3}}$ .

\subsection{The Spinor Bundle ${\cal E}_2$ over ${\bf S}^2$}

Now we would like to define the spinor bundle over the sphere ${\bf S}^2$ \cite{4,last3}. One starts first by
defining the Clifford algebra associated with the vector space  ${\bf R}^3$ . It is a complex algebra generated by
$3$ self adjoint elements ${\gamma}^{\alpha}$ which satisfy the relations, $
{\gamma}^{\alpha}{\gamma}^{\beta}+{\gamma}^{\beta}{\gamma}^{\alpha}=2{\delta}^{{\alpha}{\beta}}$, and which are
represented by  $2{\times}2$  pauli matrices .

The spin group spin$(3)$ is known to be equal to $SU(2)$ . It is the universal covering group of $SO(3)$ . It
consists of all the $2{\times}2$ transformations $S(\Lambda)$ defined by ,
${{\Lambda}^{\alpha}}_{\beta}{\gamma}^{\beta}=S^{-1} ({\Lambda}){{\gamma}^{\alpha}}S({\Lambda})$ and $det
S(\Lambda)=1$ , where ${\Lambda}$ is in $SO(3)$ . This map is clearly double valued because both $S(\Lambda)$ and
$-S(\Lambda)$ correspond to the same ${\Lambda}$ in $SO(3)$ , in other words spin$(3)$ covers $SO(3)$ twice . For
transformations ${\Lambda}$ near the identity ,
${\Lambda}^{\alpha}_{\beta}\simeq{\delta}^{\alpha}_{\beta}+{\lambda}^{\alpha}_{\beta}$ , the above map will have
one solution given by $S(\Lambda)\simeq 1+\frac{1}{4}{\lambda}_{{\alpha}{\beta}}{\gamma}^{\alpha}{\gamma}^{\beta}$
.

The spinor bundle over ${\bf S}^2$ will be defined now in three steps . First one defines the principal fiber
bundle ${\cal E}=(spin(3),{\pi},{\bf R}^3)$ over the base manifold ${\bf R}^3$ by the projection map
\begin{eqnarray}
{\pi}&:&spin(3){\longrightarrow}SO(3)\nonumber\\
&& S({\Lambda}){\longrightarrow}{\Lambda}\label{mappi}
\end{eqnarray}
It is a trivial statement that the tangent space at each point $p$ of the base manifold is ${\bf R}^3$ and
therefore at each point $p$ we have a representation $\Lambda$ of $SO(3)$ which is acting naturally , so that the
above map (\ref{mappi}) induces essentially a projection of spin$(3)$ onto ${\bf R}^3$ \cite{4} . Putting it differently
, the fiber of the above bundle ${\cal E}$ is $spin(3){\times}{\bf Z}_2$ , i.e ${\pi}^{-1}(p)=[spin(3){\times}{\bf
Z}_2]p$ .

The second step is to associate with the above bundle ${\cal E}$ , a spinor bundle ${\cal E}_3$ over ${\bf R}_3$ .
Following \cite{nakahara} , this is done by first remarking that the structure group of ${\cal E}$ is spin$(3)$ .
Then by construction the associated bundle , ${\cal E}_3=({\bf E}_3,{\pi}_3,{\bf R}^3)$ , can have as a fiber any
space on which spin$(3)$ is acting on the left . For obvious reasons we choose the Hilbert space of spinors
\begin{equation}
{\cal H}_3={\bf C}^{\infty}({\bf R}^3){\otimes}{\bf C}^2.
\end{equation}
If one now defines the right action of $g{\in}$spin$(3)$ on the space spin$(3){{\times}}{\cal H}_3$ by
\cite{nakahara,last4}
\begin{equation}
(h,\psi)g=(hg,g^{-1}\psi);{\forall} h {\in}spin(3),\psi {\in} {\cal H}_3.
\end{equation}
Then the associated spinor bundle ${\cal E}_3$ has the total space
\begin{equation}
{\bf E}_3=[spin(3){{\times}}{\cal H}_3]/spin(3),
\end{equation}
in other words the two points $(h,\psi)$ and $(hg,g^{-1}\psi)$ are identified . The projection map is , on the
other hand , defined by
\begin{eqnarray}
{\pi}_{3}&:&{\bf E}_{3}{\longrightarrow}{\bf R}^{3}\nonumber\\
&&[(h,\psi)]{\longrightarrow}{\pi}_{3}([(h,\psi)])={\pi}(h),
\end{eqnarray}
where $[$ $]$ denotes equivalence classes . The detail structure of the spinor bundle ${\cal E}_3$ is therefore
given by
\begin{equation}
{\cal H}_3{\longrightarrow}{\bf E}_3{\longrightarrow}{\bf R}^3.
\end{equation}
Its sections are by defintion the spinors ${\psi}$ which are two components wave functions:

\begin{equation}
{\psi}= \left(
\begin{array}{c}
{\psi}_{+}\\
{\psi}_{-}
\end{array}\right)\,
\end{equation}
where both  ${\psi}_{+}$  and  $ {\psi}_{-}$  are  in  ${\bf C}^{\infty}({\bf R}^3)$ .  $ {\psi} $  itself is in
the Hilbert space $ {\cal H }_3  $ .

Finally we can view  ${\bf R}^{3}-\{0\}$  as a bundle over  ${\bf S}^{2}$  where the fibers are half lines starting
at the center\cite{last3} . Each point on the fiber is then given by its radial distance $r$ . {\it{Therefore the
spinor bundle ${\cal E}_{2}$ over ${\bf S}^{2}$  can be thought of as the subbundle of  ${\cal E}_{3}$  in which
sections are independent of the fiber coordinate $r$}} . In other words the  ${\cal E}_{2}$ fiber is the subspace
${\cal H}_2$ of ${\cal H}_3$ in which wave functions $ {\psi}$ do not depend on the radial coordinate $r$ .

\subsection{Dirac Operators From Spin Connections}
\subsubsection{Generalities}
It is a known result that the Dirac operator , in arbitrary coordinates , on a manifold $ M $ is given by
\cite{4,eguchi,last3}:

\begin{equation}
{\cal D} = - i {\gamma}^{\mu} ({\partial}_{\mu}+\frac{1}{8} {\omega}_{{\mu}ab} [{\gamma}^{a},{\gamma}^{b}] ),
\end{equation}
where  ${\gamma}^{\mu}$  are the generators of the curved Clifford algebra , namely
$\{{\gamma}^{\mu},{\gamma}^{\nu}\}=2g ^{{\mu}{\nu}}$ with  ${{\gamma}^{\mu}}^{2}=1$ and ${{\gamma}^{\mu}}^{ + }=
{\gamma}^{\mu}$ . ${\gamma}^a$'s , on the other hand , are the generators of the flat Clifford algebra which will
be defined now \cite{eguchi} . First one decomposes the metric $ g^{{\mu}{\nu}} $ into tetrads as follows ,
$g_{{\mu}{\nu}}={\eta}_{a b}e^{a}_{\mu}e^{b}_{\nu}$ and ${\eta}^{a b}=g^{{\mu}{\nu}}  e^{a}_{\mu} e^{b}_{\nu}$ ,
where ${\eta}_{ab} $ is the flat metric ${\delta}_{ab}$ . The generators ${\gamma }^{a}$ of the flat Clifford
algebra are then defined by ${\gamma}^{\mu} ={\gamma}^{a}E_{a}^{\mu}$ where $E_{a}^{\mu}$ is the inverse of
$e^{a}_{\mu}$  given  by $ E^{\mu}_{a}={\eta}_{a b} g^{{\mu}{\nu}} e^{b}_{\nu}$ . This $E^{\mu}_{a}$ satisfies
therefore  the following equations $E^{\mu}_ae^{b}_{\mu}={\delta}_{a}^{b}$ and
${\eta}^{ab}E_{a}^{\mu}E^{\nu}_b=g^{{\mu}{\nu}}$ . Putting all of this in different words one can say that ,
$e^a_{\mu}$ , is the matrix which transforms the coordinate basis $dx^{\mu}$ of the cotangent bundle $T^{*}_x(M)$
to the orthonormal basis $e^{a}=e^{a}_{\mu}dx^{\mu}$ . Whereas , $E_{a}^{\mu}$ , is the matrix transforming the
basis ${\partial}/{\partial}x^{\mu}$ of the tangent bundle $T_x(M)$ to the orthonormal basis
$E_a=E_a^{\mu}\frac{{\partial}}{{\partial}x^{\mu}}$ . The above Dirac operator can then be rewritten as
\begin{equation}
{\cal D}=-i{\gamma}^{a}E_{a}^{\mu}({\partial}_{\mu} + \frac{1}{8}{\omega}_{{\mu}ab}[{\gamma}^{a},{\gamma}^{b}] )
\end{equation}
$ {\omega}_{{\mu} a b} $ , in all the above equations , is the affine spin connection one-form which will be
defined below .

All the differential geometry of the manifold $M$ is completely coded in the two following tensors : the curvature
two-form tensor $ R^{a}_{b} $ and the torsion two form-tensor $ T^{a} $ . They are given by Cartan's structure
equations:

\begin{eqnarray}
R^{a}_{b}&=&d{\omega}^{a}_{b} + {\omega}^{a}_{c}{\wedge}{\omega}^{c}_{b} {\equiv} \frac{1}{2}   R^{a}_{ b c d}
e^{c}{\wedge}e^{d}\nonumber\\
T^{a}&=&d e^{a} + {\omega}^{a}_{b}{\wedge} e^{b} {\equiv} \frac{1}{2}  T^{a}_{b c} e^{b}{\wedge} e^{c},
\end{eqnarray}
where ${\omega}^{a}_{b}$ means ${\omega}^a_b={\omega}^{a}_{b{\mu}}dx^{\mu}$ .

The Levi-Civita connection or Christoffel symbol , ${\Gamma}^{\mu}_{{\alpha}{\beta}}$ , on the manifold $M$ is
determined by the two following conditions . First one must require that the metric is covariantly constant ,
namely
$g_{{\mu}{\nu};\alpha}={\partial}_{\alpha}g_{{\mu}{\nu}}-{\Gamma}^{\lambda}_{{\alpha}{\mu}}g_{{\lambda}{\nu}}-{\Gamma}^{\lambda}_{{\alpha}{\nu}}g_{{\mu}{\lambda}}=0$
. Secondly , one requires that there is no torsion , i.e
$T^{\mu}_{{\alpha}{\beta}}=\frac{1}{2}({\Gamma}^{\mu}_{{\alpha}{\beta}}-{\Gamma}^{\mu}_{{\beta}{\alpha}})=0$ . The
Levi-Civita connection is then uniquely determined to be
${\Gamma}^{\mu}_{{\alpha}{\beta}}=\frac{1}{2}g^{{\mu}{\nu}}({\partial}_{\alpha}g_{{\nu}{\beta}}+{\partial}_{\beta}g_{{\nu}{\alpha}}-{\partial}_{\nu}g_{{\alpha}{\beta}})$
.

In the same way the Levi-Civita spin connection is obtained by restricting the affine spin connection
${\omega}_{ab}$ to satisfy the metricity and the no-torsion conditions respectively :

\begin{eqnarray}
{\omega}_{ab} + {\omega}_{ba} &=& 0\nonumber\\
de^{a} + {\omega}^{a}_{b}{\wedge}e^{b} &=& 0\label{metricitynotorsion}
\end{eqnarray}

\subsubsection{The Levi-Civita Spin Connection on ${\bf S}^2$}

The metric is given by:

\begin{equation}
ds^{2} = {\rho}^{2} {d {\theta}}^{2} + {\rho}^{2} sin^{2}{\theta} d{\phi}^{2}.
\end{equation}
But since , $ ds^{2} = g_{{\mu} {\nu}} dx^{\mu} dx^{\nu} = {\eta}_{ a b } e^{a}_{\mu} e^{b}_{\nu} dx^{\mu} dx^{\nu}
= {\eta} _{a b} e^{a} e^{b} =  \sum ( e^{a})^{2} $ , one can easily find that $e^{1}={\rho}d{\theta}$ ,
$e^{2}={\rho} sin{\theta}d{\phi}$ and therefore $e^{1}_{\theta}={\rho}$ , $e^{1}_{\phi}=0$ , $e^{2}_{\theta}=0 $
and $e^{2}_{\phi}={\rho} sin{\theta}$ . It is not then difficult to see that $de^1=0$ and
$de^2={\rho}cos{\theta}d{\theta}{\wedge}d{\phi}$ .

Similarly since , $ E^{\mu}_{a} = {\eta}_{a b} g^{{\mu} {\nu}} e^{b}_{\nu} = g^{{\mu}{\nu}} e^{a}_{\nu} $ , we have
$E^{\theta}_{a}=g^{{\theta}{\theta}} e^{a}_{\theta} = \frac{1}{{\rho}^{2}} e^{a}_{\theta}$ which leads to
$E^{\theta}_{1}=\frac{1}{\rho}$ and $E^{\theta}_{2}=0 $ . Also we have $ E^{\phi}_{a}= g^{{\phi}{\phi}}
e^{a}_{\phi} = {\frac{1}{{\rho}^{2} sin^{2}{\theta}}} e^{a}_{\phi}$ which means  $E^{\phi}_{1}= 0$ and
$E^{\phi}_{2} = \frac{1}{{\rho} sin{\theta}}$ .

From the pair of equations (\ref{metricitynotorsion}) we can find :

\begin{eqnarray}
de^{1}& =& - {\omega}_{1b}{\wedge} e^{b}=-{\omega}_{12}{\wedge}e^{2}=-{\rho} sin{\theta} {\omega}_{12}{\wedge}d{\phi}\nonumber\\
de^{2}& =& - {\omega}_{2b}{\wedge} e^{b}=-{\omega}_{21}{\wedge}e^{1}=-{\rho} {\omega}_{21}{\wedge}d{\theta},\nonumber\\
\end{eqnarray}
from which one can immediately deduce that
\begin{eqnarray}
{\omega}_{21} &= &cos{\theta} d{\phi}\nonumber\\
&or&\nonumber\\
{\omega}_{2 1 {\phi}}& =& - {\omega}_{1 2 {\phi}} = cos {\theta}.
\end{eqnarray}

\subsubsection{The Levi-Civita Spin Connection on ${\bf R}^3$}

The metric now is given by:

\begin{equation}
ds^{2} = dr^{2} + r^{2} {d {\theta}}^{2} + r^{2} sin^{2}{\theta} d{\phi}^{2},
\end{equation}
and therefore $e^{1} = r d{\theta}$ , $e^{2}=rsin{\theta}d{\phi}$ and  $e^{3}= dr$ which means that we have the
following non-vanishing components $e^{1}_{\theta}=r$ , $e^{2}_{\phi}=r sin{\theta}$ , $e^{3}_{r}=1$ . It is also
trivial to check that $de^{1}=dr {\wedge} d{\theta}$ , $de^{2}=r cos{\theta} d{\theta}{\wedge}d{\phi} +
sin{\theta} dr {\wedge}d{\phi}$ and $de^{3}=0$ .

One can also compute that $E^{\theta}_{a}=g^{{\theta}{\theta}} e^{a}_{\theta} = \frac{1}{r^{2}} e^{a}_{\theta}$
which leads to the components $E^{\theta}_{1}=\frac{1}{r}$ , $E^{\theta}_{2}=0$ and $E^{\theta}_{3}=0$ .
$E^{\phi}_{a}=g^{{\phi}{\phi}} e^{a}_{\phi} = {\frac{1}{r^{2} sin^{2}{\theta}}} e^{a}_{\phi}$ which leads to
$E^{\phi}_{1}=0$ , $E^{\phi}_{2}=\frac{1}{r sin{\theta}}$ and $E^{\phi}_{3} = 0$ . Finally $E^{r}_{a}=g^{r r}
e^{a}_{r} =  e^{a}_{r}$ leads to $E^{r}_{1}=0$ , $E^{r}_{2} = 0$ and $E^{r}_{3} = 1$ .

In this case equations (\ref{metricitynotorsion}) will read

\begin{eqnarray}
de^{1}& =& - {\omega}_{1b}{\wedge} e^{b}=-{\omega}_{12}{\wedge}e^{2} - {\omega}_{13}{\wedge} e^{3}=-r sin{\theta} {\omega}_{12}{\wedge}d{\phi} -{\omega}_{13}{\wedge} dr\nonumber\\
de^{2}& =& - {\omega}_{2b}{\wedge} e^{b}=-{\omega}_{21}{\wedge}e^{1} -{\omega}_{23}{\wedge} e^{3}=-r {\omega}_{21}{\wedge}d{\theta} - {\omega}_{23}{\wedge} dr\nonumber\\
de^{3}&=& - {\omega}_{3b}{\wedge} e^{b}=-{\omega}_{31}{\wedge}e^{1} - {\omega}_{32}{\wedge}e^{2} = -r {\omega}_{31}{\wedge}d{\theta} -r sin{\theta} {\omega}_{32}{\wedge} d{\phi},\nonumber\\
\end{eqnarray}
from which one deduce
\begin{eqnarray}
{\omega}_{21}& =& cos{\theta} d{\phi}\nonumber\\
{\omega}_{23}& =& sin{\theta} d{\phi}\nonumber\\
{\omega}_{13}& =&d {\theta}\nonumber\\
\end{eqnarray}
In other words , ${\omega}_{2 1 {\phi}}=- {\omega}_{1 2 {\phi}} = cos{\theta}$ , ${\omega}_{2 3 {\phi}}=-
{\omega}_{3 2 {\phi}} = sin{\theta}$ , ${\omega}_{1 3 {\theta}}=- {\omega}_{3 1 {\theta}} = 1$ , and all the others
are zero.

\subsubsection{Evaluation of Dirac operators}

Now we are in the position to calculate the Dirac operator both on the sphere ${\bf S}^{2}$ and on $ {\bf R}^{3} $
.

On the sphere it reads

\begin{eqnarray}
{\cal D}_{2}& = & - i {\gamma}^{a} E^{\mu}_{a} ( {\partial}_{\mu} + {\frac{1}{4}}
{\omega}_{{\mu} a b}  {\gamma}^{a}  {\gamma}^{b} )\nonumber\\
& = & - i {\gamma}^{a} E^{\theta}_{a} ( {\partial}_{\theta} + {\frac{1}{4}} {\omega}_{{\theta} a b}  {\gamma}^{a}
{\gamma}^{b} ) - i {\gamma}^{a} E^{\phi}_{a} ( {\partial}_{\phi} + {\frac{1}{4}}
{\omega}_{{\phi} a b}  {\gamma}^{a}  {\gamma}^{b} )\nonumber\\
& = & - i {\frac{{\gamma}^{1}}{\rho}} ( {\partial}_{\theta} ) - i \frac{{\gamma}^{2}}{{\rho} sin{\theta} } (
{\partial}_{\phi} + {\frac{1}{2}}
cos{\theta}  {\gamma}^{2} {\gamma}^{1} )\nonumber\\
& = & - i {\frac{{\gamma}^{1}}{\rho}} ({\partial}_{\theta} + {\frac{1}{2}}ctg{\theta} ) -
i{\frac{{\gamma}^{2}}{{\rho} sin{\theta}}}
{\partial}_{\phi}.\nonumber\\
\end{eqnarray}
Clearly  $ ( {\cal D}_{2} )^{+} = {\cal D}_{2} $ .

On ${\bf R}^3$ , on the other hand , one must have :

\begin{eqnarray}
{\cal D}_{3}& = & - i {\gamma}^{a} E^{\mu}_{a} ( {\partial}_{\mu} + {\frac{1}{4}}
{\omega}_{{\mu} a b}  {\gamma}^{a}  {\gamma}^{b} )\nonumber\\
& = & - i {\gamma}^{a} E^{\theta}_{a} ( {\partial}_{\theta} +  {\frac{1}{4}} {\omega}_{{\theta} a b}
{\gamma}^{a}  {\gamma}^{b} ) - i {\gamma}^{a} E^{\phi}_{a} ( {\partial}_{\phi} + {\frac{1}{4}} {\omega}_{{\phi} a
b}  {\gamma}^{a}  {\gamma}^{b} ) - i {\gamma}^{a} E^{r}_{a} ( {\partial}_{r} + {\frac{1}{4}}  {\omega}_{ r a b}
{\gamma}^{a}  {\gamma}^{b} )\nonumber\\
& = & - i {\frac{{\gamma}^{1}}{r}} ( {\partial}_{\theta}  + {\frac{1}{2}} {\gamma}^{1} {\gamma}^{3} ) - i
\frac{{\gamma}^{2}}{r sin{\theta} } ( {\partial}_{\phi} +  {\frac{1}{2}}  cos{\theta} {\gamma}^{2} {\gamma}^{1} +
{\frac{1}{2}} sin{\theta} {\gamma}^{2} {\gamma}^{3} ) - i {\gamma}^{3}
{\partial}_{r}\nonumber\\
& = & - i {\frac{{\gamma}^{1}}{r}}  ( {\partial}_{\theta} + {\frac{1}{2}} ctg{\theta} ) - i
{\frac{{\gamma}^{2}}{r\sin{\theta} }}
{\partial}_{\phi} - i {\gamma}^{3} ( {\partial}_{r} + \frac{1}{r} ),\nonumber\\
\end{eqnarray}
Here also we have : $ ( {\cal D}_{3} )^{+} = {\cal D}_{3} $ .

Finally we remark that $ {\cal D}_{3} $ restricted on the sphere is simply related to $ {\cal D}_{2} $ by :

\begin{eqnarray}
{\cal D}_2&=&{\cal D}_{3}{\mid}_{r=\rho} + \frac{i {\gamma}^{3}}{\rho}\nonumber\\ \label{relationr3s2}
\end{eqnarray}

\subsection{The Dirac and Chirality Operators on the Classical Sphere}

Equation (\ref{relationr3s2}) will be always our rule for finding the Dirac operator on ${\bf S}^2$ starting from the Dirac
operator on ${\bf R}^3$ . As we will show , there is an infinite number of Dirac operators on ${\bf S}^2$ which are
all related by $U(1)$ rotations and therefore they are all equivalent . The generator of these rotations is given
by the chirality operator ${\gamma}$ on the sphere which is defined by
\begin{equation}
{\gamma}=\vec{\sigma}.\vec{n}={\gamma}^{+};{\gamma}^2=1;{\gamma}{\cal D}_{2{\theta}}+{\cal
D}_{2{\theta}}{\gamma}=0.\label{chiralitydefinition}
\end{equation}
${\cal D}_{2{\theta}}$ is the Dirac operator on the sphere which is obtained from a reference Dirac operator
${\cal D}_{2g}$ by the transformation
\begin{eqnarray}
{\cal D}_{2{\theta}}&=&exp(i{\theta}{\gamma}){\cal D}_{2g}exp(-i{\theta}{\gamma})\nonumber\\
&=&(cos2{\theta}){\cal D}_{2g}+i(sin2{\theta}){\gamma}{\cal D}_{2g}.\label{diracunitary}
\end{eqnarray}
$\vec{n}$ in (\ref{chiralitydefinition}) is defined by $\vec{n}=\vec{x}/{\rho}$ .

Now what we would like to do is to find algebraic expressions of the Dirac operator ${\cal D}_2$ , in other words
global expressions with no reference to any local coordinates on the sphere ${\bf S}^2$ . There are two different
methods to do this which lead to two seemingly different Dirac operators , ${\cal D}_{2g}$ and ${\cal D}_{2w}$ .
${\cal D}_{2g}$ stands for the Dirac operator due to Grosse et al \cite{5,6,7,8,9,last3} , whereas ${\cal D}_{2w}$
stands for the Dirac operator due to the Watamuras \cite{10,11} . These two Dirac operators can also be shown to be
unitarily equivalent .

\subsubsection{The Dirac Operator ${\cal D}_{2g}$}
We start with the standard Dirac operator on $ {\bf R}^{3} $ :

\begin{equation}
{\cal D}_{3}=- i  {\sigma}_{i}  {\partial}_{i},
\end{equation}
where  ${\sigma}_{i}$   are the Pauli matrices . Now by defining , ${\gamma}_r=\frac{\vec{\sigma}.\vec{x}}{r}$ ,
one can use the identity , ${\gamma}_r^2=1$ , to rewrite ${\cal D}_3$ as
\begin{eqnarray}
{\cal D}_{3}&=&{\gamma}_r^{2}{\cal D}_{3}
=({\frac{\vec{\sigma}.\vec{x}}{r}})({\frac{\vec{\sigma}.\vec{x}}{r}})(- i {\sigma}_{i}{\partial}_{i})=- i {\frac{{\gamma}_r}{r}}(x_{i}{\partial}_{i} + i{\epsilon}_{kij}{\sigma}_{k}x_{i}{\partial}_{j}).\nonumber\\
&&
\end{eqnarray}
Recalling that ${\cal L}_{k}=-i{\epsilon}_{kij} x_{i} {\partial}_{j}$ one can finally find
\begin{equation}
{\cal D}_{3}=- i {\gamma}_r( {\partial}_{r} - {\frac {\vec{\sigma}.\vec{\cal L}}{r}}).\label{GKP1}
\end{equation}
It is a very instructive excercise to check explicitly that the operator $ {\cal D}_{3} $, written in the form
(\ref{GKP1}), is selfadjoint . To this end one introduce the following operator, $\vec{P}=- i \vec{\partial}=- i (
{\vec{n}}. {\partial}_r + {\vec{\theta}} {\frac{1}{r}} {\partial}_{\theta} +  {\vec{\phi}} {\frac{1}{r
sin{\theta}}}{\partial}_{\phi} )$ , which satisfies the identities $ \vec{n}.\vec{P}=-i {\partial}_{r} $ and
$\vec{P}.\vec{n}=- \frac{2i}{r}$ .

Now the operator ${\cal D}_{a3}$ defined by ${\cal D}_{a3}=- i {\gamma}_r {\partial}_{r}={\gamma}_r
\vec{n}.\vec{P}$ is not self-adjoint because $( {\cal D}_{a3})^{ + }= \vec{P}.\vec{n} {\gamma}_r=\vec{P} ( \vec{n}
{\gamma}_r ) + ( \vec{n} {\gamma}_r ) \vec{P}=\vec{P}( \vec{n}) {\gamma}_r + \vec{n}.\vec{P}( {\gamma}_r)+ (
\vec{n} {\gamma}_r ) \vec{P}= -  {\frac{2 i{\gamma}_r}{r}} + {\cal D}_{a3}$ , where we have used the fact that
$\vec{n} \vec{P} ({\gamma}_r)= - i {\partial}_{r} ({\gamma}_r)=-i\vec{\sigma} {\partial}_{r} ( \vec{n} )=0$.

In the same way the operator ${\cal D}_{b3}$ defined by ${\cal D}_{b3}=i {\gamma}_r{\frac {\vec{\sigma}.\vec{\cal
L}}{r}}$ is also not self-adjoint because $( {\cal D}_{b3} )^{+}=-{\frac{i}{r}} {\sigma}_{i} {\cal L}_{i}
{\gamma}_r=-{\frac{i}{r}} {\sigma}_{i} ( {\cal L}_{i} ({\gamma}_r) + {\gamma}_r {\cal L}_{i} ) =-{\frac{i}{r}}
{\sigma}_{i} ( \frac{i{\epsilon}_{ijk}{\sigma}_jx_k}{r}+{\gamma}_r{\cal
L}_i)=\frac{2i{\gamma}_r}{r}-\frac{i{\sigma}_i}{r}{\gamma}_r{\cal L}_i$ . Then by using the identity
$-\frac{i{\sigma}_i}{r}{\gamma}_r{\cal L}_i=\frac{i}{r}{\gamma}_r\vec{\sigma}.\vec{\cal L}$ one can find that
${\cal D}_{b3}^+={\cal D}_{b3}+\frac{2i{\gamma}_r}{r}$ and therefore we get ${\cal D}_3^+=({\cal D}_{a3}+{\cal
D}_{b3})^+={\cal D}_{3}$ which is what we want .

On the sphere ${\bf S}^2$ the Dirac operator will be simply given by

\begin{eqnarray}
{\cal D}_{2} & = & {\cal D}_{3}{\mid}_{r=\rho} + i \frac{{\gamma}^{3}}{\rho}\nonumber\\
 & = & i {\gamma}{\cal D}_{2g},\nonumber\\
\end{eqnarray}
where ${\cal D}_{2g}$ is the Grosse-Klim\v{c}\'{i}k-Pre\v{s}najder Dirac operator given by
\begin{equation}
{\cal D}_{2g}=\frac{1}{\rho}(\vec{\sigma}.\vec{\cal L}+1).\label{GKP}
\end{equation}
In above equations we have chosen ${\gamma}^3={\gamma}$ .

\subsubsection{The Dirac Operator ${\cal D}_{2w}$}

Another global expression for the Dirac operator ${\cal D}_2$ on the sphere can be found as follows :

\begin{eqnarray}
{\cal D}_{3} & = & - i {\sigma}_{i} {\partial}_{i}\nonumber\\
 & = & - i \vec{\sigma}[ \vec{n} ( \vec{n}. \vec{\partial} ) -  \vec{n}  {\times} (  \vec{n}  {\times}  \vec{\partial}   ) ]\nonumber\\
& = &  - i \vec{\sigma}[ {\frac{\vec{r}}{r}} {\partial}_{r} -
{\frac{i}{r^{2}}} \vec{r} {\times} \vec{\cal L} ] \nonumber\\
 & = & - i  {\gamma}_r {\partial}_{r} - \frac{1}{r^{2}} {\epsilon}_{i j k}
{\sigma}_{i} x_{j} {\cal L}_{k}. \nonumber\\
\end{eqnarray}
A new algebraic expression for ${\cal D}_2$ will then emerge

\begin{equation}
{\cal D}_{2w} =  - {\frac{1}{{\rho}^{2}}} {\epsilon}_{i j k} {\sigma}_{i} x_{j} {\cal L}_{k} + i
{\frac{{\gamma}^{3}}{\rho}}.\label{watamura1}
\end{equation}
Since we have chosen , ${\gamma}^{3} = {\gamma} $ , and by using the identity
$\frac{i{\gamma}}{\rho}=-\frac{1}{{\rho}^2}{\epsilon}_{ijk}{\sigma}_ix_j\frac{{\sigma}_k}{2}$ one can rewrite
equation (\ref{watamura1}) in the form
\begin{equation}
{\cal D}_{2w} = - {\frac{1}{{\rho}^{2}}} {\epsilon}_{i j k} {\sigma}_{i} x_{j} ( {\cal L}_{k} +
\frac{{\sigma}_{k}}{2} ).\label{watamura}
\end{equation}
This is the Watamuras Dirac operator .

From the above construction it is obvious that , ${\cal D}_{2w}=i{\gamma}{\cal D}_{2g}$ , and therefore from
equation , (\ref{diracunitary}) , one can make the following identification ${\cal D}_{2w}={\cal D}_{2\frac{\pi}{4}}$ . A more
general Dirac operator can be obtained from ${\cal D}_{2g}$ by the general transformation (\ref{diracunitary}) . The two Dirac
operators (\ref{GKP}) and (\ref{watamura}) are clearly equivalent because one can show that both operators have the same
spectrum given by ${\pm}\frac{1}{\rho}(j+\frac{1}{2})$ where $j$ is the eigenvalue of the operator $\vec{\cal
J}^2=(\vec{\cal L}+\frac{\vec{\sigma}}{2})^2=j(j+1)$ .

This can be shown as follows : First remark that ${\cal D}_{2w}^2={\cal D}_{2g}^2$\footnote{One can also remark
that $({\rho}{\cal D}_{2g})^2=\vec{\cal L}^2+{\rho}{\cal D}_{2g}$ which does not look very much like the Lichnerowicz relation
$({\rho}{\cal D}_{2g})^2=\vec{\cal L}^2+\frac{1}{4}R$ , where $R$ is the Ricci scalar of the sphere  .}which means that ${\cal
D}_{2g}$ and ${\cal D}_{2w}$ will have the same spectrum . Next one uses the identity $\vec{\cal J}^2=\vec{\cal
L}^2+\frac{1}{4}\vec{\sigma}^2+\vec{\sigma}\vec{\cal L}$ to rewrite $D_{2g}$ in the form ${\cal
D}_{2g}=\frac{1}{\rho}[\vec{\cal J}^2-\vec{\cal L}^2-\frac{1}{4}\vec{\sigma}^2+1]$ from which the above spectrum
trivially follows .

\subsection{Projectors and Winding Numbers}

One can associate with the chirality operator defined in equations (\ref{chiralitydefinition}) two projectors ,$
P_{\pm}=\frac{1\pm{\gamma}}{2}$, which satisfy : $ P_{+}+ P_{-}=1 $, $ P_{\pm}^{2}=P_{\pm} $ and $
P_{\pm}^{+}=P_{\pm} $. More precisely , these two projectors are two idempotents of the algebra $ M_{2}({\cal A})$
which define subbundles , or more precisely projective modules, $P_{+}{\cal E}_2$, $P_{-}{\cal E}_2$ of the
spinor bundle ${\cal E}_2$ over the sphere with fibers $P_{+}{\cal H}_2$ and $P_{-}{\cal H}_2$ respectively. The
connections associated with these projectors are therefore defined, up to anything which commutes with $ P_{\pm}
$ , by

\begin{equation}
{\nabla}_{\pm}=P_{\pm}dP_{\pm}\label{connection}
\end{equation}
These connections also satisfy the following condition : $ {\forall}{\psi} $ $ {\in} $ $ P_{\pm}{\cal H}_{2} $ and
$ {\forall}f{\in} {\cal A} $ we have :

\begin{equation}
{\nabla}_{\pm}({\psi}f)={\nabla}_{\pm}({\psi})f + {\psi}df.
\end{equation}
The corresponding curvature will , on the other hand , be defined by :

\begin{equation}
{\nabla}_{\pm}^{2}=P_{\pm}dP_{\pm}P_{\pm}dP_{\pm}.
\end{equation}
By using the facts that , $[P_{\pm},dP_{\pm}]=0$ and $P_{\pm}^2=P_{\pm}$ , one can compute that
${\nabla}_{\pm}^2=P_{\pm}dP_{\pm}dP_{\pm}=P_{\pm}dP_{\pm}[d(P_{\pm})+P_{\pm}d]=P_{\pm}dP_{\pm}d(P_{\pm}) +
P_{\pm}dP_{\pm}d$ . Now the facts , $d^2=0$ and $dP_{\pm}dP_{\pm}=d(P_{\pm}dP_{\pm})-P_{\pm}dP_{\pm}d$ , will
finally lead to
\begin{equation}
{\nabla}_{\pm}^2=P_{\pm}d(P_{\pm})d(P_{\pm}).\label{curvature}
\end{equation}
Next we define the Chern class as :

\begin{eqnarray}
c_{1}&=&\frac{1}{2i{\pi}} \int Tr{\nabla}_{\pm}^{2}\nonumber\\
&=&\frac{1}{2i{\pi}} \int TrP_{\pm}d(P_{\pm})d(P_{\pm}),\nonumber\\ \label{chernclass}
\end{eqnarray}
where the integral is over the surface of the sphere and the trace is taken over the spin indices . One can
actually calculate these numbers for the spinor bundles $P_{\pm}{\cal E}_2$ and find them to be given by:
\begin{equation}
c_1=\pm 1.\label{c}
\end{equation}
The computation goes as follows . First we have $d(P_{\pm})={\pm}\frac{1}{2\rho}{\sigma}_idx_i$ and therefore
$c_1={\pm}\frac{1}{8{\pi}{\rho}^3}\int_{S^2}{\epsilon}_{kij}x_kdx_i{\wedge}dx_j$ where we have used the identities
$Tr{\sigma}_i{\sigma}_j=2{\delta}_{ij}$ and $Tr{\sigma}_k{\sigma}_i{\sigma}_j=2i{\epsilon}_{ijk}$ . Finally by
using equation (\ref{twoform}) we conclude that $c_1={\pm}1$. $ c_{1}=\pm 1 $ are exactly the winding numbers of the maps
$ \vec{\xi}_{\pm}=({\xi}_{\pm}^{1},{\xi}_{\pm}^{2},{\xi}_{\pm}^{3})={\pm}\vec{n} $ respectively . In terms of Euler
angles these maps are written as ${\xi}_{\pm}^{1}={\pm}sin{\theta}cos{\phi}$,
${\xi}_{\pm}^{2}={\pm}sin{\theta}sin{\phi}$ and ${\xi}_{\pm}^{3}={\pm}cos{\theta}$ .

In a similar way higher winding numbers ${\pm}n $ are defined by the maps $
\vec{\xi}_{{\pm}n}=({\xi}_{{\pm}n}^{1},{\xi}_{{\pm}n}^{2},{\xi}_{{\pm}n}^{3}) $ given by
${\xi}_{{\pm}n}^{1}={\pm}sin{\theta}cosn{\phi}$ , ${\xi}_{{\pm}n}^{2}={\pm}sin{\theta}sinn{\phi}$ and
${\xi}_{{\pm}n}^{3}={\pm}cos{\theta}$. The corresponding projectors $ P_{{\pm}n} $ are , on the other hand ,
given by :

\begin{equation}
P_{{\pm}n} = \frac{1 + \vec{\sigma}.\vec{\xi}_{{\pm}n}}{2}.
\end{equation}
We can easily check that : $ P_{{\pm}n}^{+}=P_{{\pm}n} $ , $ P_{{\pm}n}^{2}=P_{{\pm}n} $ . These projectors define
subbundles $ P_{{\pm}n}{\cal E}_{2} $ of the spinor bundle , ${\cal E}_2$ , over the sphere with fibers given by
the Hilbert spaces $ P_{{\pm}n}{\cal H}_{2} $ . Connections $ {\nabla}_{{\pm}n} $ , curvatures $
{\nabla}_{{\pm}n}^2 $ and Chern class are still given by equations (\ref{connection}) , (\ref{curvature}) and (\ref{chernclass}) respectively
, of course with the substitution $P_{\pm}{\longrightarrow}P_{{\pm}n}$ .

It is instructive to compute the Chern classes $c_1$ for the spinor bundles, $ P_{{\pm}n}{\cal E}_{2} $,
respectively :

\begin{eqnarray}
c_{1}&=&\frac{1}{2i {\pi}} \int Tr P_{{\pm}n}d(P_{{\pm}n})d(P_{{\pm}n})\nonumber\\
&=&\frac{1}{2i{\pi}} \int Tr (\frac{ 1 + {\sigma}_{i}{\xi}_{{\pm}n}^{i}}{2}
)\frac{{\sigma}_{j}d{\xi}_{{\pm}n}^{j}}{2}{\wedge}\frac{{\sigma}_{k}d{\xi}_{{\pm}n}^{k}}{2}\nonumber\\
&=& \frac{1}{8 {\pi}} \int_{S^{2}} {\epsilon}_{i j
k}{\xi}_{{\pm}n}^{i}d{\xi}_{{\pm}n}^{j}{\wedge}d{\xi}_{{\pm}n}^{k}\nonumber\\
&=& \frac{1}{8{\pi}} \int_{Vol(S^{2})} {\epsilon}_{i j
k}d{\xi}_{{\pm}n}^{i}{\wedge}d{\xi}_{{\pm}n}^{j}{\wedge}d{\xi}_{{\pm}n}^{k}\nonumber\\
&=&\frac{1}{8{\pi}} \int_{Vol(S^{2})} {\epsilon}_{i j k}{\epsilon}^{i j k
} d{\xi}_{{\pm}n}^{1}{\wedge}d{\xi}_{{\pm}n}^{2}{\wedge}d{\xi}_{{\pm}n}^{3}\nonumber\\
&=&\frac{6}{8{\pi}} \int_{Vol(S^{2})} d{\xi}_{{\pm}n}^{1}{\wedge}d{\xi}_{{\pm}n}^{2}{\wedge}d{\xi}_{{\pm}n}^{3}\nonumber\\
&=&{\pm}n,\nonumber\\ \label{c1}
\end{eqnarray}
where we have used in the last equation the fact that ,
$d{\xi}_{{\pm}n}^{1}{\wedge}d{\xi}_{{\pm}n}^{2}{\wedge}d{\xi}_{{\pm}n}^{3}={\pm}ndn^1{\wedge}dn^2{\wedge}dn^3$.

\subsubsection{Additivity of Winding Numbers}

Another way of obtaining higher winding numbers is  by taking the tensor product, $P_{k}\otimes P_{l}$ , of two
projectors $P_{k}$ and $P_{l}$ . The product $P_{k}{\otimes}P_{l}$ is also a projector since $
(P_k{\otimes}P_l)^{2}=(P_k{\otimes}P_l) $ and $ (P_k{\otimes}P_l)^{+}=P_k{\otimes}P_l $ . It corresponds to the
spinor bundle $P_{k}{\cal E}_2{\otimes}P_{l}{\cal E}_2$ with the Hilbert space , $P_{k}{\cal
H}_2{\otimes}P_{l}{\cal H}_2$ , as a fiber . A general spinor of $P_{k}{\cal H}_2{\otimes}P_{l}{\cal H}_2$ will be
of the form $P_{k}{\psi}{\otimes}P_{l}{\phi}$ .

The Chern class of the spinor bundle $P_{k}{\cal E}_2{\otimes}P_{l}{\cal E}_2$ will be $k+l$ where $k$ and $l$ are
the winding numbers corresponding to the projectors $P_k$ and $P_l$ respectively . The proof goes as follows .
First one have the following identity ,
 $d(P_k{\otimes}P_l){\wedge}d(P_k{\otimes}P_l)=d(P_k){\wedge}d(P_k){\otimes}P_l+d(P_k)P_k{\wedge}{\otimes}P_ldP_{l}-P_kd(P_k){\wedge}{\otimes}d(P_l)P_l+P_k{\otimes}d(P_l){\wedge}d(P_{l})=
d(P_k){\wedge}d(P_k){\otimes}P_l+P_k{\otimes}d(P_l){\wedge}d(P_{l})$ , from which we obtain ,
$TrP_k{\otimes}P_ld(P_k{\otimes}P_l){\wedge}d(P_k{\otimes}P_l)=TrP_kd(P_k){\wedge}d(P_k)+TrP_ld(P_l){\wedge}d(P_{l})$
, and hence $c_1(P_k{\otimes}P_{l})=k+l=c_1(P_k)+c_1(P_l)$ .

\section{The Fuzzy Sphere ${\bf S}_F^2=({\bf A},{\bf H},D,\Gamma)$}

\subsection{The Complex Structure on ${\bf T}{\bf S}^2$}
It is very helpful to use the complex structure defined on the sphere to rewrite the Dirac operators (\ref{GKP}) and
(\ref{watamura}) and the chirality operator (\ref{chiralitydefinition}) in a form more suitable for fuzzification . As we will see shortly ,
this complex structure will provide essentially a volume form as well as a metric on the tangent space ${\bf T}{\bf
S}^2$ of the sphere .

The complex structure ${\cal J}$ on the space ${\bf T}{\bf S}_{\vec{n}}^2$ , which is tangent to ${\bf S}^2$ at
the point $\vec{n}$ , is introduced by the formula
\begin{equation}
{\cal J}_{ij}={\epsilon}_{ijk}n_k.
\end{equation}
Next one can construct from the above complex structure a projector ${\cal P}_{ij}$ defined by
\begin{eqnarray}
{\cal P}_{ij}&=&-{\cal J}_{ik}{\cal J}_{kj}\nonumber\\
&=&{\delta}_{ij}-n_in_j.
\end{eqnarray}
It can also be rewritten in the form , ${\cal P}_{ij}=(n^iAd {\cal L}_i )^2$ , where $Ad {\cal L}_i$ are the
generators of the $l=1$ adjoint representation of $SU(2)$ defined by $(Ad {\cal L}_i)_{jk}=i{\epsilon}_{ijk}$. A
simple calculation leads to the following basic identities among ${\cal J}_{ij}$ and ${\cal P}_{ij}$
\begin{eqnarray}
{\cal J}_{ij}{\cal J}_{jk}&=&-{\cal P}_{ik}\nonumber\\
{\cal P}_{ij}{\cal P}_{jk}&=&{\cal P}_{ik}\nonumber\\
{\cal P}_{ij}{\cal J}_{jk}&=&{\cal J}_{ik}.
\end{eqnarray}
The ${\cal P}$ is actually a projector on the tangent space ${\bf T}{\bf S}_{\vec{n}}^2$ , in other words the
vector ${\cal P}\vec{\xi}$ is always in ${\bf T}{\bf S}_{\vec{n}}^2$ where $\vec{\xi}$ is any vector in the vector
space ${\cal A}^3={\cal A}{\otimes}{\bf C}^3$ . This can be seen from the fact that ${\cal P}\vec{\xi}$ is , by
construction , perpendicular to the normal vector $\vec{n}$ , i.e $\vec{n}.{\cal P}\vec{\xi}={\cal
P}_{ij}n_i{\xi}_j=-({\cal J}_{ik}n_i){\cal J}_{kj}{\xi}_j=0$ . The scalar product of any two tangent vectors
${\cal P}\vec{\xi}$ and ${\cal P}\vec{\eta}$ will be given by ${\cal P}\vec{\xi}.{\cal P}\vec{\eta}=({\cal
P}_{ij}{\xi}_j)({\cal P}_{ik}{\eta}_k)={\cal P}_{jk}{\xi}_j{\eta}_k$ . This result illustrates the fact that ${\cal
P}_{ij}$ plays the role of a metric on ${\bf T}{\bf S}_{\vec{n}}^2$ . From this and from the identity
$\frac{1}{8{\pi}}\int_{S^2}{\epsilon}_{ijk}n_kdn_i{\wedge}dn_j=1$ , one can see that ${\cal J}_{ij}$ contains much
information on the metric aspects of ${\bf T}{\bf S}_{\vec{n}}^2$ .

More involved calculations show that
\begin{equation}
[{\cal L}_i,{\cal J}_{jk}]=i{\cal J}_{il}{\epsilon}_{ljk}
\end{equation}
and
\begin{equation}
 [{\cal L}_i,{\cal P}_{jk}]=i{\epsilon}_{ijl}{\cal P}_{lk}+i{\epsilon}_{ikl}{\cal P}_{lj}.
\end{equation}
The last commutation relations follow from the Jacobi identity $[{\cal L}_j,[{\cal L}_i,{\cal J}_{kl}]]+[{\cal
J}_{kl},[{\cal L}_j,{\cal L}_i]]+[{\cal L}_i,[{\cal J}_{kl},{\cal L}_j]]=0{\Longrightarrow}{\epsilon}_{ilk}{\cal
J}_{kj}+{\epsilon}_{jik}{\cal J}_{kl}+{\epsilon}_{ljk}{\cal J}_{ki}=0$. It is proven by first rewriting it in the
form , $[{\cal L}_i,P_{jk}]=-i{\cal J}_{il}{\cal J}_{hk}{\epsilon}_{ljh}+i{\cal J}_{jl}{\cal
J}_{hi}{\epsilon}_{lkh}$. Then Jacobi identity gives , $[{\cal L}_i,P_{jk}]=i{\epsilon}_{ijl}P_{lk}+i{\cal
J}_{jl}{\cal J}_{hk}{\epsilon}_{ilh}+i{\cal J}_{jl}{\cal J}_{hi}{\epsilon}_{lkh}$. Jacobi identity is used once
more to recombine the last two terms in this last equation and obtain finally the desired result .

From the complex structure ${\cal J}_{ij}$ and the projector ${\cal P}_{ij}$ one can construct the projectors
${\cal P}_{ij}^{+}$ and ${\cal P}_{ij}^{-}$, on the holomorphic and antiholomorphic parts respectively, of the
tangent space ${\bf T}{\bf S}_{\vec{n}}^2$ . They are given by
\begin{equation}
{\cal P}^{\pm}_{ij}=\frac{1}{2}({\cal P}_{ij}{\pm}i{\cal J}_{ij}),
\end{equation}
It is easy to check that ${\cal P}^{\pm}_{ij}{\cal P}^{\pm}_{jk}={\cal P}^{\pm}_{ik}$ , ${\cal P}^{\pm}_{ij}{\cal
P}^{\mp}_{jk}=0$ , ${\cal P}^{\pm}_{ij}{\cal P}_{jk}={\cal P}^{\pm}_{ik}$ and ${\cal P}^{\pm}_{ij}{\cal
J}_{jk}={\mp}i{\cal P}^{\pm}_{k}$ .

Finally the chirality operator (\ref{chiralitydefinition}) and the Dirac operators (\ref{GKP}) and (\ref{watamura}) can easily be expressed in
terms of the complex structure ${\cal J}_{ij}$ and the projector ${\cal P}_{ij}$ as
\begin{equation}
{\gamma}=-\frac{i}{2!}{\cal J}_{ij}{\sigma}_i{\sigma}_j.\label{chiralitydefinitioncs}
\end{equation}
and
\begin{eqnarray}
{\cal D}_{2g}&=&\frac{1}{\rho}{\sigma}_i{\cal P}_{ij}({\cal L}_j+\frac{1}{2}{\sigma}_j)\nonumber\\ 
{\cal D}_{2w}&=&\frac{1}{\rho}{\cal J}_{ij}{\sigma}_i({\cal L}_j+\frac{{\sigma}_j}{2}).\label{GKPwatamuracs}
\end{eqnarray}
\subsection{The Classical Sphere ${\bf S}^2$ as the K-cycle $({\cal A},{\cal H},{\cal D},{\gamma})$}

In this section we will briefly summarize all our results so far concerning the classical sphere ${\bf S}^2$ . This
will be done through the introduction of the K-cycle $({\cal A},{\cal H},{\cal D},{\gamma})$ defining ${\bf S}^2$ .
It is a theorem due to Connes \cite{1} that all the geometry of ${\bf S}^2$ is encoded in this K-cycle . The first
element of this K-cycle is the algebra ${\cal A}$ of smooth functions on ${\bf S}^2$ defined by equations (\ref{calgebra1})
and  (\ref{calgebra2}) which can be rewritten differently as

\begin{eqnarray}
{\cal A}&=&\{ f({\hat{x}})=\sum_{i_1,...,i_n}f_{i_1...i_k}\hat{x}_{i_1}...\hat{x}_{i_k}\},\nonumber\\
\end{eqnarray}
where
\begin{eqnarray}
\sum_{i=1}^3\hat{x}_i^2={\rho}^2,\nonumber\\
\end{eqnarray}
and
\begin{eqnarray}
[\hat{x}_i,\hat{x}_j]&=&0.
\end{eqnarray}
The operators $\hat{x}_i$ act on the Hilbert space ${\cal H}$ , which is generated by the vectors $\{|\vec{x}>\}$ ,
in such a way that we have
\begin{equation}
\hat{x}_i|\vec{x}>=x_i|\vec{x}> .
\end{equation}
In other words ${\cal H}$ provides the representation space of the algebra ${\cal A}$ .

The other two elements of the K-cycle $({\cal A},{\cal H},{\cal D},{\gamma})$ are the chirality operator
${\gamma}$ and the Dirac operator ${\cal D}$ which are given , in the case of ${\bf S}^2$ , by equations (\ref{chiralitydefinitioncs}) and
(\ref{GKPwatamuracs}) respectively\footnote{For odd dimensional manifolds , a chirality operator does not exist and the K-cycle
describing the manifold will consist of only three elements : an algebra ${\cal A}$ , a representation space ${\cal
H}$ and a Dirac operator $D$ .}.

\subsection{Quantization : $({\cal A},{\cal H},{\cal D},{\gamma}){\longrightarrow}({\bf A},{\bf H},D,{\Gamma})$}

Now what we would like to do is to discretise the sphere ${\bf S}^2$ a la fuzzy . In other words quantize ${\bf
S}^2$ and obtain naturally a discrete sphere ${\bf S}^2_F$ containing a finite number of points . In lattice
physics , when we directly replace the sphere by a lattice of points , we generally break explicitly all the
symetries of the problem . In here such situations are completely avoided because the discretisation is achieved
by quantizing the Wess-Zumino Lagrangian (\ref{wz}) given by
\begin{equation}
L=liTr({\sigma}_3g^{-1}\dot{g});g{\in}SU(2).\label{wz1}
\end{equation}
Of course , points $\vec{n}=\vec{x}/{\rho}$ of ${\bf S}^2$ are now identified with the equivalence classes
$[gexp(i{\sigma}_3{\theta}/2)]$ of $SU(2)/U(1)$ , because of the Hopf fibration
$\vec{x}.\vec{\sigma}={\rho}g{\sigma}_3g^{-1}$ .

As we have already shown in great detail in the introduction , the quantization of the above Lagrangian
leads to the fuzzy sphere ${\bf S}^2_F$ . The observables of (\ref{wz1}) are given by equation (\ref{expansionoffuzzys2}) or
\begin{equation}
f(L)=\sum_{i_i,...,i_k}f_{i_1...i_k}L_{i_1}...L_{i_k},\label{expansionoffuzzys21}
\end{equation}
where the operators $L_i$ are the generators of the $(2l+1)-$dimensional irreducible representation $l$ of $SU(2)$
, in other words
\begin{equation}
\sum_{i=1}^3L_i^2=l(l+1),\label{casimir}
\end{equation}
and
\begin{equation}
[L_i,L_j]=i{\epsilon}_{ijk}L_k.\label{su2commutations}
\end{equation}
The operators $n_i^F$ defined by $n_i^F=\frac{L_i}{\sqrt{l(l+1)}}$ and
$[n_i^F,n_j^F]=\frac{i}{\sqrt{l(l+1)}}{\epsilon}_{ijk}n_k^F$ are the fuzzy coordinate functions on the fuzzy sphere
${\bf S}^2_F$\footnote{Recall that the $L_i$'s are the quantum operators associated with the $t_i\approx l.n_i$ ,
see Introduction.} . From here one can also see that the noncommutativity parameter ( or the deformation
parameter ) characterizing the fuzziness of ${\bf S}^2_F$ is $ \hat{{\hbar}} = \frac{1}{\sqrt{l(l+1)}} $ . The
commutators $[n_i^F,n_j^F]$ will then approach zero when $l{\longrightarrow}{\infty}$ or equivalently
$\hat{\hbar}{\longrightarrow}0$ , which is the continuum limit .

Fuzzy functions on ${\bf S}^2_F$ are provided by equation (\ref{expansionoffuzzys21}) , whereas fuzzy points can be thought of as the
eigenvalues of the operators $L_i$ . The number of points on ${\bf S}^2_F$ is therefore finite and equal to $2l+1$
. The fact that the operators $L_i$ can not be diagonalized simultaneously reflects the property that we can not
completely localize points on ${\bf S}^2_F$ .

Given the algebra ${\bf A}$ of all operators $f(L)$ , one can define the regular representation ${\Pi }({\bf A})$  to be generated by a set of elements $L_i^L$ defined by
\begin{equation}
{\forall} f {\in} {\bf A} : L^{L}_{i} f = L_{i}f.\label{leftaction}
\end{equation}
These elements will clearly satisfy equations (\ref{casimir}) and (\ref{su2commutations}) . From the definition (\ref{leftaction}) one also see
that ${\Pi}({\bf A})$ provides a representation  of ${\bf A}$ which is acting on the left of ${\bf A}$.

 By construction the algebra ${\bf A}$ , which is the first element of the K-cycle $({\bf A},{\bf H},D,{\Gamma})$ defining the fuzzy sphere ${\bf S}^2_F$ ,
acts on the representation space ${\bf H}$ . The latter, which is the second element of the
K-cycle $({\bf A},{\bf H},D,{\Gamma})$ , was found in the introduction to be the Hilbert space of all the physical
wave functions (\ref{physicalwaves}) of the Lagrangian (\ref{wz1}) . It is the Hilbert space spanned by the basis of the
irreducible representation $l$ of $SU(2)$ , namely by
\begin{equation}
L^{2} | lm > = l(l + 1)| lm >  ,  L_{3} | lm > = m | lm >
\end{equation}

A smooth global vector field on $ {\bf S}^{2} $ is  a derivation of the algebra $ {\cal A}$ . Similarly , a smooth
global vector field on $ {\bf S}_F^{2} $ is a derivation of the algebra  $ {\bf A}$ , in other words a linear map
$X : {\bf A} {\longrightarrow} {\bf A}$ which satisfies the Leibnitz rule, $ X ( f g ) = X (f ) g + f X(g) $. One
such derivation is given by the adjoint action of the group $SU(2)$ , which is generated by $adL_{i}$ , on the
algebra ${\bf A}$ . It is defined by

\begin{eqnarray}
ad{L_{i}} ( f )& =&[ L_{i},f ]\nonumber\\
&=& L_{i} f - f L_{i}\nonumber\\
&=& ( L_{i} - L^{R}_{i} ) f,\nonumber\\ \label{adjointaction}
\end{eqnarray}
where $ L^{R}_{i} $ are the generators of the opposite representation , ${\Pi}^0({\bf A})$ , of $ {\bf A }$ which
act by right multiplication on ${\bf A}$ ,

\begin{equation}
{\forall} f {\in} {\bf A} : L^{R}_{i} f = f L_{i}.\label{rightaction}
\end{equation}
One can easily check that they satisfy
\begin{eqnarray}
\sum_{i=1}^3(L_i^{R})^2&=&l(l+1)\nonumber\\
\end{eqnarray}
and
\begin{eqnarray}
[L_i^R,L_j^R]&=&-i{\epsilon}_{ijk}L_k^R.\label{rsu2commutations}
\end{eqnarray}
$ad L_i$ defined in equation (\ref{adjointaction}) are the fuzzy analogue of the classical derivations ${\cal L}_i$ given by
${\cal L}_i=-i{\epsilon}_{ijk}x_j{\partial}_k$ .

\subsection{The Fuzzy Dirac Operators and The Fuzzy Chirality Operator}

We have already shown that in the continuum , spinors ${\psi}$ belong to the fiber ${\cal H}_2$ of the spinor
bundle ${\cal E}_2$ over the sphere . The precise definitions were given in section $(3.1.2)$ . ${\cal H}_2$ is
essentially a left ${\cal A}-$module , in other words if $f{\in}{\cal A}$ and ${\psi}{\in}{\cal H}_2$ then
$f{\psi}{\in}{\cal H}_2$ . ${\cal H}_2$ can also be thought of as the vector space ${\cal H}_2={\cal
A}{\otimes}{\bf C}^2$ . The noncommutative analogue of the projective module ${\cal H}_2$ is the projective module
${\bf H}_2={\bf A}{\otimes}{\bf C}^2$ . This is clearly an ${\bf A}-$bimodule since there is a left as well as a
right action on the space of spinors ${\bf H}_2$ by the elements of the algebra ${\bf A}$ . The exact definition of
fuzzy spinors will be presented in the next two chapter . In here we only uses these observations to conclude that
the fuzzy Dirac operators and the fuzzy chirality operator must be defined in such a way that they act on the
Hilbert space $ {\bf H}_2 $ . The Dirac operators must , on the other hand , anticommute with the fuzzy chirality
operator . They must be , of course , selfadjoint and reproduce the continuum Dirac operators in the limit
$l{\longrightarrow}{\infty}$ . In the same way , the fuzzy chirality operator should reduce to the classical one
in the continuum limit . We then must have the following requirements
\begin{eqnarray}
&a)&({\Gamma})^{2}=1\nonumber\\
&b)& {\Gamma}D_{2}=-D_2{\Gamma}\nonumber\\
&c)&({\Gamma})^{+}={\Gamma}\nonumber\\
&d)&(D_2)^{+}=D_2,
\end{eqnarray}
and
\begin{eqnarray}
&e)&Lim(D_2)_{l{\longrightarrow}{\infty}}={\cal D}_2\nonumber\\
&f)&Lim({\Gamma})_{l{\longrightarrow}{\infty}}={\gamma}.
\end{eqnarray}
\subsubsection{Fuzzy ${\gamma}$ (or ${\Gamma})$}

To get the discrete version of ${\gamma}$ one first simply replaces ${\vec n}$ in equation (\ref{chiralitydefinition}) by $
{\vec{n}}^F=\frac{\vec{L}}{\sqrt{l(l+1)}}$ to get

\begin{equation}
{\gamma}^F = {\frac{\vec{\sigma}.\vec{L}}{\sqrt{l(l+1)}}}.
\end{equation}
We can check that this ${\gamma}^F$ does not square to $1$ , $({\gamma}^F)^{2}=
{\frac{1}{l(l+1)}}{\sigma}_{i}{\sigma}_{j}L_{i}L_{j}={\frac{1}{l(l+1)}} (  {\delta}_{i j} + i{\epsilon}_{i j k}
{\sigma}_{k} ) L_{i}L_{j}=1 + {\frac{i}{2l(l+1)}} {\epsilon}_{i j k}{\sigma}_{k}[L_{i},L_{j}]=1 -
{\frac{1}{2l(l+1)}} {\epsilon}_{i j k}{\epsilon}_{i j l} {\sigma}_{k}L_{l}=1 - \frac{{\gamma}^F}{\sqrt{l(l+1)}}$.
But we can notice that $({\gamma}^F)^{2} + \frac{{\gamma}^F}{\sqrt{l(l+1)}}=1$ and therefore $( {\gamma}^F +
\frac{1}{2{\sqrt{l(l+1)}}} )^{2}=\frac{1}{4l(l+1)}+1 $ which can be rewritten as :$ ( \frac{\vec{\sigma}.\vec{L}}{l
+\frac{1}{2}} + \frac{1}{2(l+\frac{1}{2})} ) ^ {2} = 1 $ . In other words the chirality operator in the discrete is
given by :

\begin{equation}
{\Gamma}^L = {\frac{1}{l+\frac{1}{2}}} ( {\vec{\sigma}.\vec{L}} + \frac{1}{2} ).\label{leftchirality}
\end{equation}
By construction this operator has the right continuum limit and it squares to one. However by inspection
${\Gamma}^L$ does not commute with functions on ${\bf S}^2_F$ given by equation (\ref{expansionoffuzzys21}) . Remark that this was
not the case in the continuum where we had ${\gamma}f=f{\gamma}$ for any $f$ in ${\cal A}$ . The property that the
chirality operator must commute with the elements of the algebra is a fundamental requirement of the K-cycle $({\bf
A},{\bf H},D,{\Gamma})$ desribing ${\bf S}^2_F$\footnote{This is in fact a general requirement in any K-cycle , one
must always have the property $[{\gamma},f]=0$ where $f$ is an arbitrary element of the algebra and ${\gamma}$ is
the chirality operator .}. To overcome this problem one simply replace $\vec{L}$ by $-{\vec{L}}^R$ defined in
equation (\ref{rightaction}) . Since these generators act on the right of the algebra ${\bf A}$ , they will commute with
anything which act on the left and therefore the chirality operator will commute with the algebra elements as
desired. The minus sign is due to the minus sign in equation (\ref{rsu2commutations}) . The fuzzy chirality operator is then
given by

\begin{equation}
{\Gamma}= \frac{1}{l+\frac{1}{2}} (- {\vec{\sigma}.{\vec{L}}^{R}} + \frac{1}{2} )\label{fuzzychirality}
\end{equation}

\subsubsection{Fuzzy ${\cal D}_{2w}$ ( or $D_{2w} )$}
The fuzzy version of Watamuras's Dirac operator (\ref{watamura}) is simply given by

\begin{eqnarray}
D_{2w}&=&-{\frac{1}{{\rho}{\sqrt{l(l+1})}}}{\epsilon}_{i j k}{\sigma}_{i}L_{j} ( L_{k} - L_{k}^{R} + \frac{{\sigma}_{k}}{2} ),\nonumber\\
\end{eqnarray}
where we have substituted the fuzzy derivations ${ad L_k}=L_k-L_k^R$ for the classical derivations ${\cal L}_k$ and
the fuzzy coordinates $n^F_j$ for the classical coordinates $n_j$ . By construction this Dirac operator has the
right continuum limit . It can also be rewritten as , $ D_{2w}=-{\frac{1}{2 {\rho} {\sqrt{l(l+1)}}}}{\epsilon}_{i j
k}{\sigma}_{i}[L_{j}, L_{k}] + {\frac{1}{{\rho} {\sqrt{l(l+1)}}}}{\epsilon}_{i j k}{\sigma}_{i} L_{j}L_{k}^{R}-
{\frac{1}{4 {\rho} {\sqrt{l(l+1)}}}} {\epsilon}_{i j k}L_{j}[{\sigma}_{i},{\sigma}_{k}]=-{\frac{i}{{\rho}
\sqrt{l(l+1)}}}{\vec{\sigma}.{\vec{L}}} + {\frac{1}{{\rho} {\sqrt{l(l+1)}}}}{\epsilon}_{i j
k}{\sigma}_{i}L_{j}L_{k}^{R} + {\frac{i}{{\rho} \sqrt{l(l+1)}}}{\vec{\sigma}.{\vec{L}}}$ , and therefore one
obtains
\begin{equation}
D_{2w} ={\frac{1}{{\rho}}}{\epsilon}_{i j k}{\sigma}_{i}n_{j}^FL_{k}^{R}.\label{fuzzywatamura}
\end{equation}
From this expression it is obvious that this Dirac operator is selfadjoint . The last thing needed to be checked is
the claim that this Dirac operator $D_{2w}$ anticommutes with the chirality operator $ {\Gamma}$ . First one
computes

\begin{eqnarray}
D_{2w} {\Gamma}&=&{\frac{1}{{\rho} {\sqrt{l(l+1)}}}}({\epsilon}_{i j k}{\sigma}_{i}L_{j}L_{k}^{R}) {\frac{1}{l+\frac{1}{2}}} ( \frac{1}{2} - {\sigma}_{l} L_{l}^{R} )\nonumber\\
&=&\frac{1}{{\rho}(l+\frac{1}{2})\sqrt{l(l+1)}}[-{\epsilon}_{ijk}{\sigma}_i{\sigma}_lL_jL_k^RL_l^R+\frac{1}{2}{\epsilon}_{ijk}{\sigma}_iL_jL_k^R]\nonumber\\
&=&\frac{1}{{\rho}(l+\frac{1}{2})\sqrt{l(l+1)}}[{\epsilon}_{ijk}{\sigma}_l{\sigma}_iL_jL_k^RL_l^R-2{\epsilon}_{ijk}L_jL_k^RL_i^R+\frac{1}{2}{\epsilon}_{ijk}{\sigma}_iL_jL_k^R]\nonumber\\
&=&\frac{1}{{\rho}(l+\frac{1}{2})\sqrt{l(l+1)}}[{\epsilon}_{ijk}{\sigma}_l{\sigma}_iL_jL_l^RL_k^R-i{\epsilon}_{ijk}{\epsilon}_{klm}{\sigma}_l{\sigma}_iL_jL_m^R-2{\epsilon}_{ijk}L_jL_k^RL_i^R+\nonumber\\
&&\frac{1}{2}{\epsilon}_{ijk}{\sigma}_iL_jL_k^R].\nonumber\\
\end{eqnarray}
Then one computes,
\begin{eqnarray}
{\Gamma}D_{2w}&=&\frac{1}{{\rho}(l+\frac{1}{2})\sqrt{l(l+1)}}[-{\epsilon}_{ijk}{\sigma}_l{\sigma}_iL_jL_k^RL_l^R
+\frac{1}{2}{\epsilon}_{ijk}{\sigma}_iL_jL_k^R].\nonumber\\
&&
\end{eqnarray}
Taking the sum , one gets
\begin{eqnarray}
D_{2w}{\Gamma}+{\Gamma}D_{2w}&=&\frac{1}{{\rho}(l+\frac{1}{2})\sqrt{l(l+1)}}[-i{\epsilon}_{ijk}{\epsilon}_{klm}{\sigma}_l{\sigma}_iL_jL_m^R-2{\epsilon}_{ijk}L_jL_k^RL_i^R+{\epsilon}_{ijk}{\sigma}_iL_jL_k^R]\nonumber\\
&=&\frac{1}{{\rho}(l+\frac{1}{2})}[-i\vec{L}.\vec{L}^R+i{\sigma}_j{\sigma}_kL_jL_k^R+{\epsilon}_{ijk}{\sigma}_iL_jL_k^R]\nonumber\\
&=&0,
\end{eqnarray}
which shows the result .

\subsubsection{Fuzzy ${\cal D}_{2g}$ ( or $D_{2g}$ )}
The fuzzy version of the Grosse-Klim\v{c}\'{i}k-Pre\v{s}najder Dirac operator defined by equation (\ref{GKP}) is
simply given by
\begin{equation}
D_{2g}=\frac{1}{\rho}(\vec{\sigma}.ad\vec{L}+1),\label{fuzzyGKP}
\end{equation}
where we have only substituted the fuzzy derivation $ad\vec{L}=\vec{L}-\vec{L}^R$ for the classical derivation
$\vec{\cal L}$ . One can check that this Dirac operator $D_{2g}$ does not anticommute with the chirality operator
(\ref{fuzzychirality}) and therefore it is no longer unitarily equivalent to $D_{2w}$ . One therefore expect the two operators
$D_{2g}$ and $D_{2w}$ to not have the same spectrum which is indeed the case . Despite this fact $D_{2g}$ is a
much better approximation to the continuum Dirac operator than the Watamuras' Dirac opeartor $D_{2w}$
. This point will be explained in the next subsection . A thorough discussion , however , will be given in the
last section of the next chapter .

\subsection{Spectra of the Dirac Operators}
The computation of the spectra of the Dirac operators $D_{2g}$ and $D_{2w}$ will show the fact that these two
operators are no longer equivalent . Indeed , comparing the spectrum of $D_{2g}$ with the spectrum of $D_{2w}$
will allow us to see that $D_{2g}$ is really different from $D_{2w}$ . More precisely the spectrum of $D_{2g}$ is
exactly that of the continuum Dirac operators ${\cal D}_{2g}$ and ${\cal D}_{2w}$ only cut of at some top value ,
while the spectrum of $D_{2w}$ contains corrections as compared to the continuum . $D_{2g}$ is therefore a better
approximation than $D_{2w}$ . In the process of proving these claims, One will also be able to define a new
chirality operator ${\Gamma}^{'}$ which will anticommute with $D_{2g}$ .

\subsubsection{The Spectrum of $D_{2w}$}

Let us start with $D_{2w}$ . The original calculation is in \cite{10,11} . To find the spectrum one simply rewrites
the square $D_{2w}^2$ in terms of the different $SU(2)$ Casimirs , $\vec{J}^2$, $\vec{K}^2$ , $\vec{L}^2$ ,
$(\vec{L}^R)^2$ and $(\frac{\vec{\sigma}}{2})^2$ where $\vec{J}$ and $\vec{K}$ are defined by
\begin{eqnarray}
\vec{J}&=&\vec{K}+\frac{\vec{\sigma}}{2}\nonumber\\
\vec{K}&=&\vec{L}-\vec{L}^R.
\end{eqnarray}
The computation goes as follows
\begin{eqnarray}
D^2_{2w}&=&\frac{1}{{\rho}^2}({\epsilon}_{ijk}{\sigma}_in_j^FL_k^R)({\epsilon}_{lmn}{\sigma}_ln_m^FL_n^R)\nonumber\\
&=&\frac{1}{{\rho}^2}{\epsilon}_{ijk}{\epsilon}_{lmn}({\delta}_{il}+i{\epsilon}_{ilp}{\sigma}_p)(n_j^Fn_m^F)(L_k^RL_n^R)\nonumber\\
&=&\frac{1}{{\rho}^2}(\vec{n}^F)^2(\vec{L}^R)^2-\frac{1}{{\rho}^2}(n_j^Fn_k^F)(L_k^RL_j^R)+\frac{i}{{\rho}^2}({\epsilon}_{jmn}{\sigma}_k)(n_j^Fn_m^F)(L_k^RL_n^R)\nonumber\\
&-&\frac{i}{{\rho}^2}({\epsilon}_{kmn}{\sigma}_j)(n_j^Fn_m^F)(L_k^RL_n^R)\nonumber\\
&=&\frac{1}{{\rho}^2l(l+1)}\big[\vec{L}^2(\vec{L}^R)^2-L_jL_kL_k^RL_j^R-{\sigma}_kL_nL_k^RL_n^R+{\sigma}_jL_jL_mL_m^R\big]\nonumber\\
&=&\frac{1}{{\rho}^2l(l+1)}\Big[\vec{L}^2(\vec{L}^R)^2+\vec{L}.\vec{L}^R-(\vec{L}.\vec{L}^R)^2-(\vec{\sigma}.\vec{L}^R)(\vec{L}.\vec{L}^R)+(\vec{\sigma}.\vec{L})(\vec{L}.\vec{L}^R)\Big].\nonumber\\
&&
\end{eqnarray}
Now by using the two identities , $\vec{L}.\vec{L}^R=\frac{1}{2}[\vec{L}^2+(\vec{L}^R)^2-\vec{K}^2]$ and
$\vec{\sigma}.\vec{L}-\vec{\sigma}.\vec{L}^R=\vec{J}^2-\vec{K}^2-(\frac{\vec{\sigma}}{2})^2$ , the above equation
can be rewritten as
\begin{eqnarray}
D^2_{2w}&=&\frac{1}{{\rho}^2l(l+1)}\Big[\vec{L}^2(\vec{L}^R)^2+\frac{1}{2}[\vec{L}^2+(\vec{L}^R)^2-\vec{K}^2\big]\big[1-(\frac{\vec{\sigma}}{2})^2+\vec{J}^2-\frac{1}{2}\vec{L}^2-\frac{1}{2}(\vec{L}^R)^2-\frac{1}{2}\vec{K}^2\big]\Big].\nonumber\\
&&
\end{eqnarray}
One chooses to diagonalize this operator on the standard basis
\begin{eqnarray}
\vec{J}^2|jj_3>&=&j(j+1)|jj_3>\nonumber\\
J_3|jj_3>&=&j_3|jj_3>\nonumber\\
\vec{K}^2|jj_3>&=&k(k+1)|jj_3>.
\end{eqnarray}
On this basis it is obvious that $\vec{L}^2$ and $(\vec{L}^R)^2$ are both equal to $l(l+1)$ , whereas
$(\frac{\vec{\sigma}}{2})^2$ is equal to $\frac{3}{4}$ . $j$ takes the two values $j=k+\frac{1}{2}$ and
$j=k-\frac{1}{2}$  for each value of $k$ . $k$ in the other hand takes the values $k=0,1,...2l$ . The eigenvalues
of the above squared Dirac operator will then read
\begin{eqnarray}
D^2_{2w}(j)=\frac{1}{{\rho}^2}\Big[(j+\frac{1}{2})^2+\frac{[k(k+1)]^2}{4l(l+1)}-\frac{k(k+1)(j+\frac{1}{2})^2}{2l(l+1)}\Big].
\end{eqnarray}
For $j=k+\frac{1}{2}$ one can find that
\begin{eqnarray}
D^2_{2w}(j)=\frac{1}{{\rho}^2}\Big[(j+\frac{1}{2})^2+\frac{(j+\frac{1}{2})^2}{4l(l+1)}[1-(j+\frac{1}{2})^2]\Big].
\end{eqnarray}
For $j=k-\frac{1}{2}$ the same expression for $D^2_{2w}(j)$ emerges . Therefore the eigenvalues of the Dirac
operator $D_{2w}$ are given by
\begin{equation}
D_{2w}(j)={\pm}\frac{1}{{\rho}}(j+\frac{1}{2})\sqrt{[1+\frac{1-(j+\frac{1}{2})^2}{4l(l+1)}]}.\label{eigenvaluewatamura}
\end{equation}
\subsubsection{The Spectrum of $D_{2g}$}
The computation of the spectrum of the Dirac operator $D_{2g}$ is much easier than the previous one . The result of this calculation is also much simpler . The original calculation is in \cite{5,6,7,8} . It turns out
that the spectrum of $D_{2g}$ is exactly that of ${\cal D}_{2g}$ upto the eigenvalue $j=2l-\frac{1}{2}$ . Let us
prove this claim explicitly . $D_{2g}$ can be straightforwardly be rewritten as
\begin{eqnarray}
D_{2g}&=&\frac{1}{{\rho}}\big[\vec{\sigma}.(\vec{L}-\vec{L}^R)+1\big]\nonumber\\
&=&\frac{1}{{\rho}}\big[\vec{J}^2-\vec{K}^2-\frac{1}{2}(\frac{1}{2}+1)+1\big]\nonumber\\
&{\Longrightarrow}&\nonumber\\
D_{2g}(j)&=&\frac{1}{{\rho}}\big[j(j+1)-k(k+1)+\frac{1}{4}\big].
\end{eqnarray}
Again for each fixed value of $k$ , $j$ can take only two values , $j=k+\frac{1}{2}$ or $j=k-\frac{1}{2}$. For the
first value we get $D_{2g}(j)=\frac{1}{{\rho}}(j+\frac{1}{2})$ , whereas for the second value we get
$D_{2g}(j)=-\frac{1}{{\rho}}(j+\frac{1}{2})$ . The eigenvalues of the Dirac operator $D_{2g}$ then read
\begin{equation}
D_{2g}(j)={\pm}\frac{1}{{\rho}}(j+\frac{1}{2}).\label{eigenvalueGKP}
\end{equation}
\subsubsection{The Chirality Operator ${\Gamma}^{'}$}
Let us now compare the eigenvalues (\ref{eigenvaluewatamura}) and (\ref{eigenvalueGKP}) . By doing so , many key properties of the Dirac operators
$D_{2w}$ and $D_{2g}$ will then be much more obvious , and at the same time the construction of the chirality
operator ${\Gamma}^{'}$ which anticommutes with $D_{2g}$ will be better motivated and hence more natural . We have
\newpage
\begin{eqnarray}
&&\underline{\bf For~k=2l}\nonumber\\
&\bullet&j=2l+\frac{1}{2}{\longrightarrow}D_{2g}=\frac{1}{{\rho}}(2l+\frac{1}{2}+\frac{1}{2})~and~D_{2w}=0\nonumber\\
&\bullet&j=2l-\frac{1}{2}{\longrightarrow}D_{2g}=-\frac{1}{{\rho}}(2l-\frac{1}{2}+\frac{1}{2})~and~D_{2w}=-\frac{f(2l-\frac{1}{2})}{{\rho}}(2l-\frac{1}{2}+\frac{1}{2})\nonumber\\
&&\underline{\bf For~k=2l-1}\nonumber\\
&\bullet&j=2l-\frac{1}{2}{\longrightarrow}D_{2g}=\frac{1}{{\rho}}(2l-\frac{1}{2}+\frac{1}{2})~and~D_{2w}=\frac{f(2l-\frac{1}{2})}{{\rho}}(2l-\frac{1}{2}+\frac{1}{2})\nonumber\\
&\bullet&j=2l-\frac{3}{2}{\longrightarrow}D_{2g}=-\frac{1}{{\rho}}(2l-\frac{3}{2}+\frac{1}{2})~and~D_{2w}=-\frac{f(2l-\frac{3}{2})}{{\rho}}(2l-\frac{3}{2}+\frac{1}{2})\nonumber
\end{eqnarray}
\begin{eqnarray}
&.&\nonumber\\
&.&\nonumber\\
&.&\nonumber\\
&&\underline{\bf ~For~k=2}\nonumber\\
&\bullet&j=\frac{5}{2}{\longrightarrow}D_{2g}=\frac{1}{{\rho}}(\frac{5}{2}+\frac{1}{2})~and~D_{2w}=\frac{f(\frac{5}{2})}{{\rho}}(\frac{5}{2}+\frac{1}{2})\nonumber\\
&\bullet&j=\frac{3}{2}{\longrightarrow}D_{2g}=-\frac{1}{{\rho}}(\frac{3}{2}+\frac{1}{2})~and~D_{2w}=-\frac{f(\frac{3}{2})}{{\rho}}(\frac{3}{2}+\frac{1}{2})\nonumber\\
&&\underline{\bf ~For~k=1}\nonumber\\
&\bullet&j=\frac{3}{2}{\longrightarrow}D_{2g}=\frac{1}{{\rho}}(\frac{3}{2}+\frac{1}{2})~and~D_{2w}=\frac{f(\frac{3}{2})}{{\rho}}(\frac{3}{2}+\frac{1}{2})\nonumber\\
&\bullet&j=\frac{1}{2}{\longrightarrow}D_{2g}=-\frac{1}{{\rho}}(\frac{1}{2}+\frac{1}{2})~and~D_{2w}=-\frac{f(\frac{1}{2})}{{\rho}}(\frac{1}{2}+\frac{1}{2})\nonumber\\
&&\underline{\bf ~For~k=0}\nonumber\\
&\bullet&j=\frac{1}{2}{\longrightarrow}D_{2g}=\frac{1}{{\rho}}(\frac{1}{2}+\frac{1}{2})~and~D_{2w}=\frac{f(\frac{1}{2})}{{\rho}}(\frac{1}{2}+\frac{1}{2})\nonumber
\end{eqnarray}
In the above equation $f$ is defined by $f(j,l)=\sqrt{1+\frac{1-(j+\frac{1}{2})^2}{4l(l+1)}}$ .

By investigation one can immediately notice that there is a problem with the top modes , $j=2l+\frac{1}{2}$ , for
both the Dirac operators $D_{2w}$ and $D_{2g}$ . In particular , $j=2l+\frac{1}{2}$ , are zero modes for $D_{2w}$
and therefore they spoil the invertibility of this latter operator . On the other hand , the eigenvalues
$j=2l+\frac{1}{2}$ in the spectrum of $D_{2g}$ , are not paired to any other eigenvalues and this is the reason
for $D_{2g}$ not having a chirality operator . However $D_{2w}$ has the extra disadvantage of having very small
eigenvalues for large values of $j$ because $f(j,l)$ is of the order of $\frac{1}{\sqrt{l}}$ for $j$ of the order
of $l$ . In other words $D_{2g}$ is a much better Dirac operator than $D_{2w}$ if one can define for it a chirality
operator .

If we restrict ourselves to the subspace with $j{\leq}2l-\frac{1}{2}$ then clearly $D_{2g}$ must have a chirality
operator . Let us then define the projector $P$ by
\begin{eqnarray}
&P&|2l+\frac{1}{2},j_3>=0\nonumber\\
&{\rm and}&\nonumber\\
&P&|j,j_3>=|j,j_3>~,~for~all~j{\leq}2l-\frac{1}{2}.
\end{eqnarray}
Let us call $V$ the space on which $P$ projects down . The orthogonal space is $W$ .

The final thing will be now to find the chirality operator of the Dirac operator $PD_{2g}P$ . To this end one
starts by making some observations concerning the continuum . From the basic identity ${\cal D}_{2w}=i{\gamma}{\cal
D}_{2g}$ one concludes that $\{{\cal D}_{2w},{\cal D}_{2g}\}=0$ , whereas from the fact that both ${\cal D}_{2g}$
and ${\cal D}_{2w}$ share the same spectrum one obtains that $|{\cal D}_{2g}|=|{\cal D}_{2w}|$ . Obviously we have
also the fact that $|{\cal D}_{2g}|$ and $|{\cal D}_{2w}|$ commute with ${\cal D}_{2g}$ and ${\cal D}_{2w}$ . Hence
we can trivially prove the idnetity
\begin{equation}
{\gamma}=i{\cal F}_{2g}{\cal F}_{2w},\label{chiralitydefinition2}
\end{equation}
where ${\cal F}_{2g}$ and ${\cal F}_{2w}$ are the sign operators of the Dirac operators ${\cal D}_{2g}$ and ${\cal
D}_{2w}$ respectively . They are defined by
\begin{eqnarray}
{\cal F}_{2g}&=&\frac{{\cal D}_{2g}}{|{\cal D}_{2g}|}\nonumber\\
{\cal F}_{2w}&=&\frac{{\cal D}_{2w}}{|{\cal D}_{2w}|}.\nonumber\\
\end{eqnarray}
The fuzzification of (\ref{chiralitydefinition2}) is only possible if one confine ourselves to the vector space $V$ . The reason is
obvious , $F_{2w}$ which is the fuzzy version of ${\cal F}_{2w}$ will not exist on the whole space $V{\oplus}W$ .
Taking all of these matters into considerations , one end up with the new chirality operator
\begin{equation}
{\Gamma}^{'}=iF_{2g}F_{2w},\label{fuzzychirality2}
\end{equation}
where
\begin{eqnarray}
{F}_{2g}&=&\frac{{D}_{2g}}{|{D}_{2g}|},~on~V\nonumber\\
&=&0~,~on~W,\nonumber\\
\end{eqnarray}
and
\begin{eqnarray}
{F}_{2w}&=&\frac{{D}_{2w}}{|{D}_{2w}|},~on~V\nonumber\\
&=&0~,~on~W.\nonumber\\
\end{eqnarray}
By construction (\ref{fuzzychirality2}) has the right continuum limit . If it is going to assume the role of a chirality operator
on the fuzzy sphere it must also square to one on $V$ , in other words one must have on the whole space
$V{\oplus}W$ : $({\Gamma}^{'})^2=P$. It should also be selfadjoint and should anticommute with the Dirac operator
$PD_{2g}P$ . The key requirement for all of these properties to hold is the identity $\{F_{2g},F_{2w}\}=0$. The
proof which will be presented in great detail in the next chapter can be sketched as follows . First one note
that on each subspace of $V$ with fixed $j$ , the operators $D^2_{2g}$, $D^2_{2w}$ , $|D_{2g}|$ and $|D_{2w}|$
are essentially proportional to the identity and therefore they commute with the restrictions of the operators
$D_{2g}$ , $D_{2w}$ and ${\Gamma}$ to this subspace. In other words the operators $D^2_{2g}$ , $D^2_{2w}$ ,
$|D_{2g}|$ and $|D_{2w}|$ will commute with the operators $D_{2g}$ ,$D_{2w}$ and ${\Gamma}$ on the whole space $V$
. Next the basic identity , $[D_{2g},{\Gamma}]=-2i\sqrt{1-\frac{1}{(2l+1)^2}}D_{2w}$ , will lead to the result
$\{D_{2w},D_{2g}\}=0$ from which we get $\{F_{2g},F_{2w}\}=0$ .

%

\chapter{Quantum Physics on Fuzzy ${\bf S}^2$}
A fuzzy space \cite{1,2,3.1,3.2,3.3,3.4} is obtained by quantizing a manifold, treating it as a phase space. An
example is the fuzzy two-sphere ${\bf S}^2_{F}$ . It is described by operators $n_{i}^F$ subject to the relations
$\sum_{i} n_i^{F2} =1$ and $[n_{i}^F,n_{j}^F]=(i/{\sqrt{l(l+1)}}) {\epsilon}_{ijk}n_{k}^F$. Thus $L_i=\sqrt{l
(l+1)}n_i^F$ are $(2l+1)$-dimensional angular momentum operators while the canonical classical two-sphere ${\bf
S}^2$ is recovered for $l{\rightarrow}{\infty}$ . Planck's work shows that quantization creates a short distance
cut-off, therefore quantum field theories (QFT's) on fuzzy spaces are ultraviolet finite . If the classical
manifold is compact, it gets described by a finite-dimensional matrix model, the total number of states being
finite too. Noncommutative geometry \cite{1,2,3.1,3.2,3.3,3.4,4} has an orderly prescription for formulating QFT's
on fuzzy spaces so that these spaces indeed show us an original approach to discrete physics.

In this chapter , we focus our attention on the fuzzy sphere , ${\bf S}^2_{F}$ , and discuss certain of its
remarkable aspects which are entirely absent on the naively discretized ${\bf S}^2$ , namely lattice ${\bf S}^2$ .
Quantum physics on ${\bf S}^2_{F}$ is a mere matrix model in which one can coherently describe twisted topologies
like those of monopoles and solitons . Monopoles and solitons have important topological aspects like quantized
fluxes, winding numbers and curved target spaces . Naive discretizations which substitute a lattice of points for
the underlying manifolds are incapable of retaining these features in a precise way. We study in this chapter these
problems of discrete physics and matrix models and discuss mathematically coherent discretizations of monopoles
and solitons using fuzzy physics and noncommutative geometry . Traditional attempts which are usually based on
naive discrete physics have at best been awkward having ignored the necessary mathematical structures of projective
modules and cyclic cohomology . A fuzzy ${\sigma}$-model action for the two-sphere fulfilling a fuzzy
Belavin-Polyakov bound is also put forth . The last section of this chapter will be devoted to the fermion doubling
problem and its resolution on the fuzzy sphere ${\bf S}^2_F$ . One feels that this last topic is an important
contribution of fuzzy physics to discrete physics .
\section{Fuzzy Monopoles}
\subsection{Monopoles Wave Functions}
A point particle of electric charge $q$ and mass $m$ in the magnetic field of a monopole $g$ is  described by the
free Hamiltonian , $H=\frac{\vec{p}^2}{2m}$ , together with the Poisson brackets $\{x_i,x_j\}=0$ ,
$\{p_i,p_j\}=q{\epsilon}_{ijk}B_k$ and $\{x_i,p_j\}={\delta}_{ij}$. $\vec{B}$ is the magnetic field of the
monopole given by $\vec{B}=-\frac{g}{4{\pi}}\frac{\vec{x}}{r^3}$ , where $r$ is of course the radial distance
between the monopole which is assumed to be at rest at $\vec{x}=0$ and the point particle at $\vec{x}$ . It is not
difficult using the above data to find the force acting on the electric charge in the presence of the monopole , it
is given by $m\frac{d{\dot{x}}_i}{dt}=m\{\dot{x}_i,H\}=q{\epsilon}_{ijk}\dot{x}_jB_k$ . The other remark is the
fact that the canonical angular momentum $L_i={\epsilon}_{ijk}x_j(m\dot{x}_k)$ of the point particle is not
conserverd since we have,
$\frac{dL_i}{dt}={\epsilon}_{ijk}x_j(m\frac{d\dot{x}_k}{dt})=-\frac{qg}{4{\pi}}\frac{d}{dt}
(\frac{x_i}{r})$.
Therefore $J_i=L_i+\frac{qg}{4{\pi}}n_i$ where $n_i=\frac{x_i}{r}$ and $\frac{dJ_i}{dt}=0$ is what should be
interpreted as angular momentum of the point particle in the presence of a magnetic monopole , in other words the
point particle acquires an angular momentum in the direction of the line joining the particle and the monopole .

It is a known result that in the case of an electric charge in the field of a monopole , one can not find a global
system of canonical coordinates $(\vec{x},\vec{p})$ for the phase space ${\bf T}^{*}{\bf B}$ , and therefore a
global Lagrangian describing the above system can not be found by a simple Legendre transformation of the
Hamiltonian . To construct such a Lagrangian , one enlarges the configuration space ${\bf B}=\{\vec{x}\}$ to a
$U(1)$ bundle ${\bf E}$ over ${\bf B}$ given by
\begin{eqnarray}
{\bf E}&=&{\bf R}{\times}SU(2)=\{(r,g)\}\nonumber\\
&where&\nonumber\\
{\sigma}_in_i&=&g{\sigma}_3g^{-1}.
\end{eqnarray}
A global Lagrangian can then be written down as follows
\begin{eqnarray}
L&=&\frac{1}{2}m\sum_{i}\dot{x}_i^2+i\frac{qg}{4{\pi}}Tr{\sigma}_3g^{-1}\dot{g}\nonumber\\
&=&\frac{1}{2}m\dot{r}^2+\frac{1}{4}mr^2Tr[\dot{g}g^{-1},g{\sigma}_3g^{-1}]^2+i\frac{qg}{4{\pi}}Tr{\sigma}_3g^{-1}\dot{g}.\label{monopoleterm}
\end{eqnarray}
The above Lagrangian can be shown to be weakly invariant under the right $U(1)$ action ,
$g{\longrightarrow}ge^{i\frac{\theta}{2}{\sigma}_3}$ , that is to say
$L{\longrightarrow}L-\frac{qg}{4{\pi}}\dot{\theta}$ . In other words we have , like in the case considered in the
introduction , a fiber bundle structure $U(1){\longrightarrow}SU(2){\longrightarrow}{\bf S}^2$.

Following the same steps taken in the introduction to quantize the Wess-Zumino term (\ref{wz}), one can start the
quantization of (\ref{monopoleterm}) by first parametrizing the group element $g$ by a set of three real numbers
$({\xi}_1,{\xi}_2,{\xi}_3)$. The conjugate momentum ${\Pi}_i$ associated to ${\xi}_i$ is gievn by
${\Pi}_i=\frac{{\partial}L}{{\partial}\dot{\xi}^i}=i\frac{qg}{4{\pi}}Tr{\sigma}_3g^{-1}\frac{{\partial}g}{{\partial}{\xi}^i}+
\frac{1}{2}mr^2Tr[\frac{{\partial}g}{{\partial}{\xi}^i}g^{-1},g{\sigma}_3g^{-1}][\frac{{\partial}g}{{\partial}{\xi}^j}g^{-1},g{\sigma}_3g^{-1}]\dot{\xi}^j$.
The modified conjugate momentum $t_k=-{\Pi}_iN_{ik}$ can also be computed along the same lines which led to
equation (\ref{modifiedconjugatemomenta}) of the introduction . The answer is
$t_k=\frac{1}{2}\frac{qg}{4{\pi}}Tr{\sigma}_kg{\sigma}_3g^{-1}-\frac{mr^2}{2}Tr[i\frac{{\sigma}_k}{2},g{\sigma}_3g^{-1}][\dot{g}g^{-1},g{\sigma}_3g^{-1}]$. From this last equation the following constarint follows easily
\begin{equation}
P=n_kt_k-\frac{qg}{4{\pi}}\approx 0 \label{monopoleconstraint}
\end{equation}
The Hamiltonian of the system can be computed in a standard fashion , it is given by
\begin{eqnarray}
H&=&\frac{p_r^2}{2m}+\frac{1}{4}mr^2Tr[\dot{g}g^{-1},g{\sigma}_3g^{-1}]^2+vP\nonumber\\
&=&\frac{p_r^2}{2m}+\frac{1}{2mr^2}[\vec{t}^2-(\frac{qg}{4{\pi}})^2]+vP.\label{monopolehamiltonian}
\end{eqnarray}
$v$ is a Lagrange multiplier . One can next check that the first class constraint (\ref{monopoleconstraint}) have zero Poisson bracket
with the Hamiltonain (\ref{monopolehamiltonian}) , and therefore there is no secondary constraint. Observables should of course have
zero PB's with $P$, these are $r$ , $p_r$ , $t_i$ and $n_i$ or functions of them .

Wave functions of the system are ${\psi}={\psi}(r,g)$ . $p_r$ acts on ${\psi}$ as the usual differential operator
$\frac{1}{i}\frac{d}{dr}$ , while ${t}_i$ acts by left multiplication , namely
\begin{equation}
[e^{i{\theta}_it_i}{\psi}](r,g)={\psi}(r,e^{-i{\theta}_i\frac{{\sigma}_i}{2}}g).
\end{equation}
These wave functions should also satisfy the requirement
\begin{equation}
n_kt_k{\psi}=\frac{qg}{4{\pi}}{\psi}.
\end{equation}
From the above two last equations one can easily find that
\begin{equation}
{\psi}(r,ge^{-i{\theta}\frac{{\sigma}_3}{2}})=e^{i{\theta}\frac{qg}{4{\pi}}}{\psi}(r,g).
\end{equation}
Since any function of $r$ and $g$ admits the expansion
\begin{equation}
f(r,g)=\sum_{j}\sum_{m,n}c^{j}_{m,n}(r)<j,m|D^{j}(g)|j,n>,\label{monopolewavefunction}
\end{equation}
where $g{\longrightarrow}D^{j}(g)$ is the $j$ IRR of $SU(2)$ . Then under the transformation
$g{\longrightarrow}ge^{-i{\theta}\frac{{\sigma}_3}{2}}$, each term in the above expansion (\ref{monopolewavefunction}) will transform
as $<j,m|D^{j}(g)|j,n>{\longrightarrow}e^{-i{\theta}n}<j,m|D^{j}(g)|j,n>$ . In other words wave functions
${\psi}(r,g)$ should be functions of the form
\begin{equation}
\psi(r,g)=\sum_{j}\sum_{m}c^{j}_{m,n}(r)<j,m|D^{j}(g)|j,n>,
\end{equation}
with the quantization condition , $\frac{qg}{4{\pi}}=-n $ , which is the famous Dirac quantization condition . $n$
is clearly either an integer or half integer , so that
\begin{equation}
\frac{qg}{2{\pi}}=~integer.
\end{equation}
In the units where $q=2{\pi}$ we have $g={\pm}N$ where $N{\in}{\bf N}$.
\subsection{The Algebraic Formulation of Classical Monopoles }

There is an algebraic formulation of monopoles and solitons suitable for adaptation to fuzzy spaces. We first
outline it using the case of ${\bf S}^2$\cite{3.3}. It is based on the Serre-Swan theorem \cite{1} , which states that
every vector bundle ${\bf E}$ on a compact space ${\bf X}$ is isomorphic to a vector bundle ${\cal P}[{\bf C}(\bf
X){\otimes}{\bf C}^N]$, where ${\bf C}(\bf X)$ is the algebra of smooth functions on ${\bf X}$ and $N$ is some
large integer. The projector ${\cal P}$ , which is a self-adjoint idempotent , is an element of ${\bf M}_N({\bf
C}(\bf X))$ : the algebra of $N{\times}N$ matrices with entries in ${\bf C}(\bf X)$ .

Let ${\cal A}$ be the commutative algebra of smooth functions on ${\bf S}^2$. Vector bundles on ${\bf S}^2$ can be
described by projectors ${\cal P}$. ${\cal P}$ is a matrix with coefficients in ${\cal A}({\cal P}_{ij}{\in}{\cal
A}),$ and fulfills ${\cal P}^2={\cal P}$ and ${\cal P}^{\dagger}={\cal P}$. If the points of ${\bf S}^2$ are
described by unit vectors $\vec{n} {\in} {\bf R}^3$, the projector for unit monopole charge is ${\cal P}^{(1)}=(1 +
\vec{\tau}.\hat{n})/2$ where ${\tau}_i$ are Pauli matrices and $\hat{n}_i$ are coordinate functions, $\hat{n}_i
(\vec{n})=n_i$ . This calculation was carried out explicitly in section $3.1.5$ where we have computed the Chern
class for the projective module ${\cal P}^{(1)}{\cal A}^2$, where ${\cal A}^2={\cal A}{\otimes}{\bf C}^2$, and
found it to be equal to one. ${\cal P}^{(1)}{\cal A}^{2}$ are therefore sections of vector bundles for monopole
charge $1$ .

For monopole charge $\pm N $ $ (N>0)$, the corresponding projectors are
\begin{equation}
{\cal P}^{(\pm N)}=\prod _{i=1}^{i=N}\frac{(1 \pm \vec{\tau}^{(i)}.\hat{n})}{2}, \label{cprojector}
\end{equation}
where $\vec{\tau}^{(i)}$ are commuting sets of Pauli matrices . They give the following sections of vector bundles
\begin{eqnarray}
&&{\cal P}^{(\pm N)}{\cal A}^{2^{N}},\nonumber\\ \label{cprojectivemodule}
\end{eqnarray}
where ${\cal A}^{2^{N}} = {\cal A}{\otimes}{\bf C}^{2^{N}}$ consists of $2^{N}$-component vectors ${\xi}=({\xi}_1,
{\xi}_2,..., {\xi}_{2^{N}})$, ${\xi}_i{\in} {\cal A}$. $\vec{\tau}^{(i)}.\hat{n}\;{\cal P}^{(\pm N)}{\xi}=\pm
{\cal P}^{(\pm N)}{\xi}$, $\vec{\tau}^{(i)}$ acting on the $i^{th}$ ${\bf C}^2$ factor. For the trivial bundle, we
can use ${\cal P}^{0}=\frac{1 + {\tau}_3}{2}$ (or $\frac{1 - {\tau}_3}{2}$), or more simply just the identity.
Because of the additivity property of winding numbers , shown in section $(3.1.5)$ , the Chern class of the
projective module (\ref{cprojectivemodule}) is trivially equal to ${\pm}N$ [see section $(3.1.5)$ for details] .

\subsection{Fuzzy Monopoles}

The algebra ${\bf A}$ generated by $n_i^F$ is the full matrix algebra of $(2l+1){\times}(2l+1)$ matrices . Fuzzy
monopoles are described by projectors $p^{({\pm} N)}\; (p^{({\pm} N)}_{ij} {\in} {\bf A})$, which as
$l{\rightarrow}{\infty}$ approach ${\cal P}^{({\pm} N)}$. We can find them as follows .

For $N=1$, we can try $(1 +\vec{\tau}.\vec{n}^F)/2$ , but that is not an idempotent as the $n^F$'s do not commute.
We can fix that though : since $(\vec{\tau}.\vec{L})^2 = l(l+1) - \vec{\tau}.\vec{L}$,
${\gamma}_{\tau}=\frac{1}{l+1/2} (\vec{\tau}.\vec{L}+ \frac{1}{2})$ squares to $1$ as first remarked by Watamuras
\cite{10,11}\footnote{All of this was verified explicitly in section $3.2.4$ .}. Hence
\begin{equation}
p^{(1)}=\frac{1 + {\gamma}_{\tau}}{2}.
\end{equation}
There is a simple interpretation of $p^{(1)}$. We can combine $\vec{L}$ and ${\vec{\tau}}/2$ into the $SU(2)$
generator ${\vec{K}^{(1)}}=\vec{L} + {\vec{\tau}}/2$ with spectrum $k(k+1)$ ($k=l \pm1/2$). The projector to the
space with the maximum $k$ , namely $k=l+\frac{1}{2}$ , is just $p^{(1)}$. In other words we have
\begin{equation}
p^{(1)}=\frac{{K^{(1)}}^{2} - (l-1/2)(l+1/2) }{(l+1/2)(l+3/2)-(l-1/2)(l+1/2)}.
\end{equation}
where ${K^{(1)}}^2{\equiv}{\vec{K}}^{(1)}.{\vec{K}}^{(1)}$ . The proof is trivial since ${\gamma}_{\tau}$ can be
rewritten in the form ${\gamma}_{\tau}=\frac{1}{l+1/2} (\vec{\tau}.\vec{L}+
\frac{1}{2})=\frac{1}{l+1/2}(\vec{K}^{(1)2}-\vec{L}^2-(\frac{\vec{\tau}}{2})^2+\frac{1}{2})=\frac{\vec{K}^{(1)2}}{l+1/2}-(l+\frac{1}{2})$
, and therefore $p^{(1)}=\frac{{K^{(1)}}^{2} - (l-1/2)(l+1/2) }{2l+1}$.

This last remark shows the way to fuzzify ${\cal P}^{(N)}$. We substitute
\begin{equation}
\vec{K}^{(N)}=\vec{L} + \sum_{i=1}^{N} \frac{\vec{\tau}^{(i)}}{2},\label{fuzzifypn?}
\end{equation}
for $\vec{K}^{(1)}$ and consider the subspace where ${K^{(N)}}^{2}{\equiv}{\vec{K}}^{(N)}.{\vec{K}}^{(N)}$ has the
maximum eigenvalue $k_{max}(k_{max}+1),k_{max}=l+N/2$ . On this space $(\vec{L} + \vec{\tau}^{(i)}/2)^2$ has the
maximum value $(l+1/2)(l+3/2)$ and $\vec{\tau}^{(i)}.\vec{L}$ is hence $l$ . Since $\vec{\tau}^{(i)}.\vec{n}^F$
approaches $\vec{\tau}^{(i)}.\hat{n}$ on this subspace as $l \rightarrow \infty$,
$p^{(N)}$ is just its projector:
\begin{equation}
p^{(N)}\equiv p^{(+N)}={\frac{{\prod}_{k{\neq}k_{max}} [{K^{(N)}}^{2}- k(k+1) ]}{{\prod}_{k{\neq}k_{max}} [
k_{max}(k_{max}+1) - k(k+1) ]}}.
\end{equation}
$p^{(-N)}$ comes similarly from the least value $k_{min}=l-N/2$ of $k$. [We assume that $2l \geq N$.]

We remark that the limits as $l{\rightarrow}{\infty}$ of $p^{(\pm N)}$ are exactly ${\cal P}^{(\pm N)}$, and not
say ${\cal P}^{(\pm N)}$ times another projector. That is because if $\vec{\tau}^{(i)}.\vec{L}$ are all $l$, then
$\vec{\tau}^{(i)}.\vec{\tau}^{(j)}={\bf 1}$ for all $i{\neq}j$ and hence $k_{max}=l + \frac{N}{2}$. A proof goes as
follows. Vectors with ${\vec{L}}^{2}=l(l+1){\bf 1}$ can be represented as symmetric tensor products of $2l$
spinors, with components $T_{a_1...a_{2l}}$. The vectors with $\vec{\tau}^{(i)}.\vec{L}=l{\bf 1}$ as well have
components $T_{a_1...a_{2l},b_1...b_N}$ with symmetry under exchange of any $a_i$ with $a_j$ or $b_k$. So they are
symmetric under all exchanges of $b_i$ and $b_j$ and have $(\frac{\vec{\tau}^{(i)}}{2} +\frac{\vec{\tau}^{(j)}}{2}
)^2=2$, $\vec{\tau}^{(i)}.\vec{\tau}^{(j)}={\bf 1}$.

A much more straightforward proof starts by remarking that
$K^{(N)2}=l(l+1)+\sum_{i=1}^N\vec{L}.\vec{\tau}^{(i)}+\sum_{i,j=1}^N\frac{\vec{\tau}^{(i)}}{2}\frac{\vec{\tau}^{(j)}}{2}$
, from which one concludes that $\vec{K}^{(N)2}$ is maximum when all the products $\vec{L}.\vec{\tau}^{(i)}$ ,
$i=1,N$ and $ \frac{\vec{\tau}^{(i)}}{2}\frac{\vec{\tau}^{(j)}}{2}$ , $i{\neq}j$ , $i,j=1,N$ are maximum. But
$\vec{L}.\vec{\tau}^{(i)}=(\vec{L}+\frac{\vec{\tau}^{(i)}}{2})^2-l(l+1)-\frac{3}{4}$ can take only the two
different values $l$ and $-l-1$ , whereas
$\frac{\vec{\tau}^{(i)}}{2}\frac{\vec{\tau}^{(j)}}{2}=\frac{1}{2}(\frac{\vec{\tau}^{(i)}}{2}+\frac{\vec{\tau}^{(j)}}{2})^2-\frac{3}{4}$
for $i{\neq}j$ can take the two values $1/4$ and $-3/4$ , so that $K^{(N)2}$ is maximum for
$\vec{L}.\vec{\tau}^{(i)}=l$ and $\vec{\tau}^{(i)}.\vec{\tau}^{(j)}=1$ and it is given by
$K^{(N)2}=(l+N/2)(l+N/2+1)$ .

Having obtained $p^{(\pm N)}$, we can also write down the analogues of ${\cal P}^{(\pm N)}{\cal A}^{2^{N}}$: they
are the ``projective modules''
\begin{equation}
p^{(\pm N)}{\bf A}^{2^{N}}, {\bf A}^{2^{N}}= \langle (a_1,a_2,..,a_{2^{N}}):a_i{\in} {\bf A}\rangle,
\end{equation}
 and are the noncommutative
substitutes for sections of vector bundles.

If $(a_1,a_2,...,a_{2^N}) $ is regarded as a column, then column dimension of $p^{(\pm N)}{\bf A}^{2^{N}}$ is $L+1
{\equiv} 2(l{\pm}{\frac{N}{2}})+1$ , as $p^{(+ N)}$ and $p^{(-N)}$ project down to the subspaces with
$k_{max}=l+N/2$ and $k_{min}=l-N/2$ respectively , and its row dimension is $M + 1 {\equiv} 2l+1$ , as all the
$a_i$'s are in ${\bf A}=Mat_{2l+1}$ . Their difference is $\pm N$. This means that $p^{(\pm N)}{\bf A}^{2^{N}}$
can be identified with ${\hat{\cal H}}_{L,M}$ of ref\cite{6} where of course $L - M ={\pm} N$. In particular
angular momentum acts on $p^{(\pm N)}{\bf A}^{2^{N}}$ via $\vec{K}^{(N)}$ on left (they commute with $p^{(\pm
N)}$) and $-\vec{L}$ on right, while there are similar actions of angular momentum on ${\hat{\cal H}}_{L,M}$ [see
ref \cite{6}].

\section{Fuzzy ${\sigma}-$Models}

The projectors (\ref{cprojector}) also describe nonlinear $\sigma$-models. To see this, consider the projector ${\cal
P}^{(0)}=(1+{\tau}_3)/2$ and its orbit $\{h{\cal P}^{(0)}h^{-1} : h{\in}SU(2)\}$ . This orbit is clearly ${\bf
S}^{2}$ , as ${\cal P}^{0}$ is ${\tau}_3$ up to a constant matrix . If now we substitute for $h$ a field $g$ on
${\bf S}^{2}$ with values in $SU(2)$, each $g{\cal P}^{(0)}g^{-1}$ describes a map from ${\bf S}^2 $ to ${\bf S}^2$
. The second ${\bf S}^2$ is the above orbit . This $g$ is a $\sigma$-model field on ${\bf S}^2$ with target space
${\bf S}^2$ and zero winding number. (Winding number is zero as $g$ can be deformed to a constant map).

For winding number $1$, it is appropriate to consider the orbit of ${\cal P}^{(1)}=(1+\vec{\tau}\hat{n})/2$ under
$g$. For fixed $\vec{n}$, as $g(\vec{n})$ is varied, the orbit $\{g(\vec{n}) \frac{ 1 +
\vec{\tau}\hat{n}(\vec{n})}{2} g(\vec{n})^{-1}\}$ is still ${\bf S}^2$ because its points $g(\vec{n}) \frac{ 1 +
\vec{\tau}\hat{n}(\vec{n})}{2} g(\vec{n})^{-1}$ are still invariant under the right $U(1)$ action
$g(\vec{n}){\longrightarrow}g(\vec{n})h$, of course $h{\in}U(1)$ is now generated by the Pauli matrix in the fixed
direction $\vec{n}$ , namely $h=exp(i{\theta}{\tau}_n/2)$ with ${\tau}_n=\vec{\tau}\hat{n}(\vec{n})$. Finally as
$\vec{n}$ is varied , we get a map ${\bf S}^2{\rightarrow}{\bf S}^2$. (More correctly we get the section of an
${\bf S}^2$ bundle over ${\bf S}^2$).

For winding number $\pm{N}$, we can consider the orbit of ${\cal P}^{(\pm N)}$ under conjugation by
$g^{\tilde{\otimes}N}$'s where
\begin{equation}
g^{\tilde{\otimes}N} (\vec{n})= g(\vec{n}){\otimes}g(\vec{n}){\otimes} \cdots {\otimes}g(\vec{n})~(N
{\rm{~factors}}).\label{csigmafield}
\end{equation}
Here the $i$th $g(\vec{n})$ acts only on $\vec{\tau}^{(i)}$.

In the fuzzy versions of $\sigma$-models on ${\bf S}^2$ with target ${\bf S}^2$ , $g$ becomes a $2{\times}2$
unitary matrix $u$ with $u_{ij}{\in}{\bf A}$. Therefore $u{\in}U(2(2l+1))$ . ( We can impose $\det u=1$ , that
makes no difference ). An appropriate generalization $ u^{\tilde{\otimes}N} $ of $g^{\tilde{\otimes}N}$ can be
constructed as follows . If $C$ and $D$ are $2{\times}2$ matrices with entries $C_{ij}$, $D_{ij}{\in}{\bf A}$, we
can define $Ca$ and $aD$ by $(Ca)_{ij}=C_{ij}a$ and $(aD)_{ij} = aD_{ij}$. Let $C{\otimes}_{\bf A}D$ denote the
tensor product of $C$ and $D$ over ${\bf A}$ where by definition $Ca{\otimes}_{\bf A}D = C{\otimes}_{\bf A}aD$ .
This definition can be extended to more factors . For example, $C{\otimes}_{\bf A}D{\otimes}_{\bf A}E$ has the
properties $Ca{\otimes}_{\bf A}D{\otimes}_{\bf A}E = C{\otimes}_{\bf A}aD{\otimes}_{\bf A}E$, $C{\otimes}_{\bf
A}Da{\otimes}_{\bf A}E = C{\otimes}_{\bf A}D{\otimes}_{\bf A}aE$ . Then :
\begin{equation}
u^{\tilde{\otimes}N}=u{\otimes}_{\bf A}u{\otimes}_{\bf A} \ldots {\otimes}_{\bf A}u ~(~N \rm{~factors}).\label{fsigmafield}
\end{equation}
(\ref{fsigmafield}) is the fuzzification of (\ref{csigmafield}). We can understand this construction in familiar terms by writing
$u={\bf 1}_{2{\otimes}2}a_{0}$ $+ {\tau}_{j}a_{j} = {\tau}_{\mu}a_{\mu}(a_{\mu}{\in}{\bf A})$ where ${\tau}_0={\bf
1}_{2{\otimes}2}$. $[$Greek subscripts run from $0$ to $3$, Roman ones from $1$ to $3]$. Unitarity requires that
\begin{eqnarray}
{\tau}_{\mu}{\tau}_{\nu}a_{\mu}^{*}a_{\nu}&=&{\bf 1},\nonumber\\
a_{\mu}^{*}& \equiv& a_{\mu}^{\dagger}. \label{unitarya}
\end{eqnarray}
In this notation, $u{\otimes}_{\bf A}u={\tau}_{\mu}{\otimes}{\tau}_{\nu}a_{\mu}a_{\nu}$,
${\otimes}({\equiv}{\otimes}_{\bf C})$ denoting Kronecker product. It is also
${{\tau}_{\mu}}^{(1)}a_{\mu}{{\tau}_{\nu}}^{(2)}a_{\nu}$ in an evident notation. Proceeding in this way, we find,
\begin{equation}
u{\otimes}_{\bf A}u{\otimes}_{\bf A} \ldots {\otimes}_{\bf A}u = {\tau}^{(1)}_{{\mu}_1}a_{{\mu}_1}
{\tau}^{(2)}_{{\mu}_2}a_{{\mu}_2} \ldots {\tau}^{(N)}_{{\mu}_N}a_{{\mu}_N}.
\end{equation}
It is unitary in view of (\ref{unitarya}).

The significant point here is that $u^{\tilde{\otimes}N}$ is a matrix with coefficients in ${\bf A}$ and not ${\bf
A}{\otimes}{\bf A}{\otimes} \ldots {\otimes}{\bf A}$ as is the case for $u{\otimes}u{\otimes}u \ldots {\otimes}u$.
We remark that $g^{\tilde{\otimes}N}$ can also be written as $g{\otimes}_{\cal A}g{\otimes}_{\cal A}g \ldots
{\otimes}_{\cal A}g$ ($N$ factors). It is then a function only of $\vec{n}$ and has the meaning stated earlier.

The orbits of $p^{(\pm N)}$ under conjugation by $u^{\tilde{\otimes}N}$ are fuzzy matrix versions of $\sigma$-model
fields with winding numbers $\pm{N}$. [Here we take $p^{(0)}$ to be $(1+\tau_3)/2$ say and $u^{\tilde{\otimes}N}$
to be $u$ itself. Henceforth our attention will be focussed on $N \neq 0$.]

\section{Chiral Anomaly On ${\bf S}^2_F$}

\subsection{Winding Numbers for the Classical Sphere}

What about formulas for invariants like Chern character and winding number? The ideal way is to follow Connes
\cite{1,2,3.1,3.2,3.3,3.4} and introduce the Dirac and chirality operators
\begin{eqnarray}
{\cal D}_{2w}&=&-\frac{1}{\rho}{\epsilon}_{ijk} {\sigma}_{i} \hat{n}_j {\cal J}_{k},
\nonumber\\
{\gamma}&=&{\sigma}.\hat{n} \label{contoprs}
\end{eqnarray}
where ${\sigma}_i$ are Pauli matrices, $\vec{\cal
J}=-i(\vec{r}{\times}{\vec{\bigtriangledown}})+\frac{\vec{\sigma}}{2}$ is the total angular momentum and
$\hat{n}=\vec{r}/|\vec{r}|$. The two equations in (\ref{contoprs}) are essentially equations (\ref{watamura}) and (\ref{chiralitydefinition}) 
respectively which were derived in the last chapter . The important points to keep in mind here are the following :

i){\it ${\gamma}$ commutes with elements of ${\cal A}$ and anti-commutes with ${\cal D}_{2w}$.}

ii){\it ${\gamma}^2 = 1$ and ${\gamma}^{\dagger}={\gamma}$.}

The Chern numbers (or the quantized fluxes) for monopoles then are
\begin{equation}
\pm N= -\frac{1}{4 \pi}\int d (\cos{\theta}){\wedge}d{\phi}\;{\rm Tr}\;{\gamma}{\cal P}^{(\pm N)}\;[{\cal D}_{2w},
{\cal P}^{(\pm N)}]\;[{\cal D}_{2w}, {\cal P}^{(\pm N)}](\vec{n}). \label{wind0}
\end{equation}
To prove equation (\ref{wind0}) one can first remark that $[{\cal D}_{2w},{\cal
P}^{({\pm}N)}]=-i{\sigma}_i{\partial}_i({\cal P}^{({\pm}N)})$ where we have crucially used the fact
${\partial}_r({\cal P}^{({\pm}N)})=0$ . Next one shows that $Tr{\gamma}{\cal P}^{({\pm}N)}[{\cal D}_{2w},{\cal
P}^{({\pm}N)}]^2=-2i{\epsilon}_{ijk}n_kTr{\cal P}^{({\pm}N)}{\partial}_i({\cal P}^{({\pm}N)}){\partial}_j({\cal
P}^{({\pm}N)})$  by computing the trace over the Pauli matrices ${\sigma}_j$. The remaining trace is only over
the Pauli matrices ${\tau}_j^{(i)}$ . Putting this last equation in (\ref{wind0}) gives
\begin{equation}
\pm N=\frac{1}{2{\pi}i}\int Tr{\cal P}^{({\pm}N)}d{\cal P}^{({\pm}N)}{\wedge}d{\cal P}^{({\pm}N)},\label{wind1}
\end{equation}
which is exactly equation (\ref{chernclass}) of chapter $3$. In arriving at (\ref{wind1}) we have used the identity
$-{\epsilon}_{ijk}n_k d(cos{\theta}){\wedge}d{\phi}=dn_i{\wedge}dn_j$ . In equation (\ref{c}) we have already
computed explicitly the RHS of (\ref{wind1}) , for the case $N=1$ , and found it to be equal ${\pm}1$ . For the case
$N>1$ , the projector ${\cal P}^{(\pm N)}$ is given by equation (\ref{cprojector}) . It is the tensor product of $N$
projectors ${\cal P}^{({\pm}1)}$ and therefore by using the additivity of Chern classes , shown in section
$(3.1.5)$, the RHS of (\ref{wind1}) must be equal to ${\pm}N$ in a trivial manner.

These numbers do not change if ${\cal P}^{(\pm N)}$ are conjugated by $g^{\tilde{\otimes}N}$ and therefore can be
thought of as soliton winding numbers. The proof is that of the stability of the cyclic cocycle (\ref{wind0}) under
deformation of ${\cal P}^{(\pm N)}$, which was given in detail in chapter $2$ .

\subsection{Winding Numbers for the Fuzzy Sphere}

The fuzzy Dirac operator ${\bf D}$ and chirality operator ${\Gamma}$ are important for writing formulae for the
invariants of projectors. There are proposals for ${\bf D}$ and ${\Gamma}$ in \cite{5,6,7,8,10,11}, we briefly
describe those in \cite{10,11} . They were already studied in great detail in chapter $3$. There is a left and
right action (``left'' and ``right'' `` regular representations'' ${\bf A}^{L}={\Pi}(\bf A)$ and ${\bf
A}^{R}={\Pi}^0(\bf A)$) of ${\bf A}$ on ${\bf A}$: $b^{L}a = ba$ and $b^{R}a=ab, (b, a \in{\bf A},
b^{L,R}{\in}{\bf A}^{L,R})$ with corresponding angular momentum operators $L_{i}^{L}=L_i$ and $L_{i}^{R}$ and
fuzzy coordinates $\frac{L_i}{\sqrt{l(l+1)}}=n_{i}^{F}$ and $\frac{L_{i}^{R}}{\sqrt{l(l+1)}}$\footnote{All of these
notations were introduced in chapter $3$.}. ${\bf D}$ and ${\Gamma}$ are given by equations (\ref{fuzzywatamura}) and (\ref{fuzzychirality})
respectively, namely
\begin{eqnarray}
D&=&D_{2w}=\frac{1}{\rho}{\epsilon}_{ijk}{\sigma}_{i}n_{j}^{F}L_{k}^{R}, \nonumber\\
{\Gamma}&=&-\frac{{\sigma}.\vec{L}^{R} -1/2}{l+1/2}. \label{fuzzyoprs}
\end{eqnarray}
Identifying ${\bf A}^L$ as the representation of the fuzzy version of ${\cal A}$, we have as before,
\begin{eqnarray}
\Gamma b^L&=& b^L \Gamma, \nonumber \\
{\Gamma}D + D{\Gamma}&=&0, \nonumber\\
{\Gamma}^2&=&1, \nonumber\\
{\Gamma}^{\dagger}&=&{\Gamma}.
\end{eqnarray}
The carrier space of ${\bf A}^{L, R}$, ${\bf D}_{2w}$ and ${\Gamma}$ is ${\bf A}^{2}$. When $p^{(\pm N)}$ are also
included, it gets expanded to $A^{2^{(N+1)}}$ as $\vec{\tau}^{(i)}$ commute with $\vec{\sigma}$. Note that $
p^{(\pm N)}$ commute with ${\Gamma}$, as the $n$'s in (\ref{cprojector}) become $n^{F}$'s under fuzzification or equivalently
the $L$'s in (\ref{fuzzifypn?}) are being identified with $L^{L}$'s .

We now construct a certain generalization of (\ref{wind0}) for the fuzzy sphere. It looks like (\ref{wind0}), or
rather the following expression
\begin{eqnarray}
\pm N &=& -Tr_{\omega} \left( \frac{1}{|{\cal D}_{2w}|^2}\gamma \;{\cal P}^{(\pm N)}\; [{\cal D}_{2w}, {\cal
P}^{(\pm N)}] \;[{\cal D}_{2w}, {\cal P}^{(\pm N)}]\;
\right),\nonumber\\
|{\cal D}_{2w}|&=&{\rm{Positive~ square~ root~ of}}~ {\cal D}_{2w}^{\dagger}{\cal D}_{2w},\label{wind2}
\end{eqnarray}
where ${\cal F}_{2w}={\cal D}_{2w}/|{\cal D}_{2w}|$ \cite{1,2,3.1,3.2,3.3,3.4}. It is equivalent to (\ref{wind0}). It
involves a Dixmier trace $Tr_{\omega}$ and furthermore the inverse of $|{\cal D}_{2w}|$ . The highly non trivial
fact that (\ref{wind2}) is exactly equation (\ref{wind0}) is shown explicitly in example $6$ of chapter $2$ .

But the massless Dirac operator on ${\bf A}^{2^{(N+1)}}$ has zero modes and therefore $|{\bf D}_{2w}|$ has no
inverse. An easy proof is as follows. We can write the elements of $ {\bf A}^{2^{(N+1)}} $ as rectangular matrices
with entries ${\xi}_{{\lambda}j}{\in}{\bf A}$ $({\lambda}=1,2,...,2^{N};j=1,2)$ where $\lambda$ carries the action
of ${\vec{\tau}}^{(i)}$'s and $j$ carries the action of ${\vec{\sigma}}$. The dimensions of the subspaces $U_\pm$
of ${\bf A}^{2^{(N+1)}}$ with $\Gamma =\pm 1$ are
\begin{equation}
[2(l \pm 1/2)+1][(2l+1)2^{N}],
\end{equation}
which can be proven as follows . The first factor is the row dimension of $U_\pm$ in the sense that it corresponds to
the coupling of ${\vec{\sigma}}/2$ and $-\vec{L}^R$ , in other words to the index $j$ of ${\xi}_{{\lambda}j}$ . It
is deduced from the fact that
${\Gamma}=\frac{1}{l+1/2}\Big[(-\vec{L}^R+\frac{\vec{\sigma}}{2})^2-(l+1/2)^2\Big]=\frac{1}{l+1/2}\Big[j(j+1)-(l+1/2)^2\Big]$
where $j$ is the eigenvalue of $(-\vec{L}^R + \vec{\sigma}/2)^2$ . But since $j=l{\pm}1/2$ , we have
$\Gamma|_{j=l\pm 1/2}=\pm 1$ which defines the subspaces $U_{\pm}$ with row dimensions $2(l\pm 1/2)+1$ . The
second factor in $(4.28)$ is the column dimension of $U_\pm$ which corresponds to the coupling of
$\sum_{i=1}^N{\vec{\tau}}^{(i)}/2$ and $\vec{L}$.

${\bf D}_{2w}$ anticommutes with $\Gamma$. So if ${\bf D}^{(+)}_{2w}$ is the restriction of ${\bf D}_{2w}$ to the
domain $U_+$, ${\bf D}^{(+)}_{2w} = {\bf D}_{2w}|_{U_+} : U_+ \rightarrow U_-$, its index is ${\rm dim}\; U_+ -
{\rm dim}\; U_- = 2[(2l+1)2^{N}]$. This is the minimum number of zero modes of ${\bf D}_{2w}$ in $U_+$.
Calculations \cite{10,11} show this to be the exact number of zero modes, ${\bf D}_{2w}$ having no zero mode in
$U_-$.

In any case, ${\bf D}^{(+)}_{2w}$ and so ${\bf D}_{2w}$ have no inverse. So we work instead with the massive Dirac
operator ${\bf D}_{2wm}={\bf D}_{2w}+m\Gamma\;(m \neq 0)$ with the strictly positive square ${\bf D}^2_{2wm}={\bf
D}^2_{2w}+m^2$ and form the operator
\begin{equation}
f_{m}= \frac{{\bf D}_{2wm}}{|{\bf D}_{2wm}|}
\end{equation}
where
\begin{equation}
|{\bf D}_{2wm}|={\rm{Positive~ square~ root~ of}}~ {\bf D}_{2wm}^{\dagger}{\bf D}_{2wm}~, \quad
f_{m}^{\dagger}=f_{m}~, \quad f^2_{m} = {\bf1}.
\end{equation}
Consider $\frac{1-{\Gamma}}{2}p^{(N)}f_{m} p^{(N)}\frac{1+{\Gamma}}{2}$ where we pick $p^{(N)}$ and not $p^{(-N)}$
for specificity.  It anticommutes with ${\Gamma}$. Let ${\hat{V}}_{\pm} =p^{(N)} U_{\pm}$ be the subspaces of
monopoles wave functions with winding number $N$ and chirality ${\pm}1$ . It then follows that the index of the
operator
\begin{equation}
{\hat{f}}^{(+)}_{m} = \frac{1-{\Gamma}}{2} p^{(N)} f_{m} p^{(N)}\frac{1+{\Gamma}}{2}
\end{equation}
restricted to $\hat{V}_{+}$ is $2[2l+1+N]$. The proof starts by remarking that , by construction, only the matrix
elements $<p^{(N)}U_{-}|{\hat{f}}^{(+)}_{m}|p^{(N)}U_{+}>$ exist and therefore ${\hat{f}}^{(+)}_{m}$ is a mapping
from $\hat{V}_{+}=p^{(N)}U_{+}$ to $\hat{V}_{-}=p^{(N)}U_{-}$ . Hence
$Index{\hat{f}}^{(+)}_{m}=dim\hat{V}_{+}-dim\hat{V}_{-}$ . But since $p^{(N)}$ projects down to the subspace with
maximum eigenvalue $k_{max}=l+N/2$ of the operator $\vec{K}^{(N)}=\vec{L}+\sum_{i=1}^N{\vec{\tau}}^{(i)}/2$ ,
$\hat{V}_{\pm}$ has dimension $[2(l{\pm}1/2)+1][2(l+N/2)+1]$ and so the index is
\begin{equation}
Index{\hat{f}}^{(+)}_{m}=2(2l+N+1).\label{indexoff1}
\end{equation}
In the same way , the index of the adjoint
\begin{equation}
{\hat{f}}^{(+)\dagger}_m \equiv {\hat{f}}^{(-)}_m = \frac{1+{\Gamma}}{2} p^{(N)} f_m p^{(N)}\frac{1-{\Gamma}}{2},
\end{equation}
which is clearly a mapping from $\hat{V}_{-}=p^{(N)}U_{-}$ to $\hat{V}_{+}=p^{(N)}U_{+}$ , can be computed to be
$-2[2l+1+N]$,i.e
\begin{equation}
Index{\hat{f}}^{(-)}_{m}=-2(2l+N+1).
\end{equation}

We may try to associate the index of $ {\hat{f}}^{(+)}_m $ say with the winding number $N$. But that will not be
correct: this index is not zero for $N=0$. The source of this unpleasant feature is a set of unwanted zero modes.

The presence of these zero modes can be established by looking at ${\hat{f}}^{(\pm)}_m$ more closely. $
{\hat{f}}^{(\pm)}_m$ and $\Gamma$ commute with ``total angular momentum'' $\vec{J} = {\vec{L}}^{L} - {\vec{L}}^{R}
+ \sum_{i}\frac{\vec{\tau}^{(i)}}{2} + \frac{\vec{\sigma}}{2}$ while $\Gamma$ anticommutes by construction with
${\hat{f}}^{(\pm)}_m$ . So if an irreducible representation of $\vec{J}$ with $\vec{J}^2=j(j+1){\bf 1}$ occurs an
odd number of times in $\hat{V}_{+} + \hat{V}_{-}$, ${\hat{f}}^{(+)}_m + {\hat{f}}^{(-)}_m$ must vanish on at
least one of these $(2j+1)-$ dimensional eigenspaces. The remaining $(2j+1)-$ dimensional eigenspaces can pair up
so as to correspond to eigenvalues ${\pm}{\lambda}{\neq}0$ and get interchanged by $\Gamma$. The proof lies in  the
fact that if ${\lambda}_{j}{\neq}0$ is the eigenvalue of the operator ${\hat{f}}^{(+)}_m + {\hat{f}}^{(-)}_m$
associated with the eigenfunction ${\psi}_j$ , then $-{\lambda}_j$ is also an eigenvalue of ${\hat{f}}^{(+)}_m
+{\hat{f}}^{(-)}_m$ but with the eigenfunction ${\Gamma}{\psi}_j$ . In other words the $j$ IRR of $\vec{J}$ occurs
twice corresponding to the non zero modes ${\lambda}_j$ and $-{\lambda}_j$ respectively . Hence if this $j$ IRR
comes unpaired , it must only correspond to zero modes .

There are two such $j$, both in $ \hat{V}_{+}=p^{(N)}U_{+} $. They label IRR's with multiplicity $1$ and are its
maximum and minimum $j^{(N)}=2l +\frac{N+1}{2}$ and $\frac{N-1}{2}$. We can see that their eigenspaces have
${\Gamma}=+1$ as follows: the angular momentum value of $\vec{L}^{L} + \sum_{i} {\frac{\vec{\tau}^{(i)}}{2}}$ in
$p^{(N)}A^{2^{(N+1)}}$ is $l+N/2$ so that the angular momentum value of $-\vec{L}^{R}+\frac{\vec{\sigma}}{2}$ must
be $l+1/2$ to attain the $j-$ values $j^{(N)}$ and $\frac{N-1}{2}$.

A further point is that since $( 2j^{(N)} +1 ) + [2(\frac{N-1}{2})+1]=2(2l+N+1)$ is exactly the index of
${\hat{f}}^{(+)}_m$ found earlier, we can conclude that there are no other obligatory zero modes. Indeed every
other $j$ labels IRR's of multiplicity $2$, one with ${\Gamma}=+1$ and the other with $\Gamma=-1$.

The zero modes for $j^{(N)}$ are unphysical as discussed by Watamuras\cite{10,11}: there are no similar modes in
the continuum. If we can project them out, the index will shrink to $2 \frac{N-1}{2} +1 = N$, just what we want. So
let ${\pi}^{(j^{(N)})}$ be the projection operator for $j^{(N)}$, constructed in the same fashion as $p^{(N)}$,
namely
\begin{equation}
{\pi}^{(j^(N))}={\frac{{\prod}_{j{\neq}j^{(N)}} [{\vec{J}}^{2}- j(j+1) ]}{{\prod}_{j{\neq}j^{(N)}} [
j^{(N)}(j^{(N)}+1) - j(j+1) ]}}.
\end{equation}
It commutes with $p^{(N)}$ since $p^{(N)}$ commutes with $ \vec{J}$. In fact, $p^{(N)}{\pi}^{(j^{(N)})} =
{\pi}^{(j^{(N)})}$ since if $j$ is maximum, then so is $k$. We thus find that the projector
\begin{equation}
{\Pi}^{(N)}=p^{(N)} [ {\bf 1} - {\pi}^{(j^{(N)})} ]=p^{(N)} - {\pi}^{(j^{(N)})}
\end{equation}
essentially projects out the subspace with $j=j^{(N)}$ . It commutes with $\Gamma$ too. Let
\begin{eqnarray}
V_{\pm}&=&{\Pi}^{(N)}U_{\pm}, \nonumber\\
f^{(\pm)}_{m} &=& \frac{1{\mp}\Gamma}{2} {\Pi}^{(N)} f_{m} {\Pi}^{(N)} \frac{1{\pm}\Gamma}{2}\label{operatorf}
\end{eqnarray}
where ${f^{(+)}_m}^{\dagger}=f^{(-)}_m$. Then $f^{(+)}_m$ (restricted to $V_{+}$) has the index $N$ we want. This
is obvious from the construction . It is however very instructive to compute this index explicitly .

By equation (\ref{operatorf}) , the operators $f_m^{(\pm)}$ admit only the matrix elements
$<{\Pi}^{(N)}\frac{1{\mp}\Gamma}{2}{\bf A}^{2^{(N+1)}}|f^{(\pm)}_m|{\Pi}^{(N)}\frac{1{\pm}\Gamma}{2}{\bf
A}^{2^{(N+1)}}>$ and therefore $f_m^{(\pm)}$ are mappings from $V_{\pm} $ to $V_{\mp}$ . Hence the operators
$f_{m}^{(-)}f_{m}^{(+)}$ and $f_{m}^{(+)}f_{m}^{(-)}$ are mappings from $V_{+}{\longrightarrow}V_{+}$ and
$V_{-}{\longrightarrow}V_{-}$ respectively . It is next not difficult to show
\begin{eqnarray}
Tr_{V_{+}}f_{m}^{(-)}f_{m}^{(+)}-Tr_{V_{-}}f_{m}^{(+)}f_{m}^{(-)}&=&\nonumber\\
Tr\frac{1+\Gamma}{2}{\Pi}^{(N)}f_{m}^{(-)}f_{m}^{(+)}-Tr\frac{1-\Gamma}{2}{\Pi}^{(N)}f_{m}^{(+)}f_{m}^{(-)}&=&\nonumber\\
Tr\frac{1+\Gamma}{2}{\Pi}^{(N)}f_{m}\frac{1-\Gamma}{2}{\Pi}^{(N)}f_{m}-Tr\frac{1-\Gamma}{2}{\Pi}^{(N)}f_{m}\frac{1+\Gamma}{2}{\Pi}^{(N)}f_{m}&=&0.\nonumber\\
\end{eqnarray}
One can infer from this result that both $f_{m}^{(-)}f_{m}^{(+)}$ and $f_{m}^{(+)}f_{m}^{(-)}$ have the same non
zero eigenvalues . $f_{m}^{(-)}f_{m}^{(+)}$ has further the zero eigenvalues corresponding to the minimum
$\frac{N-1}{2}$ of the total angular momentum $\vec{J}$ . The other zero modes corresponding to the maximum
$j^{(N)}=2l+\frac{(N+1)}{2}$ of $\vec{J}$ are being removed by construction from $V_{+}$ . Hence the index $N$ can
be put in the form

\begin{eqnarray}
Tr\frac{1+\Gamma}{2}{\Pi}^{(N)}[{\bf 1} - f^{(-)}_m f^{(+)}_m]-Tr\frac{1-\Gamma}{2}{\Pi}^{(N)}[{\bf 1} -
f^{(+)}_m f^{(-)}_m]&=&\nonumber \\
Tr\frac{1+\Gamma}{2}{\Pi}^{(N)} - Tr\frac{1-\Gamma}{2}{\Pi}^{(N)}&=&\nonumber\\
Tr_{V_{+}}{\bf 1}-Tr_{V_{-}}{\bf 1}&=& N=~Index ~of\quad f^{(+)}_m.\nonumber\\
\label{ncindex}
\end{eqnarray}
This is because $Tr_{V_{+}}{\bf 1}-Tr_{V_{-}}{\bf
1}=dimV_{+}-dimV_{-}=[dim\hat{V}_{+}-(2j^{(N)}+1)]-dim\hat{V}_{-}$ from which the desired result follows trivially
by using equation (\ref{indexoff1}).

\subsection{Fredholm Module and Chiral Anomaly}
 We want to be able to write (\ref{ncindex}) as a cyclic cocycle coming from a Fredholm module
\cite{1,2,3.1,3.2,3.3,3.4}. The latter for us is based on a representation $\Sigma$ of ${\bf A}^{L}{\otimes}{\bf
A}^{R}$ on a Hilbert space, and operators ${\bf F}$ and $\epsilon$ with the following properties:
\begin{eqnarray}
(i)\; {\bf F}^{\dagger} &=& {\bf F}, \quad {\bf F}^2 = {\bf 1}. \\
(ii)\; \epsilon^{\dagger} &=& \epsilon, \quad \epsilon^2 = 1, \quad \epsilon \Sigma(\alpha) =
\Sigma(\alpha)\epsilon, \quad \epsilon {\bf F} = - {\bf F} \epsilon
\end{eqnarray}
where ${\alpha}{\in}{\bf A}^{L}{\otimes}{\bf A}^{R}$.  [This gives an {\it even} Fredholm module, there need be no
$\epsilon$ in an odd one.] We choose for $\Sigma$ the representation
\begin{equation}
\Sigma:\alpha \rightarrow \Sigma(\alpha) = \left( \begin{array}{cc}
                                      \alpha & 0 \\
                                       0  & \alpha
                                   \end{array}
                             \right)\label{representationsigma}
\end{equation}
on ${\bf A}^{2^{(N+1)}} \oplus {\bf A}^{2^{(N+1)}}$ and set
\begin{equation}
{\bf F} = \left( \begin{array}{cc}
             0  & f_m \\
             f_m & 0
           \end{array} \right), \quad \epsilon=\left( \begin{array}{cc}
                                                    {\bf 1} & 0 \\
                                                        0  & -{\bf 1}
                                                 \end{array} \right).
\end{equation}
Introduce the projector
\begin{equation}
P^{(N)}=\left( \begin{array}{cc}
           \frac{1+\Gamma}{2} {\Pi}^{(N)} & 0 \\
                    0  & \frac{1-\Gamma}{2} {\Pi}^{(N)}
         \end{array} \right). \label{correctedprojector}
\end{equation}
Then
\begin{equation}
(P^{(N)}{\bf F}P^{(N)})^2 = \left( \begin{array}{cc}
                   f^{(-)}_m f^{(+)}_m & 0 \\
                      0  & f^{(+)}_m f^{(-)}_m
                 \end{array}
          \right).
\end{equation}
Therefore,
\begin{equation}
~Index ~of ~f^{(+)}_m = {\rm Tr}\; \epsilon \;[P^{(N)} - (P^{(N)}{\bf F}P^{(N)})^2]. \nonumber
\end{equation}
But since \cite{1}
\begin{eqnarray}
P^{(N)}-(P^{(N)}{\bf F}P^{(N)})^2 &=& -P^{(N)}[{\bf F}, P^{(N)}]^2 P^{(N)}, \\
~Index ~of ~f^{(+)}_m &=& N = -{\rm Tr}\; \epsilon P^{(N)}\;[{\bf F}, P^{(N)}]\;[{\bf F}, P^{(N)}].
\label{dncindex}
\end{eqnarray}
This is the formulation of (\ref{ncindex}) we aimed at and is the analogue of (\ref{wind0}). It is worth remarking
that we can replace $\epsilon$ by $\left(\begin{array}{cc} {\Gamma} & 0 \\ 0 & {\Gamma}
\end{array} \right)$ here since ${\epsilon}P^{(N)} =
\left(\begin{array}{cc} {\Gamma} & 0 \\ 0 & {\Gamma} \end{array} \right) P^{(N)}$.

In $p^{(-N)}{\bf A}^{2^{(N+1)}}$ as well, the unwanted zero modes correspond to the top value $j^{(-N)}=2l
-\frac{N-1}{2}$ of ``total angular momentum ''. This can be seen by recalling that fuzzy monopole wave functions
belonging to $p^{(-N)}{\bf A}^{2^{(N+1)}}$ have the minimum eigenvalue $k_{min}=l-\frac{N}{2}$ of the operator
$\vec{K}^{(N)}=\vec{L}+\sum_{i=1}^{N}\frac{\vec{\tau}^{(i)}}{2}$. Hence by coupling $-\vec{L}^R +
\frac{\vec{\sigma}}{2}$ to $\vec{K}^{(N)}$ , one obtains the positive chirality eigenvalues $2l-\frac{N-1}{2}$ ,
$2l-\frac{N+1}{2}$ ,..., $\frac{N+2}{2}$ ,$\frac{N+1}{2}$ and the negative chirality eigenvalues $2l-\frac{N+1}{2}$
,..., $\frac{N+1}{2}$ , $\frac{N-1}{2}$. By inspection all these eigenvalues occur twice , corresponding to both
chiralities , except for the first and the last . Once the top eigenvalue $j^{(-N)}=2l -\frac{N-1}{2}$ is
suppressed, the remaining obligatory zero modes , $\frac{N-1}{2}$, have multiplicity $N$ and $\Gamma=-1$. Let
${\pi}^{(j^{(-N)})}$ be the projector for the top angular momentum. Then $j^{(-N)}$ can be projected out by
replacing $p^{(-N)}$ by
\begin{equation}
{\Pi}^{(-N)}=p^{(-N)} [ {\bf 1} - {\pi}^{(j^{(-N)})}],
\end{equation}
but now $p^{(-N)}{\pi}^{j^{(-N)}}{\neq}{\pi}^{j^{(-N)}}$, since $j^{(-N)}$ can be obtained in this case in a
variety of ways , by adding $k_{min}$ and $l+\frac{1}{2}$ like we did above , or say , by adding $k_{min}+1$ and
$l-\frac{1}{2}$ which is not in $p^{(-N)}{\bf A}^{2^{(N+1)}}$ but have a total angular momentum
$j^{(-N)}$\footnote{In the case of ${\pi}^{j^{(N)}}$ considered above, $j^{(N)}$ was obtainable only
in one way , namely by adding $k_{max}$ and $l+\frac{1}{2}$ and hence $p^{(N)}{\pi}^{j^{(N)}}={\pi}^{j^{(N)}}$.} .
Substituting ${\Pi}^{(-N)}$ for ${\Pi}^{(N)}$ in (\ref{correctedprojector}), we define $P^{(-N)}$ and then by using
(\ref{dncindex}) can associate $-N$ too with an index.

\subsection{More About Fuzzy $\sigma$-Fields}

There is the topic of fuzzy $\sigma$-fields yet to be discussed in this section. We first note that in $u$ defined
earlier, $a_{\mu}$ is to be identified with $a_{\mu}^{L}$. Let us extend $u^{{\tilde{\otimes}}N}$ and
$g^{{\tilde{\otimes}}N}$ from ${\bf A}^{2^{N}}$ and ${\cal A}^{2^N}$ to ${\bf A}^{2^{(N+1)}} = {\bf A}^{2^{N}}{\otimes}{\bf C}^2$ and
${\cal A}^{2^{N+1}} = {\cal A}^{2^{N}}{\otimes}{\bf C}^2$ so that they act as identity on the last ${\bf C}^2$'s .
In this extension we are simply including the effects of spin $\frac{\vec{\sigma}}{2}$ . We also extend them
further to ${\bf A}^{2^{(N+1)}}{\oplus}{\bf A}^{2^{(N+1)}}\equiv {\bf A}^{2^{(N+1)}}{\otimes}{\bf C}^2$ and ${\cal
A}^{2^{(N+1)}}{\oplus}{\cal A}^{2^{(N+1)}}\equiv {\cal A}^{2^{N+1}}{\otimes}{\bf C}^2$ so that they act as
identity on these last ${\bf C}^2$'s. This last extension is to take care of the effects of the representation
$\Sigma$ defined in equation (\ref{representationsigma}). Define also
\begin{equation}
Q(u) = u^{{\tilde{\otimes}}N} Q (u^{{\tilde{\otimes}}N})^{-1}
\end{equation}
for an operator $Q ( = Q({\bf 1}))$ on ${\bf A}^{2^{N+1}} {\oplus} {\bf A}^{2^{N+1}}$. The right hand side of
(\ref{dncindex}) is invariant under the substitution $P^{(N)}{\rightarrow}P^{(N)}(u)$ without changing ${\bf F}$.
So $P^{(N)}(u)$ is a candidate for a fuzzy winding number $N$ $\sigma$-field in the present context whereas
previously it was $p^{(N)}(u)$.

But we must justify this candidacy by looking at the continuum limit. In that limit, $u^{{\tilde{\otimes}}N}
{\rightarrow} g^{{\tilde{\otimes}}N}$, ${\pi}^{(j^{(N)})} {\rightarrow} {\pi}^{(j^{(N)})}_{\infty}$ say and
${\Pi}^{(N)} {\rightarrow} {\Pi}^{(N)}_{\infty}={\cal P}^{(N)} - {\pi}^{(j^{(N)})}_{\infty}$. The stability group
of ${\cal P}^{(N)}$ under conjugation by $g^{{\tilde{\otimes}}N}$ is as before $U(1)$ at each $\vec{n}$. Now it is
obvious that ${\pi}^{(j^{(N)})}$ projects down to those states where any one of $(\vec{L}^{L} +
\frac{{\vec{\tau}}^{(i)}}{2})^2$, $(\vec{L}^{L} + \frac{\vec{\sigma}}{2})^2$, $(\vec{L}^{L}-\vec{L}^{R})^2$ has
the maximum value. By this observation it is not difficult to find
\begin{eqnarray}
{\vec{\tau}}^{(i)}. {\vec{n}^{F}}{\pi}^{(j^{(N)})}&=&\frac{1}{\sqrt{1+\frac{1}{l}}}{\pi}^{(j^{(N)})}\nonumber\\
{\vec{\sigma}}.{\vec{n}^{F}}{\pi}^{(j^{(N)})}&=&\frac{1}{\sqrt{1+\frac{1}{l}}}{\pi}^{(j^{(N)})}\nonumber\\
&and&\nonumber\\
{\vec{n}}^{F}. {\vec{n}}^{F,R}{\pi}^{(j^{(N)})}&=&\frac{1}{1+\frac{1}{l}}{\pi}^{(j^{(N)})}.\label{limitdifferentobjects}
\end{eqnarray}
$\vec{n}^{F}=\frac{\vec{L}^L}{\sqrt{l(l+1)}}$ and $\vec{n}^{F,R}=-\frac{\vec{L}^R}{\sqrt{l(l+1)}}$ . They both tend
to $\vec{n}$ in the limit $l{\longrightarrow}{\infty}$, so that the last equation in (\ref{limitdifferentobjects}) trivially reduces to
the defining equation of ${\bf S}^2$. On the other hand , the other two equations in (\ref{limitdifferentobjects}) reduce to
${\vec{\tau}}^{(i)}\vec{n}{\pi}^{(j^{(N)})}_{\infty}={\vec{\sigma}}\vec{n}{\pi}^{(j^{(N)})}_{\infty}={\pi}^{(j^{(N)})}_{\infty}$
and hence ${\pi}^{(j^{(N)})}_{\infty}$ can be identified with ${\cal P}^{(N+1)}$ , namely
\begin{equation}
{\pi}^{(j^{(N)})}_{\infty}=(\prod_{i=1}^{N}\frac{1+\vec{\tau}^{(i)}.\hat{n}}{2} )\frac{1+\vec{\sigma}.\hat{n}}{2},
\end{equation}
and therefore
\begin{equation}
{\Pi}^{(N)}_{\infty}={\cal P}^{(N)}\frac{1-\vec{\sigma}\vec{n}}{2}.
\end{equation}
From this result it is clear that ${\Pi}^{(N)}_{\infty}$ has the $U(1)$ stability group at each $\vec{n}$ .
${\Pi}^{(N)}_{\infty}(g)$ is a $\sigma$-field on ${\bf S}^2$ and $P^{(N)}(u)$ is a good choice for the fuzzy
$\sigma$-field.

For winding number $-N$, we propose $P^{(-N)}(u)$ as the fuzzy $\sigma$-field. We can check its validity also by
going to the continuum limit. As $l{\rightarrow}{\infty}$, $p^{(-N)}{\rightarrow} {\cal P}^{(-N)}=\prod_{i=1}^{N}
\frac{1-{\vec{\tau}^{(i)}}.\hat{n}}{2}$ and has the $U(1)$ stability group at each $\vec{n}$. Next consider the
product $p^{(-N)}{\pi}^{(j^{(-N)})}$. The presence of $p^{(-N)}$ allows us to assume that
$({\vec{K}}^{(N)})^2=k_{min}(k_{min}+1)$, $k_{min}=l-N/2$. Also we can substitute for ${\pi}^{(j^{(-N)})}$ the
projector coupling ${\vec{K}^{(N)}}$ with $-\vec{L}^{R} + \frac{\vec{\sigma}}{2}$ to give maximum angular momentum.
This projector can be found following the same procedure as above .

${\pi}^{(j^{(-N)})}$ projects down to those states where $(\vec{L}^{L} + \frac{{\vec{\tau}}^{(i)}}{2})^2$ has its
minimum value whereas  $(\vec{L}^{L} + \frac{\vec{\sigma}}{2})^2$ and $(\vec{L}^{L}-\vec{L}^{R})^2$ have their
maximum values . This fact can be written as the requirement
\begin{eqnarray}
{\vec{\tau}}^{(i)}. {\vec{n}^{F}}{\pi}^{(j^{(-N)})}&=&-{\sqrt{1+\frac{1}{l}}}{\pi}^{(j^{(-N)})}\nonumber\\
{\vec{\sigma}}.{\vec{n}^{F}}{\pi}^{(j^{(-N)})}&=&\frac{1}{\sqrt{1+\frac{1}{l}}}{\pi}^{(j^{(-N)})}\nonumber\\
&and&\nonumber\\
{\vec{n}}^{F}. {\vec{n}}^{F,R}{\pi}^{(j^{(-N)})}&=&\frac{1}{1+\frac{1}{l}}{\pi}^{(j^{(-N)})}.
\end{eqnarray}
These equations reduce in the continuum limit to
${\vec{\tau}}^{(i)}\vec{n}{\pi}^{(j^{(-N)})}_{\infty}=-{\vec{\sigma}}\vec{n}{\pi}^{(j^{(N)})}_{\infty}=-{\pi}^{(j^{(N)})}_{\infty}$
and $\vec{n}^2{\pi}^{(j^{(-N)})}= {\pi}^{(j^{(-N)})}$ . Hence ${\pi}^{(j^{(-N)})}_{\infty}$ can be identified with
\begin{equation}
{\pi}^{(j^{(-N)})}_{\infty}=(\prod_{i=1}^{N}\frac{1-\vec{\tau}^{(i)}.\hat{n}}{2} )\frac{1+\vec{\sigma}.\hat{n}}{2},
\end{equation}
So $p^{(-N)} {\pi}^{(j^{(-N)})}$ as $l{\rightarrow}{\infty}$ can be identified with ${\cal
P}^{(-N)}{\pi}^{(j^{(-N)})}_{\infty}$ and so
\begin{eqnarray}
{\Pi}_{\infty}^{(-N)}&=&{\cal P}^{(-N)}-{\cal P}^{(-N)}{\pi}_{\infty}^{(j^{(-N)})}\nonumber\\
&=&{\cal P}^{(-N)}\frac{1-\vec{\sigma}.\vec{n}}{2}.
\end{eqnarray}
This clearly has the $U(1)$ stability group at each $\vec{n}$ which shows that $P^{(-N)}(u)$ is a good fuzzy
$\sigma$-field for winding number $-N$.

\section{Dynamics for Fuzzy $\sigma$-Models}

\subsection{Belavin-Polyakov Bound in the Fuzzy Setting}

The simplest action for the $O(3)$ nonlinear $\sigma$-model on ${\bf S}^2$ is
\begin{eqnarray}
&S&=\frac{\beta}{2}\int\frac{dcos{\theta}d{\phi}}{4{\pi}}({\cal
L}_i{\Phi}_a)(\vec{n})({\cal L}_i{\Phi}_a)(\vec{n}),\nonumber\\
&\sum_{a=1}^{3}&{\Phi}_a(\vec{n})^2=1,~{\beta}>0 \label{nlsm}
\end{eqnarray}
where ${\cal L}_i=-i{\epsilon}_{ijk}x_j{\partial}_k$ are the angular momentum operators on ${\bf S}^2$. It fulfills
the important bound [see page $112$ of the second reference of \cite{17} , and \cite{belpol}]
\begin{equation}
S{\geq}{\beta}N \label{bound}
\end{equation}
where $N({\geq}0)$ or$-N$ is as usual the winding number of the map $\vec{\Phi}: {\bf S}^2{\rightarrow}{\bf S}^2$:
\begin{eqnarray}
{\rm{~Winding ~number ~of}}~{\vec{\Phi}}&=&\frac{1}{2} \int_{S^2} \frac{dcos{\theta}d{\phi}}{4{\pi}}
{\epsilon}_{ijk}n_i{\epsilon}_{abc}{\Phi}_a ({\cal L}_j{\Phi}_b)({\cal L}_k{\Phi}_c)\nonumber\\
&=&\frac{1}{8{\pi}}\int {\epsilon}_{abc}{\Phi}_ad{\Phi}_b{\wedge}d{\Phi}_c,\label{windingphi}
\end{eqnarray}
where we have used the identity ${\epsilon}_{kpl}n_ld(cos{\theta}){\wedge}d{\phi}=-dn_k{\wedge}dn_p$ . By equation
(\ref{c1}) the second line in (\ref{windingphi}) is precisely equal to ${\pm}N$ , namely the winding number of the field
$\vec{\Phi}$ .

The bound (\ref{bound}) is obtained by integrating the inequality
\begin{equation}
({\cal L}_i{\Phi}_a {\pm} {\epsilon}_{ijk} n_j{\epsilon}_{abc}{\Phi}_b{\cal L}_k{\Phi}_c)^2{\geq}0.
\label{inequality}
\end{equation}
Indeed one shows $({\cal L}_i{\Phi}_a {\pm} {\epsilon}_{ijk} n_j{\epsilon}_{abc}{\Phi}_b{\cal
L}_k{\Phi}_c)^2=2\Big[({\cal L}_i{\Phi}_a)^2 {\pm} {\epsilon}_{ijk} n_i{\epsilon}_{abc}{\Phi}_a{\cal
L}_j{\Phi}_b{\cal L}_k{\Phi}_c\Big]$ and therefore $({\cal L}_i{\Phi}_a)^2{\geq} {\mp} {\epsilon}_{ijk}
n_i{\epsilon}_{abc}{\Phi}_a{\cal L}_j{\Phi}_b{\cal L}_k{\Phi}_c$ or $S{\geq}{\mp}{\beta}{\times}$winding number of
$\vec{\Phi}$ . Hence for winding number $+N$ , one chooses the positive sign of the inequality whereas for winding
number $-N$ we choose the negative sign and both cases lead to the bound (\ref{bound}) .

The inequality (\ref{inequality}) is saturated if and only if
\begin{equation}
{\cal L}_i{\Phi}_a {\pm} {\epsilon}_{ijk} n_j{\epsilon}_{abc}{\Phi}_b{\cal L}_k{\Phi}_c = 0 \label{saturation}
\end{equation}
for one choice of sign . The solutions of (\ref{saturation}) can be thought of as two dimensional instantons
\cite{belpol}.

We now propose a fuzzy $\sigma$-action using these properties of $S$ as our guide. Consider the inequality
\begin{equation}
\Big[[{\bf F},P(u)]\frac{1{\pm}{\epsilon}}{2}P(u)\Big]^{\dagger} \Big[[{\bf
F},P(u)]\frac{1{\pm}{\epsilon}}{2}P(u)\Big]{\geq}0
\end{equation}
where $P(u)$ can be $P^{(N)}(u)$ or $P^{(-N)}(u)$ and $Q{\geq}0$ here means that $Q$ is a nonnegative operator.
This is the analogue of (\ref{inequality}). Taking trace, we get the analogue of (\ref{bound}),
\begin{equation}
s_F{\equiv}-Tr P(u) [{\bf F},P(u)][{\bf F},P(u)]{\geq}N. \label{dinequality}
\end{equation}
The proof is similar .

One first checks $\Big[[{\bf F},P(u)]\frac{1{\pm}{\epsilon}}{2}P(u)\Big]^{\dagger} \Big[ [{\bf
F},P(u)]\frac{1{\pm}{\epsilon}}{2}P(u)\Big]=-P(u)\frac{1{\pm}{\epsilon}}{2}P(u)[{\bf
F},P(u)]^2\frac{1{\pm}{\epsilon}}{2}P(u)$. Next by taking the trace we obtain $-TrP(u)[{\bf
F},P(u)]^2{\geq}{\mp}\Big[-Tr{\epsilon}P(u)[{\bf F},P(u)]^2\Big]$. For $P(u)=P^{(N)}(u)$ we know from equation
(\ref{dncindex}) that $-Tr{\epsilon}P(u)[{\bf F},P(u)]^2=N$ and hence we choose the positive sign of the above inequality
and obtain the bound $-Tr P(u) [{\bf F},P(u)][{\bf F},P(u)]{\geq}N $ . For $P(u)=P^{(-N)}(u)$,
$-Tr{\epsilon}P(u)[{\bf F},P(u)]^2=-N$ and the bound is obtained by choosing the negative sign .

The bound is saturated if and only if
\begin{equation}
[{\bf F},P(u)]\frac{1{\pm}{\epsilon}}{2}P(u)=0 \label{dsaturation}
\end{equation}
for one choice of sign, just like in (\ref{saturation}). All this suggests the novel fuzzy $\sigma$-action
\begin{equation}
S_F={\beta}_Fs_{F}.
\end{equation}

\subsection{Continuum Limit for Fuzzy ${\sigma}-$Models}

Qualitative remarks about the approach to continuum of $S_F$ will now be made. The first is that $\beta_F$ and $m$
must be scaled as $l{\rightarrow}{\infty}$. As regards the scaling of $\beta_F$, we conjecture that
(\ref{dsaturation}) has no solution for finite $l$ and that (\ref{dinequality}) is a strict inequality. Choose
\begin{equation}
{\Lambda}(l)= \frac{1}{N}{\times}(~Minimum ~of ~ s_F)
\end{equation}
so that $\frac{s_F}{\Lambda(l)}=N\frac{s_F}{~Minimum ~of ~s_F}$ is $N$ at minimum. Then we suggest that we should
set
\begin{equation}
{\beta}_F=\frac{\beta}{\Lambda(l)}.
\end{equation}
In other words , $S_F={\beta}N\frac{s_F}{~Minimum ~of ~s_F}$ . It is our conjecture too that $\Lambda (l)$
diverges as $l{\rightarrow}{\infty}$ in such a way that (upto factors)
\begin{eqnarray}
S_F{\rightarrow}S_{\infty} &=& {\beta} Tr_{\omega} P_{\infty}(g)
[{\cal F}, P_{\infty}(g)][{\cal F},P_{\infty}(g)],\nonumber\\
{\cal F}&=& \left(\begin{array}{cc}
                  0  &  {\cal D}_{2w}/|{\cal D}_{2w}| \\
                  {\cal D}_{2w}/|{\cal D}_{2w}|   &  0
               \end{array}
          \right), \nonumber \\
g &=& \lim_{l{\rightarrow}{\infty}}u,\nonumber\\
P_{\infty}(g) &=& \lim_{l{\rightarrow}{\infty}}P(u)
\end{eqnarray}
where we have let $m$ become zero as ${\cal D}_{2w}$ has no zero mode. An alternative form of $S_{\infty}$ is
{\small
\begin{equation}
S_{\infty} = {\beta}\int\frac{d\cos{\theta}d{\phi}}{4{\pi}} Tr
   P_{\infty}(g) \left[\left(\begin{array}{cc}
                  0  &  {\cal D}_{2w} \\
                  {\cal D}_{2w}   &  0
                \end{array}
          \right)\!, P_{\infty}(g)\right]\!\!\left[\left(\begin{array}{cc}
                  0  &  {\cal D}_{2w} \\
                  {\cal D}_{2w}   &  0
                \end{array}
          \right)\!, P_{\infty}(g)\right]
\end{equation}
} where the trace $Tr$ is only over the internal indices.

We now argue that $P(u)$ itself must be corrected by cutting off all high angular momenta (and not just the top
one) while passing to continuum. Thus it was mentioned before that state vectors with top ``total'' angular
momentum $j^{({\pm}N)}$ are unphysical. Their characteristic feature is their divergence as $l \rightarrow
\infty$. That means that once normalized these vectors become weakly zero in the continuum limit. In fact any
sequence of vectors with a linearly divergent $j$ as $l{\rightarrow}{\infty}$ is unphysical. Such $j$ contribute
eigenvalues to the Dirac operator which are nonexistent in the continuum, as one can verify using the results of
\cite{10,11} for $N=0$: the spectrum of ${\bf D}_{2w}$ is ${\pm}(j+\frac{1}{2})[ 1 + (1 - (j +
\frac{1}{2})^2)/(4l(l+1))]^{1/2}$ , [see equation (\ref{eigenvaluewatamura})], while that of ${\cal D}_{2w}$ is
${\pm}(j+\frac{1}{2})$, $j$ being total angular momentum. The corresponding eigenvectors too if normalized are
weakly zero in the $l \rightarrow\infty$ limit. It seems necessary therefore to eliminate them in a suitable sense
during the passage to the limit.

One way to do so may be to use a double limit which we now describe. Let ${\pi}^{(J)}$ be the projection operator
for all states with $j{\geq}J$. Let us define
\begin{eqnarray}
P^{({\pm}N)(J)}&=&
 \left(\begin{array}{cc}
                  \frac{1+{\Gamma}}{2}p^{(\pm N)}({\bf 1} -
 {\pi}^{(J)})  &  0 \\
                  0   &  \frac{1-{\Gamma}}{2}p^{(\pm N)}({\bf 1} -
 {\pi}^{(J)})
                \end{array}
          \right), \nonumber\\
P^{{({\pm}N)}(J)}(u) &=& u^{{\tilde{\otimes}}N} P^{{({\pm}N)}(J)}
 [u^{{\tilde{\otimes}}N}]^{-1}.
\end{eqnarray}

We then consider the fuzzy $\sigma$-model with $P^{({\pm}N)(J)}(u)$ replacing $P^{({\pm}N)}(u)$ ${\equiv}
P^{({\pm}N)(j^{(N)})}(u)$ and thereby cutting off angular momenta ${\geq}J$. That would not affect index theory
arguments so long as $J>\frac{N-1}{2}$ as the important zero modes will then be left intact. We are thus led to
the cut-off action
\begin{eqnarray}
S_F^{(J)} &=& \frac{\beta}{{\Lambda}^{(J)}(l)}s_F^{(J)},\nonumber\\
s_F^{(J)} &=& -TrP^{({\pm}N)(J)}(u) [{\bf F},P^{({\pm}N)(J)}(u)]
[{\bf F},P^{({\pm}N)(J)}(u)],\nonumber\\
{\Lambda}^{(J)}(l) &=& {\frac{{\rm{~Minimum ~of}} ~s_F^{(J)}}{N}},
\end{eqnarray}
and the following suggestion: A good way to define the continuum partition function is to let $l$ and
$J{\rightarrow}{\infty}$ in that order in the partition function of $S_F^{(J)}$. Thus we propose the continuum
partition function
\begin{equation}
Z=\lim_{J{\rightarrow}{\infty}} \lim_{l{\rightarrow}{\infty}}\int d{\mu}~ exp (- S_F^{(J)}),
\end{equation}
$d{\mu}$ denoting the functional measure. The inner limit recovers the continuum where the contributions of
vectors with divergent $J$ should not matter, for this reason this method may eliminate the influence of unwanted
modes from $Z$. Perhaps an equivalent limiting procedure would be to let $l,J$ ${\rightarrow}{\infty}$ with
$J/l{\rightarrow}0$.

Taking the limit $ l{\rightarrow}{\infty} $ with fixed $J$ is compatible with the continuum description of the
$\sigma$-field. In that limit, $p^{({\pm}N)}$ becomes ${\cal P}^{({\pm}N)}$.  Next consider the vectors projected
by $p^{({\pm}N)}[1-{\pi}^{(J)}]$. The effect of the last factor on the projected vectors is as follows: For
$\Gamma=1$ say, we must combine the angular momentum value $l{\pm}\frac{N}{2}$ of $\vec{K}^{(N)}$ with the value
$l+1/2$ of $-\vec{L}^{R} + \frac{\vec{\sigma}}{2}$ to produce an allowed value $j<J$ of any such vector. So
$[\vec{K}^{(N)} + (-\vec{L}^{R} + \frac{\vec{\sigma}}{2})]^2 = j(j+1)$, $(\vec{K}^{(N)})^2 =
(l{\pm}\frac{N}{2})(l{\pm}\frac{N}{2} + 1)$ and $(-\vec{L}^{R} + \frac{\vec{\sigma}}{2})^2 =
(l+\frac{1}{2})(l+\frac{3}{2})$. Letting $l{\rightarrow}{\infty}$, we find that
$\vec{n}^F.\vec{n}^{F,R}{\rightarrow}-1$\footnote{See equation (\ref{limitdifferentobjects}) for the definition of $\vec{n}^F$ and
$\vec{n}^{F,R}$ respectively.} due to the factor $[1-{\pi}^{(J)}]$, where we have used the fact that
$\frac{\vec{\tau}^{(i)}}{l}$ and $\frac{\vec{\sigma}}{l} {\rightarrow}0$ as $l{\rightarrow}{\infty}$. But this is
just a rule instructing us to set $\vec{n}^{F,R}=-\hat{n}$ for large $l$ for these vectors, and therefore will not
show up in the continuum projector. The $\Gamma=-1$ case is no different in the continuum limit. Thus for
$l{\rightarrow}{\infty}$, $p^{({\pm}N)}(u)[1-{\pi}^{(J)}(u)]$ can be interpreted as ${\cal P}^{({\pm}N)}(g) =
g^{{\hat{\otimes}}N}{\cal P} ^{({\pm}N)}[g^{{\hat{\otimes}}N}]^{-1}$, the continuum $\sigma$-fields.

Let
\begin{equation}
 P^{({\pm} N)(J)}_{\infty}(g) = \lim_{l{\rightarrow}{\infty}}
P^{({\pm}N)(J)}(u) = \left(\begin{array}{cc}
                  \frac{1+{\gamma}}{2}{\cal P}^{({\pm}N)}(g)  &  0 \\
                  0   & \frac{1-{\gamma}}{2}{\cal P}^{({\pm}N)}(g)
                \end{array}
          \right).
\end{equation}
Then the naive $l{\rightarrow}{\infty}$, $m{\rightarrow}0$ limit of $S_F^{(J)}$ is expected to be (upto factors)
\begin{equation}
{\cal S}_{\infty} = {\beta}Tr_{\omega} P^{({\pm}N)(J)}_{\infty}(g) [{\cal F},P^{({\pm}N)(J)}_{\infty}(g)][{\cal
F},P^{({\pm}N)(J)}_{\infty}(g)]
\end{equation}
which can be simplified to
\begin{equation}
{\cal S}_{\infty} = {\beta}\int\frac{dcos{\theta}d{\phi}}{4{\pi}} Tr{\cal P}^{{(\pm)N}}(g)[{\cal D}_{2w},{\cal
P}^{{(\pm)N}}(g)][{\cal D}_{2w},{\cal P}^{{(\pm)N}}(g)].
\end{equation}
It seems to correspond to (\ref{nlsm}).

\section{The Fermion Doubling Problem and Noncommutative Geometry}

\subsection{The Fermion Doubling Problem on the Lattice}

The nonperturbative formulation of chiral gauge theories is  a long standing programme in particle physics. It
seems clear that one should regularise these theories with all symmetries intact. There is a major problem
associated with conventional lattice approaches  to this programme, with roots in topological features: The
Nielsen-Ninomiya theorem \cite{last2} states that {\it if we want to maintain chiral symmetry, then under plausible
assumptions, one cannot avoid the doubling of fermions in the usual lattice formulations}.

To see this , let us recall how one encounters the doubling problem on a $4-$dimensional Euclidean lattice . Our
discussion and notations will follow that of \cite{creutz} . One starts by remarking that the canonical continuum
Fermion action in Euclidean $4d$ space-time
\begin{equation}
{\cal L}=\bar{\psi}({\gamma}^{\mu}{\partial}_{\mu}+m){\psi}\label{startingfermionicaction}
\end{equation}
has the symmetry ${\psi}{\longrightarrow}e^{i{\theta}}{\psi}$ as well as the symmetry when $m{\longrightarrow}0$ of
${\psi}{\longrightarrow}e^{i{\theta}{\gamma}_5}{\psi}$. The associated conserved currents are known to be given by
$J_{\mu}=\bar{\psi}{\gamma}_{\mu}{\psi}$ and $J_{\mu}^{5}=\bar{\psi}{\gamma}_{\mu}{\gamma}_5{\psi}$ where
${\gamma}_5={\gamma}_1{\gamma}_2{\gamma}_3{\gamma}_4$ . It is also a known result that in quantum theory one
can not maintain the conservation of both of these currents simultaneously in the presence of gauge fields.

A regularization scheme , which maintains exact chiral invariance , of the above action can be achieved by
replacing the Euclidean four dimensional space-time by a $4-$d hypercubic lattice of $N^4$ sites . Points are now
being labeled by $x_{m}^{{\mu}}=am^{\mu}$ where $a$ is by definition the lattice spacing . $m^{\mu}$ is a four
component vector where each of the component is an integer in the range $-\frac{N}{2}<m^{\mu}{\leq}\frac{N}{2}$ .
The lattice is assumed to be periodic outside this range . Now to each site $x_m=am$ we associate a spinor variable
${\psi}_{m}$ and the derivative ${\partial}_{\mu}{\psi}(x)$ is replaced by
\begin{equation}
{\partial}_{\mu}{\psi}(x){\longrightarrow}\frac{1}{2a}\Big[{\psi}_{m_{\nu}+{\delta}_{{\mu}{\nu}}}-{\psi}_{m_{\nu}-{\delta}_{{\mu}{\nu}}}\Big],
\end{equation}
where ${\delta}_{{\mu}{\nu}}=1$ for ${\mu}={\nu}$ and $0$ otherwise . With this prescription the action (\ref{startingfermionicaction})
becomes

\begin{eqnarray}
S_a&=&\sum_{m,n}a^4\bar{\psi}_m\sum_{\mu}{\gamma}_{\mu}\frac{1}{2a}\Big[{\psi}_{m_{\nu}+{\delta}_{{\mu}{\nu}}}-{\psi}_{m_{\nu}-{\delta}_{{\mu}{\nu}}}\Big]+m\sum_{m,n}a^4\bar{\psi}_m{\psi}_m\nonumber\\
&{\rm or}&\nonumber\\
S_a&=&\sum_{m,n}\bar{\psi}_mM_{mn}{\psi}_n\nonumber\\
&where&\nonumber\\
M_{mn}&=&\frac{a^3}{2}\sum_{\mu}{\gamma}_{\mu}\Big[{\delta}^4_{m_{\nu}+{\delta}_{{\mu}{\nu}},n_{\nu}}-{\delta}^4_{m_{\nu}-{\delta}_{{\mu}{\nu}},n_{\nu}}\Big]+ma^4{\delta}^4_{m.n}.\label{discretetheory}
\end{eqnarray}
Next one can compute the propagator  $S_{mn}=(M^{-1})_{mn}$ of this action as follows. Let us define the Fourier
transform $\tilde{f}_k$ of an arbitrary complex function $f_m$ on the lattice by
$\tilde{f}_k=\sum_{m}f_me^{2i{\pi}\frac{km}{N}}$ where each component $k_{\mu}$ of $k$ is in the range
$-\frac{N}{2}<k_{\mu}{\leq}\frac{N}{2}$, i.e the momentum space lattice is periodic outside this range .
From this definition one can derive the identities $\sum_{k}e^{-2i{\pi}\frac{km}{N}}=N^4{\delta}^4_{m,0}$,
$f_m=\frac{1}{N^4}\sum_{k}\tilde{f}_ke^{-2i{\pi}\frac{km}{N}}$,
$\sum_{m}f^{*}_mg_m=\frac{1}{N^4}\sum_{k}\tilde{f}^{*}_k\tilde{g}_k$ and
$\sum_{m}f^{*}_{m_{\mu}+{\delta}_{{\mu}{\nu}}}g_m=\frac{1}{N^4}\sum_{k}\tilde{f}^{*}_k\tilde{g}_ke^{2i{\pi}\frac{k_{\nu}}{N}}$. Hence by writing
\begin{equation}
(M^{-1})_{mn}=\frac{1}{a^4N^4}\sum_{k}\tilde{M}_k^{-1}e^{2i{\pi}\frac{k(m-n)}{N}},
\end{equation}
and using the identity $(M)_{mn}(M^{-1})_{nl}={\delta}^4_{m,l}$ one can find that
\begin{equation}
\tilde{M}_k=m+\frac{i}{a}\sum_{\mu}{\gamma}_{\mu}sin(\frac{2{\pi}k_{\mu}}{N}).
\end{equation}
Let us now go to the continuum by letting $a{\longrightarrow}0$ and see if we get the ordinary Fermion propagator
back . In this limit we set $am^{\mu}{\longrightarrow}x_{m}^{\mu}$
,$\frac{2{\pi}k_{\mu}}{Na}{\longrightarrow}p_{\mu}$ , $\frac{1}{N^4a^4}\sum_{k}{\longrightarrow}\int
\frac{d^4p}{(2{\pi})^4}$ so that
\begin{equation}
(M^{-1})_{x_mx_n}=\int
\frac{d^4p}{(2{\pi})^4}\frac{1}{m+\frac{i}{a}\sum_{\mu}{\gamma}_{\mu}sin(ap_{\mu})}e^{ip(x_m-x_n)}.
\end{equation}
Each component $p_{\mu}$ is now in the range $-\frac{\pi}{a}<p_{\mu}{\leq}\frac{\pi}{a}$ . It is almost obvious
that the two different regions of the momentum space , the small momentum region $p_{\mu}=0$ as well as the large
momentum region $p_{\mu}=\frac{\pi}{a}$ , both give rise to the continuum propagator . To see this explicitly let
us rewrite the above propgator by separating the integral over each component $p_{\rho}$ in the following way
\begin{eqnarray}
(M^{-1})_{x_mx_n}&=&\int
{\Pi}_{{\nu}{\neq}{\rho}}dp_{\nu}\int_{-\frac{\pi}{a}}^{\frac{\pi}{a}}dp_{\rho}K(p_{\rho})e^{i\sum_{{\nu}{\neq}{\rho}}p_{\nu}(x_m^{\rho}-x_n^{\rho})}\nonumber\\
&=&\int
{\Pi}_{{\nu}{\neq}{\rho}}dp_{\nu}\Bigg[\int_{-\frac{\pi}{a}}^{-\frac{\pi}{2a}}dp_{\rho}K(p_{\rho})+\int_{-\frac{\pi}{2a}}^{\frac{\pi}{2a}}dp_{\rho}K(p_{\rho})+\int_{\frac{\pi}{2a}}^{\frac{\pi}{a}}
dp_{\rho}K(p_{\rho})\Bigg]e^{i\sum_{{\nu}{\neq}{\rho}}p_{\nu}(x_m^{\nu}-x_n^{\nu})},\nonumber\\
&&
\end{eqnarray}
where $K(p_{\rho})$ is the kernel
$K(p_{\rho})=\frac{1}{(2{\pi})^4}\Big[m+\frac{i}{a}\sum_{{\mu}{\neq}{\rho}}{\gamma}_{\mu}sin(ap_{\mu})+\frac{i}{a}{\gamma}_{\rho}sin(ap_{\rho})\Big]^{-1}e^{ip_{\rho}(x_m^{\rho}-x_n^{\rho})}$.
Now by changing the integration variable in the first term to $\tilde{p}_{\rho}=p_{\rho}+\frac{\pi}{a}$ and in the
last term to $\tilde{p}_{\rho}=p_{\rho}-\frac{\pi}{a}$ one obtains
\begin{eqnarray}
(M^{-1})_{x_mx_n} &=&\int
{\Pi}_{{\nu}{\neq}{\rho}}dp_{\nu}\Bigg[\int_{-\frac{\pi}{2a}}^{\frac{\pi}{2a}}dp_{\rho}K(p_{\rho})+\int_{-\frac{\pi}{2a}}^{\frac{\pi}{2a}}
d{\tilde{p}}_{\rho}\tilde{K}(\tilde{p}_{\rho})\Bigg]e^{i\sum_{{\nu}{\neq}{\rho}}p_{\nu}(x_m^{\nu}-x_n^{\nu})},\nonumber\\
&&
\end{eqnarray}
where
$\tilde{K}(\tilde{p}_{\rho})=K(\tilde{p}_{\rho}-\frac{\pi}{a})=K(\tilde{p}_{\rho}+\frac{\pi}{a})=\frac{cos{\pi}(m^{\rho}-n^{\rho})}{(2{\pi})^4}\Big[m+\frac{i}{a}\sum_{{\mu}{\neq}{\rho}}{\gamma}_{\mu}sin(ap_{\mu})-\frac{i}{a}{\gamma}_{\rho}sin(a\tilde{p}_{\rho})\Big]^{-1}e^{i\tilde{p}_{\rho}(x_m^{\rho}-x_n^{\rho})}$.

For small lattice spacing the first term is dominated by small momenta $p_{\rho}{\longrightarrow}0$ which leads to
the propagator $[m+i{\gamma}^{\mu}p_{\mu}+O(a^2)]^{-1}$ , whereas the second term is dominated by large momenta
$\tilde{p}_{\mu}{\longrightarrow}0$ which does also lead to a similar continuum propagator . It is a similar
propagator and not the same because the minus sign in $\tilde{K}(\tilde{p}_{\rho})$ can be absorbed by redefining
the ${\gamma}_{\rho}$  and therefore redefining the chirality operator so that it corresponds to a different
Fermion specie .

So for each space-time dimension we have two different regions in momentum space where the discrete theory (\ref{discretetheory})
gives the continuum Fermion propagator. Altogether we obtain in the continuum $2^4=16$ independent Fermion species
. This phenomenon is due to the fact that our regularization scheme preserves exact chiral invariance , i.e there
is no chiral anomaly . Putting it differently the effect of these extra Fermions is to cancel exactly the chiral
anomaly .

\subsection{The Fermion Doubling Problem and Fuzzy ${\bf S}^2$}

In this thesis a novel approach to discrete physics has been developed. It works with quantum fields on a ``fuzzy
space'' ${\cal M}_{F}$ obtained by treating the underlying manifold ${\cal M}$ as a phase space and quantizing it
\cite{frgrre,4,5,6,7,8,10,11}. Topological features, chiral anomalies and $\sigma-$ models have been successfully
developed in this framework \cite{6,12,13},using the cyclic cohomology of Connes\cite{1,2,3.1}.

In this section, we propose a solution of the fermion doubling problem for ${\cal M}={\bf S}^2$ using fuzzy
physics. An alternative approach can be found in \cite{6}. There have also been important developments
\cite{gw,luscher} in the theory of chiral fermions and anomalies in the usual lattice formulations. We will show
that there are striking relationships between our approach and these developments.

Quantisable adjoint orbits of compact semi-simple Lie groups seem amenable to the full fuzzy treatment and lead to
manageable finite dimensional matrix models for quantum fields . There are two such manifolds in dimension four,
namely ${\bf S}^2{\times}{\bf S}^2$ and $ {\bf C}{\bf P}^2 $. Our methods readily extend to ${\bf
S}^2{\times}{\bf S}^2$. They do not encounter obstructions for $ {\bf C}{\bf P}^2 $ as well. The published work of Grosse and Strohmaier \cite{14}
on $ {\bf C}{\bf P}^2 $ gives their description of fuzzy $4d$ fermions.

A sphere ${\bf S}^2$ is a submanifold of ${\bf R}^3$:
\begin{equation}
{\bf S}^2=\{\vec{x} \in {\bf R}^3: \sum_{i=1}^3 x_i^2={\rho}^2 \}.
\end{equation}
If $\hat{n}_i$ are the coordinate functions on ${\bf S}^2$, $\hat{n}_i(\vec{n}) = n_i=x_i/{\rho}$, then $\hat{n}_i$
commute and the algebra ${\cal A}$ of smooth functions they generate is commutative. In contrast, the operators
$n_i^F$ describing ${\bf S}_F^2$ are noncommutative:
\begin{equation}
[n_i^F, n_j^F] = \frac{i \epsilon_{ijk} n_k^F}{[l(l+1)]^{1/2}}, \quad \sum_{i=1}^3 n_i^{F2} = {\bf 1}, \quad l \in
\{\frac{1}{2}, 1, \frac{3}{2} \ldots \}.
\end{equation}
The $n_i^F$ commute and become $\hat{n}_i$ in the limit $l \rightarrow \infty$. If $L_i = [l(l+1)]^{1/2}n_i^F$,
then $[L_i, L_j] = i \epsilon_{ijk}L_k$ and $\sum L_i^2 = l(l+1)$, so that $L_i$ give the irreducible
representation (IRR) of the $SU(2)$ Lie algebra for angular momentum $l$. $L_i$ or $n_i^F$ generate the algebra
${\bf A}=M_{2l+1}$ of $(2l+1) \times (2l+1)$ matrices.

Scalar wave functions on ${\bf S}^2$ come from elements of ${\cal A}$. In a similar way, elements of ${\bf A}$
assume the role of scalar wave functions on $S_F^2$. A scalar product on ${\bf A}$ is $\langle \xi, \eta \rangle =
Tr {\xi}^{\dagger} \eta$. ${\bf A}$ acts on this Hilbert space by left- and right- multiplications giving rise to
the left and right- regular representations ${\bf A}^{L}={\Pi}(\bf A)$ , ${\bf A}^R={\Pi}^0(\bf A)$ of ${\bf A}$.
For each $a \in {\bf A}$, we thus have operators $a^{L, R} \in A^{L,R}$ acting on $\xi \in {\bf A}$ according to
$a^L \xi = a \xi, a^R \xi = \xi a$. [Note that $a^L b^L = (ab)^L $ while $a^R b^R = (ba)^R$.] We assume by
convention that elements of ${\bf A}^L$ are to be identified with fuzzy versions of functions on ${\bf S}^2_F$.
There are two kinds of angular momentum operators $L_{i}^{L}$ and $-L_{i}^{R}$. The orbital angular momentum
operator, which should annihilate ${\bf 1}$, is ${ad L}_i = L_i^L - L_i^R$. ${ad}\vec{L}$ plays the role of the
continuum $\vec{\cal L}=-i(\vec{r} \times \vec{\nabla})$.

The following two Dirac operators on ${\bf S}^2$ have occurred in the fuzzy literature:
\begin{equation}
{\cal D}_{2g} = \frac{1}{\rho}[\vec{\sigma}. \vec{\cal L} + {\bf 1}],\qquad {\cal D}_{2w} =
-\frac{1}{\rho}\epsilon_{ijk}\sigma_i \hat{n}_j {\cal J}_k,
\end{equation}
where
\begin{eqnarray}
{\cal J}_k& =&\vec{\cal L}_k + \frac{\sigma_k}{2}\nonumber \\
& =& \rm{Total~ angular~ momentum~ operators~} .
\end{eqnarray}
There is a common chirality operator $\gamma$ anticommuting with both:
\begin{equation}
\gamma = \vec{\sigma}.\hat{n} = \gamma^{\dagger}, \quad \gamma^2 = {\bf 1},\qquad \gamma {\cal D}_{2\alpha} + {\cal
D}_{2\alpha} \gamma =0.\label{g2}
\end{equation}
[${\alpha}=g,w$] . These Dirac operators {\it in the continuum} are unitarily equivalent,
\begin{equation}
{\cal D}_{2w} = \exp{(i \pi \gamma/4)} {\cal D}_{2g} \exp{(-i \pi \gamma/4)} ~~=i{\gamma}{\cal
D}_{2g},\label{gamma}
\end{equation}
and have the spectrum $ \{ \pm \frac{1}{\rho}(j+1/2): j \in \{1/2, 3/2,\ldots  \} \}$, where $j$ is total angular
momentum (spectrum of $\vec{\cal J}^2 =\{j(j+1)\}$).

Since $|{\cal D}_{2{\alpha}}|$ $(\equiv $ positive square root of ${\cal D}_{2{\alpha}}^2$ $)$ for both ${\alpha}$
share the same spectrum and rotational invariance, $ |{\cal D}_{2w}|=|{\cal D}_{2g}| $. Further being multiples of
unity for each fixed $j$, they commute with the rotationally invariant ${\gamma}$. As they are invertible too, we
have the important identity

\begin{equation}
\gamma~=~i{\frac{{\cal D}_{2g}}{|{\cal D}_{2g}|}}{\frac{{\cal D}_{2w}} {|{\cal D}_{2w}|}}.\label{extrachirality}
\end{equation}
This is also clear from the identity ${\cal D}_{2w}=i{\gamma}{\cal D}_{2g}$ which can be rewritten as $|{\cal
D}_{2w}|^{-1}{\cal D}_{2w}{\cal D}_{2g}|{\cal D}_{2g}|^{-1}=i|{\cal D}_{2w}|^{-1}{\gamma}{\cal D}_{2g}^2|{\cal
D}_{2g}|^{-1}$ . Then by using the facts $|{\cal D}_{2{\alpha}}|=\sqrt{{\cal D}_{2{\alpha}}^2}$ and
$[{\gamma},|{\cal D}_{2{\alpha}}|]=0$ one can see that (\ref{extrachirality}) follows easily.

 The discrete version of ${\cal D}_{2g}$ is:
\begin{equation}
{ D}_{2g} =\frac{1}{\rho}[ \vec{\sigma}. ad\vec{L} + {\bf 1}] ,
\end{equation}
while
\begin{eqnarray}
~Spectrum ~of ~{ D}_{2g} &=&  \left\{ \pm \frac{1}{\rho}(j+\frac{1}{2}): j \in
                              \{ \frac{1}{2}, \frac{3}{2}, \ldots
                              2l-\frac{1}{2} \} \right\} \nonumber \\
                        &\cup& \left\{ \frac{1}{\rho}(j+\frac{1}{2}):
                              j=2l+\frac{1}{2} \right\}.\label{specd1}
\end{eqnarray}
It is easy to derive (\ref{specd1}) by writing

\begin{eqnarray}
{\rho}{ D}_{2g}&=& \vec{J}^2 - (Ad\vec{L})^2 - \left(\frac{\vec{\sigma}}{2}\right)^2 + {\bf 1},~~
\left(\frac{\vec{\sigma}}{2}\right)^2 = \frac{3}{4} {\bf 1},\\
{Ad L}_k +\frac{{\sigma}_k}{2}&=&J_k \label{angmom}\\
&=& ~Total ~angular ~momentum ~operators.
\end{eqnarray}
We let $j(j+1)$ denote the eigenvalues of ${\vec{J}}^2$. Then for $(ad\vec{L})^2 = k(k+1), k \in \{0, 1, \ldots 2l
\}$, if $j=k+1/2$ we get $+(j+1/2)$ as eigenvalue of ${\rho}{ D}_{2g}$, while if $j=k-1/2$ we get $-(j+1/2)$. The
absence of $-(2l+1/2)$ in (\ref{specd1}) is just because $k$ cuts off at $2l$. (The same derivation works also for
${\cal D}_{2g}$).

The discrete version of ${\cal D}_{2w}$ is : ${ D}_{2w} =\frac{1}{\rho}\epsilon_{ijk}\sigma_i n_j^F L_k^R$. ${
D}_{2w}$ is no longer unitarily equivalent to ${ D}_{2g}$, its spectrum is given in \cite{10,11}.

The first operator has been used extensively by Grosse et al \cite{6,7,8} while the second was first introduced by
Watamuras \cite{10,11}. It is remarkable that the eigenvalues (\ref{specd1}) coincide {\it exactly} with those of
${\cal D}_\alpha$ upto $j=(2l-1/2)$. In contrast ${ D}_{2w}$ has zero modes when $j~=~2l~+\frac{1}{2}$ and very
small eigenvalues for large values of $j$, both being absent for ${\cal D}_{\alpha}$ . So ${ D}_{2g}$ is a better
approximation to ${\cal D}_\alpha$.

But ${ D}_{2g}$ as it stands admits no chirality operator anti-commuting with it. This is easy to see as its top
eigenvalue does not have its negative in the spectrum. Instead  ${ D}_{2w}$ has the nice feature of admitting a
chirality operator: the eigenvalue for top $j$, even though it has no pair, is  exactly zero. So the best fuzzy
Dirac operator has to combine the good properties of ${ D}_{2g}$ and ${ D}_{2w}$. We suggest it to be ${ D}_{2g}$
after projecting out its top $j$ mode. We will show that it then admits a chirality with the correct continuum
limit .

The chirality operator anticommuting with ${D}_{2w}$ and squaring to ${\bf 1}$ in the {\it entire} Hilbert space is
\begin{equation}
\Gamma = \Gamma^{\dagger} = -\frac{\vec{\sigma}.{\vec{L}}^R -1/2}{l+1/2},~ \Gamma^2 = {\bf 1}\label{gR}.
\end{equation}
An interpretation of ${\Gamma}$ is that $(1{\pm}{\Gamma})/2$ are projectors to subspaces where
$(-\vec{L}^R+\vec{\sigma}/2)^2$ have values $(l{\pm}\frac{1}{2})(l{\pm}\frac{1}{2}+1)$ \cite{12}. The following
identity is easily shown :
\begin{equation}
[{ D}_{2g},\Gamma]~=-~2~i~{\lambda}~{ D}_{2w};~~ {\lambda}~=~\sqrt{1-\frac{1}{(2l~+~1)^2}}~. \label{identity}
\end{equation}
Now $D_{2{\alpha}}^2$ and $ |D_{2{\alpha}}|(\equiv $ nonnegative square root of $D_{2{\alpha}}^2)$
[${\alpha}=g,w$]are multiples of identity for each fixed {\it j}, and ${\Gamma}$ commutes with ${\vec{J}}$. Hence
they mutually commute :

\begin{equation}
[A,B]~=~0~ ~for~ A~,B~=~D_{2{\alpha}}^2~ , |D_{2{\alpha}}|~ ~or  ~{\Gamma}~ .
\end{equation}
Therefore from (\ref{identity}),
\begin{equation}
\{D_{2g},D_{2w}\}=\frac{i}{2{\lambda}}[D_{2g}^2,{\Gamma}]=0 .
\end{equation}
In addition we can see that $[D_{2{\alpha}}^2,D_{2{\beta}}]=[|D_{2{\alpha}}|,D_{2{\beta}}]=0$. If we define
\begin{eqnarray}
\epsilon_{2\alpha} &=& \frac{D_{2\alpha}}{|D_{2\alpha}|} ,
 ~on ~the ~subspace ~V ~with ~;j \leq 2l-1/2 ,
 \nonumber \\
 &=& 0 ~on ~the ~subspace ~W ~with ~;j=2l+1/2,
\end{eqnarray}
it follows that
\begin{equation}
e_1={\epsilon}_{2g}~ , e_2={\epsilon}_{2w}~ , e_3=i{\epsilon}_{2g}{\epsilon}_{2w} \label{e1e2e3}
\end{equation}
generate a Clifford algebra on $V$. That is, if $P$ is the orthogonal projector on $V$,
\begin{eqnarray}
P \xi &=& \xi, \quad \xi \in V, \nonumber \\
      &=& 0,  \quad \xi \in W,
\label{pdef}
\end{eqnarray}
then $\{e_{a},e_{b}\}=2{\delta}_{ab}P$.

All this allows us to infer that $\{e_3,D_{2g}\}=0$ so that $e_3$ is a chirality operator for either $D_{2g}$ or
its restriction $PD_{2g}P$ to $V$. In addition, {\it in view of (\ref{extrachirality}), it  has the correct continuum limit
as well} so that it is a good choice for chirality in that respect too. This $e_3$ is exactly ${\Gamma}^{'}$
introduced in equation (\ref{fuzzychirality2}) .

A unitary transformation of $e_3={\Gamma}^{'}$ and $D_{2g}$ will not disturb their nice features. Such a
transformation bringing $e_3={\Gamma}^{'}$ to ${\Gamma}$ on $V$ is convenient. It can be constructed as follows.
$e_{a}$ and ${\Gamma}$ being rotational scalars leave the two-dimensional subspaces in $V$ with fixed values of
$\vec{J}^2$ and $J_3$ invariant. On this subspace, $e_{a}$ and unity form a basis for linear operators, so
${\Gamma}$ is their linear combination. As $e_1$,$e_3={\Gamma}^{'}$ and ${\Gamma}$ anticommute with $e_2$,and all
square to ${\bf 1}$, in this subspace, we infer that ${\Gamma}$ is a transform by a unitary operator
$U=exp(i{\theta}e_2/2)$ of $e_3={\Gamma}^{'}$ in each such subspace. And ${\theta}$ can depend only on $\vec{J}^2$
by rotational invariance. Thus we can replace $PD_{2g}P$ and $e_3={\Gamma}^{'}$ by the new Dirac and chirality
operators
\begin{eqnarray}
D&=& e^{(i \theta (J^2) \epsilon_2)/2} (P D_{2g} P) e^{(-i \theta (J^2)
\epsilon_2)/2},\nonumber\\
{\Gamma}&=&e^{(i \theta (J^2) \epsilon_2)/2}
{\Gamma}^{'}e^{(-i \theta (J^2) \epsilon_2)/2} \nonumber \\
&=&cos{\theta}(J^2)(i{\epsilon}_{2g}{\epsilon}_{2w})+sin{\theta}(J^2){\epsilon}_{2g}.
\end{eqnarray}
The coefficients can be determined by taking traces with ${\epsilon}_{2g}$ and $i{\epsilon}_{2g}{\epsilon}_{2w}$.

We have established that chiral fermions can be defined on $S^2_F$ with no fermion doubling, at least in the
absence of fuzzy monopoles. It is , however, easy to include them as well.

\subsection{The Ginsparg-Wilson Relation on the Fuzzy Sphere}

The Ginsparg-Wilson chiral fermion has a Dirac operator $D_{gw}$ and a hermitian chirality operator $\Gamma_{gw}$
squaring to unity. $D_{gw}$ and $\Gamma_{gw}$ fulfill the relations
\begin{equation}
D_{gw}^{\dagger}~=~\Gamma_{gw}~D_{gw}~\Gamma_{gw},~\{~\Gamma_{gw},~D_{gw}~\}~=~aD_{gw}\Gamma_{gw} D_{gw}
\end{equation}
$a$ being lattice spacing in suitable units. Now if $\Gamma'_{gw}~=~ \Gamma_{gw}(a~D_{gw})~-~\Gamma_{gw}$, then
\begin{equation}
\Gamma_{gw}^{'\dagger}~=~\Gamma_{gw}^{'},~ \Gamma_{gw}^{'2}~=~1~ ~and
~aD_{gw}~=~\Gamma_{gw}(\Gamma_{gw}~+~\Gamma_{gw}^{'}).
\end{equation}
Conversely given two idempotents $\Gamma_{gw}$ and $\Gamma'_{gw}$, we have a Ginsparg-Wilson pair
$D_{gw}~=~\frac{1}{a}~\Gamma_{gw}(\Gamma_{gw}~+~\Gamma'_{gw})$ and $\Gamma_{gw}$.

Our fermion on ${\bf S}_F^2$ admits such a formulation, except that we choose $\Gamma~+~\Gamma'$ as the Dirac
operator. Thus just like (\ref{gR}), we can also construct a left- chirality operator anticommuting with $D_{2w}$:
\begin{equation}
\Gamma^L = \Gamma^{L{\dagger}} = \frac{\vec{\sigma}.{\vec{L}}^L +1/2}{l+1/2},~~ \Gamma^{L2} = {\bf 1}.
\end{equation}
[see equation (\ref{leftchirality})] . Then the Ginsparg-Wilson Dirac operator on the fuzzy sphere is
\begin{eqnarray}
D_{gw}[{\bf S}^2_F]&=&{\Gamma}D_{2g}\nonumber\\
aD_{gw}[{\bf S}^2_F]&=&{\Gamma}({\Gamma}+{\Gamma}^{L})\nonumber\\
&where&\nonumber\\
a&=&\frac{\rho}{l+\frac{1}{2}}.
\end{eqnarray}
We have also made the identifications
\begin{eqnarray}
{\Gamma}_{gw}[{\bf S}^2_F]&=&{\Gamma}\nonumber\\
{\Gamma}_{gw}^{'}[{\bf S}^2_F]&=&{\Gamma}^{L}.
\end{eqnarray}

There is a beautiful structure underlying the algebra ${\bf K}$ of the idempotents $\Gamma, \Gamma^{L}$. It lets us
infer certain salient features of $D_{2g}$, but the results transcend ${\bf S}_F^2$. We shall now briefly describe
it.

Introduce the hermitian operators
\begin{equation}
\Gamma_{1,2}~=~\frac{1}{2}(\Gamma \pm \Gamma^{L}),\Gamma_3~=~\frac{1}{2}[\Gamma,\Gamma^{L}],\Gamma_0~=~\frac{1}{2}
\{\Gamma,\Gamma^{L}\}.
\end{equation}
Then one can directly check the following assertions

(a) $\Gamma_m~(m\neq 0)$ mutually anticommute .

(b) $\Gamma_0$ and $\Gamma_m^2$ commute with all $\Gamma_\lambda$ and are in the center of ${\bf K}$.

(c) $\Gamma_1^2~+~\Gamma_2^2 = {\bf 1} =\Gamma_3^2~ +~\Gamma_0^2$ so that $\Gamma_\lambda$ do not have eigenvalues
exceeding $1$  in modulus.

(d) $\Gamma_1^2~=~\frac{1}{2}({\bf 1}+\Gamma_0), \Gamma_2^2~=~\frac{1}{2}({\bf 1}-\Gamma_0)$.

Since
\begin{equation}
{\Gamma}_0=\frac{1}{2(l+\frac{1}{2})^2}[\vec{J}^2-2l(l+1)-\frac{1}{4}],
\end{equation}
the vector space we work with can be split into the direct sum $\oplus V_j$ of eigenspaces $V_j$ of $\Gamma_0$ with
distinct eigenvalues $\cos 2\theta_j$.  Then ${\bf K}V_j~=~ V_j$ because from property $(b)$ above ${\Gamma}_0$
commutes with all ${\Gamma}_{\lambda}$ and therefore commutes with ${\bf K}$  . On $V_j$, the possible eigenvalues
of $\Gamma_3$ which is given explicitly by
\begin{equation}
{\Gamma}_3=-\frac{{\rho}\sqrt{l(l+1)}}{(l+\frac{1}{2})^2}D_{2w},
\end{equation}
are of the form ${\pm}\sin 2{\theta}_j$ by (c) and those of ${\Gamma}_{1,2}$ are
${\pm}cos{\theta}_j,{\pm}sin{\theta}_j$ by (d). Both signs do occur on $V_j$ unless $|{\Gamma}_0|$ is ${\bf 1}$
and hence ${\Gamma}_3=0$.

For if ${\Gamma}_3{\neq}0$ on $V_j$, $|{\Gamma}_0|{\neq}1$ and hence ${\Gamma}_{1,2}$ have no zero eigenvalue and
${\Gamma}_{m}/|{\Gamma}_{m}| $ generate a Clifford algebra there. The result follow from the identity
\begin{equation}
\frac{{\Gamma}_{n}}{|{\Gamma}_{n}|} {\Gamma}_{m} \frac{{\Gamma}_{n}}{|{\Gamma}_{n}|}=-{\Gamma}_{m}, m{\neq}n~ on~
V_j~ for~ m,n=1,2,3.
\end{equation}
But if $|{\Gamma}_0|={\bf 1}$ and ${\Gamma}_{3}=0$ for $j=j_0$, then ${\Gamma}_1$ or ${\Gamma}_2$ is also zero on
$V_{j_{0}}$ and we cannot infer that the nonzero ${\Gamma}_{m}$ has eigenvalues of both signs there. We can only
say that its modulus $|{\Gamma}_{m}|$ has its maximum value ${\bf 1}$ there.

The spectrum and eigenstates of the Ginsparg-Wilson $D_{gw}={\Gamma}D_{2g}$ can also be found. Thus
$aD_{gw}=2({\Gamma}_{1}^2 + {\Gamma}_{2}{\Gamma}_{1})$. It can be diagonalised since $[{\Gamma}_1^2 ,
{\Gamma}_2{\Gamma}_1]=0$ and therefore $[{\Gamma}_1^2,D_{gw}]=0$ . On $V_j$,
${\Gamma}_1^2=[\frac{a}{2}D_{2g}]^2=cos^2{\theta}_j {\bf 1}$ and hence
${\Gamma}_2^2=[-\frac{1}{2l+1}\vec{\sigma}(\vec{L}^L+\vec{L}^R)]^2=sin^2{\theta}_j$ by property $(c)$ above . So
$({\Gamma}_2{\Gamma}_1)^2=-{\Gamma}_2^2{\Gamma}_1^2=-\cos^2{\theta}_{j} \sin^2{\theta}_j{\bf 1}$ or spectrum of
$aD_{gw}$ on $V_j$ is $={{1} + exp({\pm}2i{\theta}_j)} $.

As for its eigenvectors,, we can proceed as follows. On $V_{j_0}$, ${\Gamma}_1$ or ${\Gamma}_2$ $=0$ so that
$aD_{gw}$ is in any case diagonal. On $V_j$ $(j{\neq}j_0)$ , $|{\Gamma}_2|{\neq}0$ and
\begin{equation}
aD_{gw}=e^{-i\frac{{\Gamma}_2}{|{\Gamma}_2|}\frac{{\pi}}{4}}2\Big[{\Gamma}_1^2 +
i|\sin{\theta}_j|{\Gamma}_1\Big]e^{i\frac{{\Gamma}_2}{|{\Gamma}_2|}\frac{{\pi}}{4}}.
\end{equation}
So if ${\Gamma}_1{\psi}_{j}^{\pm}={\pm}(\cos{\theta}_j){\psi}_{j}^{\pm}$, then
\begin{equation}
aD_{gw}\Big[e^{-i\frac{{\Gamma}_2}{|{\Gamma}_2|}\frac{{\pi}}{4}}{\psi}_{j}^{\pm}\Big]= 2cos{\theta}_je^{
{\pm}i\frac{\sin{\theta}_j}{|{\sin \theta}_j|}
{\theta}_j}\Big[e^{-i\frac{{\Gamma}_2}{|{\Gamma}_2|}\frac{{\pi}}{4}}{\psi}_{j}^{\pm}\Big].
\end{equation}

%

\chapter{Fuzzy ${\bf C}{\bf P}^2$}

The central claim of this thesis as we have already stated is that regularization of quantum field theories (QFT's)
can be achieved by quantizing the underlying manifold , in other words replacing it by a non-commutative matrix
model or a ``fuzzy manifold''. Such discretization by quantization is remarkably successful in preserving
symmetries and topological features, and altogether overcoming the fermion-doubling problem . In the case of ${\bf
S}^2$ treated in the last two chapters , all of these properties were proven explicitly to hold .

In this chapter however , we will work out in detail the ``fuzzification'' of the four-dimensional ${\bf C}{\bf
P}^2$ and its QFT's . Commutative ${\bf C}{\bf P}^2$ is known to be not a spin manifold , but rather a spin${}_c$
manifold which introduces new unique features on its Dirac operator . The precise meaning of being spin${}_c$
manifold and these new features will be explained in the following . Fuzzy ${\bf C}{\bf P}^2$ will also be
formulated .

${\bf C}{\bf P}^2$ , like ${\bf S}^2$ , elegantly escapes (almost) all obstructions against fuzzification by
quantization . This will also be the case for higher ${\bf C}{\bf P}^N={\bf SU}(N+1)/{\bf U}(N)$ , and for that
matter for all co-adjoint orbits of Lie groups as was mentioned in the introduction . ${\bf C}{\bf P}^2$ is
quantized in a standard fashion by quantizing certain WZW Lagrangians of nonlinear fields with ${\bf C}{\bf P}^2$
as target spaces . The resultant fuzzy spaces are described by linear operators acting on finite dimensional
irreducible representations (IRR's) of $SU(3)$ . In addition, the elements of the Lie algebra define natural
derivations, and that helps to find Laplacian and the Dirac operator. In principle one can even define chirality
with no fermion doubling and represent monopoles and instantons as we did for the case of ${\bf S}^2$ .

In the literature, there are several studies of the fuzzy physics of ${\bf C}{\bf P}^1=S^2$
\cite{4,5,6,7,8,9,10,11,12,13,15} while there is also a rigorous and beautiful treatment of ${\bf C}{\bf
P}^2$ by Grosse and Strohmaier \cite{14}. The present work develops an alternative formulation for ${\bf C}{\bf
P}^2$. It is close to earlier treatments of $S^2$ \cite{12,13} and seems to generalize to other quantizable
orbits. It is eventually equivalent to that of \cite{14} as we show, so that the first study of ${\bf C}{\bf P}^2$
is of that reference.

Throughout this chapter , we treat ${\bf C}{\bf P}^2$ as Euclidean spacetime even though the possibility of
treating it as spacial slice is also available. Section $1$  explains the basic properties of ${\bf C}{\bf P}^2$.
We quantize it in Section $2$ to produce the fuzzy ${\bf C}{\bf P}^2$. ${\bf C}{\bf P}^2$ is not a spin , but a
spin${}_c$ manifold , and that has exotic consequences for the $SU(3)$ spectrum : left and right chiral modes
transform differently under $SU(3)$ . Section $3$ will be devoted to the formulation of the precise meaning of the
statement that ${\bf C}{\bf P}^2$ is a spin$_{c}$ manifold  .

In non-commutative geometry (NCG), a central role is assumed by the (massless) Dirac operator . Section $4$ reviews
it for ${\bf S}^2={\bf C}{\bf P}^1$ while Section $5$ studies our approach to it in detail for ${\bf C}{\bf P}^2$ .
Analysis shows its equivalence to the Dirac-K{\"a}hler operator \cite{14}. Section $6$ studies the fuzzy analogue
of the Dirac operator. This work is greatly facilitated by coherent states and star ($\star$) products. The
necessary material, contained in \cite{9,15}, is reviewed and used to discretise the continuum material for both
${\bf S}^2={\bf C}{\bf P}^1$ and ${\bf C}{\bf P}^2$. Incidentally the $\star$ product is particularly useful for
formulating fuzzy analogues of important continuum quantities like correlation functions . Last section is a breif
description of fuzzy actions.

\section{Elementary Definitions of ${\bf C}{\bf P}^2$}

${\bf C}{\bf P}^2$ is a K{\"a}hler manifold describable in different ways . It is a K{\"a}hler manifold because it
admits a metric $ds^2=g_{a\bar{b}}dz^ad{\bar{z}}^b$ which is K{\"a}hler , i.e the two-form
$K=\frac{i}{2}g_{a\bar{b}}dz^a{\wedge}d{\bar{z}}^b$ is real ( $\bar{K}=K$ ) and closed ( $dK=0$ ) . This metric
can be chosen to be the Fubini-Study metric , defined by the real-two form \cite{eguchi}
\begin{equation}
K=\frac{i}{2}{\partial}_a\bar{\partial}_{\bar{b}}Ln\big[1+\sum_{\alpha=1}^2z^{\alpha}\bar{z}^{\alpha}\big]dz^a{\wedge}d\bar{z}^b,
\end{equation}
where $z^1=x+iy$ , $z^2=z+it$. This leads to the metric
\begin{equation}
ds^2=\frac{dr^2+r^2{\sigma}_z^2}{(1+r^2)^2}+\frac{r^2({\sigma}_x^2+{\sigma}_y^2)}{1+r^2},
\end{equation}
where ${\sigma}_i$ , $i=x,y,z$ are left-invariant $1-$forms on ${\bf S}^3$ defined by the following equations
${\sigma}_i=e^i/r$ and satisfy the realtions $d{\sigma}_x=2{\sigma}_y{\wedge}{\sigma}_z $, cyclic . The $e^i$ in
here are the vierbeins of the $4-$dimensional euclidean space-time in polar coordinates. For example ,
$e^0=dr=\frac{1}{r}(xdx+ydy+zdz+tdt)$ , where $r$ is defined by
$z^1=rcos\frac{\theta}{2}exp\frac{i}{2}(\psi+\phi)$ and $z^2=rsin\frac{\theta}{2}exp\frac{i}{2}(\psi-\phi)$ ,
${\theta}$ , ${\psi}$ , ${\phi}$ are the other polar coordinates . Similar expressions for $e^i$ , $i=1,2,3$ can
also be written \cite{eguchi}.

As mentioned above , ${\bf C}{\bf P}^2$ is describable in a variety of different but equivalent ways . For example
it can be described as the orbit of the Lie group $SU(3)$ through the hypercharge operator $Y$ or its multiples
(the group $SU(3)$ has eight generators $t_i$ which satisfy $[t_i,t_j]=if_{ijk}t_k$ ; the hypercharge is
$Y=\frac{2}{\sqrt{3}}t_8$ ; in the fundamental representation of $SU(3)$ ,(${\bf 3}$), the generators are
$t_i=\frac{1}{2}{\lambda}_i$ , where the ${\lambda}_i$ are the eight Gell-Mann matrices ). This means that
\begin{equation}
{\bf C}{\bf P}^2=\{gYg^{-1};~for ~all ~g{\in}SU(3)\}.\label{definitioncp21}
\end{equation}
As the stability group of $Y$ ( $Y$ in the fundamental representation is given by $Y=\frac{1}{3}(1,1,-2)$ ) is
clearly $U(2)$:
\begin{equation}
 U(2)=\left\{\left(
  \begin{array}{cc}
    u & 0\\
    0 & {det\,u^{-{1}}}\\
  \end{array}
   \right) \in SU(3)\right\},
\end{equation}
One can check that $u^{\dagger}=u^{-1}{\Leftrightarrow}u^{\dagger}u=uu^{\dagger}=1$ and therefore $u$ is in the defining
representation of $U(2)$ . The phase $ {det\,u^{-{1}}} $ provides for us a special embedding of $U(2)$ in $SU(3)$
. We have then
\begin{equation}
{\bf C}{\bf P}^2=SU(3)/U(2). \label{s21}
\end{equation}
As its name reveals, it is also a projective complex\footnote{It is complex because its transition functions are
holomorphic.} space , in other words the space of all complex lines $C^1$ in $C^3$ which pass through the origin.
If $\xi\in C^3-\{0\}$, a point of ${\bf C}{\bf P}^2$ is defined as the equivalence class
$\langle\xi\rangle=\langle\lambda\xi\rangle$ for all $\lambda\in C^1-\{0\}$.
\begin{equation}
{\bf C}{\bf P}^2=\{\langle\xi\rangle=\langle\lambda\xi\rangle,~for ~all ~{\xi}{\in}C^3-\{0\} ~and ~all
~{\lambda}{\in}C^1-\{0\}\}
\end{equation}
One can pick from each class a representative element ${\xi}_r$ which is obtained from the corresponding ${\xi}$
by the choice $\lambda=\left(\sum_{i=1}^6|\xi_i|^2\right)^{-{1\over 2}}$ . One can then easily verify that
$\sum_{i=1}^{6}|{\xi}_{ri}|^2=1$ which is clearly still invariant under a $U(1)$ action . In other words the above
choice of ${\lambda}$ will not select a unique point ${\xi}_r$ from the class $<\xi>$ but rather it will select a
whole subclass $ <e^{i{\theta}}{\xi}_r>$  . We have then
\begin{eqnarray}
{\bf C}{\bf P}^2&=&\{\langle{\xi}_r\rangle=\langle{\xi}_r e^{i\theta}\rangle: \sum_{i=1}^6|{\xi}_{ri}|^2=1\},\nonumber\\
&{\Longleftrightarrow}\nonumber\\
{\bf C}{\bf P}^2&=&S^5/U(1).
\end{eqnarray}
In (\ref{s21}), we can first quotient $SU(3)$ by $SU(2)$. That is just the above $S^5$. This can be seen as
follows . $g{\in}SU(3)$ acts on $z=(z_1,z_2,z_3){\in}C^3$ in a natural way : $z^{'}_i=\sum_{j=1}^3g_{ij}z_j$ .
This $SU(3)$ acts also transitively\footnote{Because any two points of $S^5$ can be related by an element
$g{\in}SU(3)$.}on the sphere $S^5 = \{z \in C^3 :\sum_{i=1}^3 |z_i|^2=1\}$ of $C^3$  . At $(1,0,0)\in S^5$, the
stability group is clearly $SU(2)$ which shows
 the result. In this way we see that
\begin{equation}
{\bf C}{\bf P}^2=[SU(3)/SU(2)]/U(1)=S^5/U(1). \label{s22}
\end{equation}
The eight Gell-Mann matrices form the basis for the real vector space of traceless hermitian matrices
$\{\sum_{i=1}^8\xi_i\lambda_i, \xi=(\xi_1,...,\xi_8)\in R^8\}$ . So ${\bf C}{\bf P}^2$ is a submanifold of $R^8$ in
the sense that any element $gYg^{-1}$ of (\ref{definitioncp21}) can be clearly put in the form
$gYg^{-1}=\sum_{i=1}^{8}{\xi}_i{\lambda}_i $ .

There is a beautiful algebraic equation for this submanifold . It goes as follows: Let $d_{ijk}$ be the
totally symmetric $SU(3)$-invariant tensor defined by
\begin{equation}
\{{\lambda}_k,{\lambda}_l\}=\frac{4}{3}{\delta}_{kl}+2d_{klm}{\lambda}_m.
\end{equation}
Then
\begin{equation}
\xi\in {\bf C}{\bf P}^2\ \ \ \Longleftrightarrow\ \ \ \sum_{i,j}d_{ijk}\xi_i\xi_j=-\frac{1}{3}\,\xi_k. \label{s24}
\end{equation}

\paragraph{Proof}

A pleasant manner to demonstrate this result is as follows . The symmetric $SU(3)$ invariant product , $M \vee N $
, of any two traceless hermitian matrices $M,N$ is defined by
\begin{equation}
MVN = \frac{1}{2}\{M,N\}-\frac{1}{6}{\rm Tr}\left(\{M,N\}\right), \label{s25}
\end{equation}
For
\begin{equation}
M=\sum\chi_j\lambda_j\ \ \ N=\sum\eta_j\lambda_j,
\end{equation}
equation (\ref{s25}) will reduce to
\begin{equation}
({\chi}\vee{\eta})_i=d_{ijk}{\chi}_j{\eta}_k,
\end{equation}
where now $M\vee N=\sum_{i}({\chi}\vee{\eta})_i{\lambda}_i$ .

For $M=N=Y$ the above product ,(\ref{s25}), fulfills
\begin{equation}
M\vee M=-\frac{1}{{3}}M \label{s26}
\end{equation}
Now by $SU(3)$ invariance of the $\vee$ product\footnote{For $M^{'}=gMg^{-1}$ and $N^{'}=gNg^{-1}$ we have
$M^{'}{\vee} N^{'}=g(M{\vee} N)g^{-1}$.} , equation (\ref{s26}) is also valid at the point $M=N=gYg^{-1}$ . In other
words (\ref{s26}) is valid for all points of  ${\bf C}{\bf P}^2=\{gYg^{-1}\}$ . Parameterizing ${\bf C}{\bf P}^2$ as
$M=gYg^{-1}=\sum_{i}{\xi}_i{\lambda}_i$ , one can check that (\ref{s26}) reduces to
$\sum_{i,j}d_{ijk}{\xi}_i{\xi}_j=-\frac{1}{3}{\xi}_k$ , that is to (\ref{s24}) .

Conversely, any matrix $M$ satisfying equation (\ref{s26}) should be in the orbit of $Y$ . This can be checked by
first diagonalizing $M$ by an $SU(3)$ transformation $g$ while keeping (\ref{s26}) , then scaling the result
$M_D=g\,M\,g^{-1}$ to $D=-3M_D$ to reduce $-\frac{1}{{3}}$ to $1$, we have then $D\vee D=D$ and  $D= {\rm
diag}(a,b,-a-b)$. Comparing the difference  of the first two rows on both sides, we get $a-b=(a+b)(a-b)$. If
$a=b$, then $D=3\,a\,Y$. If $a\neq b$, then $a+b=1$. Substituting back in the first row, we get $a^2-a-2=0$, or
$a=2$ or $-1$. So $D= {\rm diag}(2,-1,-1)$ or $D= {\rm diag}(-1,2,-1)$. Both become proportional to $Y$ after Weyl
reflections, establishing the result.

\section{Quantizing ${\bf C}{\bf P}^2$}

\subsection{The Symplectic Two-Form Over ${\bf C}{\bf P}^2$ And Its Quantization}

A particular approach to quantizing coadjoint orbits was developed many years ago in \cite{17}. According to that
method, the Lagrangian giving fuzzy ${\bf C}{\bf P}^2$ is
\begin{equation}
L=i\,{\bar N}{\rm Tr} Y\,g(t)^{-1}\,{\dot g(t)},\label{s31}
\end{equation}
where $g(t){\in}SU(3)$ and ${\bar N}$ is an arbitrary constant which is yet to be determined . A point $\xi(t) \in
{\bf C}{\bf P}^2$ is related to $g(t)$ by $\xi(t)_i\lambda_i = g(t)Yg^{-1}(t)$, while the symplectic form on ${\bf
C}{\bf P}^2$ is $i{\bar N}d\Big[{\rm Tr}Yg^{-1}dg\Big]= -i\,{\bar N}{\rm Tr}Y\left[g^{-1}dg\wedge g^{-1}dg\right]$.

Let us parametrize the group element $g(t)$ by a set of eight real numbers ${\theta}_i$ , $i=1,...,8$ , and write
$g(t)=exp(i{\theta}_i{\lambda}_i/2)$ . The conjugate momenta ${\pi}_i$ associated with ${\theta}_i$ are given by
${\pi}_i=\frac{{\partial}L}{{\partial}{\dot{\theta}}_i}=i{\bar
N}TrYg^{-1}\frac{{\partial}g}{{\partial}{\theta}^i}$ . These equations are essentially providing a set of
constraints and therefore one should rewrite them as ${\pi}_i-i{\bar
N}TrYg(t)^{-1}\frac{{\partial}g}{{\partial}{\theta}^i}\approx 0$ . Now if we change the local parametrization
${\theta}_i{\longrightarrow}f_i(\epsilon)$ such that $g(f(\epsilon))=exp(i{\epsilon}_i{\lambda}_i/2)g(\theta)$
where $f(0)=\theta$ . Then one can show that the new conjugate momenta ${\Lambda}_i=-{\pi}_jN_{ji}$ are given by
\begin{equation}
{\Lambda}_i \approx\frac{{\bar N}}{2}Tr[gYg^{-1}{\lambda}_i],\label{constraintscp2}
\end{equation}
where we have used the very useful identity
$i\frac{{\lambda}_i}{2}g(\theta)=N_{ji}(\theta)\frac{{\partial}g}{{\partial}{\theta}^j}$ , with
$N_{ji}(\theta)=\frac{{\partial}f_j(\epsilon)}{{\partial}{\epsilon}_i}|_{{\epsilon}=0}$. Here ${\approx}$ denotes weak equality in the sense of Dirac.

Using $\{{\pi}_i,{\pi}_j\}=\{{\theta}_i,{\theta}_j\}=0$ and $\{{\theta}_i,{\pi}_j\}={\delta}_{ij}$ one can also
prove the following identities
\begin{eqnarray}
\{{\Lambda}_i,g\}&=&i\frac{{\lambda}_i}{2}g\nonumber\\
\{{\Lambda}_i,{\Lambda}_j\}&=&f_{ijk}{\Lambda}_k,
\end{eqnarray}
In other words ${\Lambda}_i$ are the generators of $SU(3)$ transformations which act naturally on the left of
$g(t)$ . They generate symmetries of the Lagrangian (\ref{s31}) as $L$ does not change under the transformations
$g{\longrightarrow}hg$ where $h$ is any constant element in $SU(3)$ .

$SU(3)$ can also act on the right of $g(t)$ . The generators of this action can be given in terms of ${\Lambda}_i$
by ${\Lambda}_j^R=-{\Lambda}_iU_{ij}^{(1,1)}(g)$ where $U(g)^{(1,1)}$ is the adjoint representation of the element
$g$ of $SU(3)$ defined by $U_{ij}^{(1,1)}(g){\lambda}_i=g{\lambda}_jg^{-1}$ . Similarly , these new generators
${\Lambda}_i^R$ satisfy the identities
\begin{eqnarray}
\{{\Lambda}_i^R,g\}&=&-ig\frac{{\lambda}_i}{2}\nonumber\\
\{{\Lambda}_i^R,{\Lambda}_j^R\}&=&f_{ijk}{\Lambda}_k^R.\label{identitylambdar}
\end{eqnarray}
In terms of these right generators , the constraints (\ref{constraintscp2}) take the simpler form
\begin{equation}
{\Lambda}_i^R \approx -\frac{{\bar N}}{\sqrt{3}}{\delta}_{i8},
\end{equation}
These are primary constraints . As in the case of the sphere there are no secondary constraints since the
Hamiltonian commutes with (\ref{constraintscp2}) .

From (\ref{identitylambdar}) , it is obvious that the constarints ${\Lambda}_i^R \approx -\frac{{\bar N}}{\sqrt{3}}{\delta}_{i8}$
, for $i=1,2,3,8$ are first class constraints , whereas for $i=4,5,6,7$ they are second class
constraints\footnote{Indeed one can easily check that the Poisson brackets $\{{\Lambda}_i^R,{\Lambda}_j^R\}$ do
weakly vanish on the surface ${\Lambda}_j^R \approx -\frac{{\bar N}}{\sqrt{3}}{\delta}_{j8}$ , only for $i=1,2,3,8$
.} . We can make a set of first class constraints , which is classically equivalent to all the constraints , by
taking appropriate complex combinations of the above second class constraints , namely
\begin{eqnarray}
{\Pi}_8&=&Y^R=\frac{2}{\sqrt{3}}{\Lambda}_8^R \approx -\frac{2}{3}{\bar N}\nonumber\\
{\Pi}_i&=&I_{i}^R \approx 0\nonumber\\
&and&\nonumber\\
{\Pi}_{45}&=&{\Lambda}_4^R-i{\Lambda}_5^R \approx 0\nonumber\\
{\Pi}_{67}&=&{\Lambda}_6^R-i{\Lambda}_7^R \approx 0,\nonumber\\
\end{eqnarray}
for ${\bar N}{\geq}0$ . For ${\bar N}{\leq}0$ , the two above last equations are replaced with
${\Lambda}_4^R+i{\Lambda}_5^R \approx 0$ and ${\Lambda}_6^R+i{\Lambda}_7^R \approx 0$ . Now one can check that the
Poisson brackets $\{{\Pi}_i,{\Pi}_j\}$ vanish weakly on the surface ${\Pi}_j\approx 0$ , for all $i$ .

These constraints can be realized on functions on $SU(3)$ . As all isospin singlets ( for example the s-quark or
the ${\Omega}^{-}$ ) have hypercharge in integral multiples of $\frac{2}{3}$, we find that ${\bar N}\in {\bf Z}$ .
With ${\bar N}$ fixed accordingly , the constraints together mean that for right action , we have {\it{highest
weight isospin singlet states of  hypercharge $-\frac{2}{3}{\bar N}$}} .

An IRR of $SU(3)$ is labeled by $(n_1,n_2)$ where $n_i\in {\bf N}$ . It comes from the symmetric product of $n_1$
{\bf 3}'s and $n_2$ ${\bf {\bar 3}}$'s : A tensor $T^{i_1...i_{n_1}}_{j_1...j_{n_2}}$ for  $(n_1,n_2)$ has $n_1$
upper indices, $n_2$ lower indices and is traceless, $T^{i_1i_2...i_{n_1}}_{i_1j_2...j_{n_2}}=0$ . Within an IRR ,
the orthonormal basis can be written as $|(n_1,n_2), I^2,I_3,Y\rangle$ where $I^2,I_3$ and $Y$ are square of
isospin , its third component and the hypercharge.

Let $g\rightarrow U^{(n_1,n_2)}(g)$ define the representation $(n_1,n_2)$ of $SU(3)$. Then the functions given by
$\{<(n_1,n_2),I^2,I_3,Y|U^{(n_1,n_2)}(g)|(n_1,n_2),0,0, -\frac{2}{3}{\bar N}>\}$ fulfill the constraints . By the
Peter-Weyl theorem, their linear span
\begin{equation}
\sum\xi^{(n_1,n_2)}_{I^2,I_3,Y} \langle(n_1,n_2),I^2,I_3,Y|U^{(n_1,n_2)}(g)|(n_1,n_2),0,0,
-{\textstyle\frac{2}{3}}{\bar N}\rangle \nonumber
\end{equation}
gives all the functions of interest .

If ${\bar N}=N\ge 0$ , that requires that $(n_1,n_2)=(N,0)$ . These are just the symmetric products of $N$ ${\bf
3}$'s . If  ${\bar N}=-N\le 0$ , $(n_1,n_2)=(0,N)$ or we get the symmetric product of $N$ ${\bf{\bar 3}}$'s . The
representations that we get by quantizing the Lagrangian (\ref{s31}) are thus $(N,0)$ or $(0,N)$ .

\subsection{Fuzzy ${\bf C}{\bf P}^2$ As a ``Fuzzy'' Algebraic Variety}

The coordinate functions ${\hat\xi}_i$ on ${\bf C}{\bf P}^2$ , which are defined by ${\hat\xi}_i(\xi)=\xi_i$ ,
were already shown to satisfy the algebraic relations
\begin{equation}
\sum_{ij}d_{ijk}\hat{\xi}_i\hat{\xi}_j=-\frac{1}{3}\hat{\xi}_k.\label{av1}
\end{equation}
However , since points of ${\bf C}{\bf P}^2$ are given by $gYg^{-1}=\sum_{i=1}^8\hat{\xi}_i{\lambda}_i$ , one can
easily check that the coordinate functions ${\hat\xi}_i$ also satisfy
\begin{equation}
\sum_{i=1}^8\hat{\xi}_i^2=\frac{1}{3}.\label{av2}
\end{equation}
Fuzzy ${\bf C}{\bf P}^2$ is defined by two similar equations which will reduce to (\ref{av1}) and (\ref{av2}) in a
certain limit . These equations are expected to be written in terms of ${\Lambda}_i^L$ , since under quantization
the coordinate functions ${\hat\xi}_i$ become the operators $\Lambda^L_i$ . The continuum limit is , on the other
hand , given by $N{\longrightarrow}{\infty}$ where $(N,0)$, or $(0,N)$ , are the only possible representations we
can get by quantizing the Lagrangian (\ref{s31}) .

The symmetric representations $(N,0)$ of $SU(3)$ can be constructed using $3$ creation operators ${a_i}^\dagger$
and their adjoints $a_i$ . We have
\begin{equation}
\left[a_i,{a_j}^\dagger\right]=\delta_{ij},\ \ i,j=1,2,3. \label{comm1}
\end{equation}
For the representations $(0,N)$ , we need 3 more creation operators $b^\dagger_i$ and their adjoints $b_i$ . We
concentrate below on $(N,0)$ , the treatment of $(0,N)$ being similar .

The $SU(3)$ generators are ${\Lambda}_a^L=a^\dagger\,t_a\, a,\ \ t_a=\frac{1}{2}\lambda_a$ . They fulfill
\begin{equation}
\left[\Lambda_a^L,\Lambda_b^L\right]=i\,f^{abc}\,\Lambda_c^L. \label{comm2}
\end{equation}
By using the definition $d_{ijk} =2\,{\rm Tr}(t_i\{t_j,t_k\})$ , let us compute
$\sum_{ij}d_{ijk}\Lambda_i^L\Lambda_j^L$ which should tend in the continuum limit to (\ref{av1}) . We then have
\begin{eqnarray}
\sum_{ij}d_{ijk}\Lambda_i^L\Lambda_j^L&=& \sum_{ij}2{\rm Tr}(t_i\{t_j,t_k\})\Lambda_i^L\Lambda_j^L\nonumber\\
 &=&
\sum_{ij}2(t_i)_{ab}\Big[(t_j)_{bc}(t_k)_{ca}+(t_k)_{bc}(t_j)_{ca}\Big] \Big[a^\dagger_m (t_i)_{mn} a_{n}\Big]
\Big[a^\dagger_q (t_j)_{qp} a_{p}\Big]\nonumber\\
&=&2\Big[\sum_{i}(t_i)_{ab}(t_i)_{mn}\Big]\Big[\sum_{j}(t_j)_{bc}(t_j)_{qp}\Big](t_k)_{ca}a^{+}_ma_na^{+}_qa_p\nonumber\\
&+&2\Big[\sum_{i}(t_i)_{ab}(t_i)_{mn}\Big]\Big[\sum_{j}(t_j)_{ca}(t_j)_{qp}\Big](t_k)_{bc}a^{+}_ma_na^{+}_qa_p.
\end{eqnarray}
Taking advantage of the Fierz identity
\begin{eqnarray}
\sum_\alpha (t_\alpha)_{ij}\,(t_\alpha)_{kl}= \frac{1}{2}\delta_{il}\delta_{jk}- \frac{1}{6}\delta_{ij}\delta_{kl},
\end{eqnarray}
to reduce the summations over the $i$ and $j$ indices , after a somewhat tedious but straigtforward computation ,
one gets
\begin{eqnarray}
\sum_{ij}d_{ijk}\Lambda_i^L\Lambda_j^L &=&2(t_k)_{ca}\left[ \frac{1}{4}a^\dagger_b a_a a^\dagger_{c} a_b
-\frac{1}{6}a^\dagger_b a_b a^\dagger_{c} a_a -\frac{1}{6}a^\dagger_c a_a a^\dagger_p a_p +\frac{1}{4}a^\dagger_c
a_b a^\dagger_{b} a_a \right]\nonumber\\
&=&2{\Lambda}_k^L\big[\frac{1}{4}+\frac{1}{6}a^{+}_ba_b\big].
\end{eqnarray}
Now the operators , $\sum_{ij}d_{ijk}\Lambda_i^L\Lambda_j^L$ , act on the the Hilbert space ${\cal H}_{(N,0)}$
which corresponds to the representation $(N,0)$ . This ${\cal H}_{(N,0)}$ is spanned by vectors of the form

\begin{eqnarray}
|n_1,n_2,n_3>&=&{a_1^\dagger}^{n_1}{a_2^\dagger}^{n_2}{a_3^\dagger}^{n_3} |0> , ~n_1+n_2+n_3=N\nonumber\\
a_i|0>&=&0, ~i=1,2,3,
\end{eqnarray}
and it is of dimension $\frac{1}{2}(N+1)(N+2)$ . From this last definition one will clearly have

\begin{equation}
\sum_i a^\dagger_i a_i |n_1,n_2,n_3\rangle=\sum_i n_i|n_1,n_2,n_3\rangle= N|n_1,n_2,n_3\rangle ,
\end{equation}
and hence we get
\begin{equation}
\sum_{ij}d_{ijk}\Lambda_i^L\Lambda_j^L={\Lambda}_k^L[\frac{1}{2}+\frac{N}{3}].\label{av1f}
\end{equation}
In the same way one computes the Casimir operator $\sum_{i}{\Lambda}_i^L{\Lambda}_i^L$ , which will reduce in the
continuum limit to (\ref{av2}) , as follows :
\begin{eqnarray}
\sum_{i}{\Lambda}_i^L{\Lambda}_i^L&=&\sum_{i}\big[a^{+}t_ia\big]\big[a^{+}t_ia\big]\nonumber\\
&=&\big[\sum_{i}(t_i)_{kl}(t_i)_{mn}\big]a^{+}_ka_la^{+}_ma_n\nonumber\\
&=&a^{+}_ka_k+\frac{1}{3}(a^{+}_ka_k)^2.
\end{eqnarray}
Hence on the Hilbert space ${\cal H}_{(N,0)}$ we obtain
\begin{equation}
\sum_{i}{\Lambda}_i^L{\Lambda}_i^L=N+\frac{1}{3}N^2.\label{av2f}
\end{equation}
From (\ref{av1}) , (\ref{av2}) , (\ref{av1f}) and (\ref{av2f}) one can define the fuzzy coordinate functions $\hat{\xi}_i^F$ to
be given by
\begin{equation}
\hat{\xi}_i^F=-\frac{{\Lambda}_i^L}{\sqrt{N^2+3N}}.
\end{equation}
They satisfy the following identities
\begin{eqnarray}
\sum_{ij}d_{ijk}{\hat{\xi}}_i^F{\hat{\xi}}_j^F&=&-\frac{1}{3}{\hat{\xi}}_k^F\frac{1+\frac{3}{2N}}{\sqrt{1+\frac{3}{N}}}\nonumber\\ \label{av1F}
\end{eqnarray}
\begin{eqnarray}
\sum_{i}\hat{\xi}_i^{F2}&=&\frac{1}{3},\label{av2F}
\end{eqnarray}
and
\begin{equation}
[{\hat{\xi}}_i^F,{\hat{\xi}}_j^F]=-\frac{if_{ijk}}{\sqrt{N^2+3N}}\hat{\xi}_k^F.
\end{equation}
It is a remarkable fact that  ${\hat\xi}_i^F$ fulfill essentially (\ref{av1}) and (\ref{av2}) ; for large enough $N$
equations (\ref{av1F}) and (\ref{av2F}) are very good approximations to equations (\ref{av1}) and $(\ref{av2})$ respectively .
${\hat\xi}_i^F$ will reduce exactly to ${\hat{\xi}}_i$ , and therefore will commute, at the limit
$N{\longrightarrow}{\infty}$ .

The algebra ${\bf A}_{{\bf C}{\bf P}^2}$ generated by ${\hat \xi}_i^F$ is what substitutes for the algebra of
functions ${\cal A}_{{\bf C}{\bf P}^2}=C^{\infty}({\bf C}{\bf P}^2)$ . By Burnside's theorem \cite{19} , it is the
full matrix algebra in the IRR . Fuzzy ${\bf C}{\bf P}^2$ is just the algebra  ${\bf A}_{{\bf C}{\bf P}^2}$ .

\subsection{The Fuzzy Algebra ${\bf A}_{{\bf C}{\bf P}^2}$ }

The following point , emphasised by \cite{14} is noteworthy . If ${f}\in {\cal A}_{{\bf C}{\bf P}^2}$ , it has the
partial-wave expansion

\begin{eqnarray}
f(\hat{\xi})&=&\sum_{n_1=n_2=n}{f}^n_{I^2,I_3,Y} <(n_1,n_2),I^2,I_3,Y|U^{(n_1,n_2)}(g)|(n_1,n_2),0,0,0>,\nonumber\\
gYg^{-1}&=&\sum_{i=1}^8\hat{\xi}_i{\lambda}_i.\label{noteworthy}
\end{eqnarray}
The ket $|(n_1,n_2),0,0,0> $ exists only if $n_1=n_2$ so that the sum in (\ref{noteworthy}) is restricted to $n_1=n_2=n$ .
From the above expression , it is obvious that $<(n_1,n_2),I^2,I_3,Y|U^{(n_1,n_2)}(g)|(n_1,n_2),0,0,0>$ , and
therefore $f(\hat{\xi})$ , is invariant under the $U(2)$ right action on $g$ , namely $g{\longrightarrow}gh$ where
$h{\in}U(2)$ .

If $F\in {\bf A}_{{\bf C}{\bf P}^2}$ , then $F$ too has an expansion like (\ref{noteworthy}) where the series is cut-off at
$n=N$ . This is because of the following . The $SU(3)$ Lie algebra $su(3)$ has the two following actions on $F$ ,
$F\rightarrow {\Lambda}^L_\alpha\,F=\Lambda_\alpha\,F$ and $F\rightarrow
{\Lambda}^R_\alpha\,F\,=F\,\Lambda_\alpha$ . The derivation $F\rightarrow ad\,
{\Lambda}_\alpha\,F=\left[\Lambda_\alpha,F\right]=({\Lambda}_{\alpha}^L-{\Lambda}_{\alpha}^R)F$ is the action which
annihilates ${\bf I}$ and which corresponds to the $su(3)$ action on ${\bf C}{\bf P}^2$ .

Now since $F{\in}{\bf A}_{{\bf C}{\bf P}^2}$ transforms as $(N,0)$ under the left action of $su(3)$ , while under
the right action it transforms as $(0,N)$ , ${\bf A}_{{\bf C}{\bf P}^2}$ decomposes into the direct sum of the IRR
$(n,n),n=0,...,N$ , namely : $(N,0)\otimes (0,N)=\oplus^N_{n=0}(n,n)$ . If $\{|(n,n),I^2,I_3,Y>\}$ furnishes a
basis for $(n,n)$ , then

\begin{eqnarray}
F&=&\sum {\alpha}_{I^2,I_3,Y}|(N,0),I^2,I_3,Y><(N,0),I^2,I_3,Y|\nonumber\\
&=&\sum_{n=0}^N{F}^n_{I^2,I_3,Y} T^{(n,n)}_{I^2,I_3,Y}\nonumber\\ \label{noteworthyf}
\end{eqnarray}
Identifying the expansion (\ref{noteworthyf}) with the one in (\ref{noteworthy})  for $n\leq N$ , we see that $F$ transforms like a
function on ${\bf C}{\bf P}^2$ with a terminating partial wave expansion .

A more precise statement is as follows \cite{14} . We can put a scalar product on ${\cal A}_{{\bf C}{\bf P}^2}$
using the Haar measure on $SU(3)$ and complete ${\cal A}_{{\bf C}{\bf P}^2}$ into a Hilbert space ${\cal H}_{{\bf
C}{\bf P}^2}$ . On ${\cal H}_{{\bf C}{\bf P}^2}$ , elements $f$ of ${\cal A}_{{\bf C}{\bf P}^2}$ act as linear
operators by point-wise multiplication . Let ${\cal H}^{(N,0)}_{{\bf C}{\bf P}^2}$ be the subspace of ${\cal
H}_{{\bf C}{\bf P}^2}$ carrying the IRR $(N,0)$ and $P_{(N,0)}: {\cal H}_{{\bf C}{\bf P}^2}\rightarrow {\cal
H}^{(N,0)}_{{\bf C}{\bf P}^2}$ the corresponding projector . Then we have a map ${\cal A}_{{\bf C}{\bf
P}^2}\rightarrow P_{(N,0)}\,{\cal A}_{{\bf C}{\bf P}^2}\, P_{(N,0)}$ ; $f\rightarrow P_{(N,0)}\,f P_{(N,0)}$ which
is onto ${\bf A}_{{\bf C}{\bf P}^2}$ . Thus elements of ${\bf A}_{{\bf C}{\bf P}^2}$ approximate functions in a
good sense.

\section{On the Spin$_{c}$ Structure of Classical ${\bf C}{\bf P}^2$}

It is a standard result that ${\bf C}{\bf P}^2$ does not admit a spin structure , but does admit a spin$_c$
structure . We plan to explain this result here adapting an argument of Hawking and Pope \cite{ref19} . 

{\it{The
reasoning shows that ${\bf C}{\bf P}^{N}$ for any even $N{\geq}2$ is not spin whereas it is spin if $N$ is odd}} .

The obstruction to the ${\bf C}{\bf P}^2$ spin structure comes from noncontractibile two-spheres in ${\bf C}{\bf
P}^2$. Since ${\bf C}{\bf P}^2{\simeq}SU(3)/U(2)$ we have
\begin{eqnarray}
{\pi}_2({\bf C}{\bf P}^2)&=&{\pi}_1[U(2)]={\bf Z}\nonumber\\
&and&\nonumber\\
{\pi}_1({\bf C}{\bf P}^2)&=&\{0\},
\end{eqnarray}
so that Hurewitz's theorem \cite{ref24} leads to $H^{2}({\bf C}{\bf P}^2,{\bf Z})={\bf Z}$ . Its mod $2$ reduction
is $H^{2}({\bf C}{\bf P}^2,{\bf Z}_2)={\bf Z}_2$ . 

The absence of spin structure means that the tangent bundle is
associated with the non-trivial element of ${\bf Z} _2$.

Consider a smooth map $g$ of the square $\{(s,t):0{\leq}s;t{\leq}1\}$ into $SU(3)$ which obeys the following
conditions:
\begin{eqnarray}
&g(s,0)=g(0,t)=g(1,t)={\bf 1},&\nonumber\\
&g(s,1)=e^{i{\pi}s({\lambda}_3+{\sqrt{3}}{\lambda}_8)}.&
\end{eqnarray}
[See Fig.1.]
\newpage
\bigskip
\begin{center}

\setlength{\unitlength}{0.5mm}

\begin{picture}(100,80)

\thicklines

\put(-12,12){\makebox(0,0){$(s,t)=(0,0)$}}

\put(-12,85){\makebox(0,0){$(0,1)$}}

\put(112,12){\makebox(0,0){$(1,0)$}}

\put(112,85){\makebox(0,0){$(1,1)$}}

\put(0,20){\line(1,0){100}}

\put(0,20){\line(0,1){60}}

\multiput(0,80)(10,0){10}{\line(1,0){5}}

\put(100,20){\line(0,1){60}}


\put(-5,45){\vector(0,1){15}}

\put(-5,65){\makebox(0,0){t}}

\put(50,15){\vector(1,0){15}}

\put(70,15){\makebox(0,0){s}}

\put(50,5){\makebox(0,0){$g(s,0)=1$}}

\put(-10,50){\makebox(0,0)[r]{$g(0,t)=1$}}

\put(110,50){\makebox(0,0)[l]{$g(1,t)=1$}}

\put(50,95){\makebox(0,0){$g(s,1)=exp(i\pi s(\lambda_3+\sqrt{3}\lambda_8))$}}

\put(50,-10){\makebox(0,0){\bf Fig.1}}

\end{picture}

\end{center}

\bigskip

The curve $g:(s,1){\longrightarrow}g(s,1)$ is a loop in $U(2)=\{{\rm stability ~group ~of}
~{\xi}^0\}$\footnote{${\xi}^0$ denotes here the ``north pole'' of ${\bf C}{\bf P}^2$  , that is the point given by
$[1]=\{h{\in}U(2)\}$ or $Y$ . It has the coordinates $({\xi}_1^0=0,{\xi}_2^0=0,...,{\xi}_8^0=\frac{1}{\sqrt{3}}$).}
not contractible to identity while staying within $U(2)$ . It is the generator of ${\pi}_1(U(2))$ and is associated
with nonabelian $U(2)$ monopoles \cite{extra1} . But since ${\pi}_1(SU(3))=\{0\}$, $g$ can be defined smoothly in
the entire square.

Now $U(2)$ being the stability group of ${\xi}^0$ is contained in the tangent space group $SO(4)$ at ${\xi}^0$. If
$x=(x_\mu:\mu=1,2,3,4)$ is a tangent vector at $\xi^0$ , we can map it to a $2{\times}2$ matrix $M(x)$ given by
\begin{equation}
M(x)=x_4 +i\vec{\tau}\cdot\vec{x},
\end{equation}
where ${\tau}_i$ are the Pauli matrices . The matrix $M(x)$ satsify the reality property
\begin{equation}
M(x)^{*}={\tau}_ 2M(x){\tau}_2.
\end{equation}
$SO(4)=[SU(2){\times}SU(2)]/{\bf Z}_2$ acts on $M(x)$ according to
\begin{equation}
M(x){\longrightarrow}M(x)^{'}=h_1M(x)h_2^{\dagger},\label{so4action}
\end{equation}
where $h_{i}{\in}SU(2)$ . This action preserves both the reality property and the determinant , in other words
\begin{eqnarray}
M(x)^{'*}&=&{\tau}_2M^{'}(x){\tau}_2\nonumber\\
&and&\nonumber\\
detM^{'}(x)&=&detM(x)=\sum_{{\mu}=0}^{4}x_{{\mu}}^2.
\end{eqnarray}
Hence the action (\ref{so4action}) induces an $SO(4)$ transformation on $x$ . $U(2)$ is imbedded in this $SO(4)$ , acting
on $M(x)$ as follows: $M(x){\longrightarrow}M^{'}(x)=h_1M(x)e^{-i{\tau}_3{\theta}}$.

The spin group $SU(2){\times}SU(2)=\{(h_1,h_2)\}$ is a two-fold cover of the rotation group $SO(4)$ . The inverse
image of $U(2)$ in $SU(2){\times}SU(2)$ is $SU(2){\times}U(1)$ , also a two-fold cover of $U(2)$ . In this cover
the loop
\begin{equation}
g:(s,1){\longrightarrow}g(s,1){\in}U(2)
\end{equation}
becomes
\begin{equation}
s{\longrightarrow}(e^{i{\pi}s{\tau}_3},e^{i{\pi}s{\tau}_3}){\in}SU(2){\times}U(1).
\end{equation}
It is no longer a loop , but runs from $({\bf I,\bf I})$ to $(-\bf I,-\bf I)$ . It is this that obstructs the spin
structure , as the following reasoning encountered in \cite{ref19} shows .

Let $SU(3){\longrightarrow}{\bf C}{\bf P}^2{\simeq}SU(3)/U(2)$ be the map
\begin{equation}
h{\in}SU(3){\longrightarrow}hYh^{-1}=\sum_{i=1}^8{\lambda}_{i}{\xi}_{i}.
\end{equation}
$U(2)$ here has generators ${\lambda}_i(i=1,2,3)$ and ${\lambda}_8$ . This map clearly takes the entire boundary of
the square $\{g(s,t)\}$ to ${\xi}^0$ and the square itself to a $2-$sphere ${\bf S}^2$ .

\begin{center}

\setlength{\unitlength}{0.5mm}

\begin{picture}(100,100)

\thicklines

\put(0,20){\line(1,0){100}}

\put(0,20){\line(0,1){60}}

\multiput(0,80)(10,0){10}{\line(1,0){5}}

\put(100,20){\line(0,1){60}}




\put(-10,85){{\bf P}}

\put(110,85){{\bf Q}}

\put(50,10){\makebox(0,0){\bf I}}

\put(-10,50){\makebox(0,0)[r]{\bf II}}

\put(110,50){\makebox(0,0)[l]{\bf III}}

\put(50,90){\makebox(0,0){\bf IV}}

\put(50,-5){\makebox(0,0){\bf Fig.2}}

\end{picture}

\end{center}

\bigskip

Now the tangent space at ${\xi}^0$ of ${\bf C}{\bf P}^2$ is spanned by the four $SU(3)$ Lie algebra directions
$K^{+},K^{0},\bar{K}^{0},K^{-}$ (in a complex basis) . If we write ${\bf C}{\bf P}^2$ as $\{hYh^{-1}\}$, a basis of
tangents (a frame) at $\xi^0$ is ${\lambda}_a(a=4 ,5,6,7)$ . Clearly $g(s,t){\lambda_a}g(s,t)^{-1}$ gives a frame
at $g(s,t)Yg(s,t)^{-1}$ of ${\bf C}{\bf P}^2$. This gives us a rule for transporting this frame ( and hence any
frame) smoothly over ${\bf S}^2{\in}{\bf C}{\bf P}^2$ along curves .

If $\{(s(\tau),t(\tau)),0{\leq} \tau{\leq}1\}$ is a curve on the square , then
\begin{equation}
g(s(\tau),t(\tau))Yg(s(\tau),t(\tau))^{-1}
\end{equation}
is the corresponding curve in ${\bf S}^2$ of ${\bf C}{\bf P}^2$ . The transport of the frame along the curve $
g(s(\tau),t(\tau))Yg(s(\tau),t(\tau))^{-1}$ is
\begin{equation}
g(s(\tau),t(\tau)){\lambda}_ag(s(\tau),t(\tau))^{-1}.
\end{equation}
In this rule , for the three sides I , II , III (see Fig.2) , we have the constant curve $g(s,t)Yg(s,t)^{-1}=Y$ in
${\bf S}^2$ of ${\bf C}{\bf P}^2$ , and we have the frame $g(s,t){\lambda}_ag(s,t)^{-1}={\lambda}_a$. In other
words for I , II and III we are at ${\xi}^{0}$ with the frame held fixed and equal to ${\lambda}_a$ . Along the
side IV , however , we are still at $Y$ or ${\xi}^0$ but we are rotating ${\lambda}_a$ according to
\begin{equation}
exp\{{i{\pi}s({\lambda}_3+{\sqrt{3}}{\lambda}_8)}\}\,{\lambda}_a\,exp\{{-i{\pi}s({\lambda}_3+{\sqrt{3}}{\lambda}_8)}\}.
\end{equation}
This is a $2{\pi}-$rotation of the frame as $s$ varies from $0$ to $1$ . This can be seen from the fact that
${\lambda}_3+{\sqrt{3}}{\lambda}_8=2Diag(1,0,-1)$ which will lead to
$exp[i{\pi}s({\lambda}_3+{\sqrt{3}}{\lambda}_8)]=Diag(exp(2i{\pi}s),1,exp(-2i{\pi}s))$ , and therefore the frame
rotates from ${\lambda}_a$ back to ${\lambda}_a$ as $s$ changes from $0$ to $1$ .

If spinors can be defined on ${\bf C}{\bf P}^2$ , this transport of frames will consistently lead to their
transport as well . Thus along sides I , II , III , we should be able to pick a suitable constant spinor ${\psi}$ .
But then , along IV , as $s$ increases to $1$ , we will arrive at $Q$ with $-{\psi}$ as $(-{\bf 1},-{\bf 1})$ of
$SU(2){\times}SU(2)$ flips the sign of a spinor . As we had ${\psi}$ along III , we lose continuity at $Q$ and find
that spinors do not exist for ${\bf C}{\bf P}^2$ .

It is possible to show that this conclusion is not sensitive to our choice of rule of transport of frames (that is
, connection in the frame bundle) .

{\it The spin$_c$ structure on ${\bf C}{\bf P}^2$ is achieved by introducing an additional $U(1)$ connection
for spinors which amounts to adding a hypercharge of magnitude $|Y|=1$ }. That would give an additional phase
$exp(i{\pi}{\sqrt{3}}{\lambda}_8s)$ along IV and an extra minus sign at $s=1$ cancelling the above unwanted minus
sign . Note that $1)$ this connection and extra hypercharge cancels out for frames which contain a spinor and a
complex conjugate spinor , $2)$ there is no vector bundle with this extra connection as its existence gives a
contradiction just as does the existence of the spin bundle .

Let us see what all this means for $SU(3)$ . Under $U(2)$ , at ${\xi}^0$ , the tangents transform as $(K^{+},K^0)$
and $(K^{-},\bar{K}^0)$ , that is as the IRR's $(I,Y)=(\frac{1}{2},1)$ and $(\frac{1}{2},-1)$ respectively . From
the way $M(x)$ transforms under $U(2)$ ,
\begin{equation}
M(x){\longrightarrow}M^{'}(x)=hM(x)e^{-i{\tau}_3{\theta}},h{\in}SU(2),
\end{equation}
we can see that $Y$ corresponds to ${\tau}_3$ acting on the right
of $M(x)$.

The $SU(2){\times}SU(2)$ IRR's of the non-existent spinors are as follows :

$1)$Left-handed spinors : $(1/2,0)$ with quantum numbers $(I,Y)=(1/2,0)$ under the action of the two-fold cover
$SU(2){\times}U(1)$ of $U(2)$ .

$2)$Right-handed spinors : $(0,1/2)$ with quantum numbers $(I,Y)=(0,1)$ and $(0,-1)$ .

The quantum numbers in the ${\rm spin}_c$ case follows by adding an additional hypercharge which we can take to be
$-1$ :

$1)$ Left-handed ${\rm spin}_c$ : $(I,Y)=(1/2,-1)$.

$2)$ Right-handed ${\rm spin}_c$ : $(I,Y)=(0,0)$ and $(0,-2)$ .

These are precisely the $U(2)$ quantum numbers of the representation space of tangent ${\gamma}'s$ which will be
found in Section $5.5$ . The $SU(3)$ IRR's have to contain these $U(2)$ IRR's . They are not symmetric between
left- and right-handed spinors.

The ${\rm spin}_c$  structures are not unique . Thus we have the freedom to add additional hypercharge $2n$
$(n{\in}Z)$ to the ${\rm spin}_c$  spinors , that is , tensor the ${\rm spin}_c$  bundle with any $U(1)$ bundle .
The choice of ${\rm spin}_c$ in our text is natural for our Dirac operator .

\subsection{On General ${\bf C}{\bf P}^N$}

${\bf C}{\bf P}^N$ for all odd $N$ admits a spin structure whereas those for even $N$ admit only a ${\rm spin}_c$
structure \cite{ref25} . We can understand this result too by pursuing the preceding arguments .

Let $Y^{(N+1)}=\frac{1}{N+1}\,{\rm diag}\,(1,1,...1,-N)$ be the $SU(N+1)$ ``hypercharge''. The previous $Y$ is
$Y^{(3)}$. We can represent ${\bf C}{\bf P}^{N}$ as
\begin{equation}
{\bf C}{\bf P}^{N}=SU(N+1)/U(N)=\{hY^{(N+1)}h^{-1}:h{\in}SU(N+1)\},
\end{equation}
the stability group is
\begin{equation}
U(N)=\{u{\in}SU(N+1):uY^{(N+1)}u^{-1}=Y^{(N+1)}\}.
\end{equation}
For all $N{\geq}1$ , the square of $Fig.1$ and $Fig.2$ and the map
\begin{equation}
g ~:~(s,t){\longrightarrow}g(s,t){\in}SU(N+1)
\end{equation}
can be constructed so that it is constant on sides I , II and III while
\begin{equation}
g~:~(s,1){\longrightarrow}g(s,1)
\end{equation}
gives a generator of ${\pi}_1 (U(N))$ . {\it There is an obstruction to spin structure if this loop when it acts on
a frame at $Y^{(N+1)}$ rotates it by $2{\pi}$ , in other words it acts as the noncontractible loop of $SO(2N)$} .

Let $(q_1,q_2,...,q_{N+1})$ be the ``quarks'' of $SU(N+1)$ . The hypercharge $Y^{(N)}$ of $SU(N)$ acts as the
generator $\bar{Y}^{(N)}=\frac{1}{N}(1,1,..,-(N-1),0)$ on these quarks . We can choose the loop
$g:(s,1){\longrightarrow}g(s,1)$ according to
\begin{eqnarray}
g(s,1)&=&e^{i\frac{2{\pi}s}{N}(N\bar{Y}^{(N)})}e^{-i\frac{2{\pi}s}{N}(N+1)Y^{(N+1)}}\nonumber\\
&=& \left[\begin{array}{ccccc}
                          1&0&.&.&0\\
                          0&1&0&.&0\\
                           0&.&.&1&0\\
                           0&.&.&e^{-i2{\pi}s}&0\\
                           0&0&0&0&e^{i2{\pi}s}
           \end{array}
     \right].
\end{eqnarray}
The tangent vectors at $Y^{(N+1)}$ transform like $\bar{q}^{(i)}q^{(N+1)}$ and
$\bar{q}^{(N+1)}q^{(i)}(1{\leq}i{\leq}N)$. So under $g(s,1)$,
\begin{eqnarray}
\bar{q}^{(i)}q^{(N+1)}&{\longrightarrow}&e^{i2{\pi}s}\bar{q}^{(i)}
q^{(N+1)},i\leq N-1\nonumber\\
\bar{q}^{(i)}q^{(N+1)}&{\longrightarrow}&e^{i4{\pi}s}\bar{q}^{(i)}
q^{(N+1)},i=N\nonumber\\
\bar{q}^{(N+1)}q^{(i)}&{\longrightarrow}&e^{-i2{\pi}s}\bar{q}^{(N+1)}
q^{(i)},i{\leq}N-1\nonumber\\
\bar{q}^{(N+1)}q^{(i)}&{\longrightarrow}&e^{-i4{\pi}s}\bar{q}^{(N+1)} q^{(i)},i=N.
\end{eqnarray}
Each $i$ , $i{\leq}N-1$ , gives a plane in $2N$ dimensions and each factor $e^{i2{\pi}s}$ gives a
$2{\pi}-$rotation . For $i=N$ , we have a $4{\pi}-$ rotation corresponding to $e^{i4{\pi}s}$ . Thus we have a
total product of $(N-1)+2=(N+1)$ $2{\pi}-$rotations . For $N$ odd , they are contractible in $SO(2N)$ , and for
$N$ even , they are not , showing the result we were after .

\section{The Dirac Operator On ${\bf S}^2$ Revisited : Towards Fuzzy Spinors}
This section is a warm up for what follows on ${\bf C}{\bf P}^2$ next . It contains a partial-wave analysis for
the eigenstates of the ${\bf S}^2$ Dirac operator ${\cal D}_{2g}$ which can be generalised to ${\bf C}{\bf P}^2$ .

In this section , we will essentially construct all the spinors describing a Dirac particle on ${\bf S}^2_F$ . The
answer found in this section for the continuum ${\bf S}^2$ will be simply cut-off at the top value
$j=2l-\frac{1}{2}$ to get the answer for the fuzzy ${\bf S}^2$ . This last step will be done explicitly in section
$6$ of this chapter. Given these fuzzy spinors , the projective ${\bf A}-$ bimodule ${\bf H}_2={\bf
A}{\otimes}{\bf C}^2$ is completely determined . The action of the Dirac operator on this ${\bf H}_2$ will be also
defined  .

Let us first recall the following results
\begin{eqnarray}
{\cal D}_{2g}&=&\frac{1}{{\rho}}{\sigma}_{i}{\cal P}_{ij}\,{\cal J}_{j},\nonumber\\
{\cal P}_{ij}&=&{\delta}_{ij}-\hat{n}_{i}\hat{n}_{j},{\cal J}_{i}={\cal L}_{i}+\frac{{\sigma}_{i}}{2},\nonumber\\
{\cal L}_{i}&=&-i(\vec{\hat{x}}{\wedge}\vec{\nabla})_{i}.\label{recall}
\end{eqnarray}
${\cal P}$ projects the Pauli matrices ${\sigma}_i$ to their tangent space components ${\sigma}_{i}{\cal
P}_{{i}{j}}$ . ${\cal L}_{i}$ and ${\cal J}_{i}$ are orbital and total angular momenta respectively . $\hat{x}$ are
the coordinate functions : $\hat{x}_i(x)=x_i$ .

The algebra ${\cal A}={\bf C}^{\infty}({\bf R}^3)/{\cal I}$ of smooth complex valued functions on ${\bf S}^2$
defined in section $(3.1.1)$ is ${\bf C}^{\infty}({\bf S}^2)$ . Any element $f$ of ${\cal A}$ will then have the
partial wave expansion
\begin{eqnarray}
f(\vec{\hat{x}})&=&\sum_{kk_3}f^{k}_{k_3}<kk_3|D^{(k)}(g)|k0>\nonumber\\
{\rho}g{\sigma}_3g^{-1}&=&\vec{\hat{x}}.\vec{\sigma},
\end{eqnarray}
where $D^{(k)}:g{\longrightarrow}D^{(k)}(g)$ is the $k$ IRR of $SU(2)$ .

$f(\vec{\hat{x}})$ is clearly a function over the sphere . This can be seen from the fact that when
$g{\longrightarrow}gexp(i{\sigma}_3{\theta}/2)$ , then $D^{(k)}(g){\longrightarrow}D^{(k)}(g)exp(i{\cal
J}_3^{(k)}{\theta})$ where ${\cal J}_3^{(k)}$ is the third component of the angular momentum $\vec{{\cal
J}}^{(k)}$ . Hence one sees immediately that $<kk_3|D^{(k)}(g)|k0>$ is ivariant under the right $U(1)$ action on
$g$ and therefore it is a function of $[gexp(i{\sigma}_3{\theta}/2)]{\in}{\bf S}^2$ .

The action of ${\cal L}_i$ on these functions $f$ is defined by
\begin{equation}
{\cal L}_i(<kk_3|D^{(k)}(g)|k0>)=-<kk_3|{\cal J}_i^{(k)}D^{(k)}(g)|k0>.
\end{equation}

${\cal D}_{2g}$ acts on ${\cal A}{\otimes}{\bf C}^2{\equiv}{\cal A}^2=\{(a_1,a_2):a_i{\in}{\bf C}^{\infty}({\bf
S}^2)\}$ and it anticommutes with the chirality operator ${\gamma}={\sigma}\cdot\hat{n}$ . The action of
${\sigma}\cdot\hat{n}$ on $D^{(K)}$ is defined by
\begin{equation}
[{\sigma}\cdot\hat{n}D^{(k)}](g)={\sigma}\cdot n(g)D^{(k)}(g),\label{actionofgamma}
\end{equation}
where $n_{i}(g)$ is a function on $SU(2)$ defined as follows. For any $g{\in}SU(2)$ we have :
\begin{equation}
g{\sigma}_3g^{-1}={\sigma}\cdot n(g).
\end{equation}
Using the fact that $D^{(\frac{1}{2})}(g)=g$ one can now rewrite (\ref{actionofgamma}) as
\begin{eqnarray}
[{\sigma}\cdot\hat{n}D^{(k)}](g)&=&g{\sigma}_3g^{-1}D^{(k)}(g)\nonumber\\
&=&D^{(\frac{1}{2})}(g){\sigma}_3D^{(\frac{1}{2})-1}(g)D^{(k)}(g)\nonumber\\
&=&[D^{(\frac{1}{2})}{\sigma}_3D^{(\frac{1}{2})-1}D^{(k)}](g)\nonumber\\
&{\Longrightarrow}&\nonumber\\
{\sigma}.\hat{n}&=&D^{(\frac{1}{2})}{\sigma}_3D^{(\frac{1}{2})-1}.
\end{eqnarray}
Now by taking the standard basis $\{|+>,|->\}$ defined by the equations ${\sigma}_3|\pm>=\pm|\pm>$ , one have the
following identity
\begin{equation}
{\sigma}.\hat{n}D^{(\frac{1}{2})}|\pm>=D^{(\frac{1}{2})}{\sigma}_3|\pm>={\pm}D^{(\frac{1}{2})}|\pm>.
\end{equation}
In other words $D^{(\frac{1}{2})}|+>$ and $D^{(\frac{1}{2})}|->$ are the eigenfunctions of ${\gamma}$ with $+1$ and
$-1$ helicity respectively . These spinors can be rewritten as
\begin{equation}
|{\psi}^{(\frac{1}{2})}_{\pm}>= \left(
\begin{array}{c}
<+|D^{(\frac{1}{2})}|\pm>\\
<-|D^{(\frac{1}{2})}|\pm>
\end{array}\right)\,\label{spinorfinal0}
\end{equation}

From their definition , the spinors $|{\psi}^{(\frac{1}{2})}_{\pm}>$  are functions on $SU(2)$ . For $g{\in}SU(2)$
, the function $|{\psi}^{(\frac{1}{2})}_{\pm}(g)>$ is defined by
\begin{eqnarray}
|{\psi}^{(\frac{1}{2})}_{\pm}(g)>&=&\left(
\begin{array}{c}
<+|D^{(\frac{1}{2})}(g)|\pm>\\
<-|D^{(\frac{1}{2})}(g)|\pm>
\end{array}\right)\,\label{spinorfinal}
\end{eqnarray}
where $<m|D^{(\frac{1}{2})}(g)|\pm>=g_{m,\pm}$ . This function has the obvious equivariance property
\begin{equation}
|{\psi}^{(\frac{1}{2})}_{\pm}(ge^{i\frac{{\sigma}_3}{2}{\theta}})>=|{\psi}^{(\frac{1}{2})}_{\pm}(g)>e^{{\pm}i\frac{{\theta}}{2}}.\label{equivariancespinor}
\end{equation}
Unlike (\ref{equivariancespinor}) , elements of ${\cal A}^2$ and hence too its chirality ${\pm}1$ subspaces
$\frac{1{\pm}{\sigma}.\hat{n}}{2}{\cal A}^2$ are invariant under $g\rightarrow ge^{i{\sigma}_3{\theta}}$ . This is
because they are functions on ${\bf S}^2$ and not on $SU(2)$ .

The expansion of elements of these subspaces using the above $D$'s must thus have another $D$ in each term
transforming with the opposite phase to that in (\ref{equivariancespinor}). Accounting for this fact, we can write for $|{\psi}>\in
{\cal A}^2$ ,
\begin{eqnarray}
|\psi>&=&|{\psi}^{+}>+|{\psi}^{-}>,\nonumber\\
|{\psi}^{\pm}>&=&\Big[\sum_{j,m}{\xi}_m^{j\pm}<jm|{\psi}^{(j)}_{\mp}>\Big]|{\psi}^{(\frac{1}{2})}_{\pm}>~,~{\xi}_m^{j\pm}{\in}{\bf
C} .\label{spinorexpansionfinal}
\end{eqnarray}
$|{\psi}^{+}>$ and $|{\psi}^{-}>$ belong to the subspaces $\frac{1+{\sigma}.\hat{n}}{2}{\cal A}^2$ and
$\frac{1-{\sigma}.\hat{n}}{2}{\cal A}^2$ respectively and hence they represent left handed spinors and right handed
spinors on the sphere ${\bf S}^2$  . $|{\psi}^{(j)}_{\mp}>$ in equation (\ref{spinorexpansionfinal}) is , on the other hand , defined
by
\begin{equation}
|{\psi}^{(j)}_{\mp}>=D^{(j)}|j,{\mp}\frac{1}{2}>.\label{570}
\end{equation}
$j$ here are the eigenvalues of the total angular momentum $\vec{\cal J}$ defined in equation (\ref{recall}) . It is half
integer equal to $l+\frac{1}{2}$ or $l-\frac{1}{2}$ and therefore the states $|j,{\mp}\frac{1}{2}>$ do always exist
. $l$ are of course the eigenvalues of the orbital angular momentum $\vec{\cal L}$ . It is now not difficult to see
that under the transformation $g{\longrightarrow}gexp(i{\sigma}_3{\theta}/2)$ we have
\begin{eqnarray}
|{\psi}^{(j)}_{\mp}(g)>{\longrightarrow}|{\psi}^{(j)}_{\mp}(ge^{i\frac{{\sigma}_3}{2}{\theta}})>&=&D^{(j)}(ge^{i\frac{{\sigma}_3}{2}{\theta}})|j,{\mp}
\frac{1}{2}>\nonumber\\
&=&D^{(j)}(g)e^{i{\cal
J}_3^{(j)}{\theta}}|j,{\mp}\frac{1}{2}>\nonumber\\
&=&|{\psi}^{(j)}_{\mp}(g)>e^{{\mp}i\frac{\theta}{2}}.\nonumber\\ \label{equivariancespinorextra}
\end{eqnarray}
From (\ref{equivariancespinor}) and (\ref{equivariancespinorextra}) it is then obvious that (\ref{spinorexpansionfinal}) are invariant under $g\rightarrow
ge^{i{\sigma}_3{\theta}}$.

Now orbital angular momentum ${\cal L}_{i}$ is not defined on the individual factors in (\ref{spinorexpansionfinal}) . We must lift
${\cal L}_{i}$ to the operator ${\cal K}_{i}^{L}$ which acts on $D^{(j)}$ and $D^{(\frac{1}{2})}$ in such a manner
that the spinors (\ref{spinorfinal}) are rotationally invariant . ${\cal K}_{i}^L$ can then be defined to be $SU(2)$
generators acting by left translation:
\begin{eqnarray}
[e^{i{\theta}_{i}{\cal K}_{i}^{L}} D^{(\frac{1}{2})}_{ij}](g)&\equiv&<i|D^{(\frac{1}{2})}(e^{-i{\theta}_{i}\frac{{\sigma}_{i}}{2}}g)|j>\nonumber\\
&=&<i|e^{-i{\theta}_{i}\frac{{\sigma}_{i}}{2}}g|j>\label{transformationrule}
\end{eqnarray}
We now reinterpret ${\cal J}_{i}$ as
\begin{equation}
{\cal J}_{i}={\cal K}_{i}^{L}+\frac{{\sigma}_{i}}{2}.
\end{equation}
Because of the transformation rule (\ref{transformationrule}) we have for infinitesimal transformations ,
\begin{eqnarray}
\Big[(1+i{\theta}_i{\cal
K}_{i}^{L})D^{(\frac{1}{2})}_{nm}\Big](g)&=&<n|D^{(\frac{1}{2})}\Big[(1-i{\theta}_i\frac{{\sigma}_i}{2})g\Big]|m>,
\end{eqnarray}
and therefore
\begin{eqnarray}
[D^{(\frac{1}{2})}_{nm}](g)+i{\theta}_i[{\cal K}_{i}^{L} D^{(\frac{1}{2})}_{nm}](g)&=&<n|g|m>-i{\theta}_i<n|\frac{{\sigma}_i}{2}g|m>.\nonumber\\
\end{eqnarray}
But because $[D^{(\frac{1}{2})}_{nm}](g)=<n|g|m>$ one gets
\begin{eqnarray}
-[{\cal K}_i^L D^{(\frac{1}{2})}_{nm}](g)&=&<n|\frac{{\sigma}_i}{2}g|m>=\Big[\frac{{\sigma}_i}{2}D^{(\frac{1}{2})}\Big]_{nm}(g)\nonumber\\
&{\Longrightarrow}&\nonumber\\
0&=&{\cal
K}_i^LD^{(1/2)}_{nm}+\Big[\frac{{\sigma}_i}{2}D^{(\frac{1}{2})}\Big]_{nm}\nonumber\\
&or&\nonumber\\
0&=&({\cal J}_{i}D^{(1/2)})_{nm}.\label{identity1}
\end{eqnarray}
From (\ref{spinorfinal0}) and (\ref{identity1}) we get
\begin{equation}
{\cal J}_{i}|{\psi}_{\pm}^{(\frac{1}{2})}>=0,\label{identity2}
\end{equation}
in other words the eigenfunctions of the chirality operator ${\gamma}$ are rotationally invariant .

The action of the Dirac operator on the spinors (\ref{spinorexpansionfinal}) will now be computed. First one has
\begin{eqnarray}
{\cal D}_{2g}<jm|{\psi}_{\mp}^{(j)}>|{\psi}_{\pm}^{(\frac{1}{2})}>&=&\frac{1}{{\rho}}{\sigma}_a{\cal P}_{ab}({\cal
K}_b^{L}+\frac{{\sigma}_b}{2})<jm|{\psi}_{\mp}^{(j)}>|{\psi}_{\pm}^{(\frac{1}{2})}>\nonumber\\
&=&\frac{1}{{\rho}}{\sigma}_a{\cal P}_{ab}{\cal K}_b^{L}<jm|{\psi}_{\mp}^{(j)}>|{\psi}_{\pm}^{(\frac{1}{2})}>
+\frac{1}{{\rho}}{\sigma}_a{\cal
P}_{ab}\frac{{\sigma}_b}{2}<jm|{\psi}_{\mp}^{(j)}>|{\psi}_{\pm}^{(\frac{1}{2})}>\nonumber\\
&=&\frac{1}{{\rho}}{\sigma}_a{\cal P}_{ab}\Bigg[{\cal
K}_b^{L}(<jm|{\psi}_{\mp}^{(j)}>)|{\psi}_{\pm}^{(\frac{1}{2})}>+<jm|{\psi}_{\mp}^{(j)}>{\cal K}_b^{L}(|{\psi}_{\pm}^{(\frac{1}{2})}>)\Bigg]\nonumber\\
&+&\frac{1}{{\rho}}{\sigma}_a{\cal
P}_{ab}<jm|{\psi}_{\mp}^{(j)}>\frac{{\sigma}_b}{2}\Big[|{\psi}_{\pm}^{(\frac{1}{2})}>\Big].\nonumber\\
&=&\frac{1}{{\rho}}\Big[{\cal K}_b^{L}<jm|{\psi}_{\mp}^{(j)}>\Big]\Big[{\sigma}_a{\cal
P}_{ab}|{\psi}_{\pm}^{(\frac{1}{2})}>\Big],\label{identity7}
\end{eqnarray}
where we have used the identity (\ref{identity2}) , to go from the third line to the last line.

Further simplification can be achieved by using the following identity:
\begin{eqnarray}
{\sigma}_{p}{\cal P}_{pk} ={D^{(\frac{1}{2})}}[{1\over 2}{\sigma}_3,[{1\over 2}{\sigma}_3,{\sigma}_{i}]]
{D^{(\frac{1}{2})}}^{-1}D^{(1)}_{ki},\label{identity6}
\end{eqnarray}
which can be proven as follows . From one hand we have
\begin{eqnarray}
[\frac{1}{2}{\sigma}.\hat{n},[\frac{1}{2}{\sigma}.\hat{n},{\sigma}_k]]&=&(\frac{1}{2}\hat{n}_i)(\frac{1}{2}\hat{n}_j)[{\sigma}_i,[{\sigma}_j,{\sigma}_k]]\nonumber\\
&=&(\frac{1}{2}\hat{n}_i)(\frac{1}{2}\hat{n}_j)(2i{\epsilon}_{jkl})(2i{\epsilon}_{ilp}{\sigma}_p)\nonumber\\
&=&{\sigma}_p{\cal P}_{pk}.\label{identity5}
\end{eqnarray}
On the other hand one has
\begin{eqnarray}
[\frac{1}{2}{\sigma}.\hat{n},[\frac{1}{2}{\sigma}.\hat{n},{\sigma}_k]]&=&\frac{1}{4}[D^{(\frac{1}{2})}{\sigma}_3D^{(\frac{1}{2})-1},[D^{(\frac{1}{2})}{\sigma}_3D^{(\frac{1}{2})-1},{\sigma}_k]]\nonumber\\
&=&D^{(\frac{1}{2})}[\frac{{\sigma}_3}{2},[\frac{{\sigma}_3}{2},D^{(\frac{1}{2})-1}{\sigma}_kD^{(\frac{1}{2})}]]D^{(\frac{1}{2})-1}.\nonumber\\
&&\label{identity3}
\end{eqnarray}
Now from the basic definition of the adjoint action , $g{\sigma}_jg^{-1}=R_{ij}(g){\sigma}_i$ , we have
$g^{-1}{\sigma}_jg=R_{ij}(g^{-1}){\sigma}_i=R_{ji}(g){\sigma}_i$ , where we have used the fact that $R$ is a real
orthogonal representation , i.e $R(g^{-1})=R^{T}(g)$ . The result can also be written as
\begin{equation}
D^{(\frac{1}{2})-1}{\sigma}_jD^{(\frac{1}{2})}={\sigma}_iD^{(1)}_{ji}.\label{identity4}
\end{equation}
Putting (\ref{identity4}) in (\ref{identity3}) and comparing to (\ref{identity5}) one obtains the identity (\ref{identity6}) .

Using the identity (\ref{identity6}) in equation (\ref{identity7}) we get
\begin{eqnarray}
{\cal D}_{2g}&=&\frac{1}{{\rho}}\Big[{\cal K}_b^{L}<jm|{\psi}_{\mp}^{(j)}>\Big]\Big[{D^{(\frac{1}{2})}}[{1\over
2}{\sigma}_3,[{1\over 2}{\sigma}_3,{\sigma}_{a}]]
{D^{(\frac{1}{2})}}^{-1}D^{(1)}_{ba}|{\psi}_{\pm}^{(\frac{1}{2})}>\Big]\nonumber\\
&=&\frac{1}{\rho}[D^{(1)}_{ba}{\cal K}_b^L<jm|{\psi}_{\mp}^{(j)}>]\Big[D^{(\frac{1}{2})}[\frac{{\sigma}_3}{2},[\frac{{\sigma}_3}{2},{\sigma}_a]]|{\pm}>\Big]\nonumber\\
&=&-\frac{1}{\rho}[{\cal K}_a^R<jm|{\psi}_{\mp}^{(j)}>]\Big[D^{(\frac{1}{2})}[\frac{{\sigma}_3}{2},[\frac{{\sigma}_3}{2},{\sigma}_a]]|{\pm}>\Big].\nonumber\\
&&
\end{eqnarray}
In the last line of the above equation we have used the fact that ${\cal K}_a^R=-D^{(1)}_{ba}{\cal K}_b^L$ are the
right $SU(2)$ generators acting by right translation , i.e
\begin{eqnarray}
[e^{i{\theta}_{i}{\cal K}_{i}^{R}} D^{(\frac{1}{2})}_{ij}](g)&\equiv&<i|D^{(\frac{1}{2})}(ge^{i{\theta}_{i}\frac{{\sigma}_{i}}{2}})|j>\nonumber\\
&=&<i|ge^{i{\theta}_{i}\frac{{\sigma}_{i}}{2}}|j>.
\end{eqnarray}
The proof goes as follows . From equation (\ref{transformationrule}) we have the identity $[exp(i{\theta}_{a}{\cal
K}_{a}^{L})D^{(\frac{1}{2})}_{ij}](g)=<i|exp(-i\frac{{\theta}_{a}}{2}{\sigma}_{a})g|j>=<i|gg^{-1}exp(-i\frac{{\theta}_{a}}{2}
{\sigma}_{a}) g |j>$ . But we can check that
$g^{-1}exp(-i\frac{{\theta}_a}{2}{\sigma}_a)g=exp(-i\frac{{\theta}_a}{2}g^{-1}{\sigma}_ag)$ and hence we obtain
$[e^{i{\theta}_{a}{\cal K}_{a}^{L}}
D^{(\frac{1}{2})}_{ij}](g)=<i|gexp(i\frac{\tilde{\theta}_b}{2}{\sigma}_b)|j>=[e^{i{\tilde{\theta}}_{b}{\cal
K}_{b}^{R}} D^{(\frac{1}{2})}_{ij}](g)$ , where we have used $g^{-1}{\sigma}_ag=R_{ab}(g){\sigma}_b$ and with
${\tilde{\theta}}_a=-{\theta}_bR_{ba}(g)$ . We then get $-R_{ba}(g){\cal K}_a^R={\cal K}_b^L$ and therefore ${\cal
K}^R_a=-R_{ba}(g){\cal K}_b^L$ or ${\cal K}_a^R=-D^{(1)}_{ba}{\cal K}_b^L$.

Now by using the identity
\begin{eqnarray}
{\cal K}_a^R<jm|{\psi}_{\mp}^{(j)}>&=&{\cal K}_a^R<jm|D^{(j)}|j,{\mp}\frac{1}{2}>\nonumber\\
&=&<jm|D^{(j)}{\cal J}_a^{(j)}|j,{\mp}\frac{1}{2}>,
\end{eqnarray}
we finally find that
\begin{equation}
{\cal
D}_{2g}<jm|{\psi}_{\mp}^{(j)}>|{\psi}_{\pm}^{(\frac{1}{2})}>=-\frac{1}{\rho}\sum_{m^{'}}<jm|D^{(j)}|jm^{'}><jm^{'}|{\cal
J}_{a}^{(j)}|j,{\mp}\frac{1}{2}>\Big[D^{(\frac{1}{2})}[\frac{{\sigma}_3}{2},[\frac{{\sigma}_3}{2},{\sigma}_a]]|{\pm}>\Big].
\end{equation}

Now we are ready to find the action of the Dirac operator ${\cal D}_{2g}$ on the full spinor (\ref{spinorexpansionfinal}) . First one
can easily find
\begin{equation}
{\cal D}_{2g}|{\psi}^{\pm}>=-\frac{1}{\rho}\sum_{j,m,m^{'},a}{\xi}_m^{j{\pm}}<jm|D^{(j)}|jm^{'}><jm^{'}|{\cal
J}_{a}^{(j)}|j,{\mp}\frac{1}{2}>\Big[D^{(\frac{1}{2})}[\frac{{\sigma}_3}{2},[\frac{{\sigma}_3}{2},{\sigma}_a]]|{\pm}>\Big].
\end{equation}
Next we obtain after some standard calculations
\begin{eqnarray}
{\cal D}_{2g}|{\psi}>&=&{\cal D}_{2g}|{\psi}^{+}>+{\cal D}_{2g}|{\psi}^{-}>\nonumber\\
&=&-\frac{1}{\rho}\sum_{j,m,m^{'}}{\xi}_m^{j+}<jm|D^{(j)}|jm^{'}><jm^{'}|{\cal
J}_{+}^{(j)}|j,-\frac{1}{2}>D^{(\frac{1}{2})}|->\nonumber\\
&-&\frac{1}{\rho}\sum_{j,m,m^{'}}{\xi}_m^{j-}<jm|D^{(j)}|jm^{'}><jm^{'}|{\cal
J}_{-}^{(j)}|j,+\frac{1}{2}>D^{(\frac{1}{2})}|+>\nonumber\\
\end{eqnarray}
In evaluating the sum over the $a$ index , we used the following identities
$[\frac{{\sigma}_3}{2},[\frac{{\sigma}_3}{2},{\sigma}_a]]={\sigma}_a$ , $a=1,2$ , ${\sigma}_{\pm}|\mp>=2|\pm>$ and
${\sigma}_{\pm}|\pm>=0$ . Of course ${\sigma}_{\pm}={\sigma}_1{\pm}i{\sigma}_2$ and ${\cal J}_{\pm}^{(j)}={\cal
J}_1^{(j)}{\pm}i{\cal J}_2^{(j)}$ .

Finally by using the fact that ${\cal J}_{\pm}^{(j)}$ are off-diagonal in each subspace with fixed $j$ , we find
\begin{eqnarray}
{\cal D}_{2g}|{\psi}>&=&-\frac{1}{\rho}\sum_{j,m}{\xi}_m^{j+}<jm|{\psi}_{+}^{(j)}><j,\frac{1}{2}|{\cal
J}_{+}^{(j)}|j,-\frac{1}{2}>|{\psi}_{-}^{(\frac{1}{2})}>\nonumber\\
&-&\frac{1}{\rho}\sum_{j,m}{\xi}_m^{j-}<jm|{\psi}^{(j)}_{-}><j,-\frac{1}{2}|{\cal
J}_{-}^{(j)}|j,+\frac{1}{2}>|{\psi}_{+}^{(\frac{1}{2})}>,\nonumber\\
&&\label{589}
\end{eqnarray}
which can also be written in the ``Dirac-K{\"a}hler'' form \cite{14}

\begin{eqnarray}
{\cal D}_{2g}^{D-K}{\psi}^{D-K}\equiv-\frac{1}{\rho}\sum_{j}
                   &&\left[ \begin{array}{cc}
                     0  & <j,-\frac{1}{2}|{\cal J}^{(j)}_{-}|j,+\frac{1}{2}>\\
                     <j,\frac{1}{2}|{\cal J}_{+}^{(j)}|j,-\frac{1}{2}>  &
                     0
                           \end{array}
                    \right]{\times}\nonumber\\
&&\left[
                            \begin{array}{c}
                  \sum_{m}\xi^{j+}_m<jm|{\psi}_{+}^{(j)}>\\
                  \sum_{m}\xi^{j-}_m <jm|{\psi}_{-}^{(j)}>
                              \end{array}
                            \right],\nonumber\\
&&\label{kahlerfinal}
\end{eqnarray}
where now the spinors are written in the basis $\{|{\psi}_{+}^{(\frac{1}{2})}> , |{\psi}_{-}^{(\frac{1}{2})}>\}$ ,
namely
\begin{equation}
{\psi}^{D-K}=\left[
                            \begin{array}{c}
                  \sum_{j,m}\xi^{j+}_m<jm|{\psi}_{-}^{(j)}>\\
                  \sum_{j,m}\xi^{j-}_m <jm|{\psi}_{+}^{(j)}>
                              \end{array}
                            \right].
\end{equation}

\section{The Dirac Operator on ${\bf C}{\bf P}^2$}

${\bf C}{\bf P}^2$ is not spin , but spin$_c$ \cite{14,ref19}. This fact introduces serious differences between the
${\bf C}{\bf P}^2$ Dirac operator and the Dirac operator for a spin manifold such as ${\bf C}{\bf P}^1$ discussed
last . The ${\bf C}{\bf P}^2$ Dirac operator and its fuzzy version have been treated in \cite{14}. Here we develop
an alternative approach which seems capable of generalisation to other coset spaces . We will first summarize just
some points relevant for us. The next section will give their fuzzy versions.

\subsection{The Projective Module For Tangent Bundle and its Complex Structure}

The generators $Ad\,{\lambda}_{i}$ of the adjoint representation $Ad :g\rightarrow Ad\,g$ of $SU(3)=\{g\}$ have
matrix elements
\begin{equation}
(Ad{\lambda}_{i})_{jk}=-2if_{ijk},\label{adjointrepresentation}
\end{equation}
where $f_{ijk}$ is totally antisymmetric , and such that
\begin{equation}
[Ad{\lambda}_{i},Ad{\lambda}_{j}]=2if_{ijk}Ad{\lambda}_{k}.
\end{equation}
As hypercharge commutes with itself and with the isospin generators , it follows that $f_{8ij}=0$ if $i$ or
$j=1,2,3,8$ . Thus the tangent vectors to ${\bf C}{\bf P}^2$ at ${\xi}^{0}=(0,...,0,\frac{1}{\sqrt{3}})$ , or
equivalently at $Y=\sum_{i}\xi^0_{i}{\lambda}_{i}$ , are $Ad{\lambda}_{i}$ where $i=4,5,6,7$ . The directions
$Ad{\lambda}_{j}$ , ${j}=1,2,3,8$ , are normals .

At any other point $\sum_{i}{\xi}_{i}{\lambda}_{i}=gYg^{-1}{\in}{\bf C}{\bf P}^2$ , the normals accordingly are $
Ad\,g(Ad\,\lambda_8)\, Ad\,g^{-1}=\sum_{i}{\xi}_{i}^{(8)}Ad{\lambda}_{i}$ and  $ Ad\,g(Ad\,\lambda_j)\,
Ad\,g^{-1}=\sum_{i}{\xi}_{i}^{(j)}Ad{\lambda}_{i}$ , $j=1,2,3$ . The hypercharge normal $Adg(Ad{\lambda}_8)
Adg^{-1}$ still commute with the isospin normals $Adg(Ad{\lambda}_j)Adg^{-1}$ , $j=1,2,3$ . In other words we have the following identity
$[Adg(Ad{\lambda}_j)Adg^{-1},Adg(Ad{\lambda}_8)Adg^{-1}]=2if_{j8k}Adg(Ad{\lambda}_k)Adg^{-1}=0$ which leads to
$f_{ikl}{\xi}_i^{(j)}{\xi}_k^{(8)}=0$ .

The four orthogonal directions $Ad g Ad{\lambda}_{i} Ad g^{-1}$ , for $i=4,5,6,7$ , in the trace norm span the
tangent space .

Since $(Ad{\lambda}_8)_{ij}=-2if_{8ij}$ , we can represent $AdY=\frac{1}{\sqrt{3}}Ad{\lambda}_8$ by an
$8{\times}8$ block diagonal matrix of the form $AdY=(0,{\sigma}_2,{\sigma}_2)$ . The zero is a $4{\times}4$ matrix
in which the indices $i$ and $j$ take the values $1,2,3,8$ . ${\sigma}_2$ is , on the other hand , the usual
$2{\times}2$ Pauli matrix . In the first ${\sigma}_2$ the indices $i$ and $j$ are in $\{4,5\}$ whereas in the
second ${\sigma}_2$ they are in $\{6,7\}$ . It is then obvious that the eigenvalues of $AdY$ are $0$ and ${\pm}1$ .

Since ${\xi}_8^0=\frac{1}{\sqrt{3}}$ , ${\xi}_{i}Ad{\lambda}_{i}$ for a generic point ${\xi}$ of ${\bf C}{\bf P}^2$
will also have the eigenvalues $0$ and ${\pm}1$ corresponding to the mesons $(K^{+},K^0)$ , $(K^{-},\bar{K}^0)$ ,
${\eta}^{0}$ and $\vec{\pi}$ in the flavor octet terminology .

If ${\chi}^{(+)}$ is an eigenvector for eigenvalue $+1$ , $({\xi}_{i}Ad{\lambda}_{i}) {\chi}^{(+)}={\chi}^{(+)}$ .
Then ${\xi}_{k}^{(j)}{\chi}_{k}^{(+)}= {\xi}_{k}^{(j)}({\xi}_{i}Ad{\lambda}_{i})_{kj}
{\chi}_{j}^{(+)}=2if_{kij}{\xi}_k^{(j)}{\xi}_i{\chi}_j^{(+)}$ . But since ${\xi}_i={\xi}_i^{(8)}$ and
$f_{ikl}{\xi}_i^{(j)}{\xi}_k^{(8)}=0$ we have ${\xi}_{k}^{(j)}{\chi}_{k}^{(+)}=0$ . In other words ${\chi}^{(+)}$
is a tangent to ${\bf C}{\bf P}^2$ at ${\xi}$ . So if ${\chi}^{(-)}$ is the eigenvector for eigenvalue $-1$ ,
${\chi}^{(\pm)}$ span the tangent space ${\bf T}_{\xi}{\bf C}{\bf P}^2$ . In the other hand , the null space of
${\xi}_{i}Ad{\lambda}_{i}$ spans the space of normals .

We can now present sections of the tangent bundle ${\bf T}{\bf C}{\bf P}^2$ as a projective module . Let ${\cal
A}_{{\bf C}{\bf P}^2}^8={\cal A}_{{\bf C}{\bf P}^2}{\otimes}{\bf C}^8=\{({\hat{\xi}}_1,...,{\hat{\xi}}_8)\}$ ,
where ${\hat{\xi}}_i$ are as usual defined by ${\hat{\xi}}_i(\xi)={\xi}_i$ . Then

\begin{equation}
{\cal P}=({\hat{\xi}}_{i}ad{\lambda}_{i})^2 \label{cp2tangentbundle}
\end{equation}
is a projector and ${\cal P}{\cal A}_{{\bf C}{\bf P}^2}^8$ is seen to consist of the sections of tangent bundle
from the above remarks .

The complex structure on ${\bf C}{\bf P}^2$ can be thought of as a splitting of the tangent space ${\bf
T}_{{\xi}}{\bf C}{\bf P}^2$ as the direct sum ${\bf T}_{{\xi}}^{(+)}{\bf C}{\bf P}^2+{\bf T}_{{\xi}}^{(-)}{\bf
C}{\bf P}^2$ for all ${\xi}{\in}{\bf C}{\bf P}^2$ in a smooth manner . The tensor ${\cal J}$ of complex analysis at
${\xi}$ is then ${{\pm}i}$ on ${\bf T}_{{\xi}}^{(\pm)}{\bf C}{\bf P}^2$ .

In the language of projective modules , we must thus split ${\cal P}$ as the sum of two orthogonal projectors
${\cal P}^{(\pm)}$ . The tensor ${\cal J}$ is ${\pm}i$ on ${\cal P}^{(\pm)}{\cal A}^8$ , that is , ${\cal
J}=i({\cal P}^{(+)}-{\cal P}^{(-)})$ . Hence also ${\cal J}{\cal P}={\cal P}{\cal J}={\cal J}$ .

$SU(3)-$covariance suggests the choice of ${\cal P}^{(\pm)}{\cal A}_{{\bf C}{\bf P}^2}^8$ as eigenspaces of
$\hat{\xi}_{i}Ad{\lambda}_{i}$ for eigenvalues ${\pm}1$ . Hence
\begin{equation}
{\cal P}^{(\pm)}=\frac{1}{2}{\hat{\xi}}_{i}Ad{\lambda}_{i} (\hat{\xi}_{i}Ad{\lambda}_{i}{\pm}1).\label{complexstructurecp2}
\end{equation}
As $Ad{\lambda}_{\alpha}$ is antisymmetric , we have that
\begin{equation}
{\cal P}^{(+)T}={\cal P}^{(-)},{\cal J}^{T}=-{\cal J}.
\end{equation}
From ${\cal J}$ , we can also write the Levi-Civita symbol in an $SU(3)-$covariant way . It is
\begin{equation}
{\epsilon}_{{\alpha}{\beta}{\gamma}{\delta}}={\cal J} _{[{\alpha}{\beta}}{\cal J}_{{\gamma}{\delta}]},~[~]:\ \
{\rm antisymmetrisation}.
\end{equation}

\subsection{The Gamma Matrices}

Since ${\bf C}{\bf P}^2$ is a submanifold of ${\bf R}^8$ , it is natural to start from the Clifford algebra on
${\bf R}^8$ . Let its basis be the $16{\times}16$ matrices ${\Gamma}_{i}$ $({i}=1,2,...8)$ with the relations
\begin{equation}
\{{\Gamma}_{i},{\Gamma}_{j}\}=2{\delta}_{ij}, \ \ \ {\Gamma}_{i}^{\dagger}={\Gamma}_{i} .
\end{equation}
The ${\gamma}$ matrices which will occur in the Dirac operator on ${\bf C}{\bf P}^2$ are not these ${\Gamma}'s$ ,
rather they will be $16{\times}16$ ${\gamma}$ matrices ${\gamma}_{\mu}$ with the same relations
\begin{equation}
\{{\gamma}_{i},{\gamma}_{j}\}=2{\delta}_{{i}{j}}, \ \ \ {\gamma}_{i}^{\dagger}={\gamma}_{i} .\label{cliffordalgebra}
\end{equation}
These ${\gamma}'s$ act by left multiplication on the algebra $Mat_{16}=\{M\}$ of $16{\times}16$ matrices which is
generated by ${\Gamma}_{i}$ . Thus we have
\begin{eqnarray}
{\gamma}_{i}M&\equiv&{{\Gamma}}_{i}M .
\end{eqnarray}
The matrices $M$ of $Mat_{16}$ have a scalar product $(M,N)={\rm Tr}\,(M^{\dagger}N)$ for which
${\gamma}_{i}^{\dagger}={\gamma}_{i}$ . From these definitions , one can already see that spinors on ${\bf C}{\bf
P}^2$ will be constructed out of the ${\Gamma}'s$ , i.e they will be elements of $Mat_{16}$ .

The ${\bf C}{\bf P}^2$ ${\gamma}'s$ are the tangent projections ${\gamma}_{i}{\cal P}_{{i}{j}}$ . There are only
four of them at each ${\xi}$ which are linearly independent . We have to find a four-dimension subspace of
$Mat_{16}$ at each ${\xi}$ on which they can act . If we fail in that , we will end up with more than one fermion .

We first find this subspace at ${\xi}^0$ . At ${\xi}^0$ , define the fermionic creation-annihilation operators
\begin{eqnarray}
{A}^{\dagger}_{1}&=&\frac{1}{2}({{\Gamma}}_4+i{{\Gamma}}_5), {A}_{1}=\frac{1}{2}({{\Gamma}}_4-i{{\Gamma}}_5),\nonumber\\
{A}^{\dagger}_{2}&=&\frac{1}{2}({{\Gamma}}_6+i{{\Gamma}}_7), {A}_{2}=\frac{1}{2}({{\Gamma}}_6-i{{\Gamma}}_7).
\end{eqnarray}
${A}^{\dagger}_{i}$ transform as $(K^{+},K^{0})$ whereas ${A}_{i}$ transform as $(K^{-},\bar{K}^0)$ . They satisfy
$\{A^{+}_i,A_j\}={\delta}_{ij}$ . Let
\begin{equation}
|0\rangle={{A}}_1{{A}}_2,\ \ |i\rangle={A}_{i}^{\dagger}|0\rangle\ ({i}=1,2),\ \
|3\rangle={A}^{\dagger}_{1}{A}^{\dagger}_{2}|0\rangle.
\end{equation}
They span a $4-$dimensional subspace of $Mat_{16}$ . ${\gamma}_{i}$ $(4{\leq}{i}{\leq}7)$ act irreducibly on this
space . The creation operators are defined by
\begin{equation}
a_1^{\dagger}=\frac{1}{2}({\gamma}_4+i{\gamma}_5),a_2^{\dagger}=\frac{1}{2}({\gamma}_6+i{\gamma}_7),
\end{equation}
while their adjoints define the annihilation operators . $|0\rangle$ is the vacuum state of these
creation-annihilation operators .

For an appropriate subspace at other points of ${\bf C}{\bf P}^2$ , we use the fact that $SU(3)$ acts transitively
on ${\bf C}{\bf P}^2$ . Thus we can regard ${\xi}{\in}{\bf C}{\bf P}^2$ as a function on $SU(3)$ with value
$\xi(g)$ at $g$ via the relation $gYg^{-1}=\sum_{i}{\lambda}_{i}{\xi}_{i}(g)$ . Then ${\xi}^0={\xi}(e)$, $e={\rm
identity}$ .

Now the Lie algebra of $SU(3)$ can be realised using the Clifford algebra ${{\gamma}}_{i}$ , the generators being
\begin{equation}
t_{i}^c=\frac{1}{4i}f_{ijk}{{\gamma}}_{j}{{\gamma}}_{k}.\label{tic}
\end{equation}
These $t_{i}^c$ are indeed the generators of $SU(3)$ as one can check by directly computing their commutation
relations ,
$[t^c_i,t^c_j]=\frac{1}{(4i)^2}f_{i{\mu}{\nu}}f_{j{\alpha}{\beta}}[{\gamma}_{\mu}{\gamma}_{\nu},{\gamma}_{\alpha}{\gamma}_{\beta}]$
. By using the identities $[AB,CD]=A\{B,C\}D-C\{D,A\}B+CA\{D,B\}-\{A,C\}BD$ and (\ref{cliffordalgebra}) one can find
$[t_i^c,t_j^c]=-\frac{4{\gamma}_{\mu}{\gamma}_{\beta}}{(4i)^2}[f_{{\mu}i{\alpha}}f_{{\beta}j{\alpha}}+f_{{\beta}i{\alpha}}f_{j{\mu}{\alpha}}]$
. Finally by invoking the Jacobi's identities $f_{kpl}f_{ijl}+f_{ipl}f_{kil}+f_{ipl}f_{jkl}=0$ we get the result
$[t_i^c,t_j^c]=if_{ijk}t_k^c$ .

The ${{\gamma}}_ {i}$ transform as the adjoint representation $\underline{8}$ under the action by derivation
${{\gamma}}_{\mu}{\rightarrow}[t_{i}^c,{{\gamma}}_{\mu}]$ of $SU(3)$ , as one can check that
\begin{eqnarray}
[t_{i}^c,{\gamma}_{\mu}]&=&\frac{1}{4i}f_{ijk}[{\gamma}_j{\gamma}_k,{\gamma}_{\mu}]\nonumber\\
&=&if_{i{\mu}k}{{\gamma}}_{k}\nonumber\\
&=&(Ad t_i)_{k{\mu}}{\gamma}_k,
 \label{s55}
\end{eqnarray}
where in the second line we have used the identity $[AB,C]=A\{B,C\}-\{A,C\}B$ while in the last line we used
equation (\ref{adjointrepresentation}) . The $16-$dimensional representation space of $t^c_i$ can , however , be split into
$\underline{8}{\oplus}\underline{8}$ using the projectors $P_{\pm}=\frac{1{\pm}{\Gamma}_9}{2}$ ,
${\Gamma}_9={\Gamma}_1{\Gamma}_2...{\Gamma}_8$ .

Let $T(g)$ be the image of $g$ in the $SU(3)$ representation given by (\ref{s55}). $T(g)$ can act on $Mat_{16}$ by
conjugation according to
\begin{equation}
Ad\,T(g)M=T(g)MT^{-1}(g).\label{adjointsu3}
\end{equation}
$Ad\,{t_{i}}$ are the infinitesimal generators for the action $Ad\,T(g)$ of $SU(3)$ .

The $4-$dimensional vector space at $g=e$ and its basis can be labelled as $V(e)$ and
$\{|{\nu};e\rangle,{\nu}=0,1,2,3:|{\nu};e\rangle=|\nu\rangle\}$ . The vector space and its basis at $g$ are then
\begin{eqnarray}
V(g)&=&Ad\,T(g) V(e)=T(g)V(e)T^{-1}(g),\nonumber\\
|{\nu};g\rangle &=& Ad\,T(g)|\nu;e\rangle=T(g)|\nu;e\rangle T^{-1}(g).
\end{eqnarray}
It is on this vector space that ${\gamma}_{i}{\cal P}_{{i}{j}}({\xi}(g))$ act by left-multiplication .

On the vector space $V(e)$ , the $U(2)$ subgroup of $SU(3)$ acts by conjugation . From the particle physics
interpretation of ${A}^{+}_{i}$ , we see that $V(e)$ decomposes into the direct sum

\begin{equation}
(I=0,Y=-2){\oplus}(I=\frac{1}{2},Y=-1){\oplus}(I=0,Y=0).\label{theparticlephysicsinterpretation}
\end{equation}
$(0,-2)$ corresponds to the vacuum $|0;e>$ , $(\frac{1}{2},-1)$ corresponds to the doublet $\{|i;e>,i=1,2\}$ while
$(0,0)$ corresponds to $|3;e>$ . As we have already shown in section $(5.3)$ , $(I=\frac{1}{2},Y=-1)$ is the
left-handed spinor $(\frac{1}{2},0)$ on ${\bf C}{\bf P}^2$ whereas $(I=0,Y=-2)$ and $(I=0,Y=0)$ correspond to the
right handed spinor $(0,\frac{1}{2})$ on ${\bf C}{\bf P}^2$ .

The result (\ref{theparticlephysicsinterpretation}) can be shown as follows . From equation (\ref{s55}) , one can compute $[Y^c,a_i^{+}]=a_i^{+}$
and $[Y^c,a_i]=-a_i$ , where $Y^c$ is the hypercharge operator in the representation (\ref{tic}) , i.e
$Y^c=\frac{2}{\sqrt{3}}t^c_8$ . Then one can easily prove that $[Y^c,a_1a_2]=-2a_1a_2$ ,
$[Y^c,a_i^{+}a_1a_2]=-a_i^{+}a_1a_2$ and $[Y^c,a_1^{+}a_2^{+}a_1a_2]=0$ from which the hypercharge content of
(\ref{theparticlephysicsinterpretation}) follows trivially .

The isospin content of (\ref{theparticlephysicsinterpretation}) can be computed in a similar fashion . First one obtains the commutation
relations $[t^c_3,a_1^{+}]=\frac{1}{2}a_1^{+}$ , $[t^c_3,a_1]=-\frac{1}{2}a_1$ , $[t^c_3,a_2^+]=-\frac{1}{2}a_2^+$
and $[t_3^c,a_2]=\frac{1}{2}a_2$ . Then we compute the results $[t^c_3,a_1a_2]=[t^c_3,a_1^{+}a_2^{+}a_1a_2]=0$,
$[t^c_3,a^+_1a_1a_2]=\frac{1}{2}a^{+}_1a_1a_2$ and $[t^c_3,a^+_{2}a_1a_2]=-\frac{1}{2}a^{+}_{2}a_1a_2$ with which
the proof of (\ref{theparticlephysicsinterpretation}) is completed .

{\subsubsection{5.5.2.1~~~A Brief Review Of $SU(3)$ Representation Theory}}

Before we go back to the Hilbert space $V(g)$ and describe its $SU(3)$ representation content , we would like to
give first some general remarks concerning $SU(3)$ representation theroy which will be very useful . These remarks
will also be used extensively in the study of both the Dirac operator in section $(5.5.3)$ and the wave functions
in section $(5.5.4)$ . The references for this section are in \cite{ref20} .

\paragraph{On the embedding of $SU(3)$ into $SO(8)$}
\noindent

$SO(8)$ admits two inequivalent $8-$dimensional spinor representations , the spinor representation
$\underline{8}_s$ and the conjugate representation $\underline{8}_c$. In $8$ dimensions , the Dirac spinor is
$16-$dimensional , $\underline{16}=\underline{8}_s+\underline{8}_c$ , corresponding to the fact that the only
IRR's of the $8-$dimensional Clifford algebra (\ref{cliffordalgebra}) are $16-$dimensionals. Each one of the spinor representions
$\underline{8}_s$ and ${\underline{8}}_c$ defines a Weyl Spinor in $8$ dimensions.

$SU(3)$ can be embedded into $SO(8)$ in two different ways . In the regular embedding , $SU(3)$ is embedded first
into $SU(4)$ , then $SU(4)$ is embedded into $SO(8)$ . In this case , the $SU(3)$ content of $\underline{8}_v$ ,
$\underline{8}_s$ and $\underline{8}_c$ are
\begin{eqnarray}
&&[\underline{1}+\underline{3}+\underline{1}+\bar{\underline{3}}]_{SU(3)}\subset[\underline{4}+\bar{\underline{4}}]_{SU(4)}\subset\underline{8}_v\nonumber\\
&&[\underline{1}+\underline{1}+\underline{3}+\bar{\underline{3}}]_{SU(3)}\subset[\underline{1}+\underline{1}+\underline{6}]_{SU(4)}\subset\underline{8}_s\nonumber\\
&&[\underline{1}+\underline{3}+\underline{1}+\bar{\underline{3}}]_{SU(3)}\subset[\underline{4}+\bar{\underline{4}}]_{SU(4)}{\subset}\underline{8}_c.\nonumber\\ \label{decomposition}
\end{eqnarray}
In (\ref{decomposition}) , $\underline{8}_v$ is the vector-like $8-$dimensional representation of $SO(8)$ .

In the special embedding of $SU(3)$ into $SO(8)$ , one simply have
\begin{eqnarray}
&&[\underline{8}]_{SU(3)}\subset\underline{8}_v\nonumber\\
&&[\underline{8}]_{SU(3)}\subset\underline{8}_s\nonumber\\
&&[\underline{8}]_{SU(3)}\subset\underline{8}_c.\nonumber\\
\end{eqnarray}
The generators (\ref{tic}) of $SU(3)$ define exactly this special emebdding since their $16-$dimensional
representation space was shown to decompose into the direct sum of two $8's$ , i.e
$\underline{16}=\underline{8}_s+\underline{8}_c$ .

\paragraph{The $SU(3)$ representation content of the algebra $Mat_{16}$}
\noindent

The $8-$dimensional Clifford algebra , (\ref{cliffordalgebra}) , provides for us a basis for the algebra $Mat_{16}$ of all
$16{\times}16$ matrices . This basis consists of the matrices
\begin{equation}
1,\{{\gamma}_{i_1}\},\{{\gamma}_{i_{1}i_{2}}=\frac{1}{2}({\gamma}_{i_1}{\gamma}_{i_2}-{\gamma}_{i_2}{\gamma}_{i_1})\},
\{{\gamma}_{i_1i_2i_3}\},...,\{{\gamma}_{i_1...i_7}\},
{\gamma}_9={\gamma}_1....{\gamma}_8.
\end{equation}
Each matrix ${\gamma}_{i_1..i_n}$ , for $2{\leq}n{\leq}8$ and with ${\gamma}_{i_1...i_8}={\gamma}_9$ , is
completely antisymmetric in the indices $i_1$...$i_n$ . For a fixed $n$ there are $\frac{8!}{n!(8-n)!}$
independent ${\gamma}$ matrices . In other words the $16{\times}16=256-$dimensional matrix algebra $Mat_{16}$
decomposes under $SO(8)$ as
\begin{equation}
(\underline{8}_s+\underline{8}_c)+(\underline{8}_s{\oplus}\underline{8}_c)=\underline{1}+\underline{8}_v+\underline{28}+\underline{56}+\underline{35}+\underline{35}+\underline{56}+\underline{28}+\underline{8}_v+\underline{1}.\label{decomposition0}
\end{equation}
The two singlet representations correspond to the identity and to the ${\gamma}_9$ . The two vector representations
correspond to $\{{\gamma}_i\}$ and $\{{\gamma}_{i_1...i_7}\}$ . $\underline{28}$ , $\underline{56}$ ,
$\underline{35}$ are , on the other hand ,  $\underline{28}=\{{\gamma}_{ij}\}$ ,
$\underline{56}=\{{\gamma}_{i_1i_2i_3}\}$ , $\underline{35}+\underline{35}=\{{\gamma}_{i_1...i_4}\}$ ,
$\underline{56}=\{{\gamma}_{i_1...i_5}\}$ and $\underline{28}=\{{\gamma}_{i_1...i_6}\}$ .

In terms of $SU(3)$ representations we have the decompositions
\begin{eqnarray}
&&[\underline{1}]_{SU(3)}{\subset}{\underline{1}}\nonumber\\
&&[\underline{8}]_{SU(3)}{\subset}{\underline{8}}_v\nonumber\\
&&[\underline{8}+\underline{10}+\underline{\bar{10}}]_{SU(3)}{\subset}{\underline{28}}\nonumber\\
&&[\underline{27}+\underline{10}+\underline{\bar{10}}+\underline{8}+\underline{1}]_{SU(3)}{\subset}{\underline{56}}\nonumber\\
&&[\underline{27}+\underline{8}]_{SU(3)}{\subset}{\underline{35}}.\nonumber\\ \label{decomposition2}
\end{eqnarray}

The first and second equations of (\ref{decomposition2}) are obvious by construction . The other equations , however , require
more work to prove . To this end one borrows the following identities from Slansky ,
$\underline{8}_s{\otimes}\underline{8}_s=\underline{1}+\underline{28}+\underline{35}_s$ ,
$\underline{8}_c{\otimes}\underline{8}_c=\underline{1}+\underline{28}+\underline{35}_c$ and
$\underline{8}_s{\otimes}\underline{8}_c=\underline{8}_{c}{\otimes}\underline{8}_s=\underline{8}_v+\underline{56}_v$
.

${\gamma}_{ij}'s$ are the generators of $SO(8)$ which are obtained by taking the antisymmetric product of two
${\gamma}'s$ , i.e two $\underline{8}_v$ . But we know that
$[\underline{8}{\times}\underline{8}]_{SU(3)}=[\underline{1}+\underline{8}+\underline{8}+\underline{10}+\underline{\bar{10}}+\underline{27}]_{SU(3)}$
, and $\underline{8}_v{\otimes}\underline{8}_v=\underline{1}+\underline{28}+\underline{35}_v$ . From the fact that
$[\underline{8}]_{SU(3)}\subset \underline{8}_v$ , one concludes that
$[\underline{8}{\otimes}\underline{8}]_{SU(3)}{\subset}\underline{8}_v{\otimes}\underline{8}_v$ and therefore
$[\underline{8}+\underline{10}+\underline{\bar{10}}]_{SU(3)} \subset \underline{28}$ . In the same way by comparing
$[(\underline{8}+\underline{8}){\otimes}(\underline{8}+\underline{8})]_{SU(3)}=[4(\underline{27})+4(\underline{10}+\underline{\bar{10}})+4(\underline{8}+\underline{8})+4(\underline{1})]_{SU(3)}$, with (\ref{decomposition0}) and using the first three equations of (\ref{decomposition2}) we get the result
$[2(\underline{27})+\underline{10}+\underline{\bar{10}}+2(\underline{8})+\underline{1}]_{SU(3)}{\subset}\underline{56}+\underline{35}$.
Clearly the $\underline{56}$ can not contain both the $[\underline{27}]_{SU(3)}'s$, so one of the
$[\underline{27}]_{SU(3)}'s$ is in $\underline{35}$. The two last equations of (\ref{decomposition2}) then follow easily .

\paragraph{States of the $SU(3)$ representations $(N_1,N_2)$}
\noindent

A general representation of $SU(3)$ is characterized by two integers $N_1$ and $N_2$. A basis for the Hilbert
space on which it acts can be given  by the set , $\{|(N_1,N_2);(I,I_3,Y)\rangle\}$ , where $I$ , $I_3$ and $Y$ are
the isospin , the third component of the isospin and the hypercharge quantum numbers respectively. The dimension
of this representation is $ d(N_1,N_2)=\frac{(N_1+1)(N_2+1)(N_1+N_2+2)}{2}$ .

Following Okubo's notes \cite{ref20} , the basic states of the representation $(N_1,N_2)$ are the components of
$T^{{\nu}_1....{\nu}_{N_2}}_{{\mu}_1...{\mu}_{N_1}}|0>$ . Each component is an eigenstate of $I_3$ and $Y$ . Since
each of the indices ${\mu}_i$ , ${\nu}_i$ can take only three values $1$,$2$ or $3$, one can characterize
$T^{{\nu}_1....{\nu}_{N_2}}_{{\mu}_1...{\mu}_{N_1}}$ by the number $N_1^{i}$ of lower indices having the value $i$
and  the number $N_2^{i}$ of upper indices having the value $i$ . We have as identities $N_1=\sum_{i=1}^3N_1^i$
and $N_2=\sum_{i=1}^3N_2^i$ . It is not then difficult to find that
\begin{eqnarray}
YT^{{\nu}_1....{\nu}_{N_2}}_{{\mu}_1...{\mu}_{N_1}}|0>&=&[\frac{1}{3}(N_1-N_2)-(N_1^3-N_2^3)]T^{{\nu}_1....{\nu}_{N_2}}_{{\mu}_1...{\mu}_{N_1}}|0>\nonumber\\
&and&\nonumber\\
I_3T^{{\nu}_1....{\nu}_{N_2}}_{{\mu}_1...{\mu}_{N_1}}|0>&=&\frac{1}{2}[N_1^1-N_2^1-N_1^2+N_2^2]T^{{\nu}_1....{\nu}_{N_2}}_{{\mu}_1...{\mu}_{N_1}}|0>.\label{okubo1}
\end{eqnarray}
Furthermore all the states with a given $N_1^i$ and $N_2^i$ have the isospin
\begin{equation}
I=\frac{1}{2}(N_1-N_1^3+N_2-N_2^3).\label{okubo2}
\end{equation}
From these two last equations one can write explicitly all the states of the representation $(N_1,N_2)$ .(\ref{okubo1}) and (\ref{okubo2}) will be used extensively next .

\subsubsection{5.5.2.2~~~The $SU(3)$ Representation Content Of The Hilbert Space $V(g)$ }

To see the $SU(3)$ representation content of $|\nu,g\rangle$, let us first focus on $|0;g\rangle$.
$|0;e\rangle\equiv|0\rangle$ is bilinear and antisymmetric in the ${\gamma}'s$ and has $I=0,Y=-2$. It is
antisymmetric in the ${\gamma}'s$ in the sense that $|0>$ can be written as
$|0>=\frac{1}{4}({\gamma}_{46}-i{\gamma}_{56}-i{\gamma}_{47}-{\gamma}_{57})$ where the ${\gamma}_{ij}$ are defined
by ${\gamma}_{ij}=\frac{1}{2}({\gamma}_{i}{\gamma}_{j}-{\gamma}_{j}{\gamma}_{i})$ .

The action $T(g)$ preserves the number of ${\gamma}'s$ . Thus its $SU(3)-$ orbit is contained in the vector space
spanned by the antisymmetric product of two ${\gamma}'s$ , that is , ${\gamma}^{ij}$ . We have already shown in
equation (\ref{decomposition2}) that this vector space transforms as $[\underline{8}+{\underline{10}}+\bar{{\underline
10}}]_{SU(3)}$ . Only the representation $[{\underline{10}}]_{SU(3)}$ contains an $I=0,Y=-2$ vector , namely
${\Omega}^{-}$, thus $|0;g\rangle{\in}\underline{10}=(N_1=3,N_2=0)$ . The proof for this last statement goes as
follows . From equations (\ref{okubo1}) and (\ref{okubo2}) we have $Y=\frac{1}{3}(N_1-N_2)-(N_1^3-N_2^3)=-2$ and
$I=\frac{1}{2}(N_1-N_1^3+N_2-N_2^3)=0$ which can be rewritten as $N_1^3=\frac{2}{3}N_1+\frac{1}{3}N_2+1$ and
$N_2^3=\frac{1}{3}N_1+\frac{2}{3}N_2-1$ . Now by using the facts that $N_1^3{\leq}N_1$ and $N_2^3{\leq}N_2$ which
are true by construction , one can deduce that $N_1-N_2{\leq}3$ and $N_1-N_2{\geq}3$ , i.e $N_1=N_2+3$ . In other
words only the representations $(N_1=N_2+3,N_2)$ do contain the vector $Y=-2$ , $I=0$ .

 A more explicit formula can be written. Let $|(3,0);(I,I_3,Y);e\rangle$ be the basis of vectors, which are
linear in ${\gamma}_{ij}$ and transforms as ${\underline{10}}$. We have:
$|(3,0);(0,0,-2);e\rangle\equiv|0;e\rangle$. Then
\begin{equation}
|0;g\rangle = Ad\,T(g)\,|0;e\rangle =\sum_{I,I_3,Y}|(3,0);(I,I_3,Y);e\rangle D^{(3,0)}_{(I,I_3,Y);(0,0,-2)}(g)
\label{s56}
\end{equation}
where $D^{(3,0)}:g{\rightarrow}D^{(3,0)}(g)$ is the IRR $\underline{10}$ of $SU(3)$ and the basis is labelled by
$(I,I_3,Y)$ .

We can analyse the $SU(3)$ content and write explicit formula for every $|\nu;g\rangle$ \cite{ref20}.

$|i;e\rangle(i=1,2)$ has ${\gamma}_{i}$'s and ${\gamma}_{ijk}$'s since one can write $|1>=a_2+a_1a_2a_1^{+}$ and
$|2>=-a_1+a_1a_2a_2^{+}$ . ${\gamma}_{i}$ transforms as an $[\underline{8}]_{SU(3)}$ while ${\gamma}_{ijk}$
transforms as $[\underline{27}+\underline{10}+\bar{\underline{10}}+ \underline{8}+\underline{1}]_{SU(3)}$ . We can
take linear combination of ${\gamma}_{i}$ and ${\gamma}_{ijk}$ to form two new $[\underline{8}]_{SU(3)}$'s such
that the $[\underline{8}]_{SU(3)}$ part of $|i;e\rangle$ is in a single $[\underline{8}]_{SU(3)}$. Also
$|i;e\rangle$ has $I=\frac{1}{2}$,$Y=-1$ and such a vector occurs only in $[\underline{8}]_{SU(3)}$,
$[\underline{10}]_{SU(3)}$ and $[\underline{27}]_{SU(3)}$. Thus $|i;g\rangle\,\,(i=1,2)$ transforms as the direct
sum $[\underline{8}+\underline{10}+\underline{27}]_{SU(3)}$. The proof proceeds as above , first we rewrite the
equations $Y=\frac{1}{3}(N_1-N_2)-(N_1^3-N_2^3)=-1$ and $I=\frac{1}{2}(N_1-N_1^3+N_2-N_2^3)=\frac{1}{2}$ in the
form $N_1-N_2{\geq}0$ and $N_1-N_2{\leq}3$ , where we have again used $N_1^3{\leq}N_1$ and $N_2^3{\leq}N_2$ .
Therefore the only representations $(N_1,N_2)$ which do contain the vector $I=\frac{1}{2}$ , $Y=-1$ are such that
$N_1-N_2=0,1,2,3$. But $N_1-N_2=1,2$ can also be shown to be not relevant because they lead to non-integer quantum
numbers $N_1^3$ and $N_2^3$ , so we are only left with $N_1-N_2=0,3$. $[\underline{8}]_{SU(3)}=(N_1=1,N_2=1)$,
$[\underline{10}]_{SU(3)}=(N_1=3,N_2=0)$ and $[\underline{27}]_{SU(3)}=(N_1=2,N_2=2)$ do clearly satisfy this
requirement .

There remains $|3;e\rangle$ with $I=Y=0$ . It is a linear combination of a constant, ${\gamma}_{ij}$ , and
${\gamma}_{ijkl}$ . This is obvious from the fact that one can write
$|3>=-1+a_2a_2^++a_1a_1^++a_1a_2a_1^{+}a_2^{+}$ . The constant part transforms as an $SU(3)-$singlet .
${\gamma}_{ij}$ was treated above , while ${\gamma}_{ijkl}$ was shown to transform as
$[\underline{27}+\underline{8}]_{SU(3)}$ . $U(2)$ singlets with $I=Y=0$ are contained only in $SU(3)$ singlet ,
$[\underline{8}]_{SU(3)}$ and $[\underline{27}]_{SU(3)}$ . One can actually show that all $SU(3)$ representations
$(N_1,N_2=N_1)$ do contain the vector $Y=0$ , $I=0$ . $|3;g\rangle$ transforms therefore as
$[\underline{1}+\underline{8}+\underline{27}]_{SU(3)}$ , the $[\underline{8}]_{SU(3)}$ being a mixture of the two
$[\underline{8}]_{SU(3)}'s$ from ${\gamma}_{ij}$, ${\gamma}_{ijk}$ .

For what follows, we also need formulas like (\ref{s56}) for $|i;g\rangle$ and $|3;g\rangle$. For $|i;g\rangle$,
the formula is
\begin{eqnarray}
|i;g\rangle&=&\sum_{I,I_3,Y}\Big[{\rm cos}\,{\theta}\,|(1,1);(I,I_3,Y);e\rangle
D^{(1,1)}_{(I,I_3,Y),(\frac{1}{2},\frac{(3-2i)}{2},-1)}(g)\nonumber\\
&+&{\rm sin}\,{\theta}\,{\rm cos}\,{\phi}|(3,0);(I,I_3,Y);e\rangle
D^{(3,0)}_{(I,I_3,Y),(\frac{1}{2},\frac{(3-2i)}{2},-1)}(g)\nonumber\\
&+&{\rm sin}\,{\theta}\,{\rm sin}\,{\phi}|(2,2);(I,I_3,Y);e\rangle
D^{(2,2)}_{(I,I_3,Y),(\frac{1}{2},\frac{(3-2i)}{2},-1)}(g)\Big],\nonumber\\
&&
\end{eqnarray}
where the angles ${\theta}$ and $\phi$ reflect the mixture of $[\underline{8}]_{SU(3)}$, $[\underline{10}]_{SU(3)}$
and $[\underline{27}]_{SU(3)}$ in $|i;g\rangle$.
\begin{eqnarray}
|3;g\rangle&=&\sum_{I,I_3,Y}\Big[{\rm cos}\,{\theta}^{'}\,|(0,0);(0,0,0);e\rangle\nonumber\\
&+&{\rm sin}\,{\theta}^{'}\,{\rm cos}\,{\phi}^{'}\,|(1,1);(I,I_3,Y);e\rangle
D^{(1,1)}_{(I,I_3,Y),(0,0,0)}(g)\nonumber\\
&+&{\rm sin}\,{\theta}^{'}\,{\rm sin}\,{\phi}^{'}\,|(2,2);(I,I_3,Y);e\rangle D^{(2,2)}_{(I,I_3,Y),(0,0,0)}(g)\Big].
\end{eqnarray}
For the precise values of the angles ${\theta}$ , ${\phi}$ , ${\theta}^{'}$ and ${\phi}^{'}$ see \cite{badis}.

\subsection{The Dirac Operator}

We require of the ${\bf C}{\bf P}^2$ Dirac operator ${\cal D}_{{\bf C}{\bf P}^2}$ that it is linear in derivatives
and anticommutes with the chirality operator ${\Gamma}_{{\bf C}{\bf P}^2}$ :
\begin{equation}
{\Gamma}_{{\bf C}{\bf P}^2}=-\frac{1}{4!}{\epsilon}_{ijkl}{\gamma}_{i}{\gamma}_{j}{\gamma}_{k}{\gamma}_{l}.\label{GAMMA}
\end{equation}
At ${\xi}={\xi}^0$, ${\Gamma}=-{\gamma}_4{\gamma}_5{\gamma}_6{\gamma}_7$ and is $+1$ on $|0;e\rangle$ and
$|3;e\rangle$, and $-1$ on $|i;e\rangle (i=1,2)$. Hence ${\Gamma}=+1$ on $|0;g\rangle,|3;g\rangle$ and $-1$ on
$|i;g\rangle$ for all $g$. The former have even chirality and the latter have odd chirality.

Let us prove all of this . First one rewrites equation (\ref{GAMMA}) in the form
${\epsilon}_{ijkl}{\gamma}_i{\gamma}_j{\gamma}_k{\gamma}_l=3({\cal J}_{ij}{\gamma}_i{\gamma}_j)^2-6{\cal J
}_{ij}{\cal J}_{jk}{\gamma}_i{\gamma}_k$ . Next from (\ref{complexstructurecp2}), we have ${\cal J}=i({\cal P}^{+}-{\cal
P}^{-})=i\hat{\xi}_iAd{\lambda}_i$, so that at the point ${\xi}^{0}=(0,...,0,\frac{1}{\sqrt{3}})$ , we get ${\cal
J}_{ij}=i(0,{\sigma}_2,{\sigma}_2)_{ij}$ where ${\sigma}_2$ is the usual second Pauli matrix\footnote{Again in the
notation $M_{ij}=(A,B,C)_{ij}$ the $i$ and $j$ indices take the values $1,2,3,8$ for the first entry , $4,5$ for
the second entry and $6,7$ for the third entry . In other words the matrix $M$ is block diagonal in the whole
space of the indices $i$ and $j$.} . Hence ${\cal J}_{ij}{\cal J}_{jk}|_{{\xi}^0}=-(0,1,1)_{ik}=-{\delta}_{ik}$ ,
$i$ , $k=4,5,6,7$ , and ${\cal J}_{ij}{\gamma}_i{\gamma}_j|_{{\xi}^0}
=2({\gamma}_4{\gamma}_5+{\gamma}_6{\gamma}_7)$ . In other words ${\Gamma}_{{\bf C}{\bf
P}^2}|_{{\xi}^0}=-{\gamma}_4{\gamma}_5{\gamma}_6{\gamma}_7$ . The computation for the eigenvectors is much easier
and goes as follows. ${\Gamma}_{{\bf C}{\bf
P}^2}|_{{\xi}^0}|0;e>=-{\gamma}_4{\gamma}_5{\gamma}_6{\gamma}_7a_1a_2=a_1a_2$, ${\Gamma}_{{\bf C}{\bf
P}^2}|_{{\xi}^0} |3;e>=-{\gamma}_4{\gamma}_5{\gamma}_6{\gamma}_7a_1^{+}a_2^{+}|0;e>=a_1^{+}a_2^{+}|0;e>$ and
${\Gamma}_{{\bf C}{\bf P}^2}|_{{\xi}^0}|i;e>=-a^{+}_i{\Gamma}_{{\bf C}{\bf P}^2}|_{{\xi}^0}|0;e>=-a^{+}_i|0;e>$ .

${\gamma}_{i}{\cal P}_{ij}$ anticommutes with ${\Gamma}|_{{\bf C}{\bf P}^2}$. This can be seen from equation
(\ref{cp2tangentbundle}), where we have the identity ${\cal P}=-{\cal J}^2$ and therefore at ${\xi}={\xi}^0$ we get ${\cal
P}_{ij}=(0,1,1)_{ij}={\delta}_{ij}$, $i$ , $j=4,5,6,7$ . In other words $\{{\gamma}_i{\cal
P}_{ij}|_{{\xi}^0},{\Gamma}_{{\bf C}{\bf P}^2}|_{{\xi}^0}\}=0$ . By rotational invariance this last equation
remains true at any other point on ${\bf C}{\bf P}^2$ .

The $SU(3)$ generators
\begin{eqnarray}
J_{i}&=&{\cal L}_{i} + Ad {t_{i}}\nonumber\\
&{\rm with}&\nonumber\\
 {\cal L}_{i}&=&-if_{ijk}{\hat\xi}_{j}\frac{\partial}{\partial {\hat\xi}_k}
\end{eqnarray}
commute with ${\Gamma}$ , and hence
\begin{equation}
{\cal D}_{{\bf C}{\bf P}^2}={\gamma}_{i}{\cal P}_{ij}J_{j} \label{s5extra1}
\end{equation}
anticommutes with ${\Gamma}_{{\bf C}{\bf P}^2}$,
\begin{equation}
\{{\Gamma}_{{\bf C}{\bf P}^2},{\cal D}_{{\bf C}{\bf P}^2}\}=0,\label{axiom}
\end{equation}
and is a good choice for the Dirac operator. The proof of (\ref{axiom}) starts by noticing the following identity
$[{\cal L}_m,{\cal J}_{jk}]=i{\cal J}_{mi}f_{ijk}$ which can be rewritten in the following form $[J_m,{\cal
J}_{jk}]=[{\cal L}_m,{\cal J}_{jk}]+[Adt_m,{\cal J}]_{jk}=0$ . By using this last equation , it is easy to show
that the chirality operator (\ref{GAMMA}) is rotationally invariant , i.e $[J_m,{\Gamma}_{{\bf C}{\bf
P}^2}]=0$\footnote{The tangent gammas , ${\gamma}_i{\cal P}_{ij}$ , are also rotationally invariant in this sense
, in other words we have $[J_l,{\gamma}_i{\cal P}_{ij}]=[{\cal L}_l,{\gamma}_i{\cal P}_{ij}]+{\gamma}_i[Adt_l,{\cal
P}]_{ij}=0$}. Therefore $\{{\Gamma}_{{\bf C}{\bf P}^2},{\cal D}_{{\bf C}{\bf P}^2}\}=\{{\Gamma}_{{\bf C}{\bf
P}^2},{\gamma}_i{\cal P}_{ij}J_j\}=\{{\Gamma}_{{\bf C}{\bf P}^2},{\gamma}_i{\cal P}_{ij}\}J_i$ . However it was
already shown that the tangent gammas , ${\gamma}_i{\cal P}_{ij}$ , do anticommute with ${\Gamma}_{{\bf C}{\bf
P}^2}$ , and hence $\{{\Gamma}_{{\bf C}{\bf P}^2},{\cal D}_{{\bf C}{\bf P}^2}\}=0$ .

${\cal D}_{{\bf C}{\bf P}^2}$ acts on ${\cal A}_{{\bf C}{\bf P}^2}{\otimes}Mat_{16}$. But there are only four
tangent gammas at each ${\xi}(g)$, so we have to reduce ${\cal A}_{{\bf C}{\bf P}^2}{\otimes}Mat_{16}$ to $V(g)$
(in an appropriate sense) at each ${\xi}(g)$. We can achieve this reduction as follows. The functions $\hat{\xi}$
are defined according to
\begin{eqnarray}
\hat{\xi}(g)&=&T(g)\,Y^c\,T^{-1}(g)\nonumber\\
&=&\sum_{i=1}^8{\lambda}^c_i\hat{\xi}_i(g), \label{s5extra}
\end{eqnarray}
where the notation means that $T$ and  $T^{-1}$ are to be evaluated at $g$ , and where ${\lambda}_i^c=2t_i^c$ .

We know that $U(2)$ is the stability group of $Y^c$ since $\hat{\xi}(gu)=\hat{\xi}(g)$ where $u{\in}U(2)$. Hence
${\cal A}_{{\bf C}{\bf P}^2} {\otimes}Mat_{16}$ consists of sections of the trivial $U(2)-$bundle over ${\bf
C}{\bf P}^ 2$. The same is the case for its left- and right- chiral projections
\begin{equation}
{\Psi}_{\pm}=\frac{1{\pm}{\Gamma}}{2}{\cal A}_{{\bf C}{\bf P}^2}{\otimes}Mat_{16}.
\end{equation}
But that is not the case for $|0;g\rangle$ and $|i;g\rangle$. Under $g{\rightarrow}g\,u$, $|0;g\rangle$ transforms
as an $SU(2)$ singlet with $Y=-2$ and $|i;g\rangle$ transforms as an $SU(2)$ doublet with hypercharge $Y=-1$.

Let $\hat g$ denote the matrix of functions on $SU(3)$ with ${\hat g}_{ij}(g)=g_{ij}, g\in SU(3)$. (${\hat g}$ is
just a simplified notation for $D^{(1,0)}$). We regard elements of ${\cal A}_{{\bf C}{\bf P}^2}{\otimes}Mat_{16}$
as functions of ${g}$, invariant under the substitution $g\rightarrow g\,u$. Accordingly, let us also introduce the
vectors $|a;\hat{g}\rangle;a=0,i,3$ which at $g$ are the vectors $|a;\hat{g}(g)\rangle=|a;g\rangle$. Note that on a
function $f$ on $SU(3)$, the left- and right- actions of $h{\in}SU(3)$ are $f{\rightarrow}h^{L,R}f$ where
$(h^{L}f)(g)=f(h^{-1}g)$ and $(h^Rf)(g)=f(gh)$.

Now consider, in the case of $|0,g\rangle$, the wave functions $D^{(N_1,N_2)}_{(II_3Y)(0,0,2)}$. They exist only
if $N_1=N_2-3$. The combination
\begin{equation}
\sum_{N,I,I_3,Y}D^{(N-3,N)}_{(I,I_3,Y)(0,0,2)}|0,\hat{g}\rangle \label{s57}
\end{equation}
is invariant under $g{\rightarrow}gu$ at each $g$ and can form constituents of a basis for the expansion of
functions in ${\cal A}_{{\bf C}{\bf P}^2}{\otimes}Mat_{16}$.

The remaining elements of a basis can be found in the same manner, being
\begin{eqnarray}
&&\sum_{n=0,3}\sum_{N,I,I_3,Y}\frac{1}{\sqrt{2}}\Big[D^{(N,N+n)}_{(I,I_3,Y)(\frac{1}{2},-\frac{1}{2},1)}
|1,\hat{g}\rangle +D^{(N,N+n)}_{(I,I_3,Y)(\frac{1}{2},+\frac{1}{2},1)}
|2,\hat{g}\rangle \Big]\nonumber\\
&&\sum_{N,I,I_3,Y}D^{(N,N)}_{(I,I_3,Y)(0,0,0)}|3,\hat{g}\rangle, \label{s58}
\end{eqnarray}
where $1\equiv(I=\frac{1}{2},I_3=+\frac{1}{2},Y=-1)$ and $2\equiv(I=\frac{1}{2},I_3=-\frac{1}{2},Y=-1)$. It is
almost obvious that (\ref{s57}) and (\ref{s58}) are invariant under the $U(2)$ action on the right of $g{\in}SU(3)$ ,
i.e $g{\longrightarrow}gu$. We skip here the explicit check.

 There is a subtlety we encounter at this point. [We also came
across it for $S^2$]. "Orbital" $SU(3)$ momentum ${\cal L}_{i}$ does not act on the individual factors in
(\ref{s57}) and (\ref{s58}), which are functions on $SU(3)$ and not just ${\bf C}{\bf P}^2$. It is thus necessary
to lift them to operators ${\cal K}_{i}^{L}$ which act on $\hat{g}$ in such a manner that when (\ref{s5extra}) is
used, $\hat{\xi}$ transform under $SU(3)$ in the way desired: ${\hat{\xi}}{\rightarrow}h^{L}{\hat{\xi}}$. Thus
${\cal K}_{i}^L$ are generators of $SU(3)_L$, the left-regular representation, and the Dirac equation is to be
reinterpreted as
\begin{eqnarray}
{\cal D}_{{\bf C}{\bf P}^2}&=&{\gamma}_{i}{\cal P}_{ij}J_{j}\nonumber\\
&where&\nonumber\\
J_{i}&=&{\cal K}_{i}^L+ Ad\,t_{i}. \label{s510}
\end{eqnarray}
Restricted to ${\cal A}_{{\bf C}{\bf P}^2}{\otimes}Mat_{16}$, (\ref{s510}) is the same as (\ref{s5extra1}).

$|a;g\rangle$ is $T(g)|a;e\rangle T^{-1}(g)$ so that $|a;\hat{g}\rangle =T|a;e\rangle T^{-1}$ . Now by using
$(h^{L}T)(g)=T\,(h^{-1}g)$ , we have $h^{L}[T\,|a;e\rangle
T^{-1}](g)=[T|a;e>T^{-1}](h^{-1}g)=T(h^{-1}g)|a;e>T^{-1}(h^{-1}g)=T(h^{-1})T(g) |a;e\rangle\, T^{-1}(g)T(h)$. Next
by using (\ref{adjointsu3}) , this last equation can be put in the form $h^{L}[T\,|a;e\rangle
T^{-1}](g)=AdT(h^{-1})[T(g)|a;e>T^{-1}(g)]$ . For infinitesimal transformations $h^{L}=1+i{\theta}_i{\cal K}_i^L$
and $AdT(h^{-1})=1-i{\theta}_iAdt_i$ we then obtain $J_{i}[T|a;e>T^{-1}](g)={\cal K}_{i}^L[T\,|a;e\rangle
T^{-1}](g)+Ad\,t_{i}\,[T(g) |a;e\rangle T^{-1}(g)]=0$. We conclude that
\begin{equation}
J_{i}|a,\hat{g}\rangle = 0.
\end{equation}
The expression for ${\cal P}$ is in (\ref{cp2tangentbundle}). Using the commutation relations (\ref{s55}) , we can write the
identity
\begin{equation}
[t_{i}^c{\hat{\xi}}_{i}, [t^c_{j}{\hat{\xi}}_{j},{\gamma}_{k}]]=\frac{1}{4}{\gamma}_i{\cal P}_{ik}.
\end{equation}
On the other hand by using equation (\ref{s5extra}) one finds
\begin{equation}
[t_{i}^c{\hat{\xi}}_{i},
[t^c_{j}{\hat{\xi}}_{j},{\gamma}_{k}]](g)=\frac{1}{4}T(g)[Y^c,[Y^c,T^{-1}(g){\gamma}_kT(g)]]T^{-1}(g).\label{5.130}
\end{equation}
Acting on an arbitrary matrix $M$ of $Mat_{16}$ , equation (\ref{5.130}) takes the form $[t_{i}^c{\hat{\xi}}_{i},
[t^c_{j}{\hat{\xi}}_{j},{\gamma}_{k}]](g)M=\frac{1}{4}AdT(g)[Y^c,[Y^c,T^{-1}(g){\gamma}_kT(g)]]AdT^{-1}(g)M$ ,
where we have used the definition $AdT(g)M=T(g)MT^{-1}(g)$ . Now by definition we have
$T(g){\gamma}_{k}T^{-1}(g)=(Adg)_{lk}{\gamma}_l$ so that $T{\gamma}_kT^{-1}=(Ad\hat{g})_{lk}{\gamma}_l$ .
$Ad\,\hat{g}(g)=Adg$ represents $g$ in the octet representation , it is real and orthogonal , i.e
$(Ad\hat{g}^{-1}(g))_{lk}=(Ad\hat{g}(g))_{kl}$ . Hence
\begin{eqnarray}
[t_{i}^c{\hat{\xi}}_{i},
[t^c_{j}{\hat{\xi}}_{j},{\gamma}_{k}]](g)&=&\frac{1}{4}AdT(g)[Y^c,[Y^c,{\gamma}_l]]AdT^{-1}(g)(Ad\hat{g}(g))_{kl}\nonumber\\
&{\Longrightarrow}&\nonumber\\
{\gamma}_i{\cal P}_{ik}&=&\{AdT[Y^c,[Y^c,{\gamma}_l]]AdT^{-1}\}(Ad\hat{g})_{kl}.
\end{eqnarray}
Since $|a,\hat{g}\rangle=Ad\,T|a;{e}\rangle$, $Ad\,T^{-1}|a;\hat{g}\rangle=|a;{e}\rangle$.
$[Y^c,[Y^c,{\gamma}_{l}]]$ consists only of tangent space ${\gamma}'s$ at $e$. The action of $\{.\}$ on
$|a;\hat{g}\rangle$ is thus
\begin{equation}
Ad\,T[Y^c,[Y^c,{\gamma}_{l}]]Ad\,T^{-1}|a;\hat{g}\rangle =Ad\,T\{[Y^c,[Y^c,{\gamma}_{l}]|a;{e}\rangle\}.
\end{equation}
The action of ${\cal D}_{{\bf C}{\bf P}^2}$ on typical basis vectors like (\ref{s57}) follows:
\begin{eqnarray}
{\cal D}_{{\bf C}{\bf P}^2}(\sum_{N,I,I_3,Y}D^{(N-3,N)}_{(I,I_3,Y)(0,0,2)}|0;\hat{g}\rangle )&=&\sum_{N,I,I_3,Y}
\{Ad{\hat{g}}_{kl}{\cal K}_{k}^LD^{(N-3,N)}_{(I,I_3,Y)(0,0,2)}\}Ad\,T[Y^c,
[Y^c,{\gamma}_{l}]]|0;{e}\rangle.\nonumber\\
&&
\end{eqnarray}
The term in braces also has a considerable simplification. Since $(h^{L}f)(g)=f(h^{-1}g)$
$=f(g(g^{-1}h^{-1}g))=[(g^{-1}h^{-1}g)^{R}f](g)$, $-Ad{\hat{g}}_{kl}{\cal K}_{k}^L$ are the generators ${\cal
K}_{l}^{R}$ for the $SU(3)$ acting on the right of $g$, they have the standard commutation relations $[{\cal
K}^{R}_{l},{\cal K}_{m}^{R}]=-if_{lmn}{\cal K}_{n}^{R}$. We thus find that
\begin{eqnarray}
{\cal D}_{{\bf C}{\bf P}^2}(\sum_{N,I,I_3,Y}D^{(N-3,N)}_{(I,I_3,Y)(0,0,2)}|0;\hat{g}\rangle)&=&
-\sum_{N,I,I_3,Y}\left\{{\cal K}_{l}^{R}D^{(N-3,N)}_{(I,I_3,Y)(0,0,2)}\right\}
\,Ad\,T\,[Y^c,[Y^c,{\gamma}_{l}]]|0;{e}\rangle.\nonumber\\
&& \label{s511}
\end{eqnarray}
The general wave function for even and odd chiralities can be written respectively as
\begin{eqnarray}
{\xi}^{(i)}_{j}D^{(i)}_{jj^{'}}|j^{"};\hat{g}>&;& ~j^{'}=(0,0,2),(0,0,0);~i=(N_1,N_2)\nonumber\\
 N_2-N_1&=&3 ~{\rm if} ~j^{'}=(0,0,2) ~{\rm and} ~ N_2=N_1 ~{\rm if} ~j^{'}=(0,0,0),\nonumber\\
 &&
\end{eqnarray}
and
\begin{equation}
{\eta}^{(i)}_{b}D^{(i)}_{bb^{'}}|b^{"};\hat{g}\rangle;\ \
b^{'}=(\frac{1}{2},-\frac{1}{2},1),(\frac{1}{2},\frac{1}{2},1);\ \ i=(N_1,N_2),\ \ N_2-N_1=0,3.
\end{equation}
Here $j^{"}$, $b^{"}$ are the state vectors for ${\gamma}$'s pairing with $j^{'}$, $b^{'}$ as in (\ref{s57}) and
(\ref{s58}). ${\xi}^{(i)}_{j}$, ${\eta}^{(i)}_{b}{\in}{\bf C}$ and repeated indices are summed.

Since ${\gamma}_{i}{\cal P}_{ij}$ anticommutes with ${\Gamma}_{{\bf C}{\bf P}^2}$, we can represent the effect of
${\cal D}_{{\bf C}{\bf P}^2}$ on wave functions in terms of the off-diagonal block
\begin{equation}
\left( \begin{array}{cc}
                                      0&d \\
                                     d^{+}&0
                                    \end{array}
                                     \right),
\label{offdiagform1}
\end{equation}
acting on
\begin{equation}
                                     \left( \begin{array}{c}
                                      {\xi}^{(i)}_{j}D^{(i)}_{jj^{'}}\\
                                       {\eta}^{(i)}_{b}D^{(i)}_{bb^{'}}
                                     \end{array}
                                     \right)
\label{offdiagform2}
\end{equation}
The result is the equation of \cite{14} for $m=0$. Ref \cite{14} has also found the spectrum of ${\cal D}_{{\bf
C}{\bf P}^2}$.

The zero modes of ${\cal D}_{{\bf C}{\bf P}^2}$ can be easily worked out from (\ref{s511}). When $j^{'}=(0,0,0)$,
$i$ can be $(0,0)$ [but not otherwise], and in that case, $D^{(0,0)}$ is a constant and is annihilated by ${\cal
K}^{R}_{l}$. Hence the index of ${\cal D}_{{\bf C}{\bf P}^2}$ is $1$ and the zero mode has even chirality .

\subsection{Wave Functions as Projective Modules}

${\cal D}_{{\bf C}{\bf P}^2}$ acts on a subspace of ${\cal A}_{{\bf C}{\bf P}^2}{\otimes}Mat_{16}$ . Here we find
the projector for the appropriate subspaces of either chirality . In Section $5.6$ , after summarizing the coherent
state formalism, we study the discrete analogues of such subspaces.

For ${\bf C}{\bf P}^2$, let $V(\underline{10},0)$ denotes the vector space carrying the representation
$\underline{10}=(N_1=3,N_2=0)$ of $SU(3)$ which is appropriate to the vector $|0;g\rangle$ . ${J}_{i}={\cal
K}_{i}^L+Ad\,t_{i}$ act on ${\cal A}_{{\bf C}{\bf P}^2}{\otimes}V(\underline{10},0)$ . At
${\xi}={\xi}^{0}=(0,...,0,\frac{1}{\sqrt{3}})$ , we want the subspace with
$\frac{1}{\sqrt{3}}{J}_{8}|_{{\xi}={\xi}^{0}}=\frac{1}{\sqrt{3}}Ad\,t_8=\frac{1}{2}AdY=-1$ . [As the stability
group $U(2)$ of ${\xi}^0$ acts trivially on $f({\xi}^0)$, $f{\in}{\cal A}_{{\bf C}{\bf P}^2}$, and the vector
$|0;g>$ has $Y=-2$ and $I=I_3=0$ in $\underline{10}$]. At a generic ${\xi}$, we want the subspace with
${\xi}^{i}{J}_{i}={\xi}^{i}Ad\,t_{i}=-1$ .

By using equations (\ref{okubo1}) and (\ref{okubo2}) one can compute the hyperchrage and the isospin quantum numbers of the
different states of the representation $\underline{10}=(3,0)$. They are found to be given by $Y=1-N_1^3+N_2^3$ and
$I=\frac{3}{2}-\frac{1}{2}N_1^3-\frac{1}{2}N_2^3$. $N_1^3$ and $N_2^3$ are the number of lower and upper indices
respectively of the tensor $T$, defined in equation (\ref{okubo1}) , which are having the value $3$. It is not
difficult to check that in $\underline{10}=(3,0)$ we have $N_2^3=N_2^2=N_2^1=0$ and $N_1^3=3,2,1$ or $0$ which
correspond to the states $(Y=-2,I=0)$ , $(Y=-1,I=\frac{1}{2})$ , $(Y=0,I=1)$ and $(Y=1,I=\frac{3}{2})$ respectively
. We have then the following subspaces ${\xi}^iJ_i=-1,-\frac{1}{2},0,+\frac{1}{2}$ . In particular , the subspace
with ${\xi}^iJ_i=-1$ is given by
\begin{equation}
{\cal P}^{(Y=-2,I=0)}{\cal A}_{{\bf C}{\bf P}^2}{\otimes}V(\underline{10};0)
\end{equation}
where ${\cal P}^{(-2,0)}$ is the appropriate projector:
\begin{eqnarray}
{\cal
P}^{(-2,0)}&=&\frac{({\xi}^i Adt_i+\frac{1}{2})}{(-1+\frac{1}{2})}\frac{({{\xi}}^iAdt_i-0)}{(-1-0)}\frac{({{\xi}}^iAdt_i-\frac{1}{2})}{(-1-\frac{1}{2})}\nonumber\\
&=&-\frac{1}{3}(2{\xi}^iAdt_i+1)({\xi}^iAdt_i)(2{\xi}^iAdt_i-1).
\end{eqnarray}
We can proceed in this manner to find the projectors for the remaining even chirality subspaces. Calling
$V(\underline{1}+\underline{8}+\underline{27};3)$ the vector space for $|3;g\rangle$, we have to find the
projector for the $U(2)$ singlet state at each ${\xi}$. We have seen that ${\cal K}_{i}^L$ does not contribute in
the projectors, so let $Ad\,t_{i}$ now act on $\underline{1}+\underline{8}+\underline{27}$. Define
$Adt_{i}t_{j}=[t_{i},t_{j}]$, ${\bf I}t_{i}=t_{i}$. At ${\xi}^{0}$, $(AdY)^2t_i=\frac{4}{3}(Adt_8)^2t_{i}$ give the
tangent space generators as $(AdY)^2t_i={\cal P}_{ik}t_k$ , more explicitly one can compute
$(AdY)^2t_i={\delta}_{ik}t_k$,$k=4,5,6,7$ . So $[1-(AdY)^2]t_{i}$ are the $U(2)$ generators . One checks that
$\{[1-(AdY)^2]t_{i}\}^2$ is the $SU(2)$ Casimir plus $(t_8)^2$. We want the null space of the operator
$\{[1-(AdY)^2]\,t_{i}\}^2$ at ${\xi}^{0}$. At ${\xi}$, we want the null space of
$\{[1-4({\xi}^{j}Ad\,t_{j})^2]\,t_{i}\}^2$ , this is because $4({\xi}^jAdt_j)^2|_{{\xi}={\xi}^0}=(AdY)^2$. This
last operator is therefore given by the formula
\begin{eqnarray}
B&=&\{[1-4({\xi}^{j}Ad\,t_{j})^2]\,t_{i}\}^2\nonumber\\
&=&\vec{t}^2+t_8^2\nonumber\\
&=&I(I+1)+\frac{3}{4}Y^2.
\end{eqnarray}
The spectrum of this operator $B$ can be computed using the same method which was described above for $|0;g>$ .
The representation $\underline{8}=(1,1)$ contains the following states
$(Y,I)=(-N_1^3+N_2^3,1-\frac{1}{2}N_1^3-\frac{1}{2}N_2^3)$ . But now $N_1^3=0,1$ and $N_2^3=0,1$ and hence we have
the following states
 $(Y,I)=(0,1),(1,\frac{1}{2}),(-1,\frac{1}{2}),(0,0)$ . Therefore the spectrum of $B$ on $\underline{8}$ is given by
\begin{eqnarray}
Spec
B_{8}&=&\{1(1+1),\frac{1}{2}(\frac{1}{2}+1)+\frac{3}{4},0\}.\nonumber\\
\end{eqnarray}
In the same way the representation $\underline{27}=(2,2)$ can be found to contain the states
$(Y,I)=(-N_1^3+N_2^3,2-\frac{1}{2}N_1^3-\frac{1}{2}N_2^3)$ . More explicitely we have
$(Y,I)=(0,2),(1,\frac{3}{2}),(2,1),(-1,\frac{3}{2}),(0,1),(1,\frac{1}{2}),(-2,1),(-1,\frac{1}{2})$ and $(0,0)$ since now
$N_1^3=0,1,2$ and $N_2^3=0,1,2$ . Hence the spectrum of $B$ is
\begin{eqnarray}
spec
B_{27}&=&\{2(2+1),\frac{3}{2}(\frac{3}{2}+1)+\frac{3}{4},1(1+1)+3,1(1+1),\frac{1}{2}(\frac{1}{2}+1)+\frac{3}{4},0\}.\nonumber\\
&&
\end{eqnarray}
On the representation $\underline{1}=(0,0)$ , the operator $B$ is identically $0$ .

Hence the remaining ${\Gamma}=1$ subspace is
\begin{eqnarray}
{\cal P}^{(I=0,Y=0)}{\cal
A}_{{\bf C}{\bf P}^2}{\otimes}V(\underline{1}+\underline{8}+\underline{27};3),\nonumber\\
{\cal P}^{(0,0)}={\cal P}^{(0,0)}_{\underline 8}+ {\cal P}^{(0,0)}_{\underline{27}},
\end{eqnarray}
${\cal P}^{(0,0)}_{8,27}$ being the projectors acting on the $\underline{8}$ and $\underline{27}$ parts:
\begin{eqnarray}
{\cal P}^{(0,0)}_{\underline 8}&=&\frac{[B-1(1+1)]}{[0-1(1+1)]}\frac{[B-\frac{1}{2}(\frac{1}{2}+1)-\frac{3}{4}]}{[0-\frac{1}{2}(\frac{1}{2}+1)-\frac{3}{4}]}\nonumber\\
&=&\frac{1}{3}(B-2)(B-\frac{3}{2})\nonumber\\
&and&\nonumber\\
{\cal
P}_{\underline{27}}^{(0,0)}&=&\frac{[B-2(2+1)]}{[0-2(2+1)]}\frac{[B-\frac{3}{2}(\frac{3}{2}+1)-\frac{3}{4}]}{[0-\frac{3}{2}(\frac{3}{2}+1)-\frac{3}{4}]}
\frac{[B-1(1+1)-3]}{[0-1(1+1)-3]}{\times}\nonumber\\
&&\frac{[B-1(1+1)]}{[0-1(1+1)]}\frac{[B-\frac{1}{2}(\frac{1}{2}+1)-\frac{3}{4}]}{[0-\frac{1}{2}(\frac{1}{2}+1)-\frac{3}{4}]}\nonumber\\
&=&-\frac{1}{405}(B-6)(B-\frac{9}{2})(B-5)(B-2)(B-\frac{3}{2}).
\end{eqnarray}

In a similar manner, we can find the projector for ${\Gamma}=-1$ too. It projects the $I=1/2$, $Y=-1$ states at
each ${\xi}$ from ${\cal A}_{{\bf C}{\bf P}^2}{\otimes}V(\underline{8}+\underline{10}+\underline{27})$, the $V$
factor denoting the space appropriate for $|i,g\rangle(i=1,2)$. Its explicit expression is
\begin{equation}
{\cal P}^{(-1,\frac{1}{2})}={\cal P}^{(-1,\frac{1}{2})}_{\underline 8}+{\cal
P}^{(-1,\frac{1}{2})}_{\underline{10}}+{\cal P}^{(-1,\frac{1}{2})}_{\underline{27}}.
\end{equation}
The explicit expressions of the projectors ${\cal P}^{(-1,\frac{1}{2})}_{\underline 8}$ , ${\cal
P}^{(-1,\frac{1}{2})}_{\underline{10}}$ and ${\cal P}^{(-1,\frac{1}{2})}_{\underline{27}}$ can be easily found
following the above method. We skip details here.

\section{ Fuzzification }

${\cal D}_{{\bf C}{\bf P}^2}$ acts on a subspace of ${\cal A}_{{\bf C}{\bf P}^2}{\otimes}Mat_{16}$. We can thus
conceive of a fuzzy Dirac operator ${\bf D}_{{\bf C}{\bf P}^2}$ which acts on a subspace of ${\bf A}_{{\bf C}{\bf
P}^2}{\otimes}Mat_{16}$, ${\bf A}_{{\bf C}{\bf P}^2}$ being obtained from ${\cal A}_{{\bf C}{\bf P}^2}$ by
restricting ``orbital'' $SU(3)$ IRR's to $(n,n)$, $n{\leq}N$. ${\bf D}_{{\bf C}{\bf P}^2}$ is then obtained from
${\cal D}_{{\bf C}{\bf P}^2}$ by projection to this subspace. ${\cal D}_{{\bf C}{\bf P}^2}$ commutes only with the
total $SU(3)$ Casimir $J_{i}^2$ and not with orbital $SU(3)$ Casimir ${\cal L}_{i}^2$. This causes edge effects
distorting the spectrum of ${\bf D}_{{\bf C}{\bf P}^2}$ for those states having $(n,n)$ near $(N,N)$ which ${\cal
D}_{{\bf C}{\bf P}^2}$ mixes with $(n^{'},n^{'})$, $n^{'}{\geq}N$. This particular edge phenomenon does not occur
for ${\bf S}^2={\bf C}{\bf P}^1$ where orbital angular momentum ${\cal L}^2_{i}$ commutes with the Dirac operator.
A way to eliminate such problems is suggested by the work of \cite{6,7,8,9}: We introduce the cut-off not on the
orbital Casimir, but on the {\it total} Casimir, retaining all states upto the cut-off. That seems the best
strategy as it will give a fuzzy Dirac operator ${\bf D}_{{\bf C}{\bf P}^2}$ with a spectrum exactly that of the
continuum operator ${\cal D}_{{\bf C}{\bf P}^2}$ upto the cut-off point, and which has chirality (chirality
${\Gamma}_{{\bf C}{\bf P}^2}$ of ${\cal D}_{{\bf C}{\bf P}^2}$ commutes with $J^2_{i}$) and no fermion doubling.
This approach is the same as the method adopted for ${\bf S}^2$ in \cite{6,7,8,9} . For ${\bf S}^2$, the edge
effect turned up as the absence of the ${-}E$ eigenvalue subspace for the maximum total angular momentum when the
cut-off is introduced in orbital angular momentum, and attendant problems with chirality.

${\bf D}_{{\bf C}{\bf P}^2}$ being just a restriction of ${\cal D}_{{\bf C}{\bf P}^2}$, we can continue to use
(\ref{s5extra1}) in calculation, just remembering the truncation of the spectrum. That means that the analysis in
Section $5.5$ can be used intact. In the final expressions like (\ref{offdiagform1}) and (\ref{offdiagform2}), $i$
labels the IRR and the Dirac operator acts in subspace with fixed $i$. So the cut-off can be introduced on
$i=(N_1,N_2)$.

\subsection{Coherent States and Star Products : The Case of ${\bf S}^2\simeq{\bf C}{\bf P}^1$}

These have been treated in \cite{ref21,9,15}. Here we summarize the main points so that we can outline the relation
of wave functions like (\ref{s57}) and those based on matrices for fuzzy physics.

Let us first consider ${\bf S}^2={\bf C}{\bf P}^1$ and its fuzzy versions. The algebra ${\bf A}$ is $Mat_{2l+1}$.
$SU(2)$ acts on ${\bf A}$ on left and right with generators $L_i^{L}$ and $-L_i^{R}$, and orbital angular momentum
is ${\cal L}_{i}=L_i^L-L_i^R$. The spectrum of ${\cal L}^2$ is $K(K+1)$, $K=0,1,..,2l$. We can find a basis of
matrices $T^{K}_{M}$ diagonal in ${\cal L}^2$ and ${\cal L}_3$(with eigenvalue $M$) and standard matrix elements
for ${\cal L}_i$. ${\bf A}$ acts on a $(2l+1)-$dimensional vector space with the familiar basis $|l,m\rangle$.
$T^{K}_{M}$ are orthogonal, $K(K+1)$ and $M$ being eigenvalues of ${\cal L}^2$ and ${\cal L}_3$:
\begin{equation}
(T^{K}_{M},T^{K^{'}}_{M^{'}}):=\,Tr\,T^{K\dagger}_{M}T^{K^{'}}_{M^{'}}= ~{\rm
constant}\times{\delta}_{KK^{'}}{\delta}_{MM^{'}}.
\end{equation}
The above suggests that there is a way to regard ${\bf A}$ as ``functions'' on ${\bf S}^2$ with angular momenta
cut-off at $2l$. Such functions are also represented by the linear span of spherical harmonics $Y_{KM}$,
$K{\leq}2l$. We want to clarify the relation of $Y_{KM}$'s to the matrices $T^{K}_{M}$ in ${\bf A}$.

Towards this end, let us introduce the coherent states
\begin{equation}
|g\rangle=U^{(l)}(g)|l,l\rangle
\end{equation}
induced from the highest weight vector $|l;l\rangle$. $g{\rightarrow}U^{(K)}(g)$ is the angular momentum $K$ IRR of
$SU(2)$. Note the identity
\begin{equation}
|ge^{i{{\sigma}_3\over 2}{\theta}}\rangle=e^{il{\theta}}|g\rangle.
\end{equation}
It is a theorem \cite{ref21} that the diagonal matrix elements $\langle g|a|g\rangle$ completely determine the
operator $a$. Further $\langle  ge^{i{{\sigma}_3\over 2}{\theta}}|a|ge^{i{{\sigma}_3\over 2}{\theta}}\rangle
=\langle g|a|g\rangle$ so that $\langle g|a|g\rangle$ depends only on
\begin{equation}
g{\sigma}_3g^{-1}={\sigma}\cdot x,\ \ \sum_{i=1}^3 x^2_{i}=1 ; x{\in}{\bf S}^2.
\end{equation}
In this way, we have the map
\begin{eqnarray}
&&{\bf A}{\rightarrow}{\bf C}^{\infty}({\bf S}^2),\nonumber\\
&&a{\rightarrow}{\tilde a};\nonumber\\
&where&\nonumber\\
&&{\tilde a}(x)=\langle g|a|g\rangle.
\end{eqnarray}
In this map, the image of $T^{K}_{M}$ is $Y_{KM}$ after a phase choice:
\begin{equation}
Y_{KM}(x)=\langle g|T^{K}_{M}|g\rangle . \label{phasechoice}
\end{equation}
For, under $g{\rightarrow}hg$ , $x{\rightarrow}R(h)x$ where $h{\rightarrow}R(h)$ is the $SU(2)$ vector
representation. Under this transformation, since
\begin{equation}
Y_{KM}(R(h)x)=\sum_{M^{'}=-K}^K D^{(K)}(h)_{MM^{'}}Y_{KM^{'}}(x)
\end{equation}
and
\begin{equation}
T^{K}_{M}{\longrightarrow}U^{(K)}(h)^{-1}T^{K}_{M}U^{(K)}(h)=\sum_{M^{'}=-K}^{K}D^{(K)}(h)_{MM^{'}}T^{K}_{M^{'}},
\end{equation}
where $h{\longrightarrow}D^{(K)}(h)$ is the angular momentum $K$ IRR of $SU(2)$  in a matrix representation, we
have the proportionality of the two sides. (\ref{phasechoice}) and phase conventions fix the constant of
proportionality.

The map $T^{K}_{M}{\rightarrow}Y_{KM}$ is an isomorphism at the level of vector spaces. It can be extended to the
noncommutative algebra ${\bf A}$ by defining a new product on $Y_{KM}$'s, the star product. Thus consider $\langle
g|T^{K}_{M}T^{L}_{N}|g\rangle$. The functions $Y_{KM}$ and $Y_{LN}$ completely determine $T^{K}_{M}$ and
$T^{L}_{N}$, and for that reason also this matrix element. Hence it is the value of a function $Y_{KM}*Y_{LN}$,
linear in each factor, at $x$:
\begin{equation}
\langle g|T^{K}_{M}T^{L}_{N}|g\rangle = [Y_{KM}*Y_{LN}](x).
\end{equation}
The product $*$ here , the star product , extends by linearity to all functions with angular momenta ${\leq}2l$.
The resultant algebra is isomorphic to the algebra ${\bf A}$.

The explicit formula for $*$ has been found by Pre\v{s}najder \cite{9} (see also \cite{15}). The image of ${\cal
L}_{i}a$ is just $-i(\vec{x}{\wedge}{\vec{\nabla}})_{i}{\tilde a}$. We will use the same symbol ${\cal L}_i$ to
denote $-i(\vec{x}{\wedge}{\vec{\nabla}})_i$ derivation. The $*$ product is covariant under the $SU(2)$ action in
the sense that
\begin{equation}
{\cal L}_i({\tilde a}*{\tilde b})=({\cal L}_i{\tilde a})*{\tilde b}+{\tilde a}*({\cal L}_{i}{\tilde b}).
\end{equation}
It depends on $l$ and approaches the commutative product of ${\bf C}^{\infty}({\bf S}^2)$ as
$l{\longrightarrow}{\infty}$. Coherent states thus give an intuitive handle on the matrix representation of
functions.

But on ${\bf S}^2$, we also have monopole bundles. Sections of these bundles for Chern  class $n$ are spanned by
the rotation matrices $D^{(j)}_{mn}$, $j{\geq}|n|$. They have the equivariance property
\begin{equation}
D^{(j)}_{mn}(ge^{i{{\sigma}_3\over 2}{\theta}})=D^{(j)}_{mn}(g)e^{in{\theta}}.\label{5157}
\end{equation}
This last equation is essentially a generalization of equation (\ref{equivariancespinorextra}) , in other words one can identify
$D^{(j)}_{mn}$ with $<j,m|{\psi}^{(j)}_n>$ where $|{\psi}^{(j)}_n>=D^{j}|j,n>$ [see equation (\ref{570})].

How do we represent them by matrices?

In the first instance, let $n{\geq}0$ and consider the coherent states (now with an additional label)
\begin{eqnarray}
|g;l+n\rangle &=&U^{(l+n)}(g)|l+n,l+n\rangle\nonumber\\
|g;l\rangle&=&U^{(l)}(g)|l,l\rangle.\label{coherentstatefinal}
\end{eqnarray}
They span vector spaces $V_{l+n}$ and $V_{l}$. We can consider the linear operators $Hom(V_{l+n},V_l)$ from
$V_{l+n}$ to $V_{l}$. They are $[2l+1]{\times}[2(l+n)+1]$ matrices in a basis of $V_{l+n}$ and $V_l$, and have
$U^{(l)}(g)$ acting on their left(with generators $L^{L}_i$) and $U^{(l+n)}(g)$ acting on their right (with
generators $-L^{R}_i$). We can decompose $Hom(V_{l+n},V_l)$ under the ``orbital'' angular momentum group
$U^{(l)}{\otimes}U^{(l+n)}$ (with generators ${\cal L}_i=L^{L}_{i}-L^{R}_i$) into the direct sum
${\oplus}_{K=n}^{2l+n}(K)$ with the IRR $K$ having the basis $T^{K}_{M}$, with ${\cal L}_{3}T^{K}_{M}=MT^{K}_{M}$.
As before, we choose $T^{K}_{M}$ so that ${\cal L}_i$ follow standard phase conventions. $T^{K}_{M}$ are orthogonal
\begin{equation}
Tr(T^{K^{'}}_{M^{'}})^{\dagger}T^{K}_{M}= ~{\rm constant}\times{\delta}_{K^{'}K}{\delta}_{M^{'}M}.
\end{equation}
Now consider

\begin{equation}
\langle g;l|T^{K}_{M}|g;l+n\rangle.
\end{equation}

It transforms in precisely the same manner as $D^{(K)}_{Mn}(g)$ under $g{\rightarrow}hg$ and
$g{\rightarrow}ge^{i{{\sigma}_3\over 2}{\theta}/2}$ and hence after an overall normalisation,
\begin{equation}
\langle g;l|T^{K}_{M}|g;l+n\rangle=D^{(K)}_{Mn}(g).
\end{equation}
Thus $Hom(V_{l+n},V_l)$ are fuzzy versions of sections of vector bundles for Chern class $n{\geq}0$. For $n<0$,
they are similarly $Hom(V_l,V_{l+|n|})$. This result is due to \cite{ref22} (see also \cite{6,7,8,9,14}). An
explicit formulae for the fuzzy version of rotation matrices can be found in \cite{9}.

It is interesting that Chern class has a clear meaning even in this matrix model: It is $|V|-|W|$ for $Hom(V,W)$,
where $|V|$ and $|W|$ are dimensions of $V$ and $W$.

There are two (inequivalent) fuzzy algebras acting on $Hom(V,W)$. $Mat_{|V|}={\bf A}_{|V|}$ acts on the right and
$Mat_{|W|}={\bf A}_{|W|}$ acts on the left, where now a subscript has been introduced on ${\bf A}$. These left and
right actions have their own $*$'s, call them $*_{|V|}$ and $*_{|W|}$: if $a{\in}{\bf A}_{V}$, $b{\in}{\bf A}_{W}$
and ${\tilde a}$ and ${\tilde b}$ are the corresponding functions, then
\begin{equation}
bT^{K}_{M}a{\longrightarrow}{\tilde b}*_{|W|}Y_{KM}*_{|V|}{\tilde a}
\end{equation}
under the map of $Hom(V,W)$ to sections of bundles. There is also a fuzzy analogue for tensor products of bundles.
Thus we can compose elements of $Hom(V,W)$ and $Hom(W,X)$ to get $Hom(V,X)$
\begin{equation}
Hom(V,X)=Hom(V,W){\otimes}_{{\bf A}_{|W|}}Hom(W,X).
\end{equation}
Its elements are $ST$, $S{\in}Hom(V,W)$, $T{\in}Hom(W,X)$. Its Chern class is $|V|-|X|$. If ${\tilde S}$ and
${\tilde T}$ are the representatives of $S$ and $T$ in terms of sections of bundles, then
$ST{\longrightarrow}{\tilde S}*{\tilde T}$.

Tensor products ${\Gamma}_{1}{\otimes}{\Gamma}_2$ of two vector spaces ${\Gamma}_1$ and ${\Gamma}_2$ over an
algebra $B$ are defined only if ${\Gamma}_1({\Gamma}_2)$ is a right-(left-) $B$-module \cite{ref23}. Hence
$Hom(V,W){\otimes}_{{\bf A}_{|W|}}Hom(W^{'},X)$ is defined only if $W=W^{'}$. So ${\tilde S}*{\tilde T}$ is rather
different in its properties from the usual tensor product of bundle sections, in particular ${\tilde T}*{\tilde
S}$ makes no sense if $V\ne X$.

\subsection{Fuzzy Dirac Spinors on ${\bf S}^2_F$}

We can now comment on the fuzzy form of (\ref{kahlerfinal}) . Elsewhere the Watamuras \cite{10,11} and following them, us
\cite{12,13}, investigated the Dirac operator as acting on ${\bf A}{\otimes}C^2={\bf A}^2$, ${\bf A}=Mat_{2l+1}$.
That led to rather an elaborate formalism because of the cut-off in orbital angular momentum. So as indicated
earlier, it seems more elegant to cut-off total angular momentum at some value $j_0$.

We can now argue such a cut-off leads to the formalism of \cite{5,6,7,8,9} and to supersymmetry. Thus let
$T^{j}_{m+}{\in}Hom(V_{l+1/2},V_l)$ with the transformation property
$<g;l|T^{j}_{m+}|g;l+\frac{1}{2}>{\longrightarrow}e^{i\frac{\theta}{2}}<g;l|T^{j}_{m+}|g;l+\frac{1}{2}>$ under
$g{\longrightarrow}ge^{i\frac{\theta}{2}{\sigma}_3}$ , where we have used the definitions (\ref{coherentstatefinal}) . One also has
the transformation property

\begin{equation}
U^{(l)}(g)^{\dagger}T^{j}_{m+}U^{(l+\frac{1}{2})}(g)=\sum_{m^{'}}D^{(j)}_{mm^{'}}(g) T^{j}_{m^{'}+}
\end{equation}
[So $j{\leq}2l+1/2$ and $j_0=2l+1/2$]. Then one can make the identification
\begin{equation}
D^{j}_{m+}(g)=\langle g;l|T^{j}_{m+}|g;l+\frac{1}{2}\rangle,
\end{equation}
since from equation (\ref{5157}) it is easy to see that
$D^{j}_{m+}(g){\longrightarrow}e^{i\frac{\theta}{2}}D^{j}_{m+}(g)$ under
$g{\longrightarrow}ge^{i\frac{\theta}{2}{\sigma}_3}$ .

{\it{The subscript $+$ in $T^{j}_{m+}$ indicates helicity $-$}} , i.e $T^{j}_{m+}$ is the fuzzy version of
$<j,m|{\psi}^{(j)}_{+}>$ of (\ref{570}) so that it will be associated with the negative helicity part of the wave
function [see equation (\ref{spinorexpansionfinal})].

For helicity $+$, but for same $j_0$, we have to consider $T^{j}_{m-}{\in}Hom(V_l,V_{l+1/2})$, with
\begin{equation}
U^{(l+\frac{1}{2})}(g)^{\dagger}T^{j}_{m-}U^{(l)}(g)=\sum_{m^{'}}D^{(j)}_{mm^{'}}(g)T^{j}_{m^{'}-}.
\end{equation}
Of course now ,
\begin{equation}
D^{j}_{m-}(g)=<g;l+\frac{1}{2}|T^{j}_{m-}|g;l>,
\end{equation}
where both sides will acquire now a phase $exp(-i\frac{\theta}{2})$ under the right $U(1)$ action, namely under
$g{\longrightarrow}gexp(i\frac{\theta}{2}{\sigma}_3)$ . $T^j_{m-}$ is then clearly the fuzzy version of
$<j,m|{\psi}^{j)}_{-}>$ .

This is the formalism of \cite{5,6,7,8,9} . As we have united $V^{(l)}$ and $V^{(l+1/2)}$, it is natural to
consider $OSp(2,1)$ or even $OSp(2,2)$ SUSY as discovered first by Grosse et al in the second paper of \cite{6}.

Because of the mixing of $l$ and $l+1/2$, we have to reconsider the action of the matrix algebra ${\bf A}$
approximating ${\cal A}={C}^{\infty}({\bf S}^2)$. $Mat_{2l+1}$ acts on $T^{j}_{m+}(T^{j}_{m-})$ on the left(right)
while $Mat_{2l+2}$ acts on $T^{j}_{m+}(T^{j}_{m{-}})$ on the right(left). So it is best to regard fuzzy functions
to act on left(say) of $T^j_{m+}$ and right of $T^j_{m-}$ as $Mat_{2l+1}$. This suggestion is slightly different
from that of \cite{5,6,7,8,9} where they regard the fuzzy algebra to be $Mat_{2l+1}$ on $T^{j}_{m+}$ and
$Mat_{2l+2}$ on $T^{j}_{m-}$, both acting on left. However, our proposal does not generalize to instanton
(monopole) sectors.

We can restore spin parts to fuzzy wave functions. The spin wave functions for helicity $\pm$ are
$T^{\frac{1}{2}}_{m_{s}{\pm}}$ , where $m_s$ denotes the two components of the spinor . The positive chirality
spinors are defined by
\begin{equation}
<g;l|T^{\frac{1}{2}}_{m_s+}|g;l+\frac{1}{2}>=D^{\frac{1}{2}}_{m_s+}=<\frac{1}{2},m_s|{\psi}_{+}^{(\frac{1}{2})}>,
\end{equation}
while the negative chirality spinors are defined by
\begin{equation}
<g;l+\frac{1}{2}|T^{\frac{1}{2}}_{m_s-}|g;l>=D^{\frac{1}{2}}_{m_s-}=<\frac{1}{2},m_s|{\psi}_{-}^{(\frac{1}{2})}>.
\end{equation}
So the two components of the total fuzzy wave functions for helicity $\pm$ are
\begin{equation}
<\frac{1}{2},m_s|{\psi}^{\pm}_F>=\Big[\sum_{j,m} {\xi}^{j\pm}_{m} T^{j}_{m{\mp}}\Big]T^{\frac{1}{2}}_{m_s{\pm}},\ \
\ {\xi}^{j\pm}_{m}{\in}{\bf C},m_s=+\frac{1}{2},-\frac{1}{2}.
\end{equation}
This is the fuzzy version of equation (\ref{spinorexpansionfinal}).

The Dirac operator ${\bf D}_{2g}$ is given by the truncated version of (\ref{589}) :
\begin{eqnarray}
&&{\rho}\sum_{m_s}({\bf D}_{2g})_{m_s^{'}m_s}\{\sum_{j,m}\xi^{j+}_{m} T^{j}_{m-}T^{1/2}_{m_s+}+\sum_{j,m}\xi^{j-}_{m}T^{j}_{m+}T^{1/2}_{m_s-}\} =\nonumber\\
&&-\{\sum_{j,m}\xi^{j+}_{m}T^{j}_{m+}(J^{(j)}_{+})_{+1/2,-1/2}\} \{T^{1/2}_{m_s^{'}-}\}
-\{\sum_{j,m}\xi^{j-}_{m}T^{j}_{m-}(J^{(j)}_{-})_{-1/2,+1/2}\}
\{T^{1/2}_{m_s^{'}+}\},\nonumber\\
&&j{\leq}2l+1/2,
\end{eqnarray}
$J^{(j)}_{i}$ being the angular momentum $j$ images of $\frac{{\sigma}_{i}}{2}$.

\subsection{ The Case of ${\bf C}{\bf P}^2$}

Coherent states for ${\bf C}{\bf P}^2$ can be defined using highest weight states. For IRR $(3,0)$, we can pick
the highest weight state with $I=I_3=0,\ \ \ Y=-2/3$, namely the $c-$quark: $|0,0,-2/3\rangle$ ${\equiv}$
$|0,0,-2/3;(3,0)\rangle$. Then if $g{\longrightarrow}U^{(3,0)}(g)$ defines the IRR,
$|g;(3,0)\rangle=U^{(3,0)}(g)|0,0,-2/3;(3,0)\rangle$. For the IRR $(N,0)$, we can simply replace
$|0,0,-2/3;(3,0)\rangle$ by its $N-$fold tensor product
\begin{equation}
|0,0,-\frac{2}{3};(3,0)\rangle{\otimes}|0,0,-\frac{2}{3};(3,0)\rangle
{\otimes}...{\otimes}|0,0,-\frac{2}{3};(3,0)\rangle=|0,0,-\frac{2N}{3};(N,0)\rangle,
\end{equation}
and set
\begin{equation}
|g;(N,0)\rangle=U^{(N,0)}(g)|0,0,-\frac{2N}{3};(N,0)\rangle.
\end{equation}
For $(0,N)$, we can use the ${\bar{c}}-$quark state $|g;(0,3)\rangle$ $=U^{(0,3)}(g)|0,0,+2/3;(0,3)\rangle$ and its
tensor product states.

The development of ideas now keep following ${\bf S}^2={\bf C}{\bf P}^1$. Full details can be found in \cite{15}.

General theory confirms that the maps $a{\longrightarrow}{\tilde a}$ from matrices in the $(N,0)$ or $(0,N)$ IRR to
functions on ${\bf C}{\bf P}^2$, defined by
\begin{eqnarray}
{\tilde a}({\xi})&=&\langle (N,0);g|a|g;(N,0)\rangle\nonumber\\
&or&\nonumber\\
{\tilde a}(\xi)&=&\langle(0,N);g|a|g;(0,N)\rangle.
\end{eqnarray}
are one-to-one so that a $*-$product on ${\tilde a}$'s exists. In this map, the $SU(3)$ generators ${\cal L}_{i}$
acting on ${\tilde a}$ become the corresponding ${\bf C}{\bf P}^2$  $SU(3)$ operators
$-if_{ijk}{\hat\xi}_{j}\frac{\partial} {{\partial}{\hat\xi}_{k}}$. We shall use the same symbol ${\cal L}_{i}$ for
these operators too. The orbital $SU(3)$ action is compatible with $*$ in the sense that ${\cal L}_{i}({\tilde
a}*{\tilde b})=({\cal L}_{i}{\tilde a})*{\tilde b} + {\tilde a}*({\cal L}_{i}{\tilde b})$. Irreducible tensor
operators of $SU(3)$ are well studied \cite{ref26}. With their help, fuzzy analogues of $D-$matrices can be
constructed, as also sections of $U(1)$ and $U(2)$ bundles.

The fuzzy ${\bf C}{\bf P}^2$ Dirac operator is the cut-off version of (\ref{s511}). It can be put in a matrix form
as in (\ref{offdiagform1}) and (\ref{offdiagform2}). We omit the details: the necessary group theory is already to
be found in \cite{14} while the rest is routine.

\section{Fuzzy Scalar Fields}

Here we briefly indicate a certain fuzzy version of the free scalar field action. It is very natural and a
generalization of fuzzy ${\bf C}{\bf P}^1$ action proposed earlier \cite{5,6,7,8,9,10,11}. Still certain less
obvious actions based on cyclic cohomology have been proposed \cite{13,15}, they have distinct topological
advantages and correct continuum limits as well.

The operators ${Ad\, L}_i=L_i^L-L_i^R$ correspond to the $SU(3)$ generators for functions on ${\bf C}{\bf P}^2$. A
Laplacian for fuzzy ${\bf C}{\bf P}^2$ is thus ${Ad\, L}_i^2$. A scalar field ${\phi}$ is a polynomial in the fuzzy
coordinate functions $\hat{{\xi}}_i$, so ${\phi}$ is just a matrix in ${\bf A}_{{\bf C}{\bf P}^2}$. The Euclidean
action for ${\phi}$ is
\begin{eqnarray}
S(\phi)&=&{\rm ~constant}\times Tr({\phi}^{+}{Ad\, L_i}^2{\phi}),\nonumber\\
{Ad\, L}_i{\phi}&=&[L_i,\phi].
\end{eqnarray}
Let ${\lambda}_K$ be the eigenvalue of the continuum operator for the IRR $(K,K)$. Ref \cite{14} gives
\begin{equation}
\lambda_K=2K(K+1)\ \
\end{equation}
If $N$ is the maximum $K$ for the fuzzy space, then ${Ad\, L}_i^2$ has the spectrum
$\{{\lambda}_0,{\lambda}_1,..{\lambda}_N\}$, it is just the cut-off spectrum of the continuum Laplacian.

\chapter{Conclusions}

As we have shown in this thesis , fuzzy physics has reached a certain level of maturity which makes it a very
strong alternative to lattice physics . It is superior to lattice gauge theory in two fundamental aspects of
physical phenomena , namely symmetry and topology . This claim is supported by the ease and sucess with which
fuzzy physics and noncommutative geometry incorporate nontrivial topological configurations of field theory such
as monopoles and instantons . By using fuzzy physics together with the appropriate tools from NCG one can also
retain all symmetries and anomalies of the continuum theory such as the difficult chiral anomaly . Fermion
doubling is elegantly avoided using a formalism which is less involved and mathematically better founded than the
formalism used in lattice gauge theory. Another advantage of fuzzy physics is its powerful structure which allows
us to put the whole paradigm of discretization by quantization on a very solid mathematical ground . Everything
one says is mathematically strictly precise.

One thinks that the two following big questions remaining to be answered
\begin{itemize}
\item[-]At the level of the formalism one needs to go beyond coadjoint orbits .
\item[-]At the level of physical applications one still has many open problems to be addressed for fuzzy spaces .
\end{itemize}
Regarding the first question , one would like to find a fuzzy version of ${\bf S}^4$ and odd dimensional manifolds
, and prove the existence of star product for such spaces. Having discrete versions of $4-$dimensional spaces allows us to write the fuzzy standard model and therefore start doing actual computational physics.

On the second question above  , one has yet to do the following
\begin{itemize}
\item[1]{\bf Fuzzy Gauge Theories and Continuum Limit}

A precise and practical formulation of general gauge theories on fuzzy ${\bf S}^2$ and fuzzy ${\bf C}{\bf P}^2$
would be highly interesting . To this end the reformulation of fuzzy physics in a form closer to the continuum by
using star products is needed . Building analogies with lattice gauge theory is also important. Such an analogy is given in
chapter $4$ where a Ginsparg-Wilson like set of identities for fuzzy ${\bf S}^2$ is found with the lattice spacing
$a$ being identified with the fuzziness parameter $1/(l+\frac{1}{2})$.

New effects will distinguish between noncommutative gauge theories and lattice gauge theories . Establishing
continuum limit will be the guide for the correct fuzzy version of gauge theories .

Once we have fuzzy gauge theories one can go and look for the different properties of such theories : confinement
, asymptotic freedom , unitarity and causality , IR-UV mixing , chiral symmetry ..

\item[2]{\bf Local Chiral Anomaly}

The global form of the anomaly was found in chapter $4$ of this thesis. The local form which was treated in
\cite{9} is extremly interesting from the physical point of view. The fuzzy sphere is acting here as our regulator.
One strongly expects that the answer found has the structure of the commutative case plus corrections of the order
of the fuzziness. Working out this problem explicitly will possibly shed new lights on the nature of chiral
anomaly . Chiral anomaly on fuzzy ${\bf C}{\bf P}^2$ is an open problem .

\item[3]{\bf Perturbative Dynamics of Fuzzy Field Theories}

The difference between fuzzy field theories and noncommutative field theories is the fact that in the former the
noncommutativity parameter ${\theta}$ is equal to the cut-off ${\Lambda}$. In other words in fuzzy field theories
there is no issue of the noncommutativity of the two limits ${\theta}{\longrightarrow}0$ and
${\Lambda}{\longrightarrow}{\infty}$, since ${\theta}=1/l$ and ${\Lambda}=l$ . Therefore , if the UV-IR phenomenon
persists in fuzzy field theories one can discard the above noncommutativity as a source of it . The problem might
be easier in this context since we are dealing with finite-dimensional matrix models . A good starting point is the work \cite{seiberg,vaidya}.
\end{itemize}

\end{document}